# Nonlinear Time-History Analysis of Soil-Structure Systems Incorporating Frequency-Dependent Impedance Functions

S. Farid Ghahari[1,a], Alborz Ghofrani[2,b], Jian Zhang[1,c], and Ertugrul Taciroglu[1,c]

[1] University of California, Los Angeles, California

[2] Google, Mountain View, California

## Executive Summary

To accurately analyze structures, Soil-Structure Interaction (SSI) effects must be taken into account. One approach to analyze SSI effects is to create and analyze a complete Finite Element Model (FEM) of the full system wherein the soil medium is represented as a semi-infinite domain. This so-called "direct" method approach is frequently adopted in research studies. But, it is typically avoided in engineering practice due to the labor-intensive finite element model development, and the high computational cost. In practice, SSI analysis is mostly carried out through a substructure approach. In this approach, the superstructure is usually modeled through a very detailed FE model and is placed on a soil-foundation substructure which is represented by a system called Impedance Function (IF). Then, the entire system is analyzed under Foundation Input Motions (FIMs) obtained from Free-Field Motions (FFMs) considering Kinematic Interaction (KI) effects. While the method is theoretically designed for linear-elastic behavior due to the superposition assumption, the substructure method can be partially applied to nonlinear systems for which the condensation process is performed only on the viscous elastic soil-foundation. Although IFs for various soil and foundation configurations can be obtained from analytical, numerical, or experimental analyses, their implementation in the time-domain is not trivial because they are frequency-dependent with unlimited bandwidth. A simple solution for this problem has been to convert these IFs to some lumped-parameter physical models with frequency-independent components, but there is no straightforward way to connect these components. More importantly, the coefficients of these components could be non-physical parameters that cannot be modeled in FE software like OpenSEES or the final lumped model could be unstable. To resolve the aforementioned problems with the physical models, various alternative approaches have been proposed in the literature. In this project, we review some of the existing solutions and verify them through numerical examples. After extensive review and evaluation, the Hybrid Time Frequency Domain (HTFD) method [1] seems a more practical solution with fewer stability issues. This method is implemented in Opensees to be used by researchers and practitioners.

[a] Research scientist (faridghahari@ucla.edu)

[b] Software engineer

[c] Professor

In Chapter 1, the soil-structure interaction and the direct and substructure approaches will be described in more detail. Chapter 2 presents a summary of the theoretical background needed to be able to build upon the existing remedies for the representation of frequency-dependency in the time domain. Also, the SSI implementation in seismic design codes will be briefly reviewed. In Chapter 3, the characteristics of soil-foundation impedance functions will be first described some of the available solutions will be discussed in detail along with numerical implementation in Matlab, and extensive verification. The chosen HTFD method will be further verified in Matlab through a series of linear and nonlinear time and frequency domain analyses in Chapter 4. This method can be employed using the current capabilities of Opensees, so Chapter 5 shows how the method can be implemented in Opensees to solve SDOF and MDOF linear and nonlinear soil-structure problems. Such an explicit implementation needs some extra coding which may not be practical in real-life applications. Therefore, HTFD is built-in within Opensees by introducing new elements and analysis methods in Opensees. The details of such implementation along with its verification are presented in Chapter 6. A complete application example will also presented in this chapter. Chapter 7 presents major conclusions and achievements made in this project as well as suggestions and recommendations for future studies.

# TABLE OF CONTENTS

## LIST OF FIGURES

# LIST OF TABLES

# CHAPTER 1: INTRODUCTION

## 1.1. SOIL-STRUCTURE INTERACTION

Soil-Foundation-Structure Interaction (SFSI) has been well studied for more than 40 years (e.g., [2]–[4]). SFSI can be classified into two distinct effects: Kinematic and Inertial effects [5]. Even in the absence of the structure, a massless foundation experiences different movement during an earthquake (called Foundation Input Motion (FIM)) from Free-Field Motion (FFM) which would be recorded at the same site if the foundation was not there (see Figure 1-1 left and middle). This effect which is dubbed kinematic interaction is due to stiffness differences between the foundation and surrounding soil. FIM is dependent on both foundation's and site's properties as well as wave fields. For example, base-slab averaging is the major source of kinematic interaction in the surface foundation where the foundation's slab experiences an average of inclined and incoherent waves [6]–[11]. For embedded foundations or piled foundations, base-slab averaging is accompanied by embedment effects to make FIM much more different from FFM [12]–[14]. Dynamic response of the structure inserts force and moments to the base and causes the foundation to have a different response from FIM. This effect is named inertial interaction (Figure 1-1 right). Due to such inertial effects, a vibrating structure can operate as a wave-propagating source and change the wave field around. Consequently, FFM recorded around the structure can no longer assume free-field.

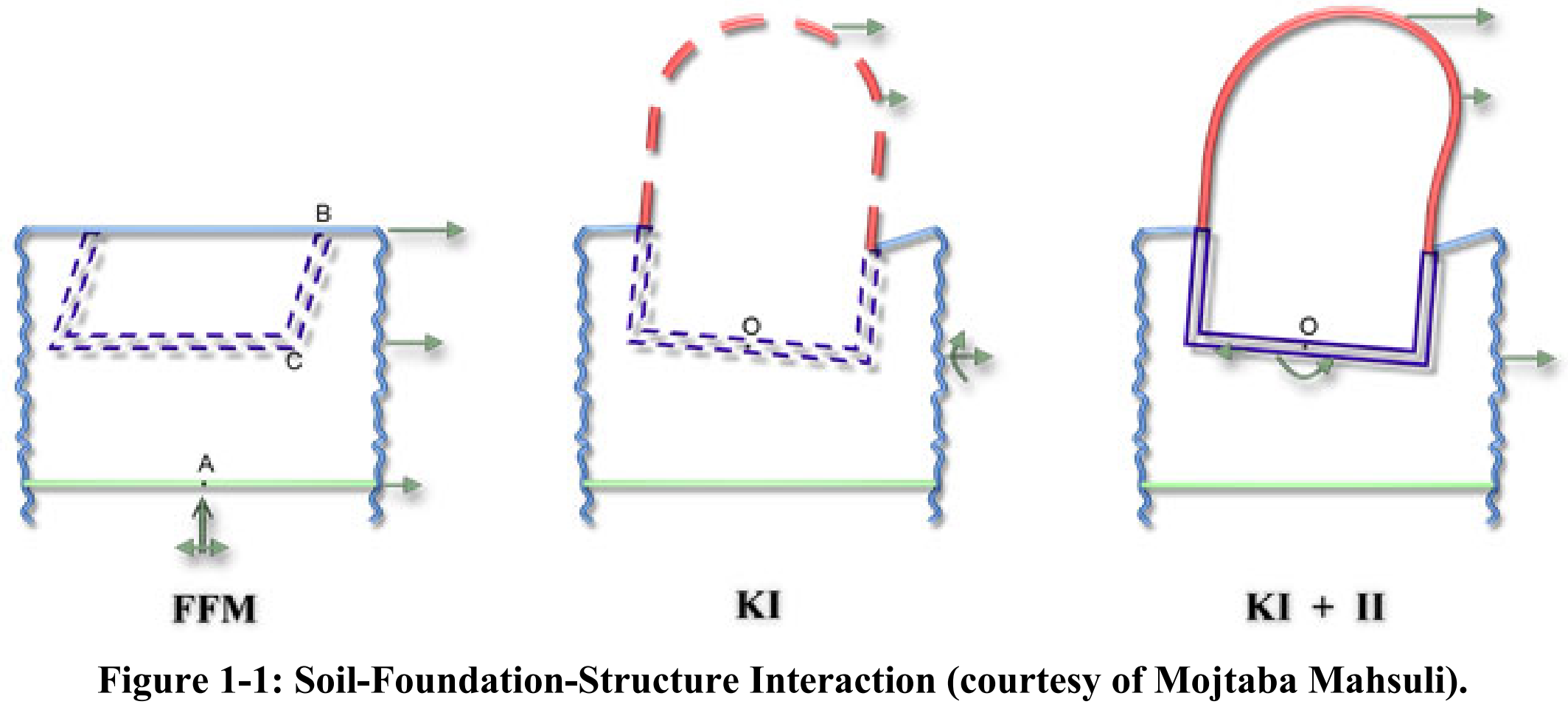

**Figure 1-1: Soil-Foundation-Structure Interaction (courtesy of Mojtaba Mahsuli).**

## 1.2. SUBSTRUCTURE APPROACH

As shown in Figure 1-2, one approach to analyze SFSI effects is to create and analyze a complete Finite Element Model (FEM) of the full system wherein the soil medium is represented as a semi-infinite domain (e.g., [15], [16]). This method of analysis is usually referred to as the *direct method*. In the direct method approach, the region of the soil containing the structure is modeled up to an artificial boundary, because the soil domain is nominally an unbounded half-space (unless of course the bedrock or a rock outcrop is nearby). In order to avoid reflections of the outbound waves at the artificial/truncated boundary, special provisions must be made—for example, frequency-independent springs and viscous dashpots proposed by Lysmer and Kuhlemeyer [17] must be attached in all directions at the remote boundary. The direct method is frequently adopted in research studies. Because superposition is not required, material nonlinearities in both the structure and the soil domain can be considered; thus, this is a quite general approach to SFSI analysis. Nevertheless, the direct method is typically avoided in engineering practice due to the labor-intensive finite element model development, and the high computational cost associated with carrying out successive simulations under multiple input motions.

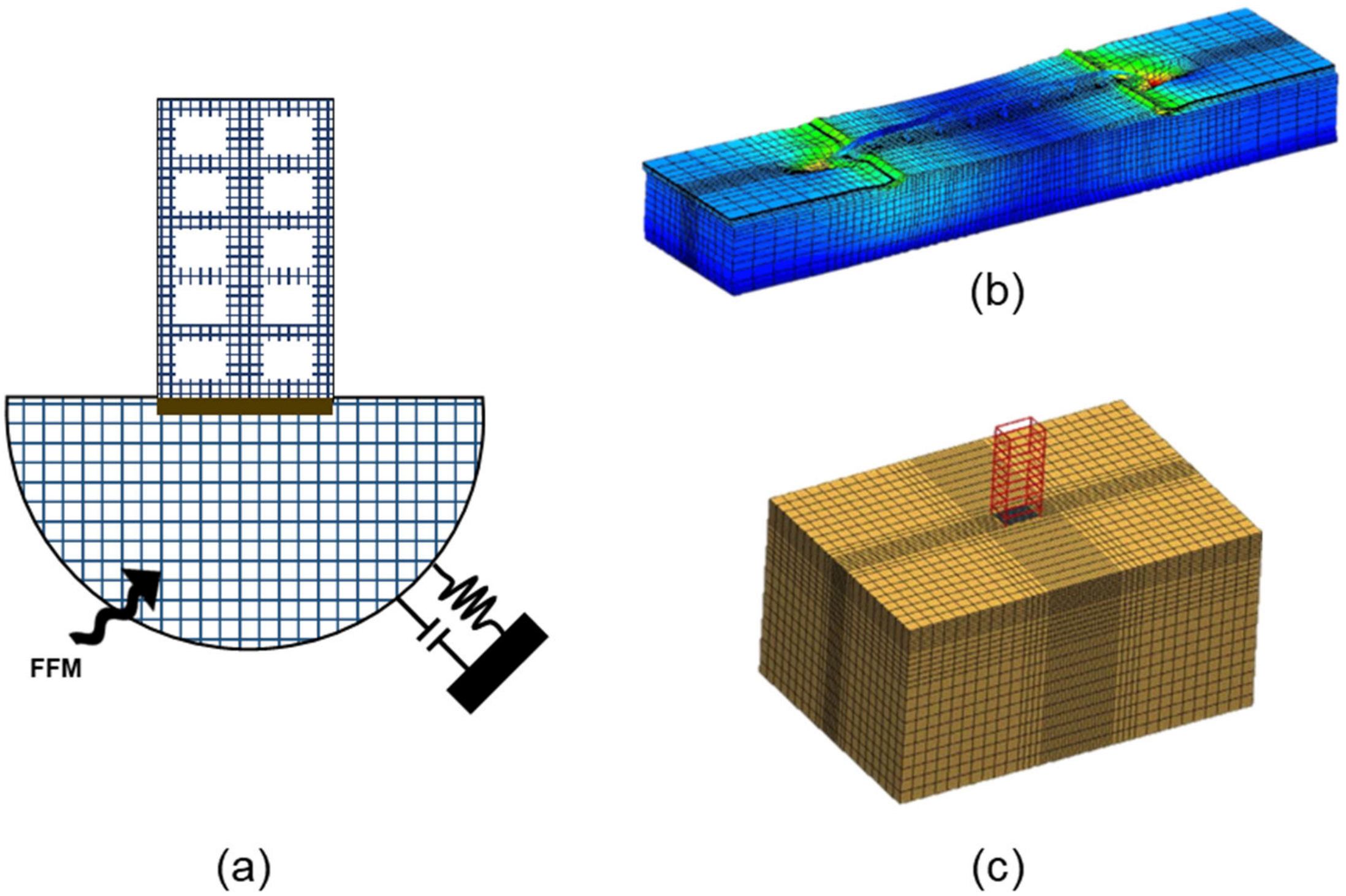

**Figure 1-2: (a) Schematic presentation of the direct method and two example studies by (b) Elgamal et al.** [16]**, and (c) Torabi and Rayhani** [15]**.**

The primary alternative approach is the so-called substructure method, wherein the SFSI problem is broken down into three distinct parts, which are combined to formulate the complete solution (Figure 1-3).

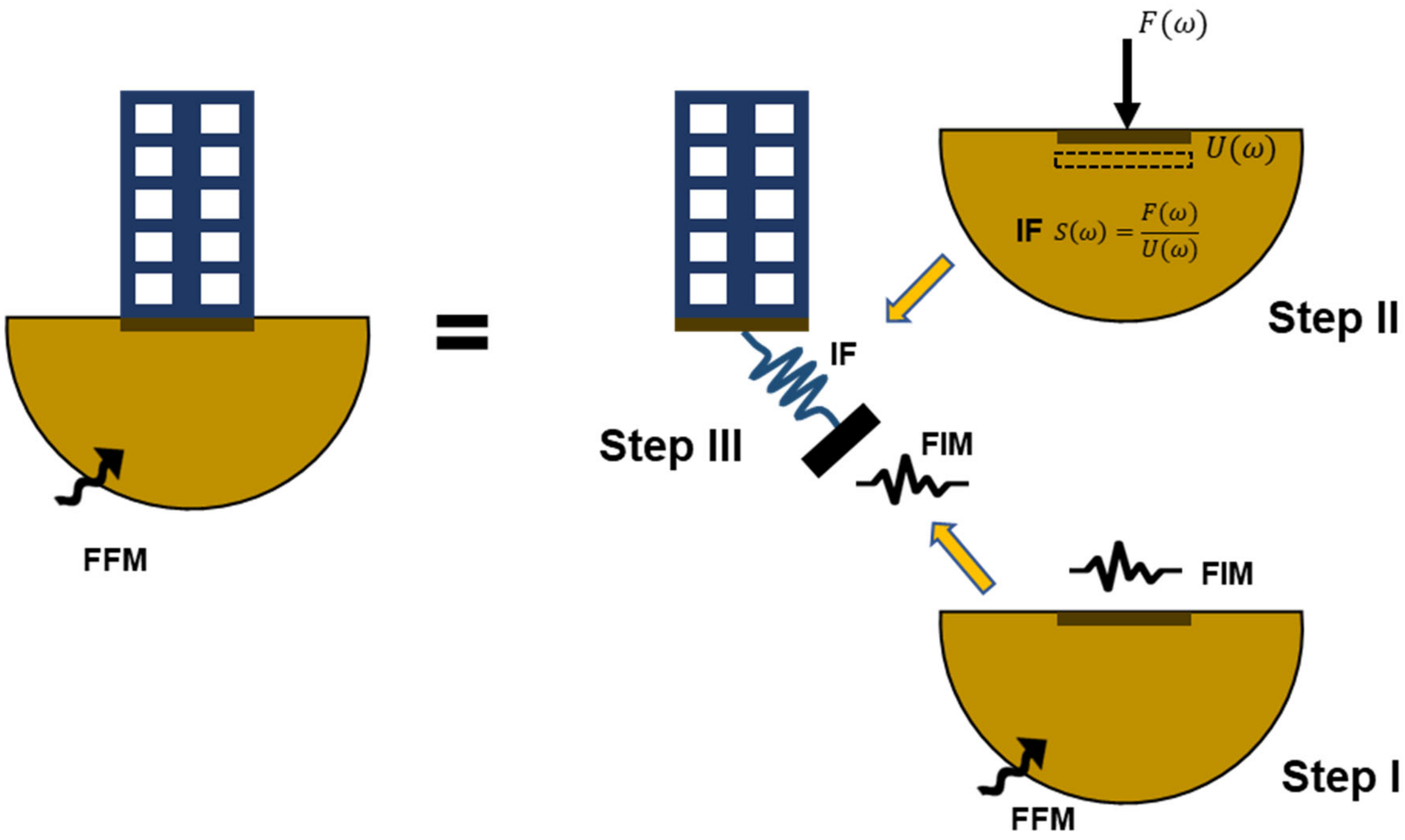

**Figure 1-3: A schematic presentation of the substructure method: (a) soil-structure response problem, (b) evaluation of FIMs, (c) evaluation of impedance function, and (d) analysis of structure on compliant base subjected to FIMs [18].**

The superposition procedure inherent to this approach requires an assumption of system linearity. The aforementioned three parts of the substructure method are as follows:

**Step I: Estimation of FIMs**. As stated above, the usual contrast between the stiffness of a (nearly rigid) foundation and the surrounding soil causes the motion experienced by the foundation (FIM) to differ from the FFM. This kinematic effect may produce rocking and torsional input motions in addition to the modification of horizontal motions. Foundations with different stiffness, shape, and embedment may produce different FIMs under the same FFM. There are several methods to calculate FIMs from FFM using foundation specification [18]–[21]; however, they are rather limited mostly because of the general understanding that KI reduces the horizontal component of FFM although it may produce rotational/rocking input excitations.

**Step II: Determination of soil-foundation Impedance Functions (IFs)**. To represent the stiffness and damping effects of the soil-foundation system, discrete springs and dashpots are attached to the structure's base as schematically shown in Figure 1-4 for a multi-story building. These springs and dashpots are collectively referred to as the Impedance Functions (IFs) (a.k.a. the dynamic stiffness). The impedance functions depend on the frequency of excitation and can be compactly represented as $s(\omega) = k(\omega) + ic(\omega)$ for each direction/mode of motion [5]. The term $k(\omega)$ represents the soil's stiffness, while $c(\omega)$ denotes the damping effects related to both small-scale material hysteresis and radiation. This step is crucial in substructure analysis, so, many analytical, numerical, and experimental methods have been developed to estimate frequency-dependent impedance functions. Naturally, analytical solutions were obtained only for very special cases. For example, Luco and Westman [22] presented the impedance function for a disk-shaped foundation on an elastic half-space by assuming frictionless contact, and for a strip foundation bonded to an elastic half-space. Thanks to ongoing advances in numerical methods at the time, several researchers obtained various solutions for impedance functions for more complex situations. For example, Day [23] obtained the frequency-dependent dynamic stiffness of an embedded cylindrical foundation. Later on, identical results were obtained by Apsel [24] who used integral equations. Similar numerical studies were also conducted for rectangular foundations [25]–[27]. In order to make the proposed solutions easier to implement and use, Pais and Kausel [28] presented a series of approximate formulae for the aforementioned, analytically or numerically obtained, impedances. One of the earliest laboratory studies on SSI was conducted by Richart and Whitman [29] and after about 20 years, another experimental attempt was made by Dobry et al. [30]. As laboratory-scale tests were eventually deemed inadequate to replicate/mimic the actual field conditions, in situ dynamic (usually harmonic excitation) tests on large to medium-scale specimens were pursued [31]–[36]. Naturally, the forced vibration tests are expensive and laborious; and field tests on actual structures are usually avoided due to service disruptions and safety concerns. Moreover, it is generally difficult to attain adequately high vibrations at the foundation levels of massive/large structures even through forced vibration testing [37]. Hence, extracting impedance functions from real data recorded during earthquakes has attracted attention recently [38]–[48].

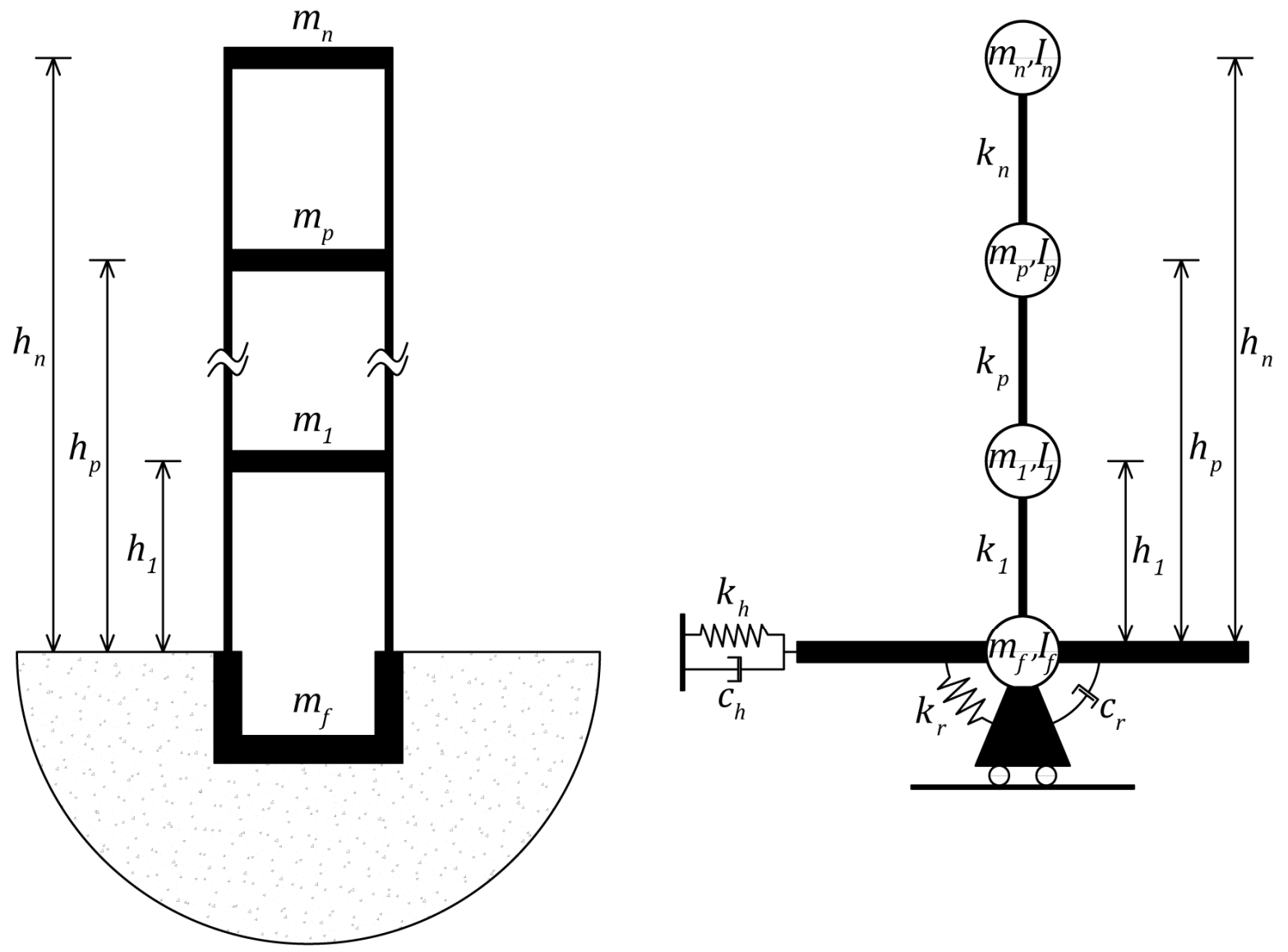

**Figure 1-4: Soil-structure system (left) and its substructure model (right).**

**Step III:** Dynamic analysis of the structure supported on a compliant base represented by the impedance function and subjected to a base excitation consisting of the FIM. Having available FIMs and frequency-dependent impedance functions, the soil-structure system must be analyzed in the frequency domain, which is quite straightforward, because of the implicit superposition assumption in the substructure approach as well as the frequency-dependency of the impedance function. However, in real-life problems, the superstructure is allowed to show nonlinear behavior during moderate earthquake events, which cannot be handled in the frequency domain. So, to be able to use the substructure method for such problems while frequency-dependent impedance function is considered, innovative solution methods need to be used.

## 1.3. OBJECTIVE OF THIS STUDY

This research study aims to:

- Review existing methods to carry out time-domain analysis of soil-structure systems in which frequency-dependency and nonlinearity are limited, respectively, to the soil-foundation and superstructure.
- Select one of the solutions and extensively evaluate its performance.
- Implement the selected method in the Opensees [49].

# CHAPTER 2: SOIL-STRUCTURE INTERACTION

## 2.1. THEORETICAL BACKGROUND

Figure 2-1 shows a general soil-structure domain that is composed of three areas [50]: 1- superstructure, which can behave nonlinearly, 2- part of the soil-foundation domain, near-field domain, which can behave nonlinearly, and 3- semi-infinite soil domain, far-field domain, which is assumed to behave linearly. All nodes within the nonlinear domain are called “s” nodes, while nodes at the boundary of the linear and nonlinear domains are called “b” nodes.

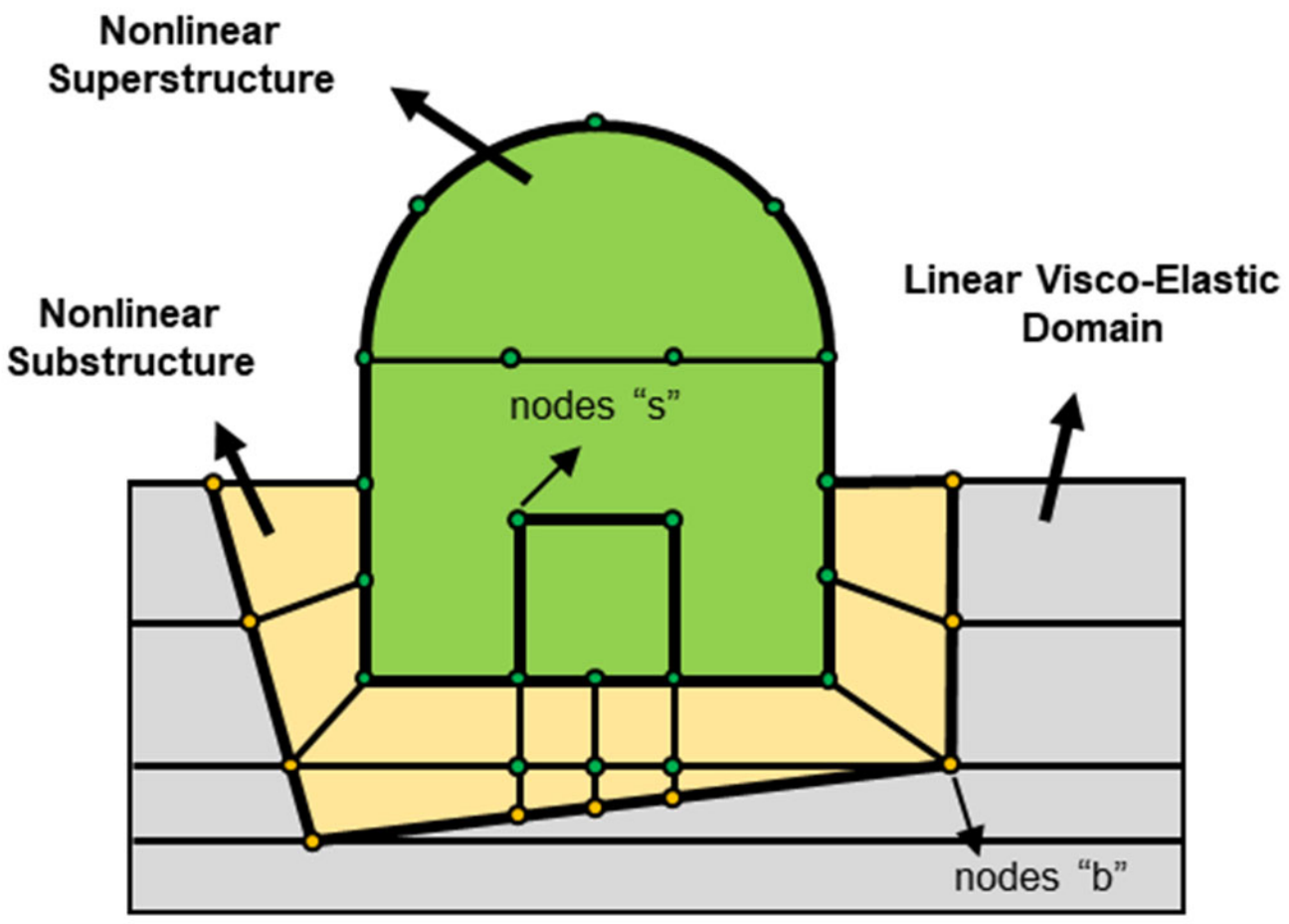

**Figure 2-1: Soil-structure domain.**

If there was no structure, we would have a free-field domain as shown in Figure 2-2(left). The motion of the nodes of this domain under ground motion excitation is called free-field motion. The free-field motion of all “b” nodes is shown as $\mathbf{u}_b^f$. The dynamic stiffness corresponding to these nodes is also shown as $\mathbf{S}_{bb}^f$. The free-field domain can be considered as a combination of a domain with excavation and an excavated part as shown on the right side of Figure 2-2. The motion of the “b” nodes at the domain with excavation is called scattered motion and is denoted by $\mathbf{u}_b^g$, which is different from the free-field motion because its associated dynamic stiffness $\mathbf{S}_{bb}^g$ is different from $\mathbf{S}_{bb}^f$.

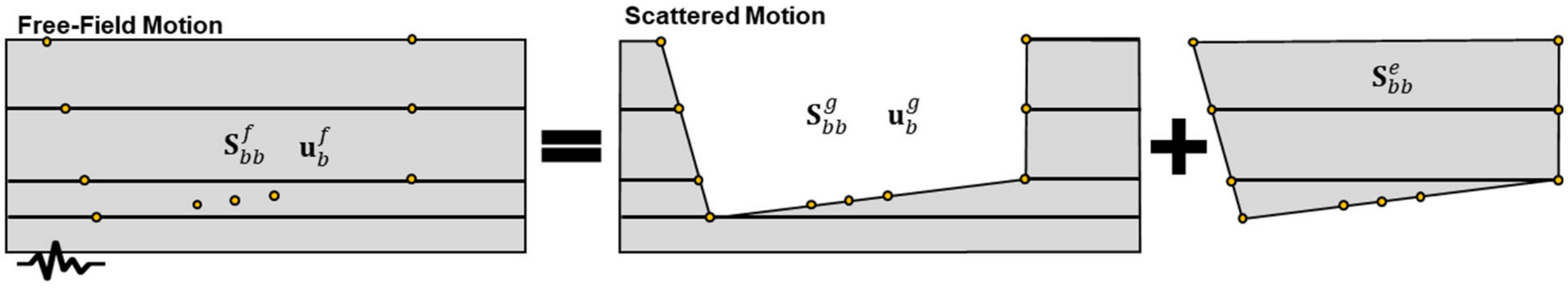

**Figure 2-2: Free-field domain.**

Now, let's write the equation of motion of the nodes "b" and "s" in Figure 2-1:

$$\begin{bmatrix} \mathbf{M}_{ss} & \mathbf{M}_{sb} \\ \mathbf{M}_{bs} & \mathbf{M}_{bb} \end{bmatrix} \begin{bmatrix} \ddot{\mathbf{u}}_s^t \\ \ddot{\mathbf{u}}_b^t \end{bmatrix} + \begin{bmatrix} \mathbf{p}_s \\ \mathbf{p}_b + \int_0^t \mathbf{S}_{bb}^g (t-\tau)\mathbf{u}_b^t d\tau \end{bmatrix} = \begin{bmatrix} \mathbf{0} \\ \int_0^t \mathbf{S}_{bb}^g (t-\tau)\mathbf{u}_b^g d\tau \end{bmatrix}. \tag{2-1}$$

In the equation above, $\mathbf{p}_s$ is the force applied to the structural nodes (strain and damping), which can be nonlinear[4], $\mathbf{p}_b$ is the part of the forces applied to the "b" nodes from the superstructure (strain and damping), which can be nonlinear too[5], and $\int_0^t \mathbf{S}_{bb}^g (t-\tau)\mathbf{u}_b^t d\tau$ is the linear forces applied to the "b" nodes from the semi-infinite domain. Note that this force is not the interaction force because it is written using the total displacement of "b" nodes. As seen, due to the frequency-dependency of the dynamic stiffness, this force is represented through a convolution integral[6]. Similarly, the external force $\int_0^t \mathbf{S}_{bb}^g (t-\tau)\mathbf{u}_b^g d\tau$ is represented through a convolution integral too[7].

An equation similar to (2-1) can be written for the free-field domain as below where the excavated part plays the role of the superstructure with no nodes within the "s" domain

$$\mathbf{M}_{bb}\ddot{\mathbf{u}}_b^f + \mathbf{p}_b + \int_0^t \mathbf{S}_{bb}^g (t-\tau)\mathbf{u}_b^f d\tau = \int_0^t \mathbf{S}_{bb}^g (t-\tau)\mathbf{u}_b^g d\tau. \tag{2-2}$$

Assuming that the excavated part in the free-field domain remains linear, we have

$$\mathbf{M}_{bb}\ddot{\mathbf{u}}_b^f + \mathbf{p}_b = \int_0^t \mathbf{S}_{bb}^e (t-\tau)\mathbf{u}_b^f d\tau. \tag{2-3}$$

By combining Eqs. (2-2) and (2-3), we have

$$\int_0^t [\mathbf{S}_{bb}^e(t-\tau) + \mathbf{S}_{bb}^g(t-\tau)]\mathbf{u}_b^f d\tau = \int_0^t \mathbf{S}_{bb}^g (t-\tau)\mathbf{u}_b^g d\tau. \tag{2-4}$$

---

[4] In the case of a linear superstructure, $\mathbf{p}_s = \mathbf{C}_{ss}\dot{\mathbf{u}}_s^t + \mathbf{C}_{sb}\dot{\mathbf{u}}_b^t + \mathbf{K}_{ss}\mathbf{u}_s^t + \mathbf{K}_{sb}\mathbf{u}_b^t$.

[5] In the case of a linear superstructure, $\mathbf{p}_b = \mathbf{C}_{bs}\dot{\mathbf{u}}_s^t + \mathbf{K}_{bs}\mathbf{u}_s^t$.

[6] In the case of a frequency-indepednent impedance function, $\int_0^t \mathbf{S}_{bb}^g (t-\tau)\mathbf{u}_b^t d\tau$ could be written as $\mathbf{C}_{bb}\dot{\mathbf{u}}_b^t + \mathbf{K}_{bb}\mathbf{u}_s^t$.

[7] In the case of a frequency-indepednent impedance function, $\int_0^t \mathbf{S}_{bb}^g (t-\tau)\mathbf{u}_b^g d\tau$ could be written as $\mathbf{C}_{bb}\dot{\mathbf{u}}_b^g + \mathbf{K}_{bb}\mathbf{u}_s^g$.

Also, we know that $\mathbf{S}_{bb}^{f} = \mathbf{S}_{bb}^{g} + \mathbf{S}_{bb}^{e}$, so

$$\int_{0}^{t} \mathbf{S}_{bb}^{f}(t-\tau)\mathbf{u}_{b}^{f} d\tau = \int_{0}^{t} \mathbf{S}_{bb}^{g}(t-\tau)\mathbf{u}_{b}^{g} d\tau. \tag{2-5}$$

Therefore, in Eq. (2-1), we can replace the right-hand side with $\int_0^t \mathbf{S}_{bb}^{f}(t-\tau)\mathbf{u}_{b}^{f} d\tau$, i.e.,

$$\begin{bmatrix} \mathbf{M}_{ss} & \mathbf{M}_{sb} \\ \mathbf{M}_{bs} & \mathbf{M}_{bb} \end{bmatrix} \begin{bmatrix} \ddot{\mathbf{u}}_{s}^{t} \\ \ddot{\mathbf{u}}_{b}^{t} \end{bmatrix} + \begin{bmatrix} \mathbf{p}_{s} \\ \mathbf{p}_{b} + \int_{0}^{t} \mathbf{S}_{bb}^{g}(t-\tau)\mathbf{u}_{b}^{t} d\tau \end{bmatrix} = \begin{bmatrix} \mathbf{0} \\ \int_{0}^{t} \mathbf{S}_{bb}^{f}(t-\tau)\mathbf{u}_{b}^{f} d\tau \end{bmatrix}, \tag{2-6}$$

which means the scattered motion does not need to be calculated.

Eq. (2-6) can be rewritten in a compact form as

$$\begin{bmatrix} \mathbf{M}_{ss} & \mathbf{M}_{sb} \\ \mathbf{M}_{bs} & \mathbf{M}_{bb} \end{bmatrix} \begin{bmatrix} \ddot{\mathbf{u}}_{s}^{t} \\ \ddot{\mathbf{u}}_{b}^{t} \end{bmatrix} + \begin{bmatrix} \mathbf{p}_{s} \\ \mathbf{p}_{b} \end{bmatrix} = \begin{bmatrix} \mathbf{0} \\ \mathbf{q}_{b} \end{bmatrix} - \begin{bmatrix} \mathbf{0} \\ \mathbf{r}_{b} \end{bmatrix}, \tag{2-7}$$

where $\mathbf{r}_b = \int_0^t \mathbf{S}_{bb}^{g}(t-\tau)\mathbf{u}_{b}^{t} d\tau$ and $\mathbf{q}_b = \int_0^t \mathbf{S}_{bb}^{f}(t-\tau)\mathbf{u}_{b}^{f} d\tau$. $\mathbf{r}_b$ is called the interaction force in the following although, as mentioned earlier, it is not an exact definition because it is calculated using the total displacement of the soil-foundation nodes[8].

## 2.2. SOIL-STRUCTURE INTERACTION IN DESIGN CODES

Traditionally, a simplified model like Figure 2-3 has been used in design codes. This model consists of a Single-Degree-Of-Freedom (SDOF) superstructure with height $h_s$, mass $m_s$, stiffness $k_s$, and damping $c_s$ on top of a massless sway-rocking foundation supported by translational and rotational resisting force $F$ and moment $M$, respectively, provided through the soil-foundation substructure. This model can be viewed as a one-story building or the first mode of a multi-story building in which soil-structure interaction is expected to be more dominant [51]. In the latter, the height $h_s$ represents the effective height which is the distance from the base to the centroid of the inertial forces associated with the first mode.

The resisting soil-foundation forces/moments can be expressed as

$$F(t) = F^{-1}\{\bar{S}_u(\omega)\} * u_f(t), \tag{2-8}$$

$$M(t) = F^{-1}\{\bar{S}_\theta(\omega)\} * \theta_f(t), \tag{2-9}$$

where $u_f(t)$ and $\theta_f(t)$ are relative foundation sway and rocking deformation, respectively, $F^{-1}\{\}$ stands for the Inverse Fourier Transform, and $\bar{S}_j(\omega)$ $(j = u, \theta)$[9] represents the complex-valued frequency-

---

[8] The actual interaction force is $\int_0^t \mathbf{S}_{bb}^{g}(t-\tau)[\mathbf{u}_{b}^{t} - \mathbf{u}_{b}^{g}]d\tau$.

[9] An overbar is usually used throughout this text to emphasize frequency-domain representation of a parameter.

dependent impedance function corresponding to the aforementioned DOFs. $\bar{S}_j(\omega)$ is usually represented as [52]

$$\bar{S}_j(a_0) = k_j(a_0, \nu) + i\omega c_j(a_0, \nu), \tag{2-10}$$

where $\omega$ is the excitation frequency, $\nu$ is the soil Poisson ratio, and $a_0 = \omega r/V_s$ is a dimensionless frequency with $V_s$ is soil shear wave velocity and $r$ is the equivalent radii of the foundation which could be different for each DOF ($r_u$ and $r_\theta$) resulting in two different $a_0$s. As seen, $k_j(a_0, \nu)$ and $c_j(a_0, \nu)$ can be considered, respectively, as physical spring and dashpot with frequency-dependent coefficients. To further simplify the presentation, $k_j(a_0, \nu)$ and $c_j(a_0, \nu)$ are usually expressed as

$$k_j = \alpha_j K_j, \tag{2-11}$$

$$c_j = \beta_j \frac{K_j r_j}{V_s}, \tag{2-12}$$

where $K_j$ is the static stiffness of a disk on a half-space and $\alpha_j$ and $\beta_j$ represent frequency-dependency. The explicit dependency on the frequency is dropped for simplification. Translational and rocking static stiffnesses of a disk on half-space are analytically available as [53]

$$K_u = \frac{8}{2-\nu} G r_u, \tag{2-13}$$

$$K_\theta = \frac{8}{3(1-\nu)} G r_\theta^3, \tag{2-14}$$

in which $G$ is the soil's shear modulus of elasticity. As mentioned in the previous chapter, there are analytical or numerical formulas for frequency-dependent impedance functions. A series of such formulas can be found in [28], [54]. Having frequency-dependent impedance functions, or equivalently frequency-dependent coefficients of springs and dashpots, it is easy to obtain equivalent stiffness of the system as

$$\tilde{k} = \frac{1}{\frac{1}{k_s} + \frac{1}{k_u} + \frac{h_s^2}{k_\theta}}. \tag{2-15}$$

So, the ratio of the soil-structure system's period[10] $\tilde{T}$ to the superstructure's period[11], $T = 2\pi\sqrt{m_s/k_s}$, is [55]

$$\frac{\tilde{T}}{T} = \sqrt{1 + \frac{k_s}{k_u} + \frac{k_s h_s^2}{k_\theta}}. \tag{2-16}$$

To estimate the flexible-base period $\tilde{T}$ using fixed-base period, Eq. (2-16) cannot be used directly because $k_u$ and $k_\theta$ are frequency-dependent and must be set at $\omega = 2\pi/\tilde{T}$ which is not known a priori. Therefore, an iterative solution needs to be employed [56].

---

[10] It is sometimes called flexible-base period.
[11] It is sometimes called fixed-base period.

To estimate the flexible-base damping ratio, Veletsos and Nair [57] proposed the following formula

$$\tilde{\xi} = \tilde{\xi}_0 + \frac{\xi}{\left(\tilde{T}/T\right)^3}, \tag{2-17}$$

in which $\tilde{\xi}_0$ is the hysteretic and radiation damping ratio provided through soil-foundation, and $\xi = Tc_s/4\pi m_s$ is the viscous damping in the superstructure. Expressions for $\tilde{\xi}_0$ can be found in [57], [58], and contrary to the flexible-base period, Eq. (2-17) can be used in a single step once $\tilde{T}$ is determined.

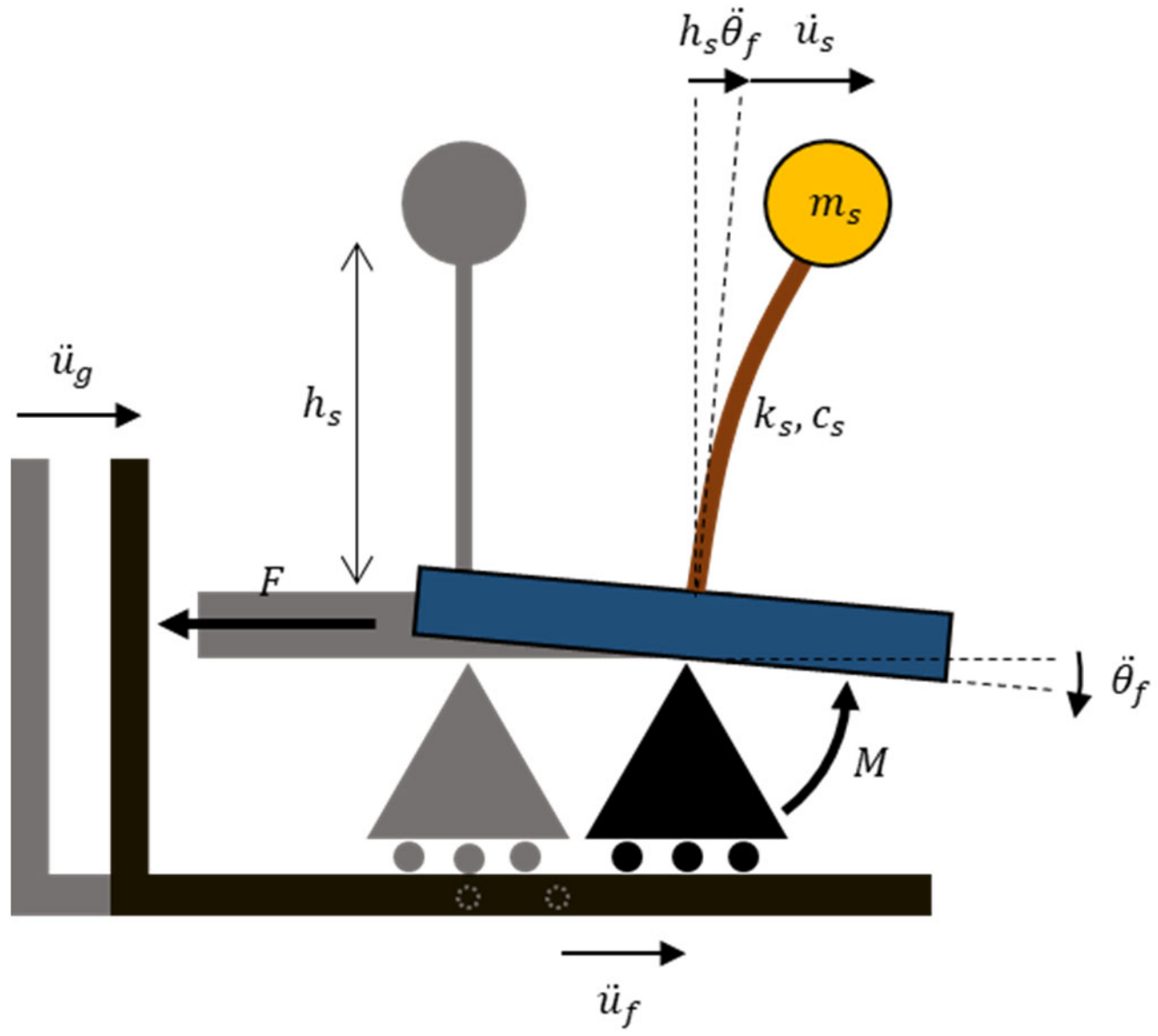

**Figure 2-3: Simplified model for SSI analysis.**

### 2.2.1. Practical Application

ATC[12] 3-06 [59] is one of the earliest seismic documents in which simplified soil-structure interaction analysis is formally included. In such documents, the SSI analysis is based on the theoretical background introduced in the previous section. That is, design shear force is calculated from the response spectrum by using the flexible-base period, which is always longer than the fixed-base period, and the flexible-base damping ratio, which is usually higher than the fixed-base value. To do so, $\tilde{T}/T$ and $\tilde{\xi}$ are estimated non-iteratively from pre-compiled graphs using the fixed-based dimensionless parameter $\sigma = V_s T/h_s$, structural aspect ratio $h_s/r$, soil Poisson ratio ν, soil hysteretic damping ratio $\beta$ (part of $\tilde{\xi}_0$), structure to soil mass $\gamma = m_s/\rho\pi r_u^2 h_s$, and foundation embedment ratio $e/r$. A sample of such graphs is shown in Figure 2-4. The schematic effect of such period elongation and damping increase on spectral demand used in traditional force-based design methods is shown in Figure 2-5. The mentioned approach has been used more or less since then in most of the seismic codes whenever soil-structure interaction is needed to be considered. For example, below are the steps used in 2000 NEHRP Provisions [60] to consider soil-structure interaction:

[12] Applied Technology Council

- Determine fixed-base fundamental period, $T$, and damping ratio, $\xi$ (Section 5.4.2).

- Determine the stiffness of the fixed-base building, $k_s$, using the effective mass, $k_s$, of the building and $T$ (Equation 5.8.2.1.1-2). The effective mass can be taken as 70% of the total mass of the building.

- Determined the effective height, $h_s$, as the height of the building if it is a single-story building or 70% of the total height if it is a multi-story building.

- Calculate mass ratio, $\gamma = m_s/\rho A h_s$, using soil's mass density $\rho$, foundation's area $A$, and superstructure's effective mass and height (Equation 5.8.2.1.1-4).

- Calculate equivalent circular disk radii, $r_u$ and $r_\theta$, to represent the foundation geometry (Equations 5.8.2.1.1-5 and -6).

- Determine soil's shear wave velocity $V_s$ (or elastic modulus of elasticity $G$) and Poisson's ratio $\nu$ using the available soil profile, and adjust these values based on the embedment and expected peak ground acceleration (Table 5.8.2.1.1).

- Calculate the foundation static stiffnesses, $K_u$ and $K_\theta$, and frequency-dependent stiffness modification factor by established principles of foundation mechanics, for example, Eqs. (2-13) and (2-14) for static stiffnesses and [57] for modification factors.

- Evaluate period lengthening using Eq. (2-16) (Equation 5.8.2.1.1-1).

- Calculate $\tilde{\xi}_0$ from Figure 5.8.2.1.2 using period lengthening ratio and soil hysteretic damping ratio (represented through the level of shaking).

- Evaluate reduction in base shear using Equation 5.8.2.1-2.

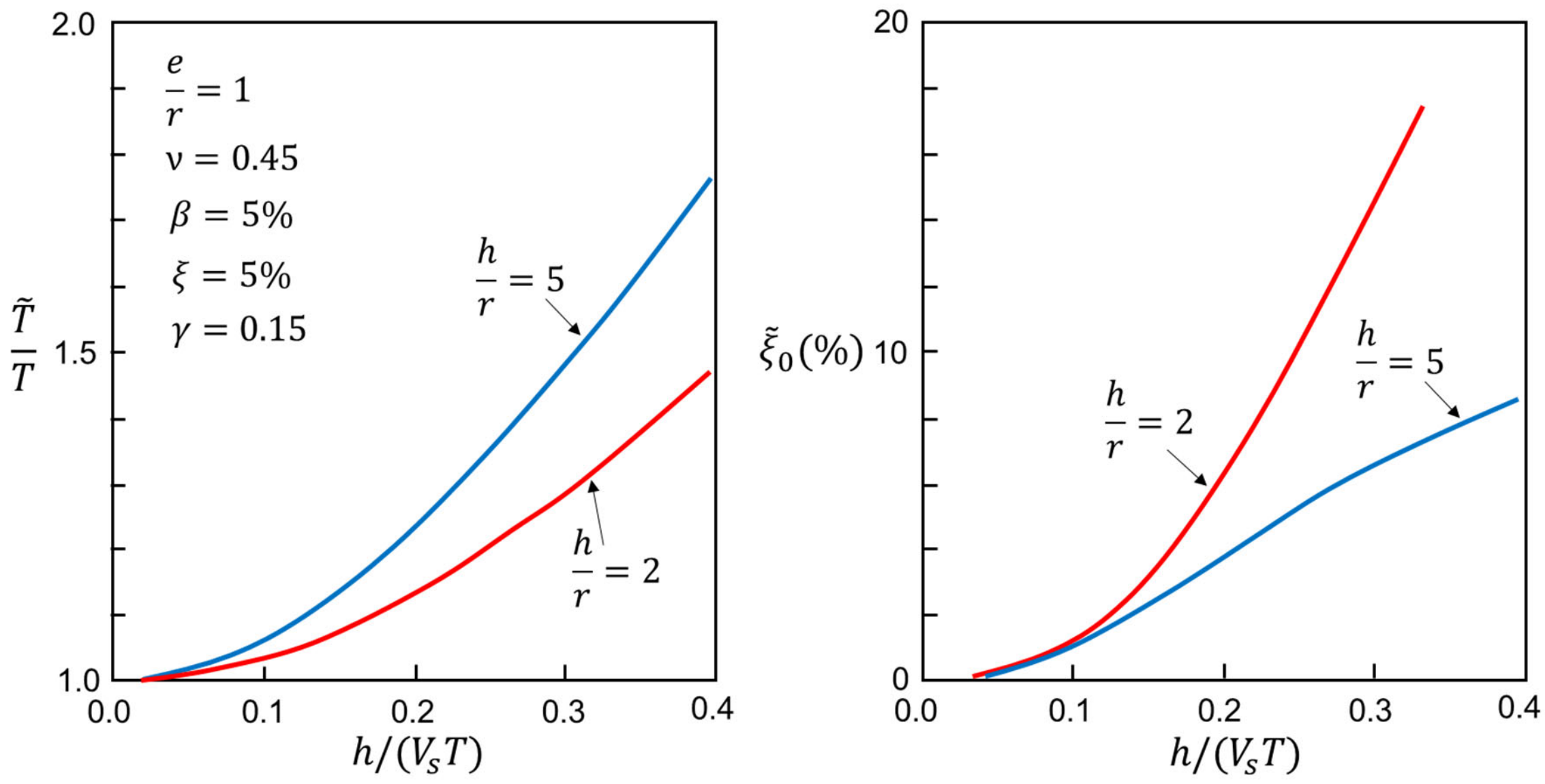

**Figure 2-4: Flexible-base period and damping for a rigid circular foundation embedded into a viscoelastic half-space obtained by Bielak** [58] **(the image is reproduced from** [52]**).**

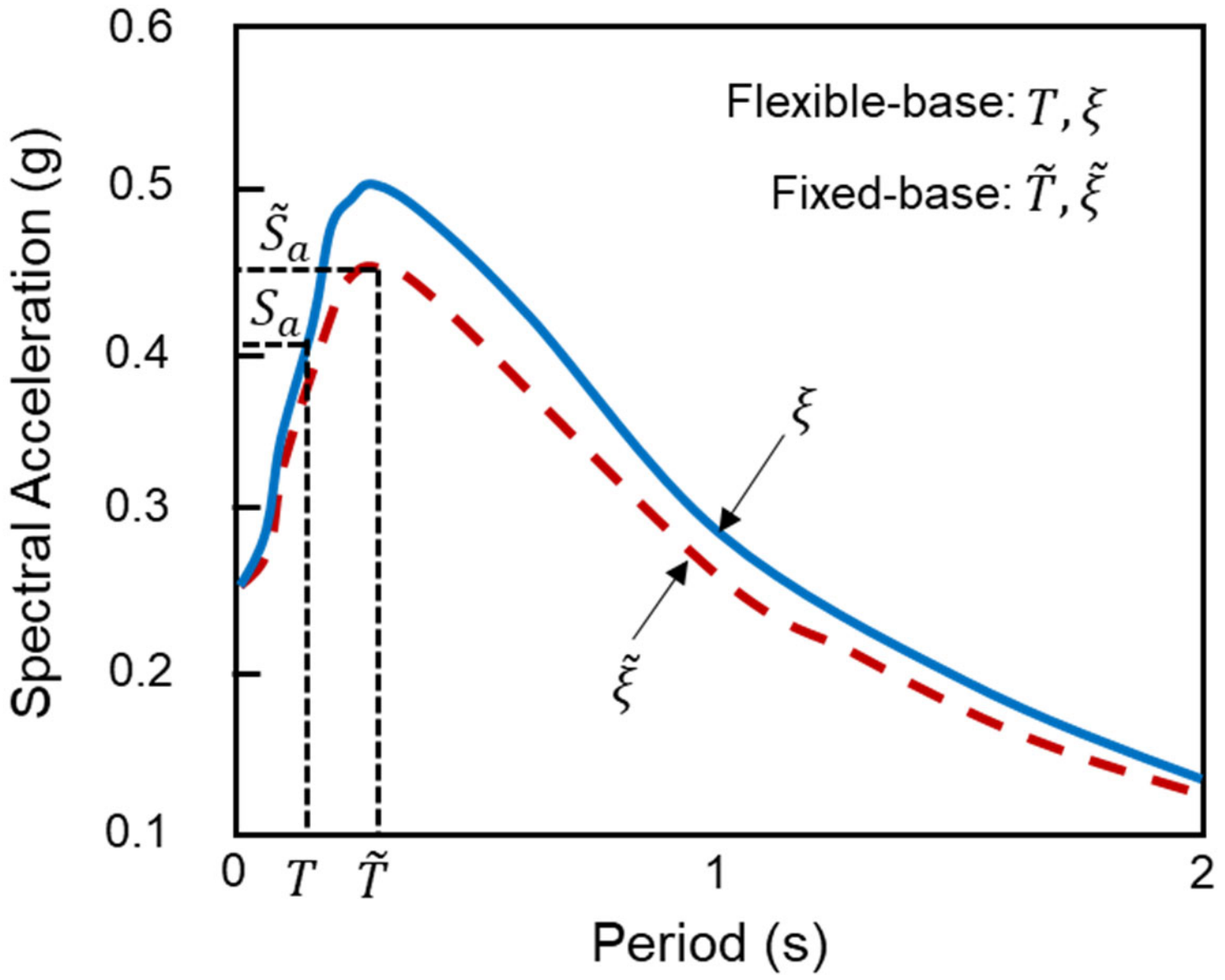

**Figure 2-5: Effects of period lengthening and damping increase on the sceptral acceleration demand (the image is reproduced from** [61]**).**

# CHAPTER 3: EXISTING METHODS

## 3.1. INTRODUCTION

As it was described in Chapter 1, the substructure approach has to be solved in the frequency domain because of the frequency-dependency of the soil-foundation impedance function. However, this method is frequently employed in the time domain to be able to analyze systems with a nonlinear superstructure. Therefore, the frequency-depednency is usually neglected by setting the impedance function at a dominant frequency (the fundamental frequency of the flexible-base system (refer to Chapter 2), or the dominant frequency of the excitation). In this chapter, available solutions to address frequency dependency in the nonlinear time-history analysis are briefly reviewed.

### 3.1.1. The Problem

To review the existing solutions, let's consider a very simple example shown in Figure 3-1(a). In this example, an SDOF system is placed on top of a sway foundation with a frequency-dependent impedance function of $\bar{S}(\omega)$. This impedance function can be, for example, obtained numerically or analytically by solving a problem shown in Figure 3-1(b).

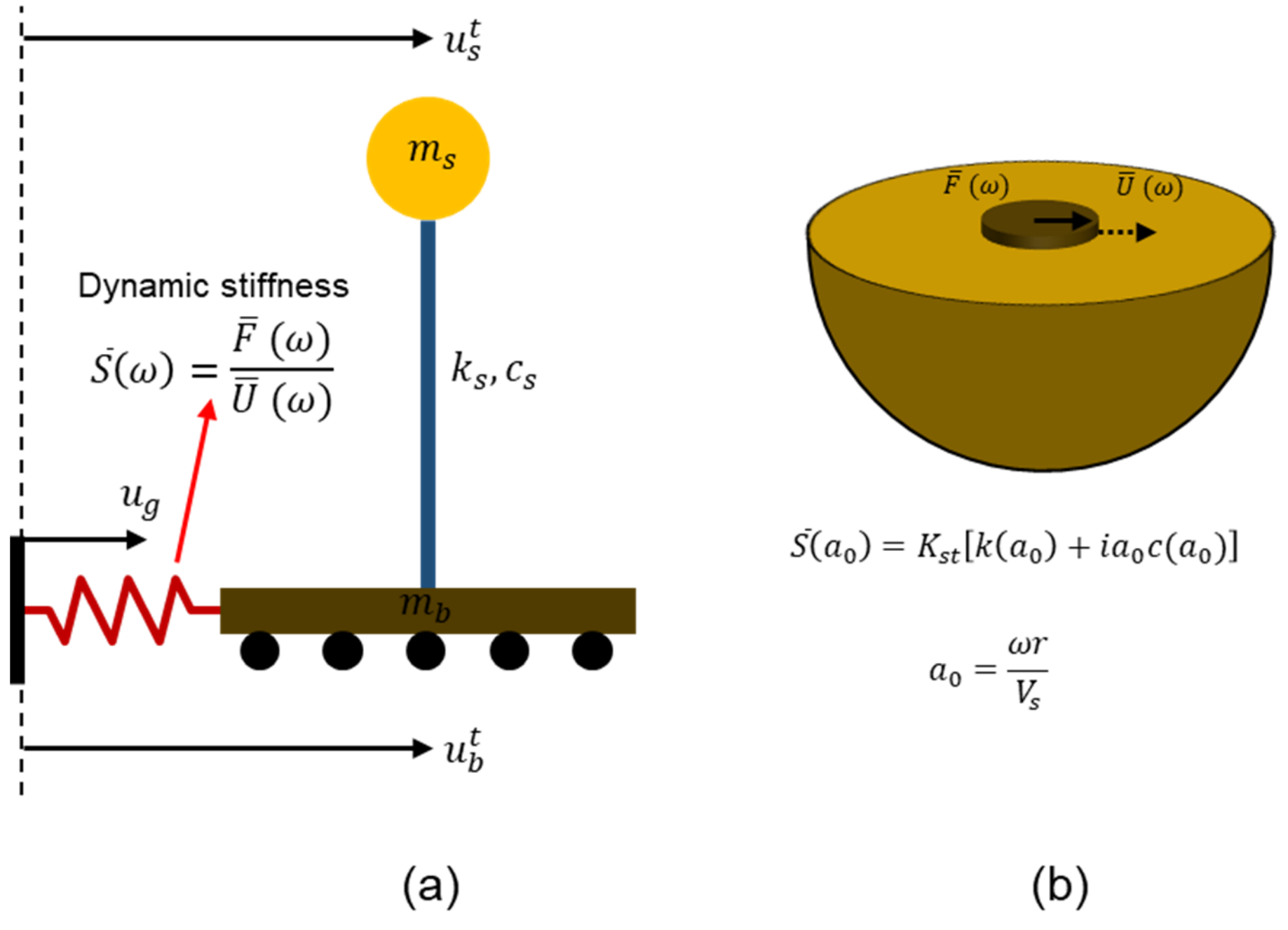

**Figure 3-1: (a) a simple SDOF on a frequency-dependent sway soil-foundation substructure, and (b) the rigid foundation on a semi-infinite soil half-space.**

One can write the equation of motion of this system in the time domain as

$$\begin{bmatrix} m_s & 0 \\ 0 & m_b \end{bmatrix}\begin{bmatrix} \ddot{u}_s^t \\ \ddot{u}_b^t \end{bmatrix} + \begin{bmatrix} c_s & -c_s \\ -c_s & c_s \end{bmatrix}\begin{bmatrix} \dot{u}_s^t \\ \dot{u}_b^t \end{bmatrix} + \begin{bmatrix} k_s & -k_s \\ -k_s & k_s \end{bmatrix}\begin{bmatrix} u_s^t \\ u_b^t \end{bmatrix} + \begin{bmatrix} 0 \\ s(t) * u_b^t \end{bmatrix} = \begin{bmatrix} 0 \\ s(t) * u_g \end{bmatrix}, \tag{3-1}$$

in which $s(t)$ is the inverse Fourier Transform of $\bar{S}(\omega)$, i.e., $s(t) = F^{-1}\{\bar{S}(\omega)\}$. As seen, the frequency-dependent impedance function appears as a convolution ($*$) in the time domain. Although there is already a computational cost here because of that integration, there is another issue too. The inverse Fourier transform of the impedance function could be non-causal [62], which makes the implementation impossible because the reaction force at any time instant would be a function of the state at future times. As an example, Veletsos and Verbic [63] approximated the rocking impedance function of a disk on half-space as

$$S(a_0) = K_{st}\left[1 - b_1\frac{(b_2a_0)^2}{1 + ib_2a_0} - b_3{a_0}^2\right], \tag{3-2}$$

where $K_{st} = \frac{8Gr^3}{3(1-\upsilon)}$, $a_0 = \frac{\omega r}{V_s}$, and $b_1$, $b_2$ and $b_3$ are some constants, and, then, they used the inverse Fourier Transform to obtain its time-domain representation and consequently reaction moment as shown below.

$$M(t) = K_{st}\left[(1 - b_1)\theta(t) + b_1b_2\frac{r}{V_s}\dot{\theta}\ (t) + b_3\left(\frac{r}{V_s}\right)^2\ddot{\theta}(t) + \frac{b_1}{b_2}\frac{V_s}{r}\int_{-\infty}^{+\infty}\theta(x)\,e^{-V_s/(b_2r)(t-x)}dx\right]. \tag{3-3}$$

As seen in Eq. (3-3), in addition to extra computational expenses due to the convolution integration, the integration term in Eq. (3-3) is not a causal term because the rotation at future time instants is needed to obtain moment at the current time.

### 3.1.2. Impedance Function Characteristic

Looking at the impedance functions, they all can be decomposed into two parts as shown in Figure 3-2: a singular part $\bar{S}_s(\omega)$ which is the part that is not zero at high frequencies (not square integrable), and a regular part $\bar{S}_r(\omega)$. In other words,

$$\bar{S}(\omega) = \bar{S}_s(\omega) + \bar{S}_r(\omega). \tag{3-4}$$

The singular part can be usually represented as a combination of mass, stiffness, and damping terms as shown below.

$$\bar{S}_s(\omega) = -m_b\omega^2 + i\omega c_b + k_b. \tag{3-5}$$

So, the total impedance function can be represented in the time domain as

$$s(t) = m_b\ddot{\delta}\ (t) + c_b\dot{\delta}\ (t) + k_b\delta(t) + \frac{1}{2\pi}\int_{-\infty}^{+\infty}\bar{S}_r(\omega)e^{i\omega t}d\omega. \tag{3-6}$$

Therefore, the soil-foundation reaction force can be written as

$$F(t) = \{m_b \ddot{u}_b^t(t) + k_b[u_b^t(t) - u_g(t)] + c_b[\dot{u}_b^t(t) - \dot{u}_g(t)]\} + \left\{\int_0^t s_r(t-\tau)\,[u_b^t(\tau) - u_g(\tau)]d\tau\right\}. \quad (3\text{-}7)$$

Eq. (3-7) is the motivation behind most of the existing solutions and methods to represent frequency-dependent soil-foundation impedance function in the time domain. For example, one may neglect the second term in comparison with the first term, to represent the impedance function by using lumped physical models [64]. Although it is possible to develop physical models to have that integral convolution too, some other solutions are more focused on the convolution part, to reduce the computational cost by using recursive formulations [65]. Note that the computational cost of the regular term could be huge because all degrees of freedom of the nodes on the structure-soil interface from the start of the excitation contribute to the forces.

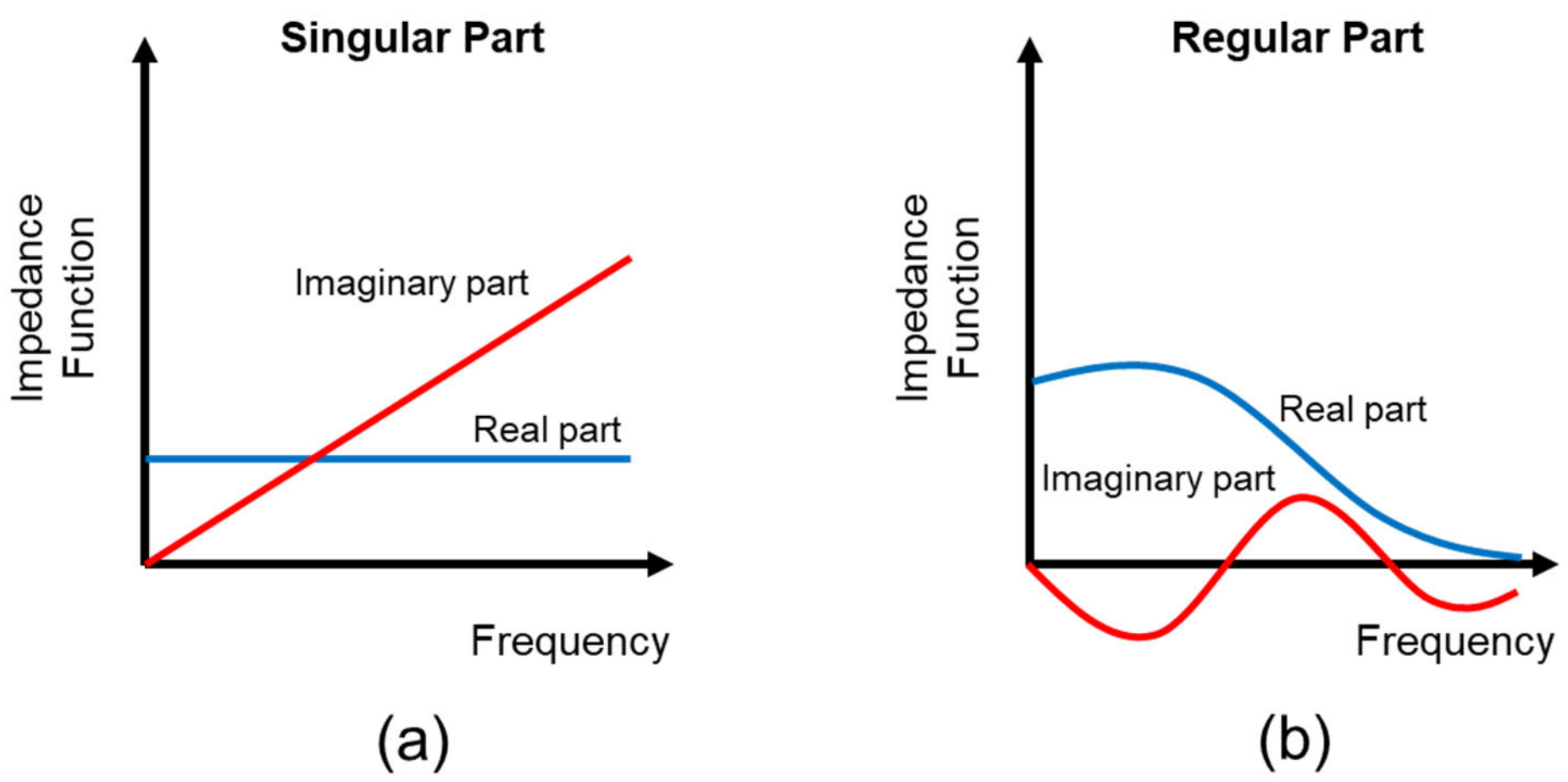

**Figure 3-2: Schematic representation of singular and regular parts of the impedance function in the frequency domain (the image is reproduced from** [66]**).**

Based on the extensive works by pioneer researchers like John Wolf, several solutions have been proposed during the last 40 years some of which will be reviewed below.

### 3.2. PHYSICAL MODELS

As mentioned earlier, physical models like the ones shown in Figure 3-3, are an obvious solution [64]. While this method is easy to use for simple problems like a single node at the soil-structure interface, it can not be applied to large-scale problems. Also, it is not trivial to make a such model for complex impedance functions. Moreover, this solution may result in some non-physical parameters, like a negative mass or dashpot coefficient which cannot be used within Finite Element software. So, we do not pursue this approach in this project, and will only use it for the verification examples wherever needed.

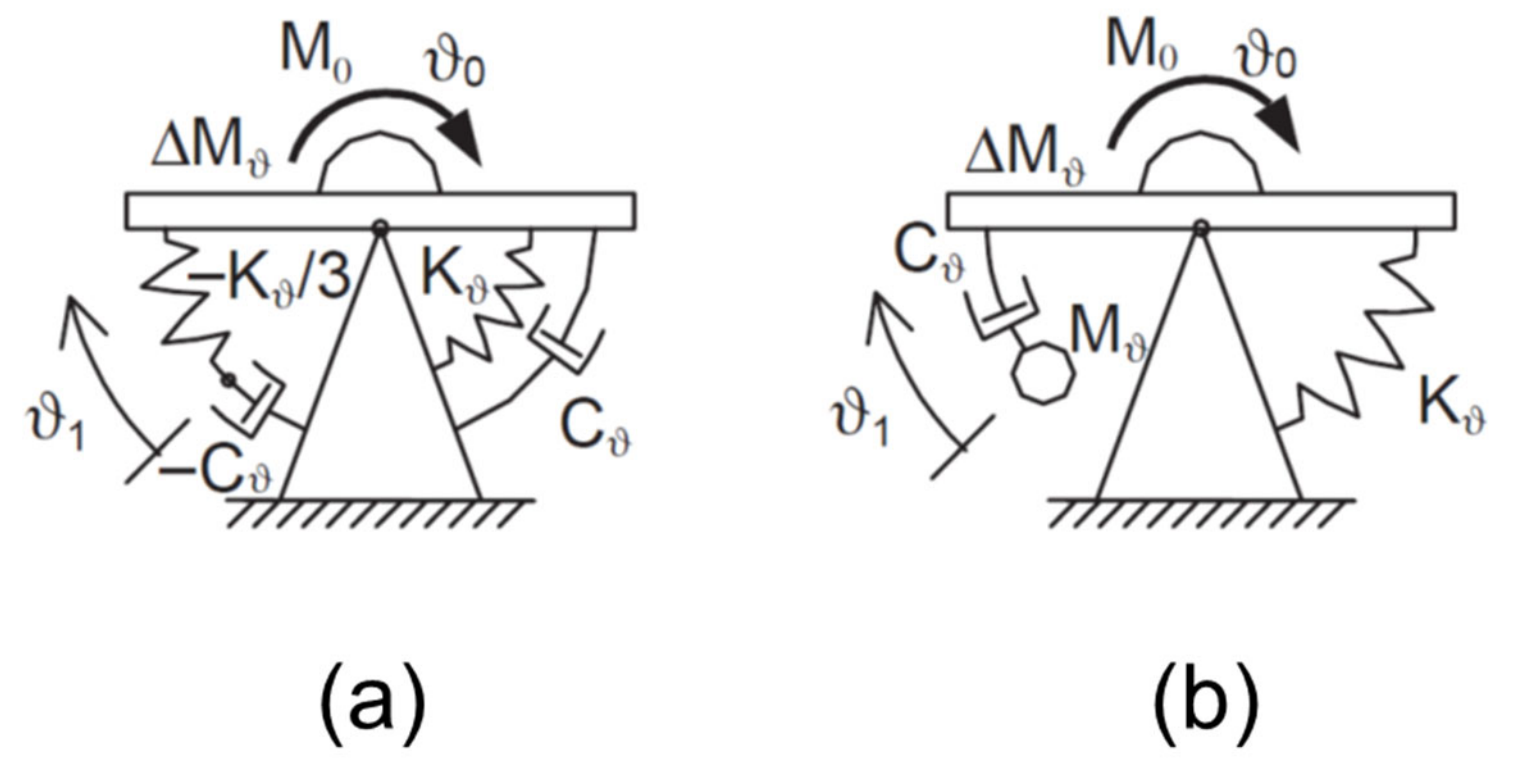

**Figure 3-3: Two physical models proposed by Wolf** [5] **for the rotational impedance function of a foundation on the surface of a homogenous half-space.**

## 3.3. TIME-DOMAIN REPRESENTATION OF IMPEDANCE FUNCTION

One avenue to solve the frequency-dependency in the time domain through a different method than the Inverse Fourier Transform (which is prone to the causality issue mentioned before) is to calculate the interaction force through Finite Impulse Response (FIR) or Infinite Impulse Response (IIR) filters connecting foundation state (displacement, velocity, and acceleration) to the reaction force. These two possibilities will be discussed in the following sections.

### 3.3.1. Finite Impulse Response (FIR) Filter Representation

In the Finite Impulse Response (FIR) representation, the reaction force is estimated using the foundation state at the current and previous time instants. Between 2005 and 2008, Nakamura published a series of papers and proposed this solution using displacement, velocity, and acceleration foundation responses [66]–[70]. This method, which will be reviewed below, is less prone to stability issues, at least at the filter level, because the filter does not have any pole.

The idea behind the FIR filter proposed by Nakamura [66] is shown in Figure 3-4 through a simple example. Figure 3-4(a) shows the cone model [5] for the translational response of a surface foundation on a multi-layered semi-infinite soil domain. As shown in Figure 3-4(b), once the unit impulse displacement is imposed, a force due to the simultaneous reaction of soil is produced. However, in addition to the simultaneous component, the reflective reactive also contributes to the total reaction force but with some time delays due to the finite wave propagation speed. That is,

$$F(t) = F_0(t) + \sum_{j=1}^{n} c_j' \, F_0\big(t - T_j\big), \tag{3-8}$$

where $F_0(t)$ is the simultaneous reaction force, while $F_0\big(t - T_j\big)$ is the reflective reaction force due to j-th reflection which is assumed to be a $c_j'$ fraction of $F_0(t)$. $T_j$ is the occurrence time of the reflection reaction, and $n$ is the total number of the considered reflections. Assuming the reaction force is related to the foundation displacement and velocity through soil's stiffness $K_0$ and damping $C_0$, respectively, Eq. (3-8) can be written as

$$F(t) = \{K_0 u(t) + C_0 \dot{u}(t)\} + \{\textstyle\sum_{j=1}^{n} c'_j \left[K_0 u(t - T_j) + C_0 \dot{u}(t - T_j)\right]\}. \tag{3-9}$$

Using Fourier Transform, Eq. (3-9) can be expressed in the frequency domain as

$$\bar{F}(\omega) = \bar{U}(\omega)\{K_0 + i\omega C_0\} + \bar{U}(\omega)\left\{\sum_{j=1}^{n} c'_j \left[K_0 e^{-i\omega T_j} + i\omega C_0 e^{-i\omega T_j}\right]\right\}, \tag{3-10}$$

which can be rewritten as

$$\bar{F}(\omega) = (K_0 + i\omega C_0)\left\{1 + \sum_{j=1}^{n} c'_j\, e^{-i\omega T_j}\right\}\bar{U}(\omega), \tag{3-11}$$

from which the corresponding impedance function can be obtained as

$$\bar{S}(\omega) = (K_1 + i\omega C_1)\left\{1 + \sum_{j=1}^{n} c'_j\, e^{-i\omega T_j}\right\}. \tag{3-12}$$

So, provided that $K_0$, $C_0$, and coefficients $c'_j$ and time delays $T_j$ are determined, the reaction force taking into account the frequency-dependent impedance function in the form of Eq. (3-12) can be calculated in the time domain using Eq. (3-9) where an Impulse Response Function (IRF) representation of the impedance function convolved with foundation response. As mentioned earlier and can be seen in Eq. (3-9), this IRF model contains two components: 1- simultaneous and 2- time-delay components that are schematically shown in Figure 3-5(a) and (b), respectively. It is interesting to note that these two time-domain components, respectively, correspond to the singular and regular parts of the impedance function in the frequency domain which are shown in Figure 3-2.

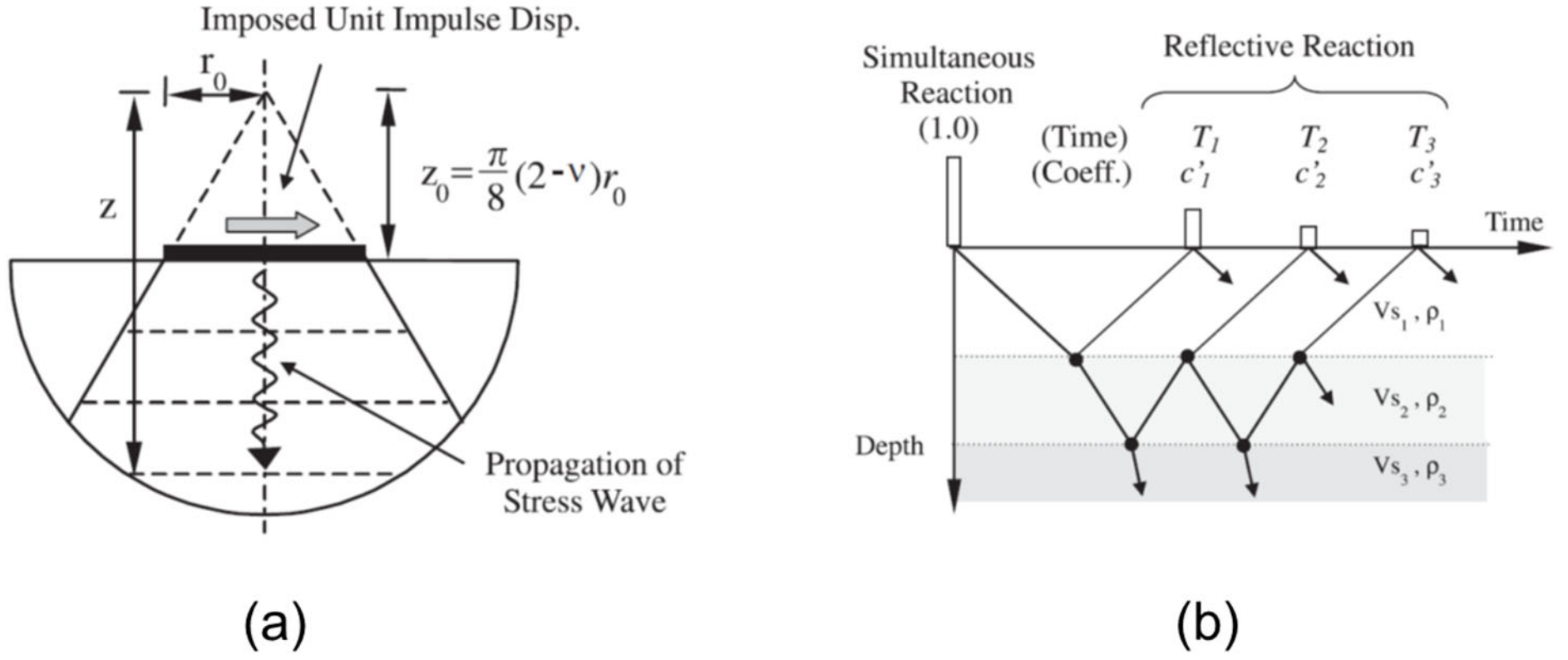

**Figure 3-4: (a) Cone model for translational foundation response, and (b) impulse response on multi-layered soil** [66]**.**

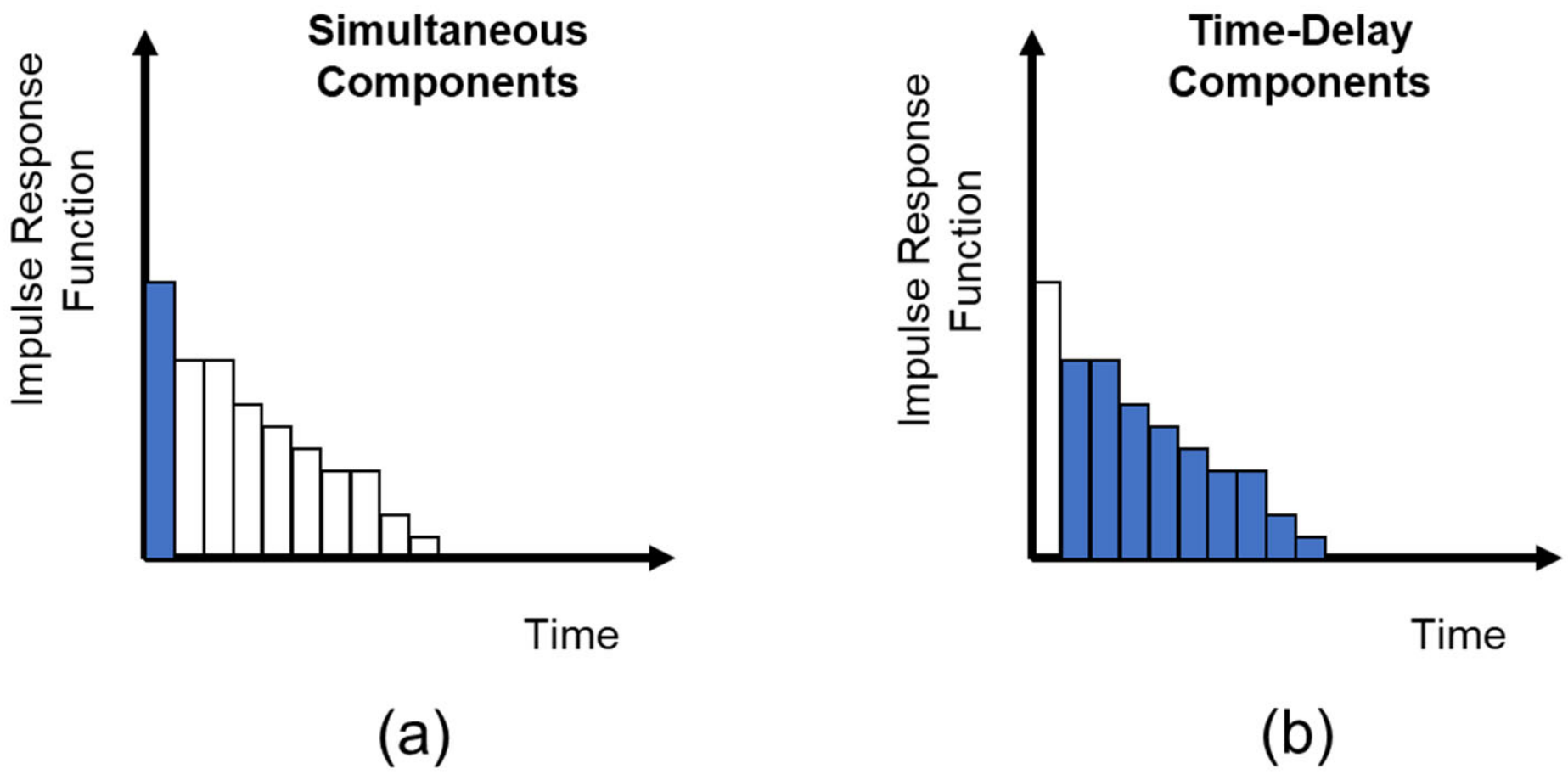

**Figure 3-5: Schematic representation of simultaneous and time-delay components of the impedance function in the time domain (the image is reproduced from** [66]**).**

To obtain the unknown parameters needed in Eq. (3-9), let's rewrite Eq. (3-9) as

$$F(t) = \sum_{j=0}^{n} k_j u\left(t - T_j\right) + c_j \dot{u}\left(t - T_j\right), \tag{3-13}$$

where $k_j$ and $c_j$ are called stiffness and damping terms of the IRF hereafter. If we discretize IRFs (stiffness and damping terms) at intervals of $\Delta t$ (which can be larger than the time interval used for time history analysis), Eq. (3-13) can be written as

$$F(t) = \sum_{j=0}^{N-1} k_j u\left(t - t_j\right) + c_j \dot{u}\left(t - t_j\right), \tag{3-14}$$

where $t_j = j\Delta t$, $k_j = k(t_j)$ and $c_j = c(t_j)$. The equation above contains $2N$ unknown coefficients to be determined, so at least $2N$ equations are needed. The frequency-dependent impedance function corresponding to Eq. (3-14) is

$$\bar{S}(\omega) = \sum_{j=0}^{N-1} k_j\, e^{-i\omega t_j} + i\omega \sum_{j=1}^{n} c_j\, e^{-i\omega t_j}. \tag{3-15}$$

So, having samples of the impedance function at $p = 1..M \geq N$ frequencies, we have

$$\begin{bmatrix} Real\{\bar{S}(\omega_p)\} \\ Imag\{\bar{S}(\omega_p)\} \end{bmatrix} = \begin{bmatrix} \sum_{j=0}^{N-1} k_j \cos\theta_{pj} + \omega_p \sum_{j=0}^{N-1} c_j \sin\theta_{pj} \\ -\sum_{j=0}^{N-1} k_j \sin\theta_{pj} + \omega_p \sum_{j=0}^{N-1} c_j \cos\theta_{pj} \end{bmatrix}, \tag{3-16}$$

from which we have

$$\begin{bmatrix}\mathbf{G}_K \\ \mathbf{G}_C\end{bmatrix} = [\mathbf{C}_K \quad \mathbf{C}_C]^{-1} \begin{bmatrix}\mathbf{D}(\omega_1) \\ \vdots \\ \mathbf{D}(\omega_M)\end{bmatrix}, \tag{3-17}$$

where

$$\mathbf{D}(\omega_p) = \begin{bmatrix} Real\{\bar{S}(\omega_p)\} \\ Imag\{\bar{S}(\omega_p)\}\end{bmatrix}, \tag{3-18}$$

$$\mathbf{G}_K = \begin{bmatrix} k_0 \\ \vdots \\ k_{N-1}\end{bmatrix}, \tag{3-19}$$

$$\boldsymbol{G}_C = \begin{bmatrix} c_0 \\ \vdots \\ c_{N-1}\end{bmatrix}, \tag{3-20}$$

$$\mathbf{C}_K = \begin{bmatrix} [\mathbf{C}_K]_{1,0} & \cdots & [\mathbf{C}_K]_{1,N-1} \\ \vdots & \ddots & \vdots \\ [\mathbf{C}_K]_{M,0} & \cdots & [\mathbf{C}_K]_{M,N-1}\end{bmatrix}, \ with\ [\mathbf{C}_K]_{p,j} = \begin{bmatrix}\cos\theta_{pj} \\ -\sin\theta_{pj}\end{bmatrix}, \tag{3-21}$$

$$\boldsymbol{C}_C = \begin{bmatrix} [\boldsymbol{C}_C]_{1,0} & \cdots & [\boldsymbol{C}_C]_{1,N-1} \\ \vdots & \ddots & \vdots \\ [\boldsymbol{C}_C]_{M,0} & \cdots & [\boldsymbol{C}_C]_{M,N-1}\end{bmatrix}, \ with\ [\mathbf{C}_C]_{p,j} = \begin{bmatrix}\omega_p\cos\theta_{pj} \\ \omega_p\sin\theta_{pj}\end{bmatrix}. \tag{3-22}$$

To solve Eq. (3-17), the following points should be taken into account:

- The number of selected frequencies $M$ should be equal to $N$ to avoid least-squares analysis;
- The time interval of the IRFs ($\Delta t$) should be as short as necessary but as long as possible. It is recommended to use $\Delta t \sim \frac{1}{f_{max}}$ where $f_{max}$ is the maximum frequency of the given impedance function data;

The Matlab implementation of this method, referred to as Method A, along with two other modified/improved versions of this method [69], is available upon request from the first author.

#### 3.3.1.1 Verification Example 1

To verify the Matlab codes, they are used to reproduce examples presented in [70]. In the first example, Method A is used to obtain a time-domain representation of the impedance function of a semi-infinite rod with exponentially increasing section area. The analytical solution to obtain its frequency-dependent impedance function is presented in Appendix A. Figure 3-6 shows real and imaginary parts of the total impedance function and its regular (frequency-dependent) part scaled by the static stiffness. The Impulse Response Function of the regular part for which a closed-form solution is available is also shown in Figure 3-7. Method A with 20 frequency points from 0 to 20 Hz is used to estimate stiffness and damping IRFs. Figure 3-8 shows a comparison between the exact and the estimated IRFs. As seen, both IRFs are estimated very accurately. Finally, the reconstructed real and imaginary parts of the impedance function are shown in Figure 3-9(a). As seen, both parts pass through the exact data points and they are very similar to their exact counterparts already shown in Figure 3-6(a) although a few ripples are observed at the high-frequency end.

The zoomed-in plot between zero and 5 Hz is shown in Figure 3-6(b) and displays a very good performance of the method in the lower frequencies.

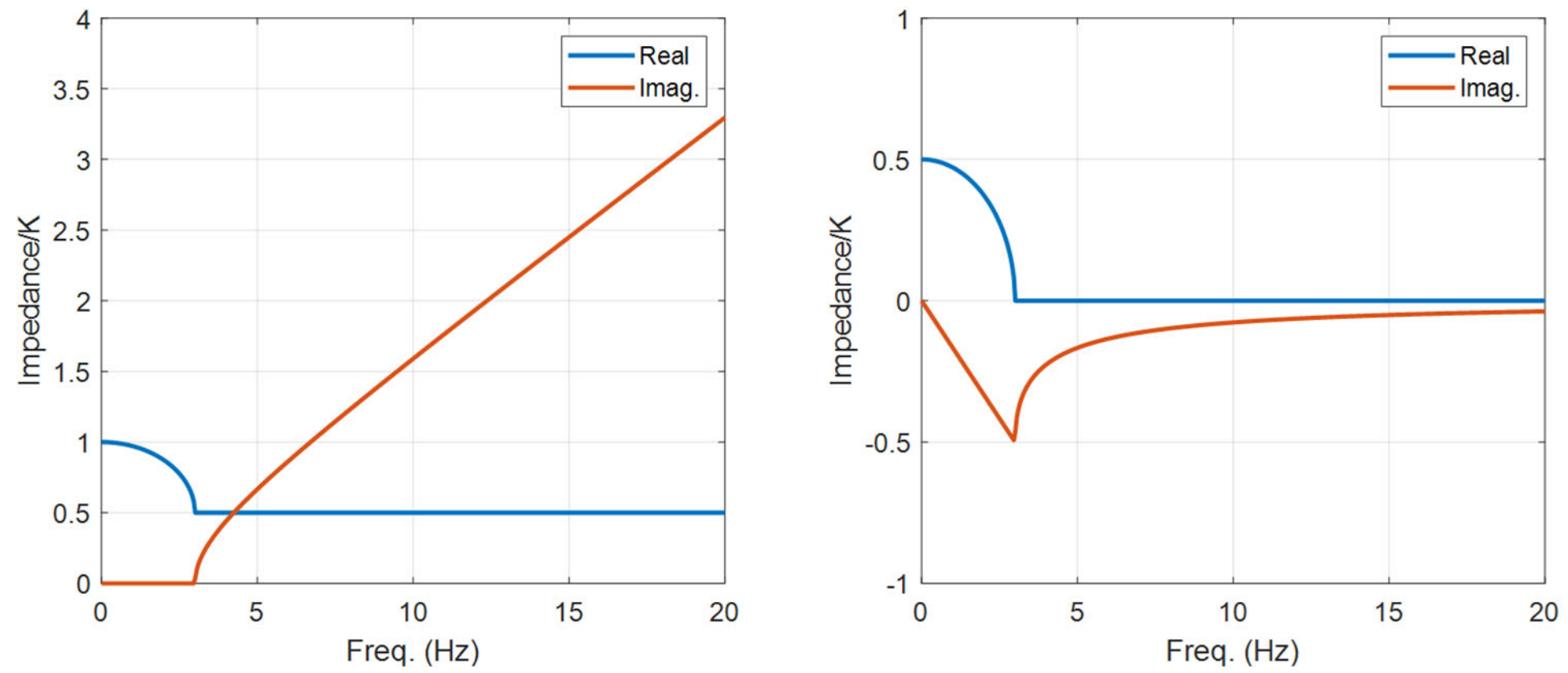

**Figure 3-6: Real and imaginary parts of the (a) total impedance function and (b) its regular part.**

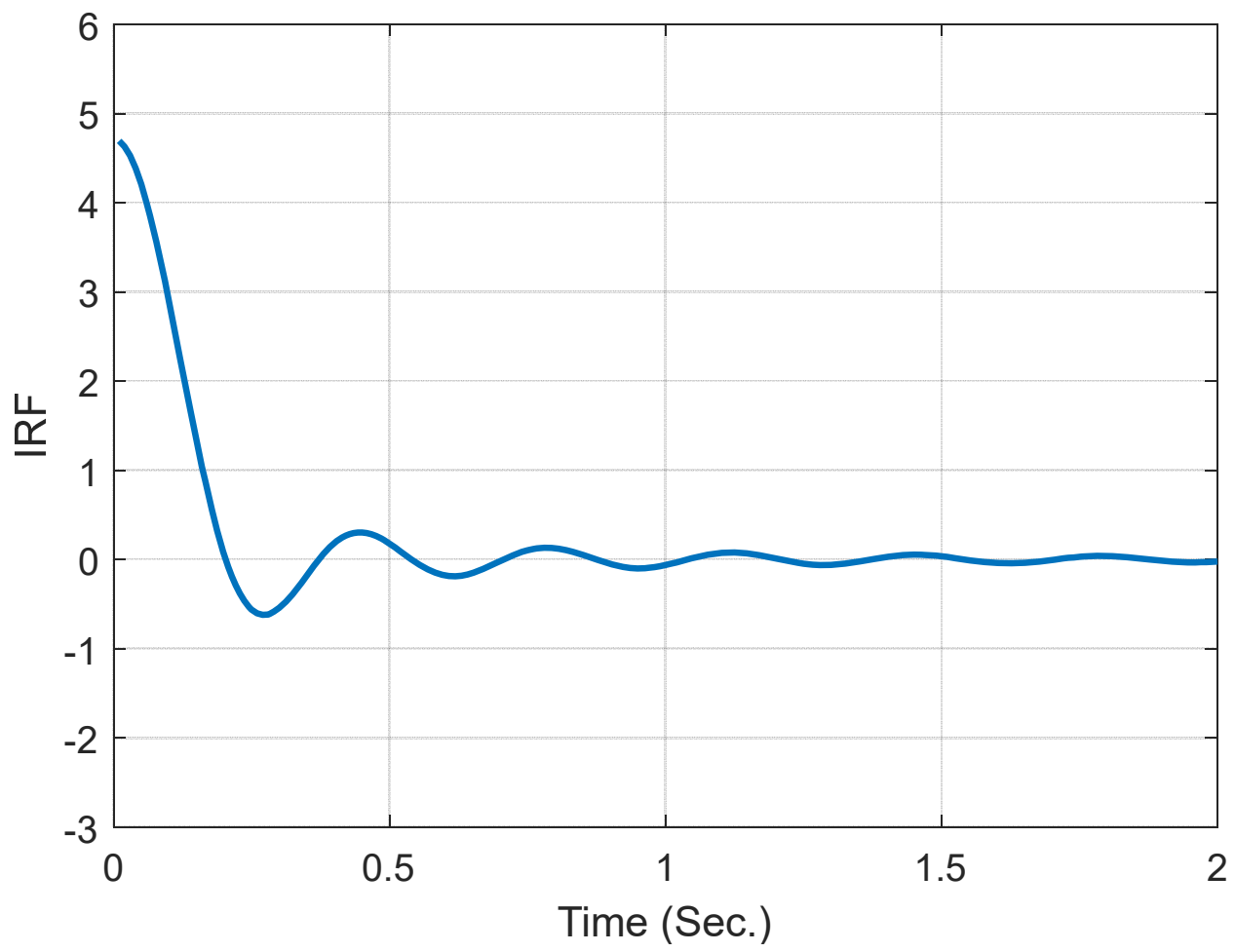

**Figure 3-7: The Impulse Response Function of the regular part of the impedance function.**

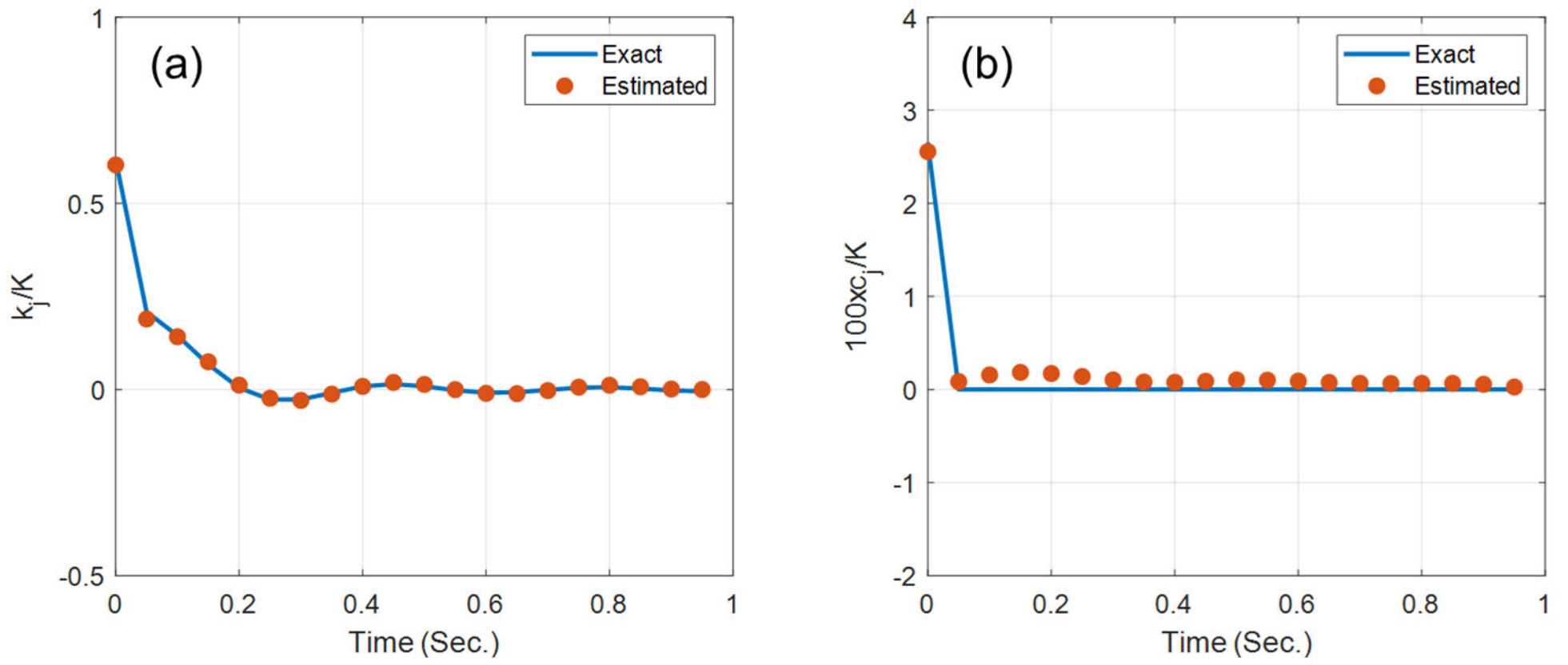

**Figure 3-8: Comparison between exact and estimated (a) stiffness and (b) damping IRFs.**

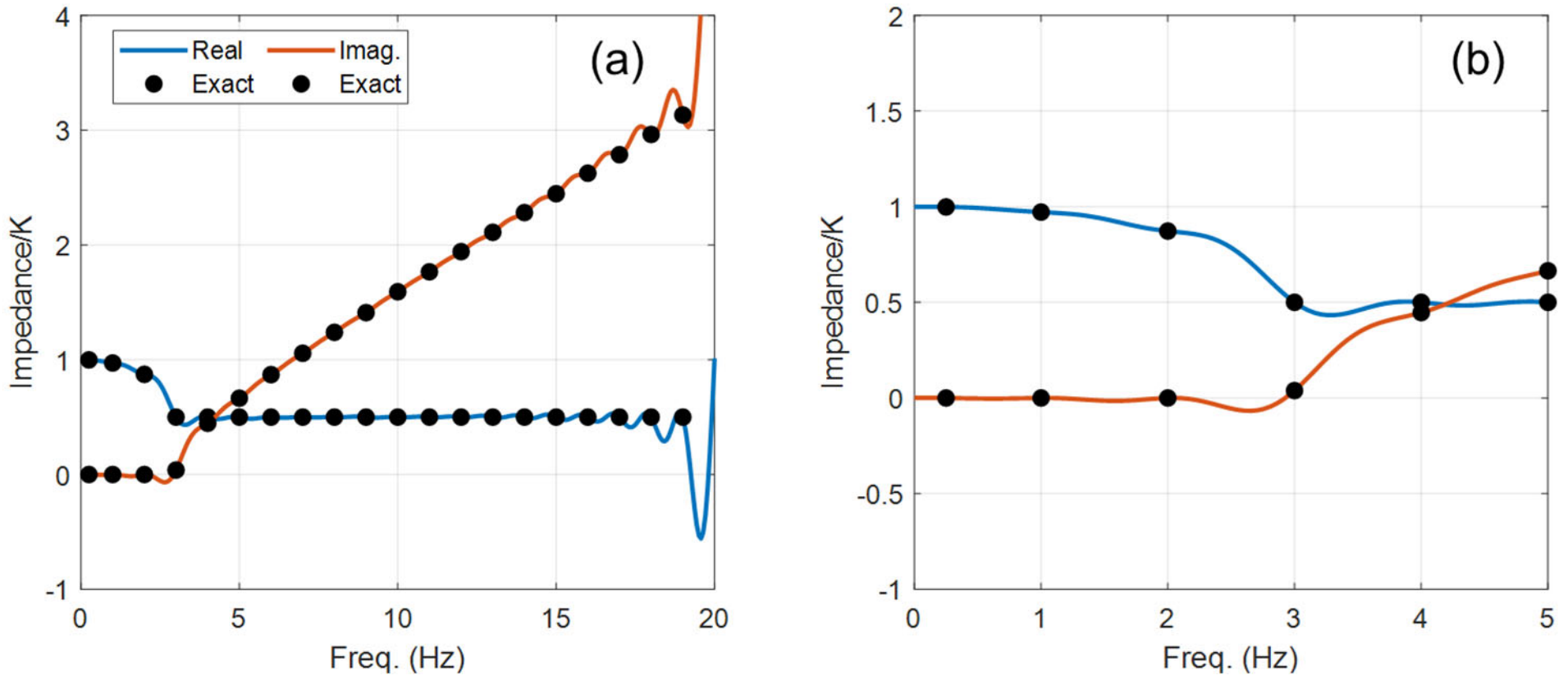

**Figure 3-9: (a) Real and imaginary parts of the reconstructed impedance function. (b) The plot zoomed between zero to 5 Hz.**

### 3.3.1.2 Verification Example 2

The second example shows the application of Method A to sway and rocking impedance functions of a rigid square foundation on a 2-layered soil as shown in Figure 3-10. The impedance functions are originally developed by Tajimi [71] and obtained from the graphs presented in [66]. Figure 3-11 shows the stiffness and damping IRFs for sway and rocking directions estimated from the mentioned impedance functions using 20 frequency points uniformly distributed from zero to 20 Hz. Note that a limited length of 0.4 seconds is considered for both IRFs to emphasize that the IRFs can be truncated after the level of amplitude drops to very small values. In Figure 3-11, the IRFs are scaled to their first values. Figure 3-12 shows the reconstructed impedance functions in sway and rocking versus data points used for the estimation. Similar to the previous example, the method is able to represent the impedance function in the entire frequency band of interest.

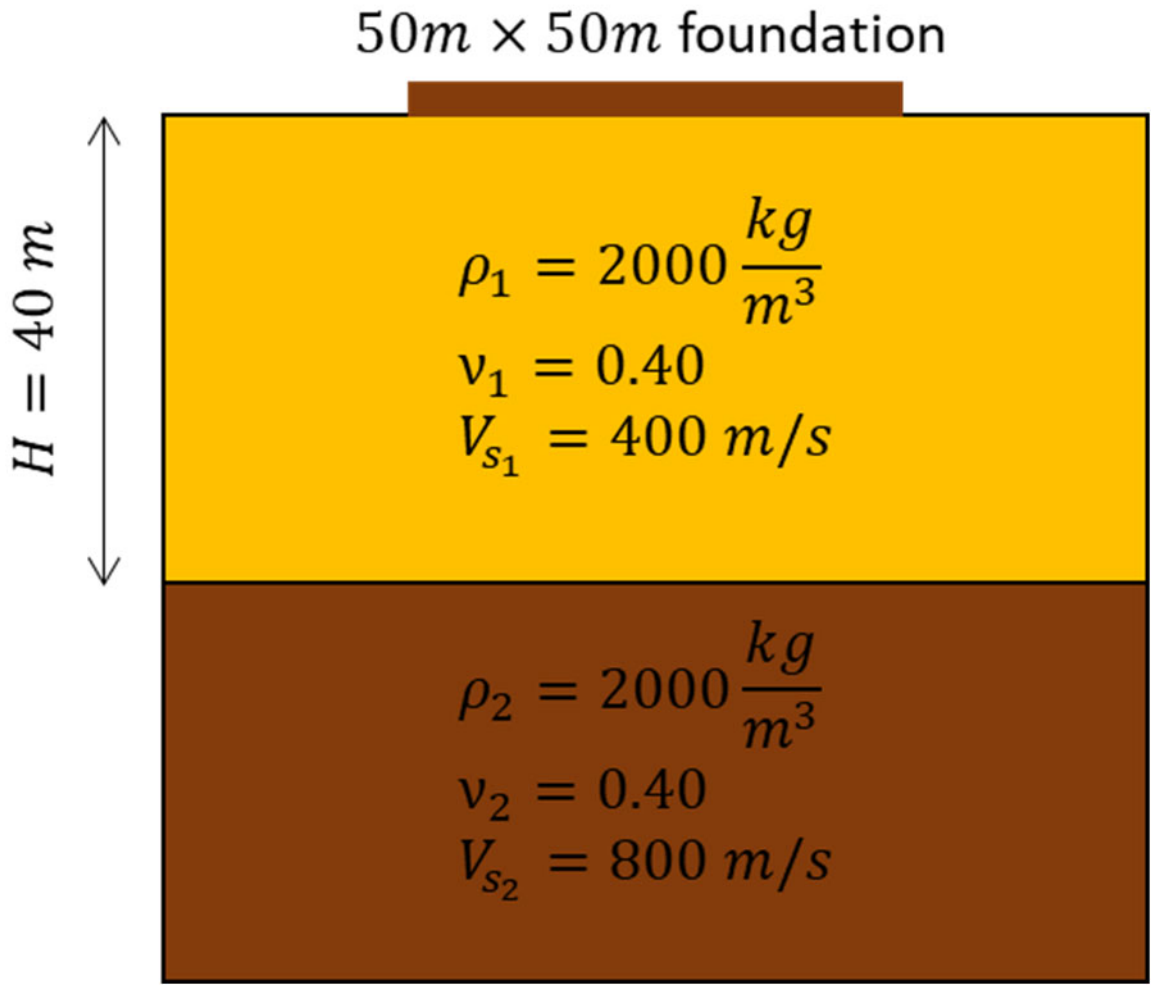

**Figure 3-10: A rigid square foundation on a 2-layered soil domain** [66].

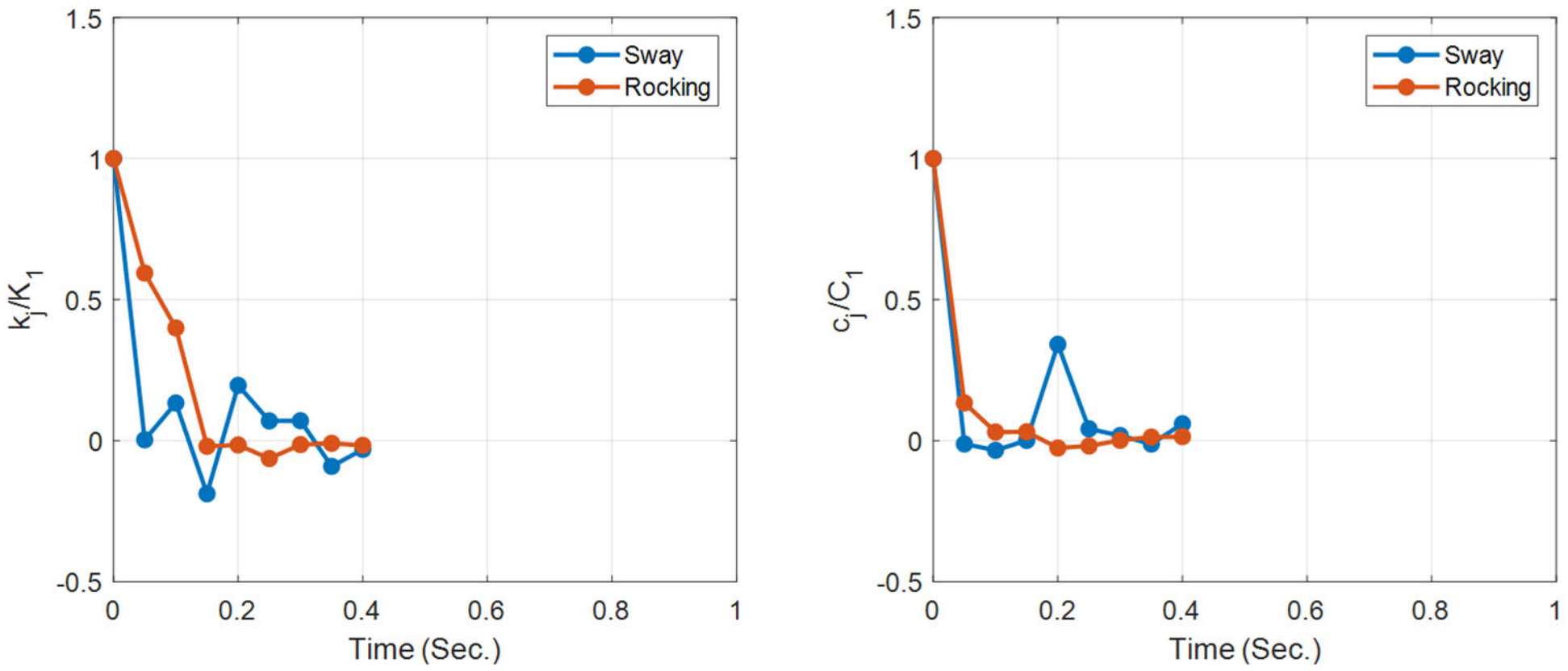

**Figure 3-11: Estimated stiffness and damping IRFs for sway and rocking directions.**

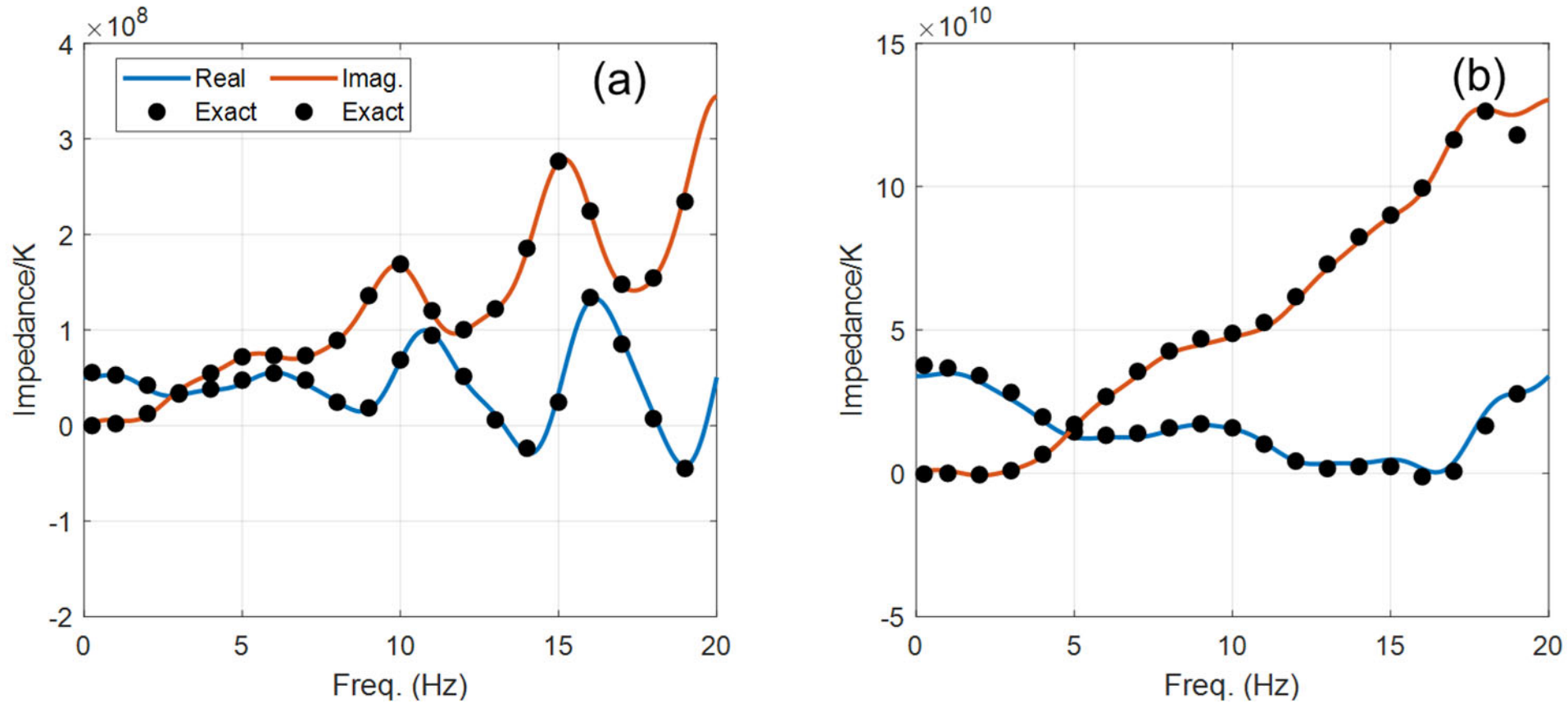

**Figure 3-12: Real and imaginary parts of the reconstructed impedance function in (a) sway and (b) rocking directions.**

### 3.3.1.3 Application Example

To show the performance of Nakamura's method in time-history analysis, a 6-DOF shear-building model is placed on the frequency-dependent sway-rocking foundation presented in the previous section with slightly different properties as shown in Figure 3-13. This example has been originally used in [72].

Figure 3-14 shows real and imaginary parts of the sway and rocking impedance functions used in this example. Using these impedance functions, the absolute acceleration Frequency Response Function (FRF) of the roof is calculated and shown in Figure 3-15. While the fixed-base natural frequency of the system is 4.51 $Hz$, the first dominant peak in the FRF is around 3.14 $Hz$ due to the inertial soil-structure interaction.

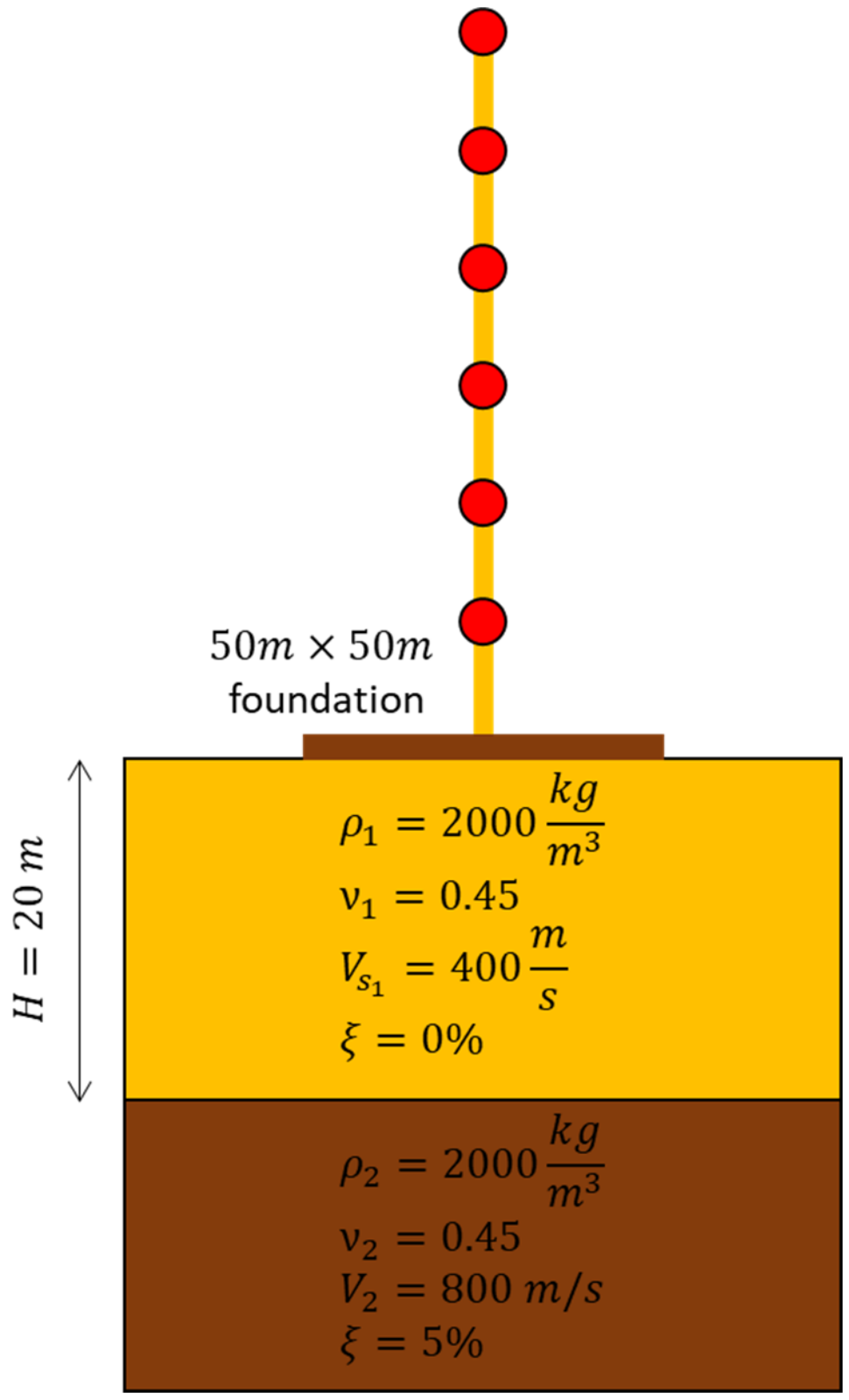

**Figure 3-13: An MDOF system on a 2-layered soil profile.**

To run the system under ground motion excitations, the impedance functions are transferred to the time domain using Nakamura's Method B. The data points used for the time-domain representation and reconstructed impedance functions are shown in Figure 3-14. As seen, the reconstructed impedance functions are almost identical to the actual impedance functions.

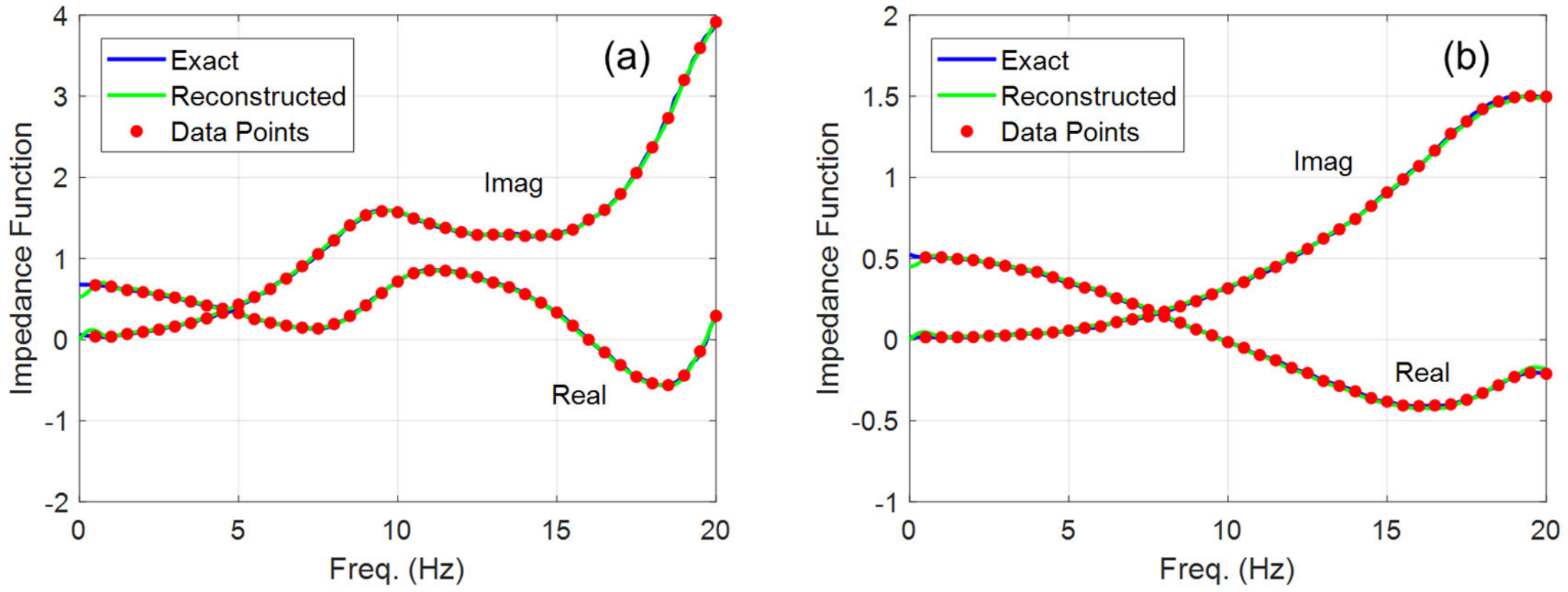

**Figure 3-14: Exact, data points, and reconstructed impedance functions in sway (a) and rocking (b) directions.**

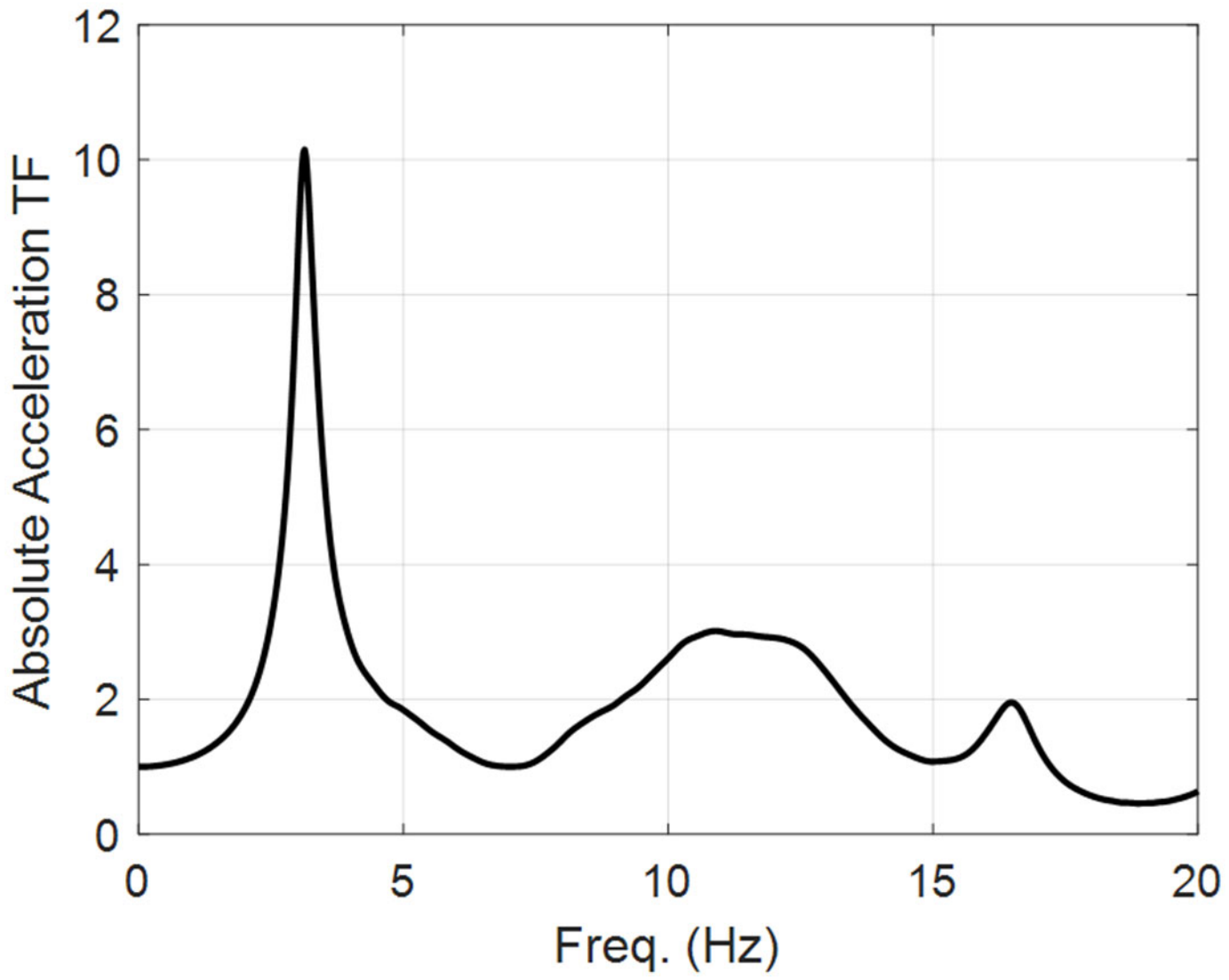

**Figure 3-15: Roof absolute acceleration FRF.**

Figure 3-16 shows the responses of the system under a synthetic ground motion. To emphasize the importance of the frequency-dependent impedance function, a 40-second chirp signal with frequency varying from 1 to 15 Hz is used as ground motion excitation (Figure 3-17). In Figure 3-16, the exact responses shown in blue are obtained by solving the equation of motions in the frequency domain to be able to include the frequency-dependency of the impedance function. The time-domain solution shown in red is the solution obtained by solving the system in the time domain using the estimated time-domain impedance function. As seen, the approximate time-domain responses all match the exact responses. To show the importance of considering the frequency-dependency of the impedance function, the solution obtained through an approach typically used in seismic design (see Chapter 2) is also presented in gray. In this solution, impedance functions are set at the fundamental frequency of the soil-structure system. As seen in Figure 3-16, as long as the frequency of the excitation is close to the fundamental mode, this solution is accurate, but it significantly deviates from the exact response in frequencies far from the fundamental frequency. To see how the time-domain representation of impedance functions is used within time-domain numerical integration, refer to Appendix C where the stability criteria of this method are also discussed. A Matlab code to reproduce the results of this example is available upon request from the first author.

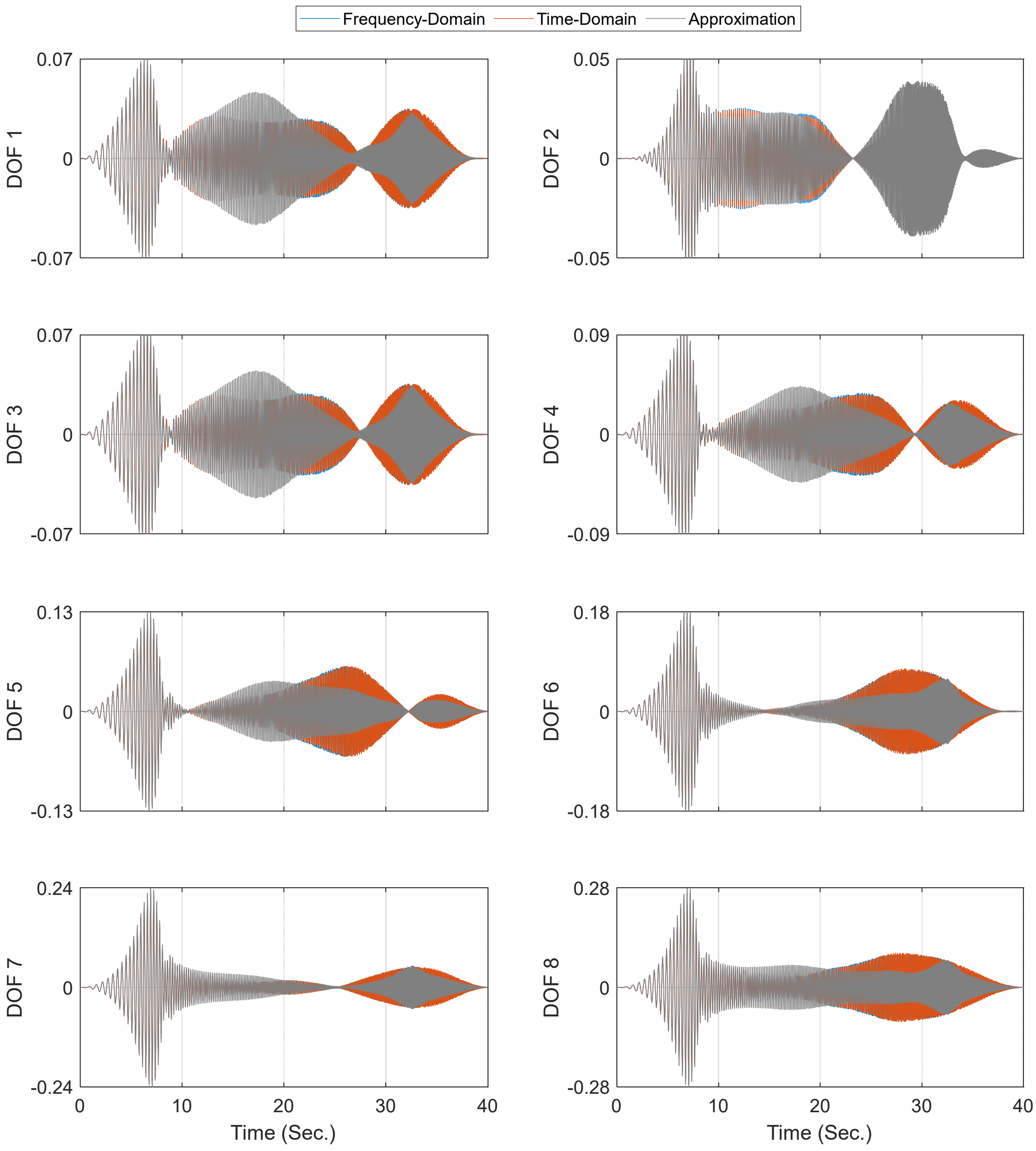

**Figure 3-16: Comparison between exact responses obtained through the frequency-domain solution with the approximate responses obtained through the time-domain representation of the impedance function and approximate responses obtained by using the frequency-independent impedance function.**

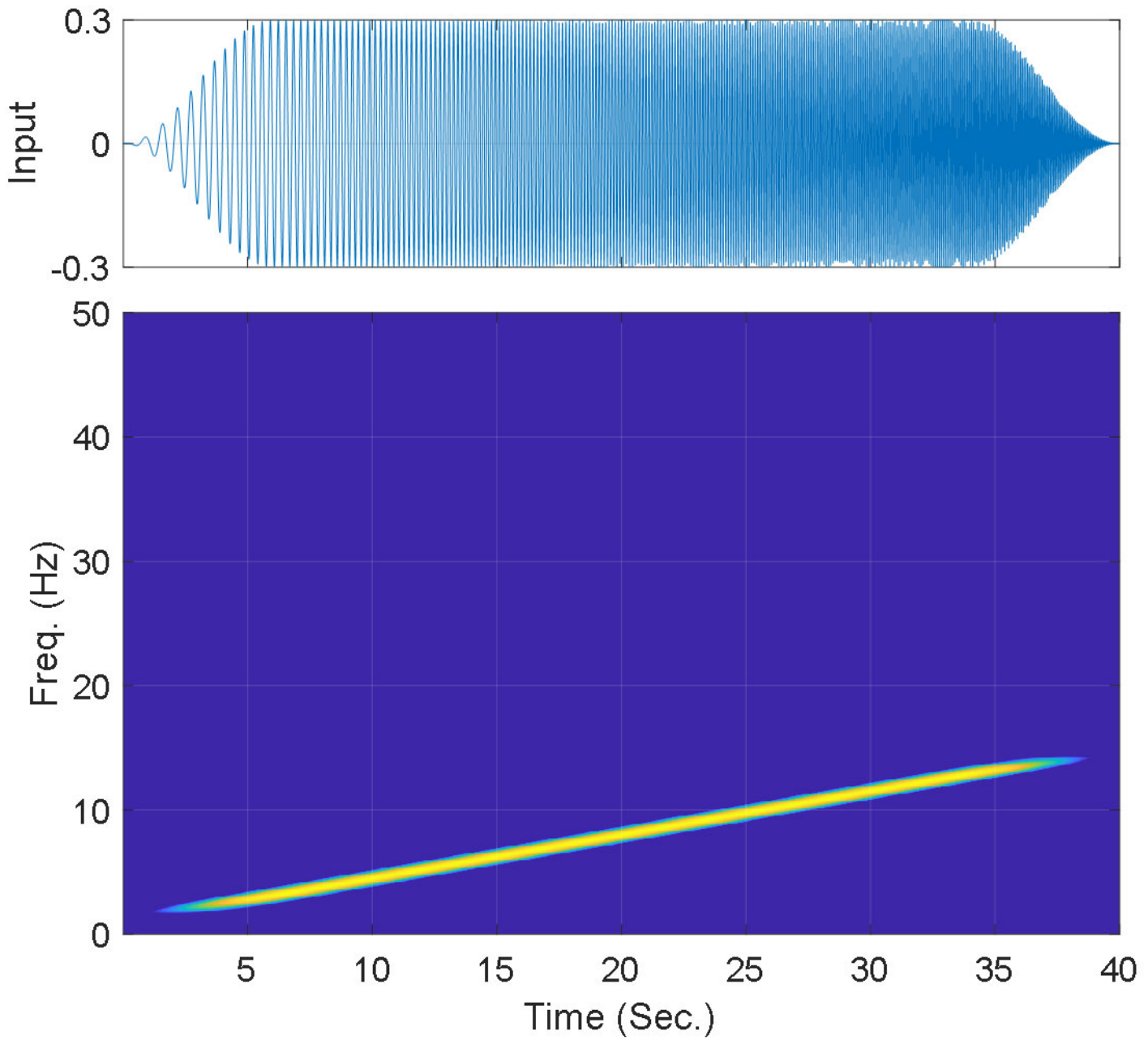

**Figure 3-17: Time and time-frequency representation of the input excitation.**

As reviewed in this section, the time-domain representation of the impedance functions using FIR filters could be a potential solution to address the frequency-dependency in the time domain. However, as shown in Appendix C, the stability of the numerical integrator is an issue. While it is possible to check the stability of the problem (it is computationally costly) before carrying out the time-history analysis, the problem with this solution is how to adjust the parameters to achieve stability. Therefore, this method is not pursued in this project.

### 3.3.2. Infinite Impulse Response (IIR) Filter Representation

The idea of calculating reaction force through a recursive process goes back to early works by Wolf and Motosaka [73], [74]. Let's assume that the Discrete Fourier Transform (DFT) of the dynamic stiffness/impedance function of the soil-foundation, $\bar{S}(\omega)$, is available at $N$ discrete frequencies. Theoretically, its time-domain representation can be calculated using Inverse DFT (IDFT) as follows[13]

$$s[n] = \frac{1}{N\Delta t}\sum_{j=0}^{N-1} \bar{S}(\omega_j)\, e^{-i\omega_j n\Delta t}, \tag{3-23}$$

from which the Z-Transform [75] of the impedance function can be written as

$$S(z) = \frac{1}{N\Delta t}\sum_{j=0}^{N-1} \bar{S}(\omega_j)\frac{1}{1 - e^{-i\omega_j\Delta t}z^{-1}}. \tag{3-24}$$

The interaction force can be calculated using the inverse Z-Transform as follows

[13] It is implicitly assumed that the impedance function only contains the regular part.

$$R[n] = \sum_{j=0}^{N-1} R^j[n], \tag{3-25}$$

where

$$R^j[n] = e^{i\omega_j \Delta t} R^j[n-1] + \frac{1}{2N}\bar{S}(\omega_j)\big(u[n-1] + u[n]\big), \tag{3-26}$$

in which a linear variation is assumed for the foundation displacement $u[n]$. To verify this approach, the force needed to apply a displacement pulse to a rod on an elastic foundation is calculated and compared to the analytical solution. As shown in Appendix B, the regular part of the impedance function in the frequency domain is

$$\bar{S}(a_0) = i\sqrt{a_0^2 - 1} - ia_0, \tag{3-27}$$

where $a_0$ is a dimensionless frequency. The rod is subject to the following displacement

$$u(t) = \frac{u_0}{2}\left[1 - \cos 2\pi \frac{t}{t_0}\right], \qquad for \quad t < t_0 \tag{3-28}$$

with $u_0 = 1$ and $t_0 = 2$. Figure 3-18(a) shows this imposed displacement. The comparison between the exact reaction force and the reaction forces calculated through Eq. (3-25) is shown in Figure 3-18(b). As seen, the recursive solution is identical to the exact solution. The Matlab code to reproduce these results is available upon request from the first author.

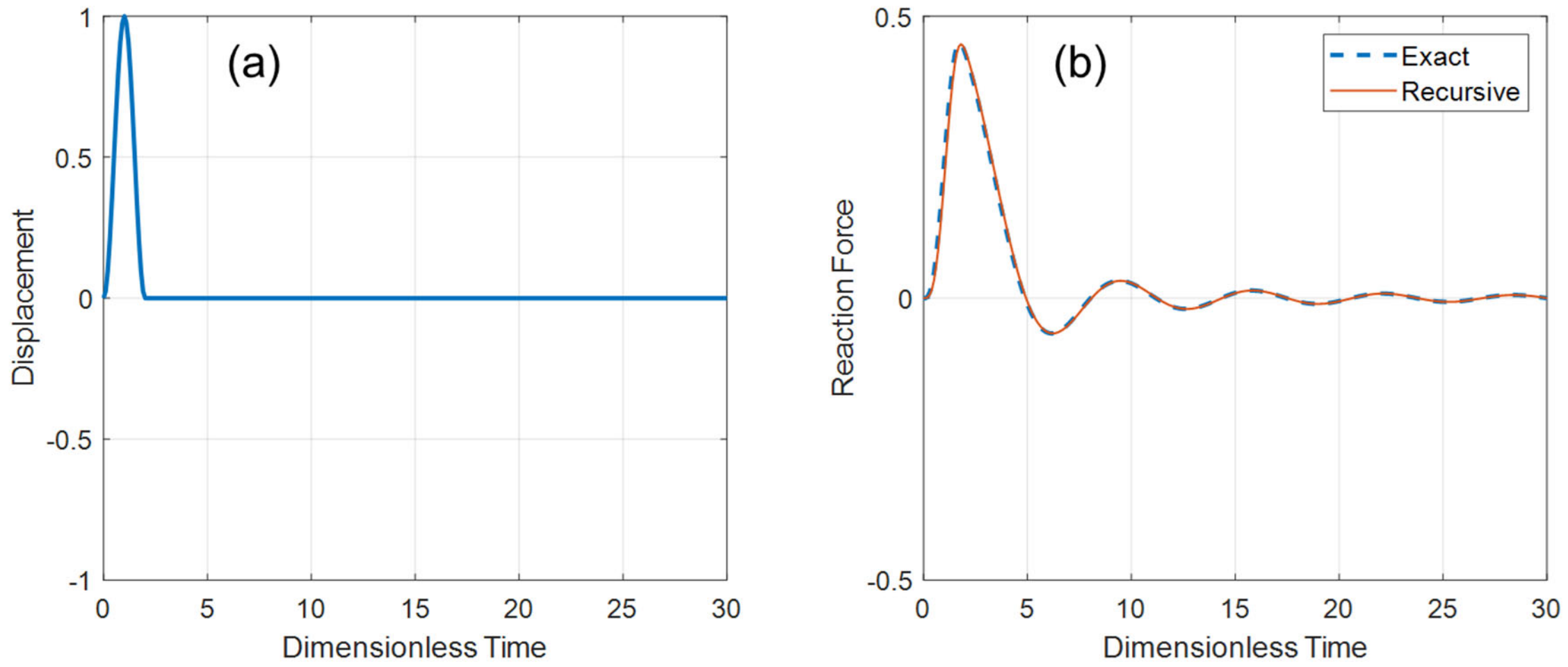

**Figure 3-18: (a) The imposed displacement, and (b) the exact and calculated reaction forces.**

Looking at Eq. (3-26), it is easy to show that employing a delay operator $q^{-1}$ where $u[n-1]=q^{-1}u[n]$, $R^j[n]$ can be rewritten as a rational expression follows,

$$R^j[n]=\frac{\frac{\bar{S}(\omega_j)}{2N}(1+q^{-1})u[n]}{1-e^{i\omega_j\Delta t}q^{-1}}. \tag{3-29}$$

This IIR representation is equivalent to modeling the impedance function at a frequency $\omega_j$ as a rational expression as

$$\bar{S}[z]=\frac{\frac{\bar{S}(\omega_j)}{2N}(1+q^{-1})}{1-e^{i\omega_j\Delta t}q^{-1}}. \tag{3-30}$$

Based on this idea and using Eq. (3-25), Wolf and Motosoka [73] suggested approximating the regular part of the impedance function in the frequency domain as a ratio of two polynomials as

$$\bar{S}(a_0)\cong\bar{S}(ia_0)=\frac{P(ia_0)}{Q(ia_0)}=K\frac{1+p_1(ia_0)+\cdots+p_{M-1}(ia_0)^{M-1}}{1+q_1(ia_0)+\cdots+q_M(ia_0)^M}, \tag{3-31}$$

where $K$ is the static stiffness, $p_j$ and $q_j$ are real-valued coefficients, and $a_0$ is the dimensionless frequency. To obtain these $2M-1$ unknown coefficients, the least-squares approach can be employed to minimize the following objective function calculated at $N$ frequencies

$$\varepsilon=\sum_{k=1}^{N}\|\bar{S}(a_{0k})Q(ia_{0k})-P(ia_{0k})\|^2, \tag{3-32}$$

where $\|.\|$ is 2-norm. To minimize $\varepsilon$, we have

$$\frac{\partial\varepsilon}{\partial p_j}=0\ \ for\, j=1,\dots,M-1, \tag{3-33}$$

$$\frac{\partial\varepsilon}{\partial q_j}=0\ \ for\, j=1,\dots,M, \tag{3-34}$$

which results in a system of $2M-1$ linear equations with real coefficients for the $2M-1$ unknown $p_j$ and $q_j$.

As an alternative approach, the ratio in Eq. (3-31) can be expressed in a partial-fraction expression as

$$\bar{S}(ia_0)=\sum_{j=1}^{M}\frac{A_j}{ia_0-s_j}, \tag{3-35}$$

where $s_j$s are poles of $\bar{S}(ia_0)$ (roots of $Q(ia_0)$) and $A_j$ are residues at the poles, i.e., $A_j=(ia_0-s_j)\bar{S}(ia_0)|_{ia_0=s_j}$ . Eq. (3-35) can be converted to the time domain using continuous-time Fourier Transform as

$$s(\hat{t}) = \sum_{j=1}^{M} A_j e^{s_j \hat{t}}, \tag{3-36}$$

where $\hat{t}$ is the dimensionless time. $s(\hat{t})$ can be discretized in the middle of the time intervals as

$$s[n] = \sum_{j=1}^{M} A_j e^{s_j \hat{t}\frac{2n+1}{2}\Delta\hat{t}}, \tag{3-37}$$

where $\Delta\hat{t}$ is the size of the time intervals. By applying right-hand Z-Transformation, we have

$$S(z) = \sum_{j=1}^{M}\sum_{n=0}^{\infty} s[n]z^{-n} = \sum_{j=1}^{M} \frac{A_j e^{\frac{s_j\Delta\hat{t}}{2}}}{1 - e^{s_j\Delta\hat{t}}z^{-1}}. \tag{3-38}$$

To have a stable $S(z)$, poles must be within the unit circle, i.e., $\left|e^{s_j\Delta\hat{t}}\right| < 1$. The equation above can be expressed as

$$S(z) = \frac{b_0 + b_1 z^{-1} + \cdots + b_{M-1} z^{-(M-1)}}{1 - a_1 z^{-1} - \cdots - a_M z^{-M}}. \tag{3-39}$$

Considering $R(z) = S(z)U(z)$, we have

$$R[n] = \sum_{j=1}^{M} a_j R[n-j] + \sum_{j=0}^{M-1} b_j u[n-j]. \tag{3-40}$$

To obtain $a_j$ and $b_j$ coefficients, a system of linear equations can be constructed similar to Eq. (3-32). The Matlab function “invfreqz” in its basic setting carries out the such operation using the complex-curve fitting approach [76]. As this IIR representation of the impedance function is prone to instability, the Matlab function “invfreqz” provides an algorithm that guarantees the stability of the resulting linear system by searching for the best fit using a numerical, iterative scheme [77].

To show the application of this approach, consider the previous example. Comparison between exact and impedance function estimated through the IIR representation is shown in Figure 3-19. In this example, the Matlab function “invfreqz” is used to estimate $S(z)$ introduced in Eq. (3-39) with $M$ =8 and by using 58 frequency points uniformly distributed from zero to $a_0 = 3$. As seen, the IIR filter is able to very accurately represent the regular part of the impedance function of the rod on an elastic foundation for $a_0 < 3$. As “invfreqz” with the iterative scheme is used, the estimated filter must be stable. To check the stability, the poles of the estimated filter are shown in Figure 3-20. As seen, all poles are within the unit circle confirming the stability of the estimated filter. Finally, the reaction force under pulse displacement is calculated using Eq. (3-40) and is compared with the exact counterpart in Figure 3-21. As seen, the estimated force is almost identical to the exact force.

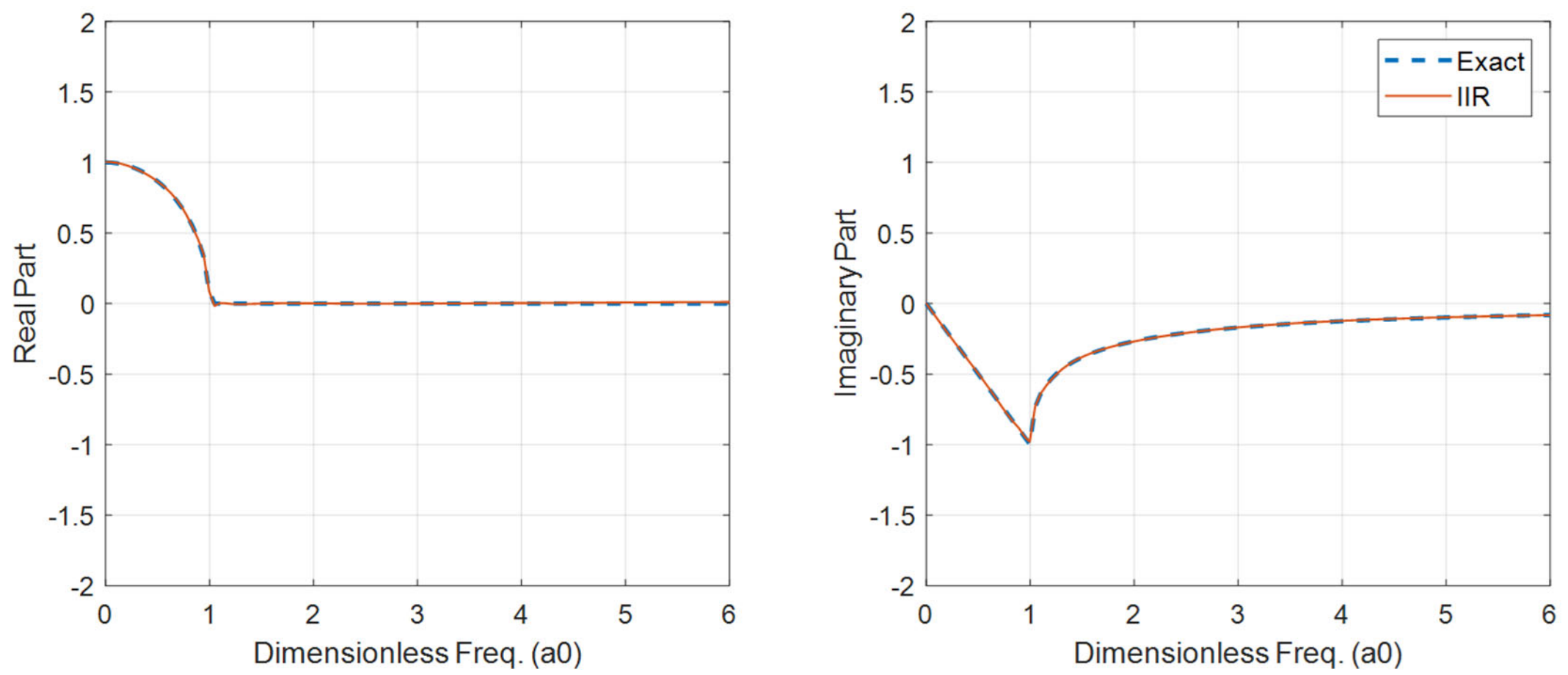

**Figure 3-19: Comparison between exact and reconstructed impedance function.**

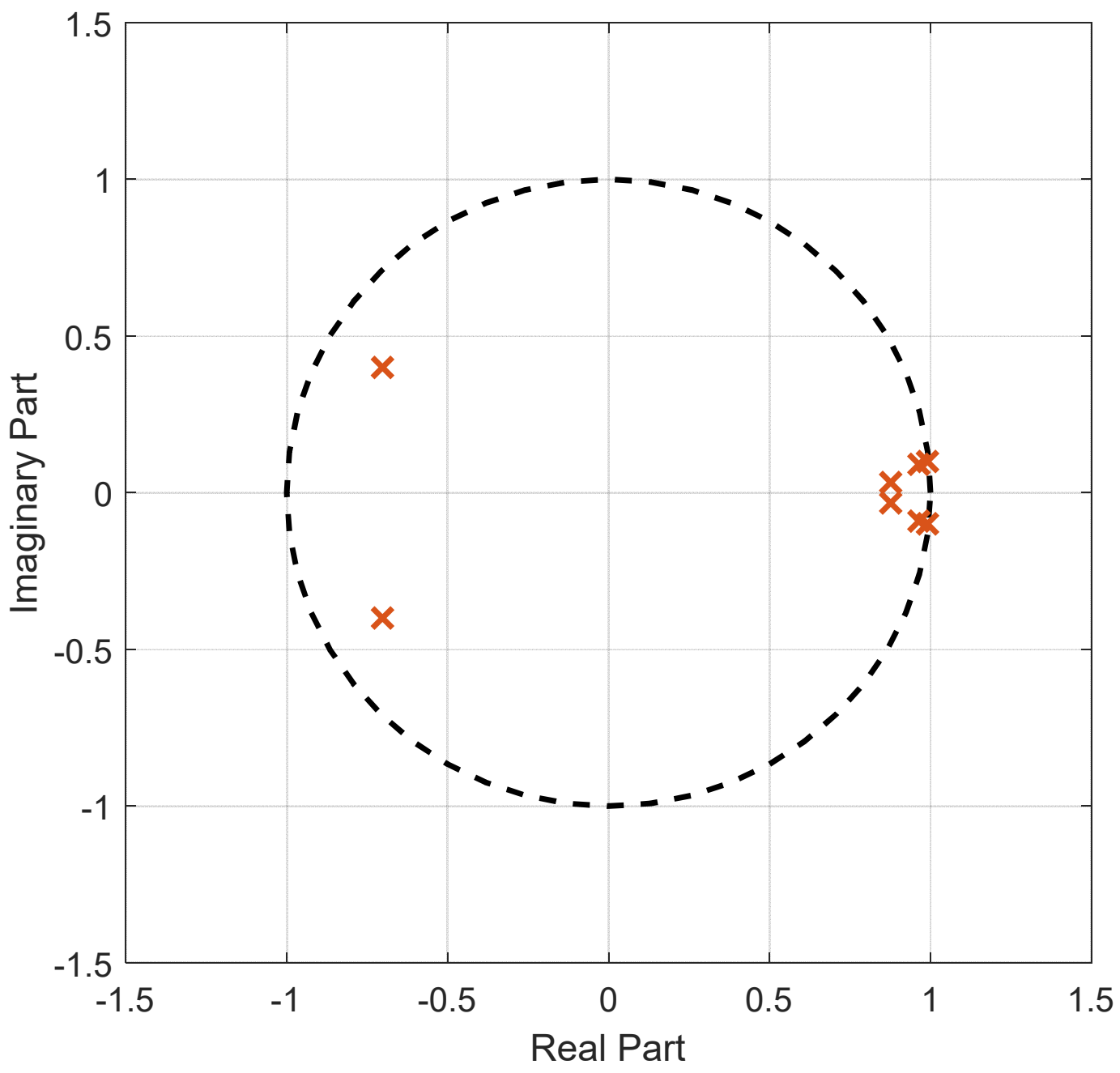

**Figure 3-20: Poles of the estimated IIR filter.**

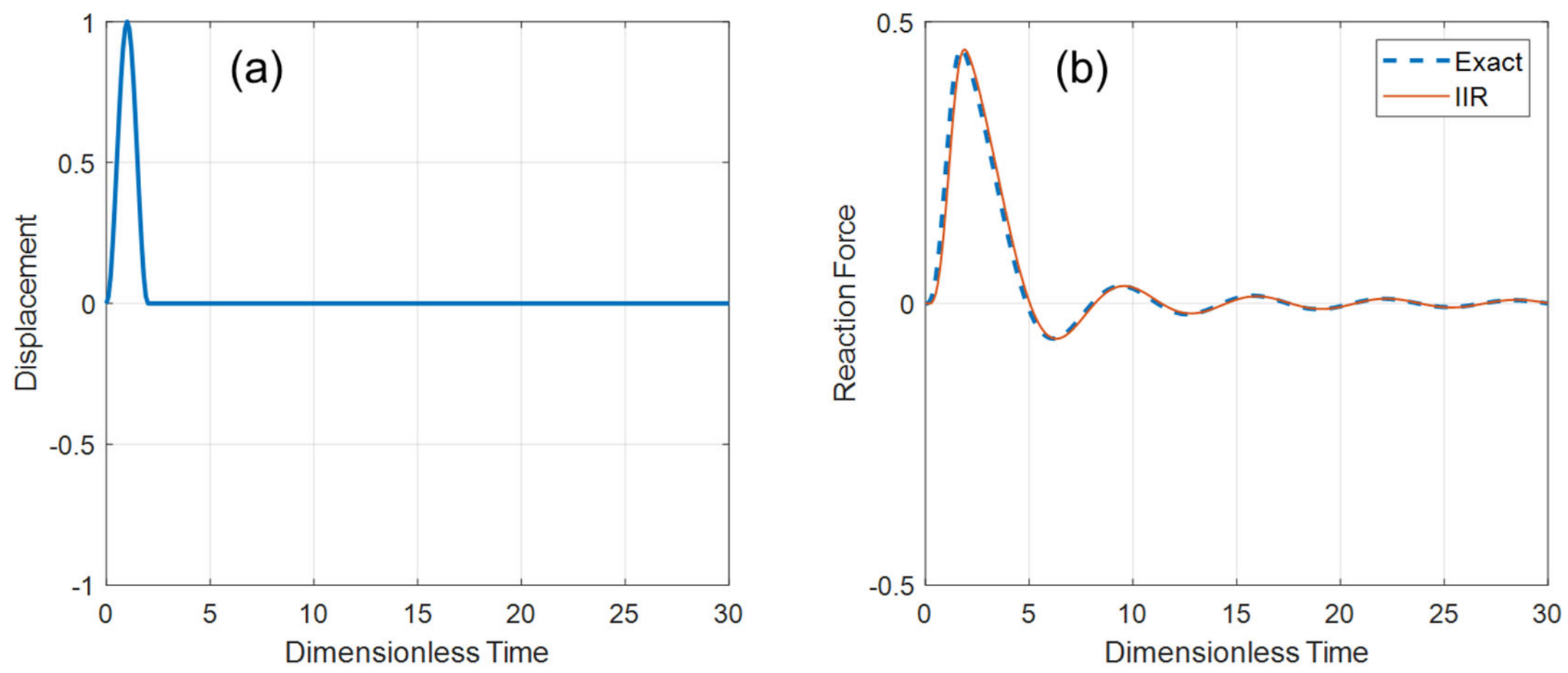

**Figure 3-21: (a) The imposed displacement, and (b) the exact and calculated reaction forces.**

Based on Wolf's and Motosaka's works which were reviewed above, Safak [78] suggested the application of the IIR filters to represent frequency-dependent impedance function in the time domain without limiting its application to the regular part. In his formulation, the reaction force can be written as

$$R[n] = \sum_{p=0}^{n_b} b_p u[n-p] - \sum_{p=1}^{n_a} a_p R[n-p], \tag{3-41}$$

which is identical to Eq. (3-40) with minor differences in sign convention. By calculating the Discrete-Time Fourier Transform of both sides, we have

$$\bar{R}(\omega) = \sum_{p=0}^{n_b} b_p\, e^{-i\omega p}\, \bar{U}(\omega) - \sum_{p=1}^{n_a} a_p\, e^{-i\omega p} \bar{R}(\omega) \qquad \omega \in [-\pi,\, \pi), \tag{3-42}$$

and the impedance function can be defined as

$$\bar{S}(\omega) = \frac{\sum_{p=0}^{n_b} b_p\, e^{-i\omega p}}{1 + \sum_{p=1}^{n_a} a_p\, e^{-i\omega p}}. \tag{3-43}$$

So, by solving a minimization problem like what is shown before (Eq. (3-32)), coefficients $b_p$ and $a_p$ can be estimated. In other words, we would like to minimize

$$\varepsilon(\omega) = \bar{S}(\omega) - \frac{N(\omega)}{D(\omega)}, \tag{3-44}$$

where $\bar{S}(\omega)$ is the impedance function and $\frac{N(\omega)}{D(\omega)}$ is the IIR representation that needs to be adjusted. While the Matlab function "invfreqz" can be used for this purpose, herein we derive the formulation to carry out the minimization by solving a system of linear equations.

By multiplying both sides of the equation above by $D(\omega)$, we get the following objective function that needs to be minimized over $N$ discrete frequencies $\omega_k$

$$E=\sum_{k=1}^{N}|D(\omega_k)\varepsilon(\omega_k)|^2=\sum_{k=1}^{N}|D(\omega_k)\bar{S}(\omega_k)-N(\omega_k)|^2, \tag{3-45}$$

where the weight of the peaks in the impedance function is reduced. To solve this minimization, $N(\omega_k)$ and $D(\omega_k)$ can be expressed as

$$N(\omega_k)=\sum_{p=0}^{n_b}b_p\,e^{-i\omega_k p}=\sum_{p=0}^{n_b}b_p\,(\cos p\omega_k-i\sin p\omega_k)=\alpha_k+i\beta_k, \tag{3-46}$$

$$D(\omega_k)=\sum_{p=0}^{n_a}a_p\,e^{-i\omega_k p}=\sum_{p=0}^{n_a}a_p\,(cos\,p\omega_k-i\,sin\,p\omega_k)=\sigma_k+i\tau_k, \tag{3-47}$$

where

$$\alpha_k=\sum_{p=0}^{n_b}b_p\,cos\,p\omega_k\quad\text{and}\quad\beta_k=\sum_{p=0}^{n_b}-b_p\,sin\,p\omega_k, \tag{3-48}$$

$$D\sigma_k=\sum_{p=0}^{n_a}a_p\,cos\,p\omega_k\quad\text{and}\quad\tau_k=\sum_{p=0}^{n_a}-a_p\,sin\,p\omega_k\,. \tag{3-49}$$

So, the objective function can be rewritten as

$$E=\sum_{k=1}^{N}E(\omega_k), \tag{3-50}$$

with

$$E(\omega_k)=(\sigma_kR_k-\tau_kI_k-\alpha_k)^2+(\sigma_kI_k+\tau_kR_k-\beta_k)^2, \tag{3-51}$$

where the reference impedance function is presented as

$$\bar{S}(\omega_k)=R_k+iI_k. \tag{3-52}$$

To minimize the objective function, we have

$$\sum_{k=1}^{N}\frac{\partial E(\omega_k)}{\partial b_p}=0\ \text{ for }\ p=0,\dots,n_b, \tag{3-53}$$

$$\sum_{k=1}^{N}\frac{\partial E(\omega_k)}{\partial a_p}=0,\ \text{ for }\ p=1,\dots,n_a\,. \tag{3-54}$$

These derivatives can be computed as

$$\frac{\partial E(\omega_k)}{\partial b_p} = -2\cos p\omega_k\,(\sigma_k R_k - \tau_k I_k - \alpha_k) + 2\sin p\omega_k\,(\sigma_k I_k + \tau_k R_k - \beta_k), \tag{3-55}$$

$$\frac{\partial E(\omega_k)}{\partial a_p} = 2(R_k\cos p\omega_k + I_k\sin p\omega_k)(\sigma_k R_k - \tau_k I_k - \alpha_k) + 2(I_k\cos p\omega_k - R_k\sin p\omega_k)(\sigma_k I_k + \tau_k R_k - \beta_k). \tag{3-56}$$

Using these derivates, we can solve the minimization and obtain $n_a + n_b + 1$unknown coefficients through solving the following system of linear equations

$$\mathbf{Ax} = \mathbf{c}, \tag{3-57}$$

where

$$\mathbf{A} = \sum_{k=1}^{N}\begin{bmatrix}\mathbf{A}_{bb}^k & \mathbf{A}_{ba}^k \\ \mathbf{A}_{ab}^k & \mathbf{A}_{aa}^k\end{bmatrix}, \tag{3-58}$$

$$\mathbf{x} = \begin{bmatrix}\mathbf{b} \\ \mathbf{a}\end{bmatrix}, \tag{3-59}$$

$$\mathbf{c} = \sum_{k=1}^{N}\begin{bmatrix}\mathbf{c}_b^k \\ \mathbf{c}_a^k\end{bmatrix}, \tag{3-60}$$

in which

$$\mathbf{A}_{bb}^k = \begin{bmatrix} -c_{0k}(-c_{0k}) + s_{0k}(s_{0k}) & -c_{0k}\,(-c_{1k}) + s_{0k}(s_{1k}) & \cdots & -c_{0k}(-c_{n_bk}) + s_{0k}(s_{n_bk}) \\ -c_{1k}(-c_{0k}) + s_{1k}(s_{0k}) & -c_{1k}\,(-c_{1k}) + s_{1k}(s_{1k}) & \cdots & -c_{1k}(-c_{n_bk}) + s_{1k}(s_{n_bk}) \\ \vdots & \vdots & \ddots & \vdots \\ -c_{n_bk}(-c_{0k}) + s_{n_bk}(s_{0k}) & -c_{n_bk}\,(-c_{1k}) + s_{n_bk}(s_{1k}) & \cdots & -c_{n_bk}(-c_{n_bk}) + s_{n_bk}(s_{n_bk}) \end{bmatrix}, \tag{3-61}$$

$$\mathbf{A}_{ba}^k = \begin{bmatrix} -c_{0k}(R_k c_{1k} + I_k s_{1k}) + s_{0k}(I_k c_{1k} - R_k s_{1k}) & \cdots & -c_{0k}(R_k c_{n_ak} + I_k s_{n_ak}) + s_{0k}(I_k c_{n_ak} - R_k s_{n_ak}) \\ -c_{1k}(R_k c_{1k} + I_k s_{1k}) + s_{1k}(I c_{1k} - R_k s_{1k}) & \cdots & -c_{1k}(R_k c_{n_ak} + I_k s_{n_ak}) + s_{1k}(I_k c_{n_ak} - R_k s_{n_ak}) \\ \vdots & \ddots & \vdots \\ -c_{n_bk}(R_k c_{1k} + I_k s_{1k}) + s_{n_bk}(I_k c_{1k} - R_k s_{1k}) & \cdots & -c_{n_bk}(R_k c_{n_ak} + I_k s_{n_ak}) + s_{n_bk}(I_k c_{n_ak} - R_k s_{n_ak}) \end{bmatrix}, \tag{3-62}$$

$$\mathbf{A}_{ab}^k = \begin{bmatrix} (R_k c_{1k} + I_k s_{1k})(-c_{0k}) + (I_k c_{1k} - R_k s_{1k})(s_{0k}) & (R_k c_{1k} + I_k s_{1k})\,(-c_{1k}) + (I_k c_{1k} - R_k s_{1k})(s_{1k}) & \cdots & (R_k c_{1k} + I_k s_{1k})(-c_{n_bk}) + (I_k c_{1k} - R_k s_{1k})(s_{n_bk}) \\ \vdots & \vdots & \ddots & \vdots \\ (R_k c_{n_bk} + I_k s_{n_bk})(-c_{0k}) + (I_k c_{n_bk} - R_k s_{n_bk})(s_{0k}) & (R_k c_{n_bk} + I_k s_{n_bk})\,(-c_{1k}) + (I_k c_{n_bk} - R_k s_{n_bk})(s_{1k}) & \cdots & (R_k c_{n_bk} + I_k s_{n_bk})(-c_{n_bk}) + (I_k c_{n_bk} - R_k s_{n_bk})(s_{n_bk}) \end{bmatrix}, \tag{3-63}$$

$$\mathbf{A}_{aa}^k = \begin{bmatrix} (R_k c_{1k} + I_k s_{1k})(R_k c_{1k} + I_k s_{1k}) + (I_k c_{1k} - R_k s_{1k})(I_k c_{1k} - R_k s_{1k}) & \cdots & (R_k c_{1k} + I_k s_{1k})(R_k c_{n_ak} + I_k s_{n_ak}) + (I_k c_{1k} - R_k s_{1k})(I_k c_{n_ak} - R_k s_{n_ak}) \\ \vdots & \ddots & \vdots \\ (R_k c_{n_ak} + I_k s_{n_ak})(R_k c_{1k} + I_k s_{1k}) + (I_k c_{n_ak} - R_k s_{n_ak})(I_k c_{1k} - R_k s_{1k}) & \cdots & (R_k c_{n_ak} + I_k s_{n_ak})(R_k c_{n_ak} + I_k s_{n_ak}) + (I_k c_{n_ak} - R_k s_{n_ak})(I_k c_{n_ak} - R_k s_{n_ak}) \end{bmatrix}, \tag{3-64}$$

$$\mathbf{c}_b^k = \begin{bmatrix} c_{0k}(R_k c_{0k} + I_k s_{0k}) - s_{0k}(I_k c_{0k} - R_k s_{0k}) \\ c_{1k}(R_k c_{0k} + I_k s_{0k}) - s_{1k}(I_k c_{0k} - R_k s_{0k}) \\ \vdots \\ c_{n_bk}(R_k c_{0k} + I_k s_{0k}) - s_{n_bk}(I_k c_{0k} - R_k s_{0k}) \end{bmatrix}, \tag{3-65}$$

$$\mathbf{c}_a^k = \begin{bmatrix} -(R_k c_{1k} + I_k s_{1k})(R_k c_{0k} + I_k s_{0k}) - (I_k c_{1k} - R_k s_{1k})(I_k c_{0k} - R_k s_{0k}) \\ \vdots \\ -\left(R_k c_{n_a k} + I_k s_{n_a k}\right)(R_k c_{0k} + I_k s_{0k}) - \left(I_k c_{n_a k} - R_k s_{n_a k}\right)(I_k c_{0k} - R_k s_{0k}) \end{bmatrix}, \tag{3-66}$$

$$\mathbf{b} = \begin{bmatrix} b_0 \\ b_1 \\ \vdots \\ b_{n_b} \end{bmatrix}, \tag{3-67}$$

$$\mathbf{a} = \begin{bmatrix} a_1 \\ \vdots \\ a_{n_a} \end{bmatrix}. \tag{3-68}$$

In the equations above, $c_{pk} = \cos p\omega_k$ and $s_{pk} = \sin p\omega_k$.

To verify the implemented filter estimation method, an IIR filter whose discrete-time version is analytically available is used. Eq. (3-69) shows the Transfer Function (TF) of the absolute acceleration response of an SDOF system under base acceleration in which $\omega_n$ and $\xi$ are natural frequency and damping ratio, respectively.

$$TF(s) = \frac{2\xi\omega_n s + \omega_n^2}{s^2 + 2\xi\omega_n s + \omega_n^2}. \tag{3-69}$$

To obtain the analytical discrete-time representation of this function, bilinear or so-called Tustin mapping is used to preserve the stability of the TF from the Laplace domain to Z-domain (see [79] for more details on various types of transformation). The transformed TF can be expressed as

$$TF(z^{-1}) = \frac{b_0 + b_1 z^{-1} + b_2 z^{-2}}{1 + a_1 z^{-1} + a_2 z^{-2}}, \tag{3-70}$$

where

$$b_0 = \frac{4\xi\omega_n\Delta t + \omega_n^2\Delta t^2}{4 + 4\xi\omega_n\Delta t + \omega_n^2\Delta t^2}, \tag{3-71}$$

$$b_1 = \frac{2\omega_n^2\Delta t^2}{4 + 4\xi\omega_n\Delta t + \omega_n^2\Delta t^2}, \tag{3-72}$$

$$b_2 = \frac{-4\xi\omega_n\Delta t + \omega_n^2\Delta t^2}{4 + 4\xi\omega_n\Delta t + \omega_n^2\Delta t^2}, \tag{3-73}$$

$$a_1 = \frac{2\omega_n^2\Delta t^2 - 8}{4 + 4\xi\omega_n\Delta t + \omega_n^2\Delta t^2}, \tag{3-74}$$

$$a_2 = \frac{4 - 4\xi\omega_n\Delta t + \omega_n^2\Delta t^2}{4 + 4\xi\omega_n\Delta t + \omega_n^2\Delta t^2}. \tag{3-75}$$

Figure 3-22 shows the zeros and poles of these two systems for a case with $\omega_n = 2\pi 5 \frac{rad}{sec}$ and $\xi = 5\%$. As seen, the bilinear transformation preserves the stability of the system, and all poles of the continuous-time

system which are on the left half of the coordinate plane in Figure 3-22a are within the unit circle in Figure 3-22b.

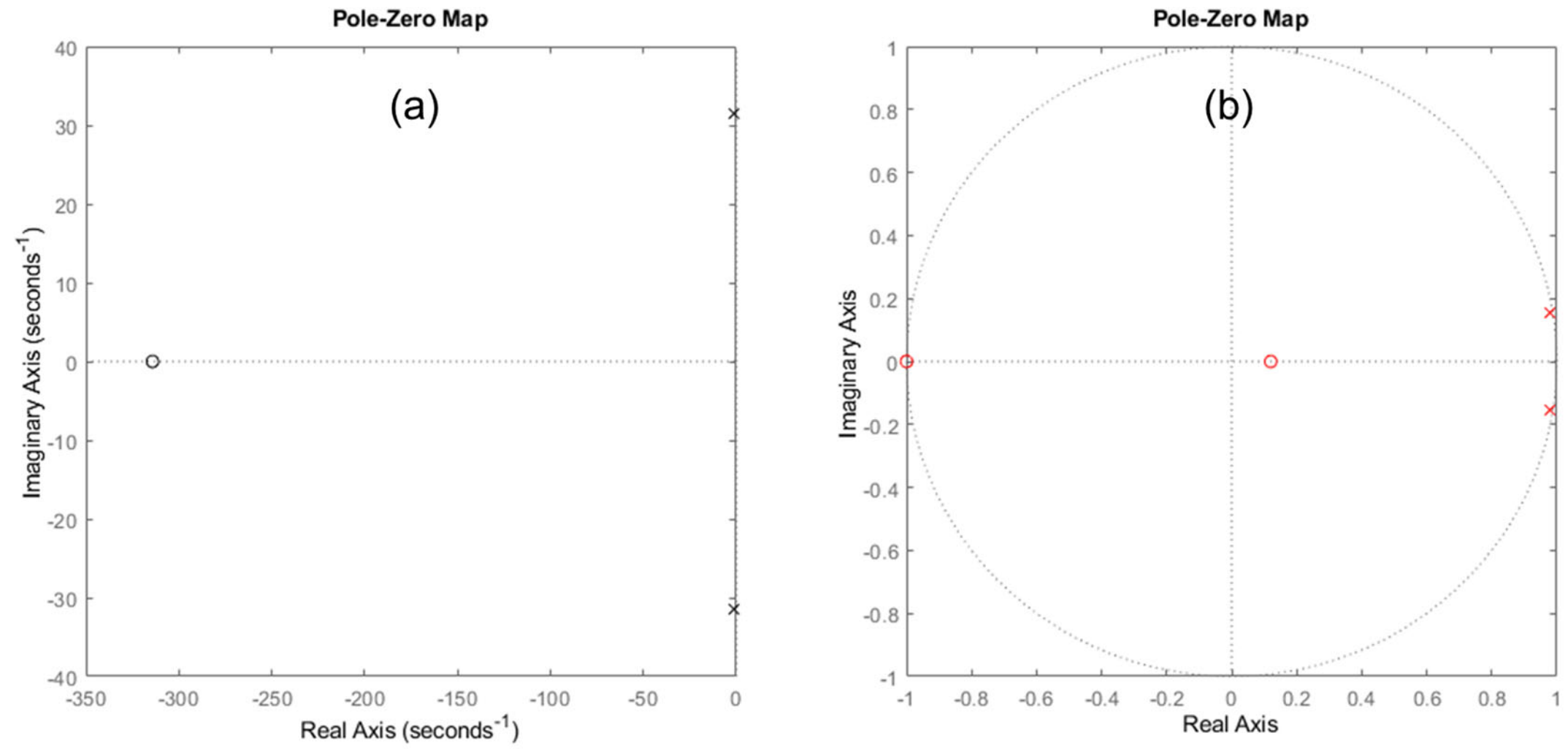

**Figure 3-22: Poles (x) and zeros (o) of the (a) continuous-time and (b) discrete-time systems.**

Figure 3-23 shows the magnitude of the Frequency Response Function (FRF) of various representations of the system. The gray curve is the analytical continuous-time domain (3-69). Matlab has built-in Continuous-to-Discrete Tustin transformation, and the blue curve shows the FRF of the discrete-time system obtained using this Matlab function considering a sampling frequency of 200 Hz. As expected, because an unlimited bandwidth is wrapped into a limited frequency, a slight distortion at high frequencies is observed. The discrete-time system shown in Eq. (3-70) is shown in red, which verifies the accuracy of the analytical mapping. Using 200 frequency points selected between zero to 50 Hz from the discrete-time FRF, the minimization approach presented above is employed to obtain an equivalent IIR filter. The FRF of this system is shown in green, and as seen, it is identical to the reference systems.

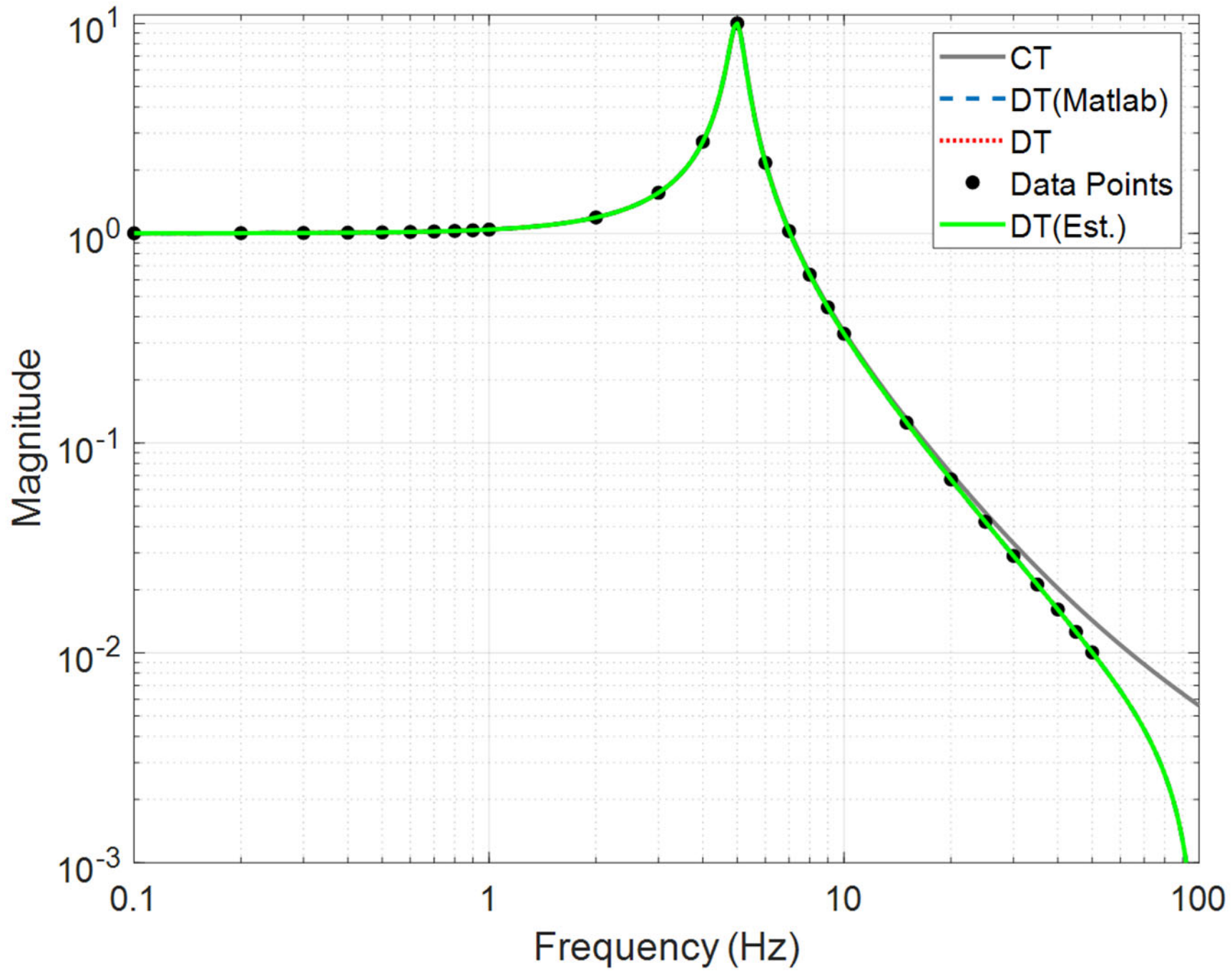

**Figure 3-23: Comparison between analytical and estimated IIR filters.**

The Matlab function that carried out the filter estimation using the real-valued least-squares formulation is available upon request from the first author. It must be emphasized that this approach does not guarantee the stability of the estimated filter. Gash [80] extensively studied the stability of the filter as well as the stability of the soil-structure system. He implemented complex-valued least-squares to solve the filter estimation. In this approach, the stability is enforced after the design by mirroring the poles of the estimated filter into the unit circle, similar to the approach used in Matlab built-in "invfreqz" filter. While the stability of the filter can be addressed through this approach or through constrained minimization (see, e.g., [81]), the stability of the entire soil-structure system is a very challenging task. Gash [80] developed criteria to check the stability of the soil-structure system for a simple SDOF superstructure, which is not enough for real-life application. Based on this fact, the application of the IIR filter representation of the impedance function is limited and will not be pursued in this project.

### 3.4. ITERATIVE TIME-FREQUENCY SOLUTIONS

Methods presented in the previous section were designed to represent frequency-dependent impedance functions in the time domain. A different approach to extend the substructure method to the time domain is to try to decompose the problem into frequency-dependent and nonlinear parts and solve them in frequency or time and check and correct in the other domain. In the following, two types of these approaches will be presented.

### 3.4.1. Hybrid Frequency Time Domain (HFTD) Method

Hybrid Frequency-Time Domain (HFTD) method is the method originally developed to solve nonlinear problems in the frequency domain to reduce computational costs, which was a crucial issue some decades ago [82]. However, because the method works in the frequency domain, the frequency-dependency can be taken into account very easily, as long as it is decoupled from the nonlinear part of the system [83]–[86]. The idea behind this method is to use an equivalent linear-elastic system, which can have a frequency-dependent part, and compensate for the error in restoring nonlinear force by using a time-varying pseudo-force which is estimated iteratively. This method is briefly reviewed in the following, but it is mostly suitable for problems dominant by frequency-dependent parts with very limited nonlinearity that is quite opposite the soil-structure problems. In the next section, a different iterative method will be presented which is more suitable for soil-structure interaction problems and will be extensively studied, verified, and implemented in this project.

The governing equation of motion of a soil-structure system in which frequency-dependency and nonlinearity are limited to the soil-foundation and superstructure, respectively, will be discussed in detail in the next chapter when the HTFD method will be reviewed. This equation is shown below

$$\begin{bmatrix} \mathbf{M}_{ss} & \mathbf{M}_{sb} \\ \mathbf{M}_{bs} & \mathbf{M}_{bb} \end{bmatrix} \begin{bmatrix} \ddot{\mathbf{u}}_s^t \\ \ddot{\mathbf{u}}_b^t \end{bmatrix} + \begin{bmatrix} \mathbf{p}_s \\ \mathbf{p}_b \end{bmatrix} + \begin{bmatrix} \mathbf{0} \\ \mathbf{r}_b \end{bmatrix} = \begin{bmatrix} \mathbf{0} \\ \mathbf{q}_b \end{bmatrix}, \tag{3-76}$$

with

$$\mathbf{r}_b = \mathbf{S}_{bb}^g * \mathbf{u}_b^t, \tag{3-77}$$

$$\mathbf{q}_b = \mathbf{S}_{bb}^f * \mathbf{u}_b^f. \tag{3-78}$$

in the equations above, $\mathbf{u}_b^t$ and $\mathbf{u}_s^t$ are displacement of the nodes at the soil-structure boundary and within the structure, respectively. $\mathbf{p}_s$ is the nonlinear force applied to the structural nodes (strain and damping), $\mathbf{p}_b$ is the part of the forces applied to the "b" nodes from the superstructure (strain and damping), which can be nonlinear too. $\mathbf{q}_b$ is the external load imposed to "b" nodes to Free-Field Motion (FFM) at "b" nodes, $\mathbf{u}_b^f$, where $\mathbf{S}_{bb}^f$ is the stiffness of the free-field domain. $\mathbf{r}_b$ represents the interaction force although it is not exactly the interaction force because it is calculated using the total displacement of the soil-foundation nodes $\mathbf{u}_b^t$ and stiffness of the excavated domain $\mathbf{S}_{bb}^g$. As seen, due to the frequency-dependency of the soil-foundation and free-field impedance function, $\mathbf{r}_b$ and $\mathbf{q}_b$ are calculated through time convolution. It is possible to solve this equation in the frequency domain provided that all parts remain linear-elastic. So, we can replace the nonlinear part with an equivalent linear-elastic and compensate for the error in the restoring force by adding a pseudo-force. In other words, we can rewrite Eq. (3-76) as

$$\begin{bmatrix} \mathbf{M}_{ss} & \mathbf{M}_{sb} \\ \mathbf{M}_{bs} & \mathbf{M}_{bb} \end{bmatrix} \begin{bmatrix} \ddot{\mathbf{u}}_s^t \\ \ddot{\mathbf{u}}_b^t \end{bmatrix} + \begin{bmatrix} \tilde{\mathbf{C}}_{ss} & \tilde{\mathbf{C}}_{sb} \\ \tilde{\mathbf{C}}_{bs} & \mathbf{0} \end{bmatrix} \begin{bmatrix} \dot{\mathbf{u}}_s^t \\ \dot{\mathbf{u}}_b^t \end{bmatrix} + \begin{bmatrix} \tilde{\mathbf{K}}_{ss} & \tilde{\mathbf{K}}_{sb} \\ \tilde{\mathbf{K}}_{bs} & \mathbf{0} \end{bmatrix} \begin{bmatrix} \mathbf{u}_s^t \\ \mathbf{u}_b^t \end{bmatrix} + \begin{bmatrix} \mathbf{0} \\ \mathbf{r}_b \end{bmatrix} = \begin{bmatrix} \mathbf{0} \\ \mathbf{q}_b \end{bmatrix} + \mathbf{f}_{ps}, \tag{3-79}$$

with

$$\mathbf{f}_{ps} = \begin{bmatrix} \tilde{\mathbf{C}}_{ss} & \tilde{\mathbf{C}}_{sb} \\ \tilde{\mathbf{C}}_{bs} & \mathbf{0} \end{bmatrix} \begin{bmatrix} \dot{\mathbf{u}}_s^t \\ \dot{\mathbf{u}}_b^t \end{bmatrix} + \begin{bmatrix} \tilde{\mathbf{K}}_{ss} & \tilde{\mathbf{K}}_{sb} \\ \tilde{\mathbf{K}}_{bs} & \mathbf{0} \end{bmatrix} \begin{bmatrix} \mathbf{u}_s^t \\ \mathbf{u}_b^t \end{bmatrix} - \begin{bmatrix} \mathbf{p}_s \\ \mathbf{p}_b \end{bmatrix}, \tag{3-80}$$

where $\mathbf{f}_{ps}$ is the pseudo-force vector and the tilde shows the properties of the assumed equivalent linear-elastic system. Eq. (3-80) can be converted to the frequency domain as

$$\left\{ -\omega^2 \begin{bmatrix} \mathbf{M}_{ss} & \mathbf{M}_{sb} \\ \mathbf{M}_{bs} & \mathbf{M}_{bb} \end{bmatrix} + i\omega \begin{bmatrix} \tilde{\mathbf{C}}_{ss} & \tilde{\mathbf{C}}_{sb} \\ \tilde{\mathbf{C}}_{bs} & \mathbf{0} \end{bmatrix} + \begin{bmatrix} \tilde{\mathbf{K}}_{ss} & \tilde{\mathbf{K}}_{sb} \\ \tilde{\mathbf{K}}_{bs} & \bar{\mathbf{S}}_{bb}^g(\omega) \end{bmatrix} \right\} \begin{bmatrix} \bar{\mathbf{u}}_s^t(\omega) \\ \bar{\mathbf{u}}_b^t(\omega) \end{bmatrix} = \begin{bmatrix} \mathbf{0} & \mathbf{0} \\ \mathbf{0} & \bar{\mathbf{S}}_{bb}^f(\omega) \end{bmatrix} \begin{bmatrix} \mathbf{0} \\ \bar{\mathbf{u}}_b^f(\omega) \end{bmatrix} + \bar{\mathbf{f}}_{ps}(\omega), \tag{3-81}$$

where the overbar shows the frequency-domain representation. This equation can be easily solved in the frequency domain considering the frequency-dependent soil-foundation impedance function provided that the pseudo-force is known. However, to have $\mathbf{f}_{ps}$, the state of the system must be completely known. Therefore, the problem is solved iteratively. Psuedo-force time-history is assumed (e.g., zero time history at the first iteration) and the response of the system is obtained in the frequency domain. Then, $\mathbf{p}_s$ and $\mathbf{p}_b$ are calculated using the actual nonlinear behavior of the system and $\mathbf{f}_{ps}$ is updated through Eq. (3-80). If the difference between the new pseudo-force and the assumed one is greater than the acceptable threshold, this new pseudo-force is used and the analysis is repeated. This process continues until convergence is achieved.

#### 3.4.1.1 Convergence Criteria

The stability of the HFTD method has been studied by Darbre and Wolf [86]. To have stability under any level of nonlinearity, a necessary condition is to have convergence at every step because this method converges progressively. So, working with one time-step corresponds to an initial-value problem for which we can write the following equation

$$f(0) = \lim_{i\omega \to \infty} F(\omega). \tag{3-82}$$

In the discrete-time domain, the maximum frequency is the Nyquist frequency, so to evaluate the Fourier Transform at infinite $i\omega$, we must evaluate it at the $\Omega = \frac{\pi N}{T}$, which can be calculated as shown below.

$$F(\omega)|_{i\omega=\Omega} = \Delta t f(0) + \Delta t \sum_{n=1}^{N-1} f(n\Delta t)\, e^{-n\pi}. \tag{3-83}$$

Assuming that function $f(t)$ does not increase with time, the summation is small and we can approximately write the following equation.

$$f(0) \approx \frac{F(\omega = -i\Omega)}{\Delta t}. \tag{3-84}$$

For a moment, let's assume that the input excitation is harmonic. The exact displacement response can be calculated using the dynamic stiffness of the system as follows

$$\bar{\mathbf{u}}(\omega) = \bar{\mathbf{S}}(\omega)^{-1} \bar{\mathbf{p}}(\omega). \tag{3-85}$$

In the HFTD method, we solve a linear-elastic system with stiffness $\tilde{\mathbf{S}}_0$ under the excitation and an extra pseudo-force which is the difference between the exact internal force and the linear-elastic force. So, we can write this equation at $j-th$ iteration.

$$\bar{\mathbf{u}}(\omega)_j = \bar{\mathbf{S}}_0(\omega)^{-1}\{\bar{\mathbf{p}}(\omega)\text{-}[\bar{\mathbf{S}}(\omega) - \bar{\mathbf{S}}_0(\omega)]\bar{\mathbf{u}}(\omega)_{j-1}\}. \tag{3-86}$$

We repeat this recursive equation down to the first iteration where the pseudo-force is zero, we get this equation

$$\bar{\mathbf{u}}(\omega)_j = \bar{\mathbf{S}}(\omega)^{-1}\bar{\mathbf{S}}_0(\omega)\left[\mathbf{I}\text{-}\bar{\mathbf{A}}(\omega)^{j+1}\right]\bar{\mathbf{S}}_0(\omega)^{-1}\bar{\mathbf{p}}(\omega), \tag{3-87}$$

where

$$\bar{\mathbf{A}}(\omega) = \mathbf{I} - \bar{\mathbf{S}}_0(\omega)^{-1}\bar{\mathbf{S}}(\omega). \tag{3-88}$$

So, the limit of $\bar{\mathbf{u}}(\omega)$ after infinite numbers of iterations can converge to the exact response provided that the

$$\lim_{j\to\infty} \bar{\mathbf{A}}(\omega)^{j+1} = 0. \tag{3-89}$$

which means the spectral radius of the matrix $\bar{\mathbf{A}}(\omega)$ must be zero. Based on the initial-value theorem shown in Eqs. (3-82) to (3-84), we can extend the stability criteria obtained for harmonic excitation to transient excitation but only for the Nyquist frequency as shown here.

$$\rho(\omega = -i\Omega) < 1. \tag{3-90}$$

where $\rho(\omega) = \max_i|\lambda_i(\omega)|$ in which $\lambda_i(\omega)$ are eigenvalues of the matrix $\bar{\mathbf{A}}(\omega)$. Darbre and Wolf [86] showed in their paper that this condition is even sufficient to ensure the stability of the HFTD procedure. As the exact dynamic stiffness of the system changes in time and the method converges in a time-progressive manner (convergence at any time is reached only after the solution has converged at all previous times), the solution must be carried out in small time segments. A very short time segment increases convergence chance but it also increases computational cost, so there is a balance between convergence and computational cost.

#### 3.4.1.2 Verification Example

To verify the HFTD method, the response of a nonlinear SDOF system on a sway-rocking soil-foundation substructure with frequency-dependent rocking impedance function is calculated under 1940 El Centro ground motion. The details of this example can be found in the next chapter where it will be used to verify the HTFD method. To obtain the ground truth response, an equivalent lumped-parameter model is used. A window size of 5 seconds is used to run this example. Figure 3-24 shows the comparison between the true response and the response obtained using the HFTD method. In the HFTD method, elastic properties of the superstructure are used for the equivalent linear-elastic model. As seen, the method is able to accurately calculate the response of the system taking into account both the frequency-dependency of the soil-foundation impedance function and the nonlinearity of the superstructure. A Matlab code to carry out the HFTD method in sequential time windows and a video showing the progress of the analysis are available upon request from the first author.

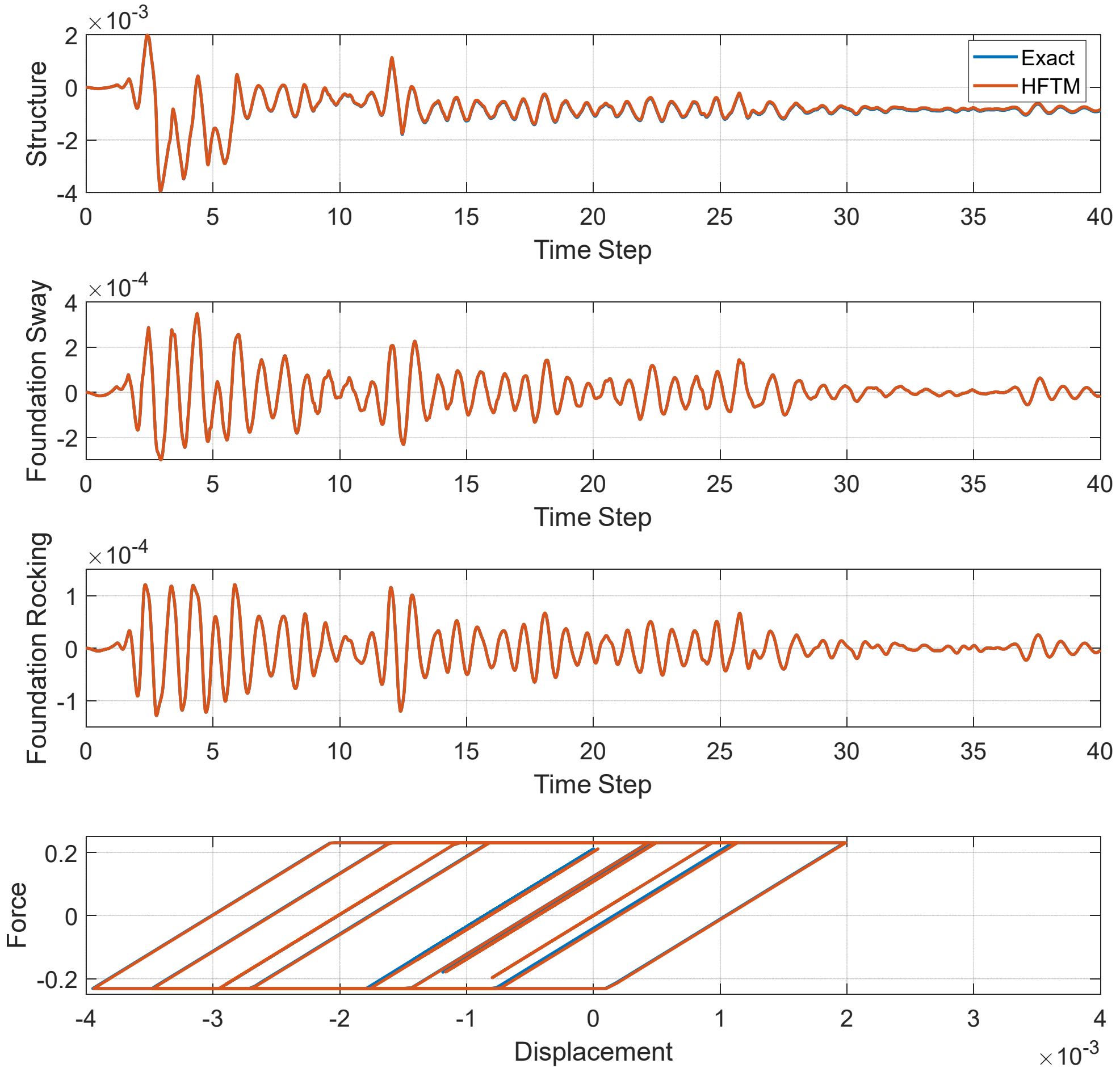

**Figure 3-24: Comparison between true responses and responses obtained using the HFTD method.**

### 3.4.2. Hybrid Time Frequency Domain (HTFD) Method

As described in the previous section, the HFTD method is not efficient for problems with large nonlinear DOFs because the number of pseudo-forces to be estimated would be large. To resolve this issue, Bernal and Youssef [1] proposed Hybrid Time-Frequency Domain (HTFD) method, to solve the system in the time domain and compensate for frequency-dependency using a pseudo-force. In this project, we extensively study this method and implement it in Opensees. The details of this method will be presented in the following sections. The implementation and verification of this method will be presented as separate chapters.

The dynamic stiffness of the soil-foundation system in Eq. (2-7), i.e., $\mathbf{S}_{bb}^{g}$, can be decomposed in the frequency domain into the regular and singular parts as below

$$\overline{\mathbf{S}}_{bb}^{g}(\omega) = \overline{\mathbf{S}}_{bb}^{g,r}(\omega) + \overline{\mathbf{S}}_{bb}^{g,s}(\omega), \tag{3-91}$$

where the overbar denotes frequency-domain representation and superscripts $r$ and $s$ represent regular and singular parts, respectively. The singular part can be well modeled using a lumped physical model with some arbitrary mass, stiffness, and damping values denoted by subscript "ref" as below

$$\overline{\mathbf{S}}_{bb}^{g,s}(\omega) = -\omega^2 \mathbf{M}_{ref} + i\omega \mathbf{C}_{ref} + \mathbf{K}_{ref}. \tag{3-92}$$

By using such decomposition, the so-called interaction force can be written as

$$\mathbf{r}_b = \mathbf{M}_{ref}\ddot{\mathbf{u}}_b^t + \mathbf{C}_{ref}\dot{\mathbf{u}}_b^t + \mathbf{K}_{ref}\mathbf{u}_b^t + \mathbf{r}_b^r, \tag{3-93}$$

where $\mathbf{r}_b^r$ represents the regular part, and we call it pseudo-force hereafter and represented by $\mathbf{f}_{ps}$. Now, we replace $\mathbf{r}_b$ into Eq. (2-7) which results in the following equation

$$\begin{bmatrix} \mathbf{M}_{ss} & \mathbf{M}_{sb} \\ \mathbf{M}_{bs} & \mathbf{M}_{bb} + \mathbf{M}_{ref} \end{bmatrix} \begin{bmatrix} \ddot{\mathbf{u}}_s^t \\ \ddot{\mathbf{u}}_b^t \end{bmatrix} + \begin{bmatrix} \mathbf{0} & \mathbf{0} \\ \mathbf{0} & \mathbf{C}_{ref} \end{bmatrix} \begin{bmatrix} \dot{\mathbf{u}}_s^t \\ \dot{\mathbf{u}}_b^t \end{bmatrix} + \begin{bmatrix} \mathbf{p}_s \\ \mathbf{p}_b + \mathbf{K}_{ref}\mathbf{u}_b^t \end{bmatrix} = \begin{bmatrix} \mathbf{0} \\ \mathbf{q}_b \end{bmatrix} - \begin{bmatrix} \mathbf{0} \\ \mathbf{f}_{ps} \end{bmatrix}. \tag{3-94}$$

To solve this equation in the time domain, an initial value is assumed for $\mathbf{f}_{pf} = \boldsymbol{f}_0$, which is the frequency-dependent component of the equation. Once the response of the system is estimated, $\boldsymbol{f}_0$ is compared with its true value, $\boldsymbol{f}_1$[14], that can be estimated in the frequency domain using the frequency-dependent dynamic stiffness and converted back to the time domain as below

$$\boldsymbol{f}_1 = F^{-1}\left\{\left[\overline{\mathbf{S}}_{bb}^{g}(\omega) - \left(-\omega^2 \mathbf{M}_{ref} + i\omega \mathbf{C}_{ref} + \mathbf{K}_{ref}\right)\right] \overline{\mathbf{U}}_b^t(\omega)\right\}, \tag{3-95}$$

where $\overline{\mathbf{U}}_b^t(\omega)$ is the Fourier Transform of the total displacement of the "b" nodes and $F^{-1}\{.\}$ denotes Inverse Fourier Transform. If the difference between the assumed pseudo-force, $\boldsymbol{f}_0$, and the true pseudo-force time history, $\boldsymbol{f}_1$, is above a considered threshold, then $\boldsymbol{f}_0 = \boldsymbol{f}_1$ is used and the time history analysis is repeated. This process is repeated, i.e., until the difference between consecutive pseudo-forces calculated below goes less than the threshold.

$$\epsilon = \frac{\|\boldsymbol{f}_1 - \boldsymbol{f}_0\|}{\|\boldsymbol{f}_0\|}. \tag{3-96}$$

In the equation above, $\|.\|$ denotes Euclidean norm. The benefit of this approach relative to the Hybrid Frequency Time Domain (HFTD) [86] that solves the equation in the frequency domain and checks in the time domain is the number of unknown forces. Here the number of unknown forces is limited to the number of frequency-dependent DOFs which is usually significantly smaller than the number of nonlinear DOFs in civil engineering problems like a bridge or building model with soil-structure interaction effects [48], [87].

[14] It is not technically a true value as long as $\mathbf{u}_b^t$ is not accurate.

#### 3.4.2.1 Convergence Criteria

Same as with any iterative solution, the convergence of the HTFD must be evaluated. Also, the properties of the reference lumped model, i.e., $\mathbf{M}_{ref}$, $\mathbf{C}_{ref}$, and $\mathbf{K}_{ref}$, need to be determined. As described in the original reference [1], this method is quite stable using some easy settings. In this section, we review the criteria that need to be met to guarantee the convergence of the method.

It is accepted that the singular part of the soil-foundation impedance function in Eq. (3-91) controls the high-frequency portion of the impedance function, so we can write

$$\mathbf{r}_b = \mathbf{M}_\infty \ddot{\mathbf{u}}_b^t + \mathbf{C}_\infty \dot{\mathbf{u}}_b^t + \mathbf{K}_\infty \mathbf{u}_b^t + \mathbf{f}_{ps}, \tag{3-97}$$

from this and Eq. (3-91), we can conclude

$$\mathbf{M}_\infty = -\frac{1}{2}\lim_{\omega\to\infty} Real\left\{\frac{d^2\bar{\mathbf{S}}_{bb}^g(\omega)}{d\omega^2}\right\}, \tag{3-98}$$

$$\mathbf{C}_\infty = \lim_{\boldsymbol{\omega}\to\infty} Imag\left\{\frac{\bar{\mathbf{S}}_{\boldsymbol{bb}}^{\boldsymbol{g}}(\boldsymbol{\omega})}{\boldsymbol{\omega}}\right\}, \tag{3-99}$$

$$\mathbf{K}_\infty = \lim_{\boldsymbol{\omega}\to\infty} Real\left\{\bar{\mathbf{S}}_{\boldsymbol{bb}}^{\boldsymbol{g}}(\boldsymbol{\omega}) + \mathbf{M}_\infty\omega^2\right\}, \tag{3-100}$$

and by having Eq. (3-97), $\mathbf{r}_b^r$ can be written as

$$\mathbf{f}_{ps} = \mathbf{M}_\infty \ddot{\mathbf{u}}_b^t + \mathbf{C}_\infty \dot{\mathbf{u}}_b^t + \mathbf{K}_\infty \mathbf{u}_b^t + \int_0^t \mathbf{S}_{bb}^{g,r}(t-\tau)\mathbf{u}_b^t d\tau - \left[\mathbf{M}_{ref}\ddot{\mathbf{u}}_b^t + \mathbf{C}_{ref}\dot{\mathbf{u}}_b^t + \mathbf{K}_{ref}\mathbf{u}_b^t\right], \tag{3-101}$$

where $\mathbf{S}_{bb}^{g,r} = F^{-1}\{\bar{\mathbf{S}}_{bb}^{g,r}(\omega)\}$. At a discrete-time instant $n$, Eq. (3-101) can be rewritten as

$$\begin{aligned}\mathbf{f}_{ps}[n] = {} & (\mathbf{M}_\infty - \mathbf{M}_{ref})\ddot{\mathbf{u}}_b^t[n] + (\mathbf{C}_\infty - \mathbf{C}_{ref})\dot{\mathbf{u}}_b^t[n] + (\mathbf{K}_\infty - \mathbf{K}_{ref} + \mathbf{S}_{bb}^{g,r}[0]\Delta t)\mathbf{u}_b^t[n] \\ & + \Delta t \sum_{k=1}^{n} \mathbf{S}_{bb}^{g,r}[k]\mathbf{u}_b^t[n-k],\end{aligned} \tag{3-102}$$

where $\Delta t$ is the sampling time. Now, let's assume that the damping in the "s" domain is linear, which is a common assumption, so we can rewrite Eq. (3-94) as

$$\begin{bmatrix}\mathbf{M}_{ss} & \mathbf{M}_{sb}\\ \mathbf{M}_{bs} & \mathbf{M}_{bb}+\mathbf{M}_{ref}\end{bmatrix}\begin{bmatrix}\ddot{\mathbf{u}}_s^t\\ \ddot{\mathbf{u}}_b^t\end{bmatrix} + \begin{bmatrix}\mathbf{C}_{ss} & \mathbf{C}_{sb}\\ \mathbf{C}_{bs} & \mathbf{C}_{ref}\end{bmatrix}\begin{bmatrix}\dot{\mathbf{u}}_s^t\\ \dot{\mathbf{u}}_b^t\end{bmatrix} + \begin{bmatrix}\mathbf{p}_s\\ \mathbf{p}_b + \mathbf{K}_{ref}\mathbf{u}_b^t\end{bmatrix} = \begin{bmatrix}\mathbf{0}\\ \mathbf{q}_b\end{bmatrix} - \begin{bmatrix}\mathbf{0}\\ \mathbf{f}_{ps}\end{bmatrix}. \tag{3-103}$$

Note that $\mathbf{p}_s$ and $\mathbf{p}_b$ here would be different from those in Eq. (3-94) because damping is explicitly presented. As the pseudo-force is limited to the "b" nodes, let's extract the related equation which is

$$\left(\mathbf{M}_{bb} + \mathbf{M}_{ref}\right)\ddot{\mathbf{u}}_b^t[n] + \mathbf{C}_{ref}\dot{\mathbf{u}}_b^t[n] + \mathbf{K}_{ref}\mathbf{u}_b^t[n] = \mathbf{z}[n] - \mathbf{f}_{ps}[n], \tag{3-104}$$

where

$$\mathbf{z}[n] = \mathbf{q}_b[n] - \mathbf{M}_{bs}\ddot{\mathbf{u}}_s^t[n] - \mathbf{C}_{bs}\dot{\mathbf{u}}_s^t[n] - \widetilde{\mathbf{K}}_{bs}(\mathbf{u}_s^t[n] - \mathbf{u}_s^t[n-1]) - \mathbf{p}_b[n-1], \tag{3-105}$$

with $\widetilde{\mathbf{K}}_{sb}$ is the part of the stiffness matrix that supports "b" nodes from the "s" domain. The Tilda sign denotes the tangent stiffness as the superstructure can behave nonlinearly. By inserting Eq. (3-102) for $\mathbf{r}_b^r[n]$ into Eq. (3-104), we have

$$\left(\mathbf{M}_{bb} + \mathbf{M}_{ref}\right)\ddot{\mathbf{u}}_b^t[n] + \mathbf{C}_{ref}\dot{\mathbf{u}}_b^t[n] + \mathbf{K}_{ref}\mathbf{u}_b^t[n] = \mathbf{w}[n] - \mathbf{f}_{ps}[n], \tag{3-106}$$

where

$$\mathbf{f}_{ps}[\boldsymbol{n}] = (\mathbf{M}_\infty - \mathbf{M}_{\boldsymbol{ref}})\ddot{\mathbf{u}}_{\boldsymbol{b}}^{\boldsymbol{t}}[\boldsymbol{n}] + (\mathbf{C}_\infty - \mathbf{C}_{\boldsymbol{ref}})\dot{\mathbf{u}}_{\boldsymbol{b}}^{\boldsymbol{t}}[\boldsymbol{n}] + (\mathbf{K}_\infty - \mathbf{K}_{\boldsymbol{ref}} + \mathbf{S}_{\boldsymbol{bb}}^{g,r}[\mathbf{0}]\Delta t)\mathbf{u}_{\boldsymbol{b}}^{\boldsymbol{t}}[\boldsymbol{n}], \tag{3-107}$$

$$\mathbf{w}[\boldsymbol{n}] = \mathbf{z}[n] - \Delta t\sum_{k=1}^{n}\mathbf{S}_{bb}^{g,r}[k]\mathbf{u}_b^t[n-k]\,. \tag{3-108}$$

As seen in Eq. (3-106), $\mathbf{w}[n]$ does not contain any soil-foundation response at time instant $n$, so it does not affect the convergence limit of the iterative solutions of Eqs. (3-106) and (3-107).

To investigate the convergence properties of Eqs. (3-106) and (3-107), let's first consider the common case in any numerical integration method where the differential equations are cast into a set of algebraic equations as

$$\mathbf{A}\Delta\mathbf{u} = \Delta\mathbf{f}, \tag{3-109}$$

where $\Delta\mathbf{u} = \mathbf{u}[n] - \mathbf{u}[n-1]$. Assuming the matrix $\mathbf{A}$ can be expressed as $\mathbf{A} = \mathbf{A}_0 + \Delta\mathbf{A}$, Eq. (4-1) can be then rewritten as

$$\mathbf{A}_0\mathbf{u}[n] = \Delta\mathbf{f} + \mathbf{A}\mathbf{u}[n-1] - \Delta\mathbf{A}\mathbf{u}[n]. \tag{3-110}$$

Therefore, after $k$ iterations to advance from step $n-1$ to $n$, the response is given by

$$\{\mathbf{u}[n]\}_k = \left[\sum_{m=1}^{k}(-1)^{m-1}\left(\mathbf{A_0}^{-1}\Delta\mathbf{A}\right)^{m-1}\right]\mathbf{A_0}^{-1}\left[\Delta\mathbf{f} + \mathbf{A}\mathbf{u}[n-1]\right]. \tag{3-111}$$

Eq. (3-111) converges to the final solution provided that the maximum of the absolute eigenvalues of the matrix ${\mathbf{A}_0}^{-1}\Delta\mathbf{A}$ is less than unity. For the problem at hand, based on Eqs. (3-106) and (3-107) and considering the Newmark-Beta numerical integration method with parameters $\gamma$ and $\beta$ and, matrices $\mathbf{A_0}$ and $\Delta\mathbf{A}$ are as follows

$$\mathbf{A}_0 = \frac{1}{\beta\Delta t^2}\left(\mathbf{M}_{bb} + \mathbf{M}_{ref}\right) + \frac{\gamma}{\beta\Delta t}\mathbf{C}_{ref} + \mathbf{K}_{ref}, \tag{3-112}$$

$$\Delta\mathbf{A} = \frac{1}{\beta\Delta t^2}\left(\mathbf{M}_\infty - \mathbf{M}_{\boldsymbol{ref}}\right) + \frac{\gamma}{\beta\Delta t}\left(\mathbf{C}_\infty - \mathbf{C}_{\boldsymbol{ref}}\right) + \mathbf{K}_\infty - \mathbf{K}_{\boldsymbol{ref}} + \mathbf{S}_{\boldsymbol{bb}}^{g,r}[\mathbf{0}]\Delta t. \tag{3-113}$$

So, if the reference damping matrix is set as

$$\mathbf{C}_{ref} = \mathbf{C}_\infty + \frac{\beta\Delta t}{\gamma}\left(\mathbf{K}_\infty - \mathbf{K}_{ref} + \mathbf{S}_{bb}^{g,r}[\mathbf{0}]\Delta t\right) + \frac{1}{\gamma\Delta t}\left(\mathbf{M}_\infty - \mathbf{M}_{ref}\right), \tag{3-114}$$

the matrix $\Delta\mathbf{A}$ would be zero and convergence is guaranteed. However, the rate of convergence would be a function of $\mathbf{M}_{ref}$ and $\mathbf{K}_{ref}$. So, it is recommended to set these matrices such that the real part of the impedance function at low frequencies is well approximated [1]. For example, as foundation mass is usually explicitly modeled, $\mathbf{M}_{ref}$=0 and $\mathbf{K}_{ref} = \mathbf{K}_{static}$ are suitable selections. It is noteworthy to mention that Eq. (3-114) is a sufficient condition of the convergence, but it is not necessary. The convergence can be achieved even under mild settings like $\mathbf{C}_{ref} = \mathbf{C}_\infty$ too.

# CHAPTER 4: VERIFICATION IN MATLAB

## 4.1. INTRODUCTION

To verify the HTFD method, preferably a soil-structure system with nonlinear superstructure resting on a semi-infinite soil domain (Figure 4-1) should be modeled and used as a ground truth model. However, to simplify the verification problem, a Single-Degree-Of-Freedom (SDOF) is placed on top of a sway-rocking lumped-parameter foundation model shown in Figure 4-2. This 3-DOF (foundation sway $u_f$; foundation rocking $\varphi$; internal DOF $\varphi_1$) discrete soil-foundation model with frequency-independent mass, spring, and dashpots is based on the concept of cone model to represent a 2-DOF soil-foundation model with frequency-independent sway and frequency-dependent rocking impedance function provided by introducing the additional internal DOF [64]. In the following sections, the model with a 2-DOF frequency-dependent soil-foundation is called the actual model, the model with the 2-DOF frequency-independent soil-foundation is called the approximate model, and the model with 3-DOF soil-foundation, representing frequency-dependency, is called the physical model.

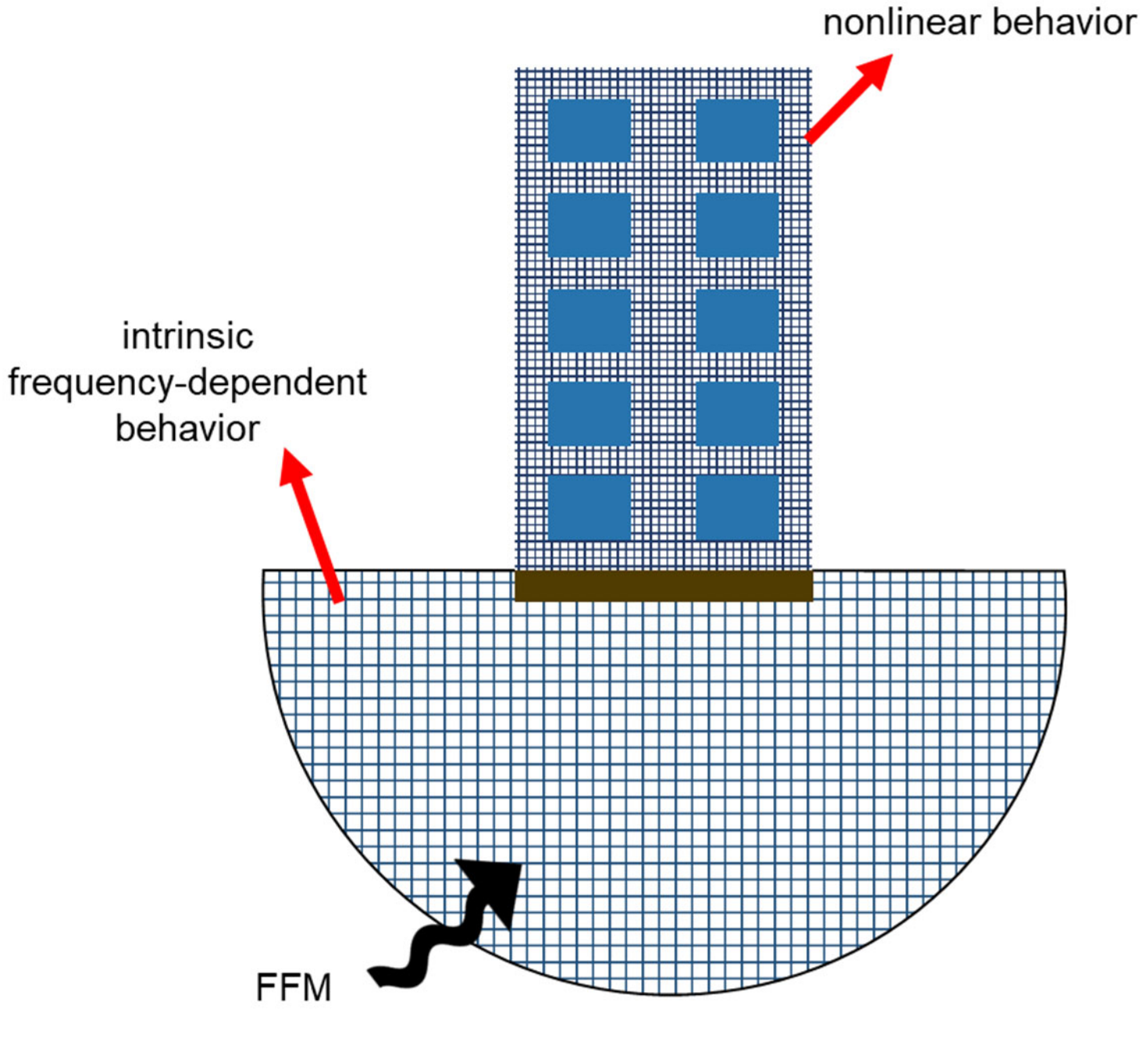

**Figure 4-1: A direct model of a soil-structure system.**

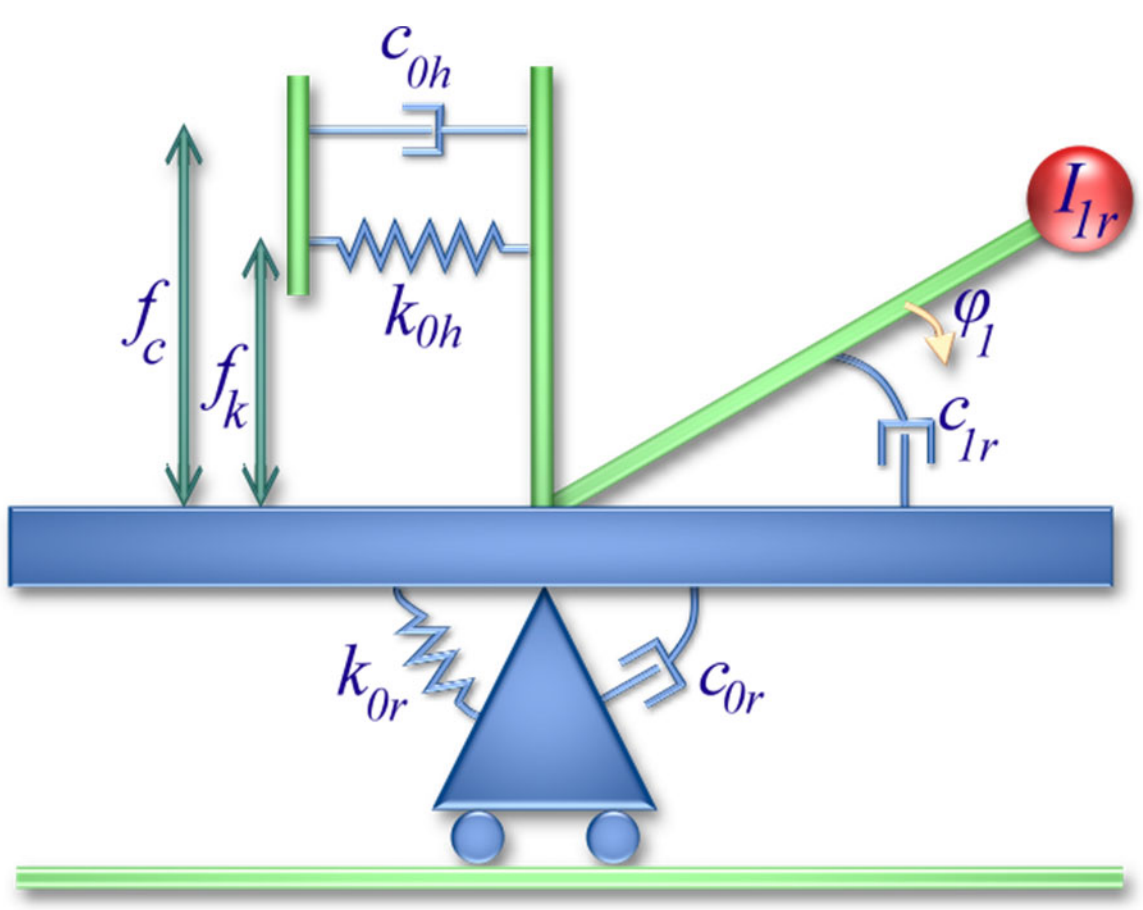

**Figure 4-2: A lumped-parameter model with frequency-independent coefficients to represent soil-foundation impedance function frequency-dependency** [88].

## 4.2. TECHNICAL BACKGROUND

The coefficients of discrete rotational mass, sway and rocking spring, and dashpot elements used in the introduced soil-foundation substructure are calculated using the following equations [64]:

$$k_{0h} = \frac{8\rho V_s^2 r}{2-\nu}\left(1+\frac{e}{r}\right), \tag{4-1}$$

$$c_{0h} = \frac{r}{V_s}\gamma_{0h} k_{0h}, \tag{4-2}$$

$$k_{0r} = \frac{8\rho V_s^2 r^3}{3(1-\nu)}\left(1+2.3\frac{e}{r}+0.58\left(\frac{e}{r}\right)^3\right), \tag{4-3}$$

$$c_{0r} = \frac{r}{V_s}\gamma_{0r} k_{0r}, \tag{4-4}$$

$$c_{1r} = \frac{r}{V_s}\gamma_{1r} k_{0r}, \tag{4-5}$$

$$I_{1r} = \left(\frac{r}{V_s}\right)^2 \mu_{1r} k_{0r}, \tag{4-6}$$

where $r$ is the radius of the cylindrical foundation, $e$ is the embedment depth, and $\rho$, $\nu$, and $V_s$ are soil's mass density, Poisson's ratio, and shear wave velocity, respectively. Dimensionless parameters $\gamma_{0h}$, $\gamma_{0r}$, $\gamma_{1r}$, and $\mu_{1r}$ are calculated using the following equations:

$$\gamma_{0h} = 0.68 + 0.57\sqrt{\frac{e}{r}}, \tag{4-7}$$

$$\gamma_{0r} = 0.15631\frac{e}{r} - 0.08906\left(\frac{e}{r}\right)^2 - 0.00874\left(\frac{e}{r}\right)^3, \tag{4-8}$$

$$\gamma_{1r} = 0.4 + 0.03\left(\frac{e}{r}\right)^2, \tag{4-9}$$

$$\mu_{1r} = 0.33 + 0.1\left(\frac{e}{r}\right)^2, \tag{4-10}$$

As shown in Figure 4-2, the sway spring and dashpot are placed with the eccentricities $f_k$ and $f_c$, respectively, with respect to the bottom of the foundation to account for the possible sway-rocking coupling existing in embedded foundations. These eccentricities can be calculated as

$$f_k = 0.25e, \tag{4-11}$$

$$f_c = 0.32e + 0.03e\left(\frac{e}{r}\right)^2, \tag{4-12}$$

Figure 4-3(left) shows an SDOF superstructure with height $h$ on top of the introduced soil-foundation substructure in which $m$, $I$, $k$, and $c$ are, respectively, translational mass, rotational mass moment of inertia, stiffness, and damping of the superstructure. The mass, $m_f$, and mass moment of inertia, $I_f$, of the foundation are lumped at the middle of the foundation depth. As shown in Figure 4-3(right), the response of this system under horizontal, $u_g$, and rotational, $\varphi_g$, ground motions can be entirely defined by already introduced three soil-foundation DOFs ($u_f$, $\varphi$, and $\varphi_f$) and the added superstructure relative DOF, $u$. That is,

$$\mathbf{u} = [u, u_f, \varphi, \varphi_1]^T. \tag{4-13}$$

Four governing equations of motions of the system can be written as

$$m\left[\ddot{u} + \ddot{u}_f + \ddot{u}_g + (h+e)\left(\ddot{\varphi} + \ddot{\varphi}_g\right)\right] + c\dot{u} + ku = 0, \tag{4-14}$$

$$\begin{aligned} m[\ddot{u} + \ddot{u}_f + \ddot{u}_g + (h+e)(\ddot{\varphi} + \ddot{\varphi}_g)] + m_f\left[\ddot{u}_f + \ddot{u}_g + \frac{e}{2}(\ddot{\varphi} + \ddot{\varphi}_g)\right] + c_{0h}(\dot{u}_f + f_c\,\dot{\varphi}) \\ + k_{0h}(u_f + f_k\,\varphi) = 0, \end{aligned} \tag{4-15}$$

$$\begin{aligned} m(h+e)[\ddot{u} + \ddot{u}_f + \ddot{u}_g + (h+e)(\ddot{\varphi} + \ddot{\varphi}_g)] + m_f\frac{e}{2}\left[\ddot{u}_f + \ddot{u}_g + \frac{e}{2}(\ddot{\varphi} + \ddot{\varphi}_g)\right] \\ + (I + I_f)(\ddot{\varphi} + \ddot{\varphi}_g) + c_{0r}\,\dot{\varphi} + c_{1r}\,(\dot{\varphi} - \dot{\varphi}_1) + c_{0h}f_c(\dot{u}_f + f_c\,\dot{\varphi}) + k_{0r}\,\varphi \\ + k_{0h}f_k(u_f + f_k\,\varphi) = 0, \end{aligned} \tag{4-16}$$

$$I_{1r}(\ddot{\varphi}_1 + \ddot{\varphi}_g) + c_{1r}\,(\dot{\varphi}_1 - \dot{\varphi}) = 0, \tag{4-17}$$

which can be rewritten in matrix/vector form as

$$\mathbf{M\ddot{u}} + \mathbf{C\dot{u}} + \mathbf{Ku} = -\mathbf{ML\ddot{u}}_g, \tag{4-18}$$

where $\ddot{\mathbf{u}}_g = [\ddot{u}_g, \ddot{\varphi}_g]^T$ and system matrices are

$$\mathbf{M} = \begin{bmatrix} m & m & m(h+e) & 0 \\ m & m + m_f & m(h+e) + m_f\dfrac{e}{2} & 0 \\ m(h+e) & m(h+e) + m_f\dfrac{e}{2} & m(h+e)^2 + m_f\left(\dfrac{e}{2}\right)^2 + I + I_f & 0 \\ 0 & 0 & 0 & I_{1r} \end{bmatrix}, \tag{4-19}$$

$$\mathbf{C} = \begin{bmatrix} c & 0 & 0 & 0 \\ 0 & c_{0h} & c_{0h}f_c & 0 \\ 0 & c_{0h}f_c & c_{0r} + c_{1r} + c_{0h}{f_c}^2 & -c_{1r} \\ 0 & 0 & -c_{1r} & c_{1r} \end{bmatrix}, \tag{4-20}$$

$$\mathbf{K} = \begin{bmatrix} k & 0 & 0 & 0 \\ 0 & k_{0h} & k_{0h}f_k & 0 \\ 0 & k_{0h}f_k & k_{0r} + k_{0h}{f_k}^2 & 0 \\ 0 & 0 & 0 & 0 \end{bmatrix}, \tag{4-21}$$

$$\mathbf{L} = \begin{bmatrix} m & m(h+e) \\ m + m_f & m(h+e) + m_f \dfrac{e}{2} \\ m(h+e) + m_f \dfrac{e}{2} & m(h+e)^2 + m_f \left(\dfrac{e}{2}\right)^2 + I + I_f \\ 0 & I_{1r} \end{bmatrix}. \tag{4-22}$$

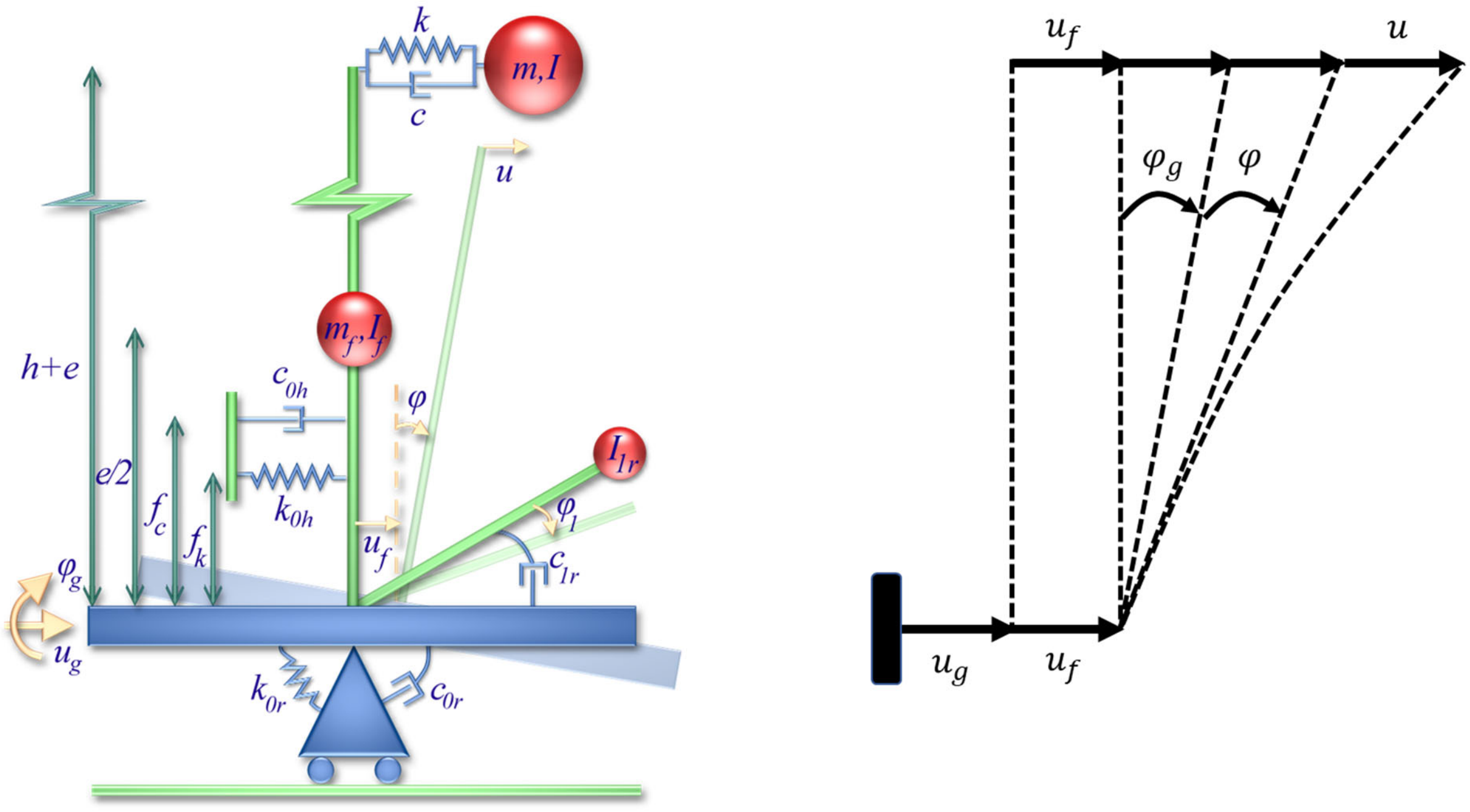

**Figure 4-3: An SDOF system on top of the frequency-dependent soil-foundation substructure (left) and degrees of freedom of the entire system (right)** [88].

To construct system matrices, properties of soil, foundation, and superstructure have to be defined. However, instead of directly specifying these properties, it is well-accepted that the dynamic response of a soil-structure system can be described using the following non-dimensional parameters [89]: $a_0^{fix} = \frac{\omega_{fix}h}{V_s}$, $\overline{m} = \frac{m}{\rho r^2 h}$, $\frac{h}{r}$, $\frac{e}{r}$, $\frac{m_f}{m}$, and $\nu$ where $\omega_{fix}$ is the superstructure's natural frequency with the fixed-based condition. The superstructure's damping is modeled as stiffness-proportional damping which is calculated as $c = 2m\omega_{fix}\xi$ where $\xi$ is the damping ratio. Here these parameters are set at values reported in Table 4-1. These values may not be realistic and are selected to show the performance of the method in the presence of significant soil-structure interaction effects.

**Table 4-1: Values of parameters set in the example.**

| Parameter | Value |
|---|---|
| $\boldsymbol{a_0^{fix}}$ | 4 |
| $\boldsymbol{\omega_{fix}}$ | 15.7 (rad/s) |
| $\boldsymbol{\frac{h}{r}}$ | 3 |
| $\boldsymbol{\frac{e}{r}}$ | 1 |
| $\boldsymbol{\frac{m}{\rho r^2 h}}$ | 0.5 |
| $\boldsymbol{\frac{m_f}{m}}$ | 0.5 |
| $\boldsymbol{\xi}$ | 0.02 |
| $\boldsymbol{\nu}$ | 0.25 |

## 4.3. SYSTEM MATRICES

Figure 4-4 shows a Matlab function to calculate soil-structure matrices using parameters set in Table 4-1. To simplify the calculation and also be able to generate structural responses as absolute values, not dimensionless values, an arbitrary value of foundation radius of $r = 8\ m$ and structural mass of $m = 1\ kg$ are used to run this Matlab function. Also, this function carries out complex eigenanalysis [90] and outputs underdamped modal properties from which $a_0^{flexible} = \frac{\omega_{flexible,1} h}{V_s}$ is computed that is 1.8 in this example.

```
function [dof,M,C,K,L,wn,xi,a0flex,ms,ks,cs,mf,h,he,Is,If,m0r,k0r,c0r,k0h,c0h,c1r,I1r,fk,fc,e,vs] =
SSIModel_01(a0fix,h_r0,e_r0,Tn,xi_s,ms,v,r0)

        dof = 4;
        wn = 2*pi/Tn;
        ks = ms*wn^2;
        cs = 2*xi_s*ms*wn;
        mratio = 0.5;
        mf = mratio*ms;
        mbar = 0.5;
        h = h_r0*r0;
        e = e_r0*r0;
        he = h+e;
        roh = (1/mbar)*ms/(h*r0^2);
        Is = 0.25*ms*r0*r0;
        If = 0.25*mf*r0*r0;
        vs = wn*h/a0fix;
        G = roh*vs^2;
        Kh = 8*G*r0/(2-v);
        Kr = 8*G*r0^3/(3*(1-v));
        gama0h = 0.68+0.57*sqrt(e_r0);
        gama0r = 0.15631*e_r0-0.08906*e_r0^2-0.00874*e_r0^3;
        gama1r = 0.4+0.03*e_r0^2;
        mu1r = 0.33+0.1*e_r0^2;
        fk = 0;
        fc = 0;
        k0h = Kh*(1+e_r0);
        kr_embedded = Kr*(1+2.3*e_r0+0.58*e_r0^3);
        k0r = kr_embedded-G*r0^3/(2*(2-v))*(1+e_r0)*e_r0^2;
        c0h = r0/vs*gama0h*k0h;
        c0r = r0/vs*gama0r*kr_embedded;
        c1r = r0/vs*gama1r*kr_embedded;
        m0r = 0;
        I1r = r0^2/vs^2*mu1r*kr_embedded;
        % System matrices
        M = [ms ms ms*he 0;ms ms+mf ms*he+mf*e/2  0;ms*he ms*he+mf*e/2
        ms*he^2+mf*e^2/4+Is+If+m0r  0;0 0 0 I1r];
        C = [cs 0 0 0;0 c0h c0h*fc 0;0 c0h*fc c0h*fc^2+c0r+c1r -c1r;0 0 -c1r c1r];
        K = [ks 0 0 0;0 k0h k0h*fk 0;0 k0h*fk k0h*fk^2+k0r 0;0 0 0 0];
        L = [ms ms*he;ms+mf ms*he+mf*e/2;ms*he+mf*e/2  ms*he^2+mf*e^2/4+Is+If+m0r;0  I1r];
        % Modal properties
        [~,D ] = eig([zeros(dof,dof),eye(dof);-M\K,-M\C]);
        D = diag(D);
        D = D(imag(D)~=0);
        wn = abs(D);
        xi = -real(D)./wn;
        [~,id] =unique(wn);
        wn = wn(id);
        xi = xi(id);
        [wn,id] = sort(wn);
        xi = xi(id);
        a0flex = min(wn)*h/vs;
```

**Figure 4-4: A Matlab function to calculate soil-structure properties using controlling parameters.**

## 4.4. LINEAR-ELASTIC RESPONSE OF THE PHYSICAL MODEL IN THE TIME DOMAIN

Once the system matrices of the physical model are constructed, the time-history response of the model can be computed under horizontal and rotational Foundation Input Motions (FIMs). Herein, we neglect kinematic interaction effects and analyze the model only under the North-South component of the ground motion recorded in the 1940 El Centro earthquake (see Figure 4-5). The linear-elastic response history of the model is carried out in the time domain using the Newmark method.

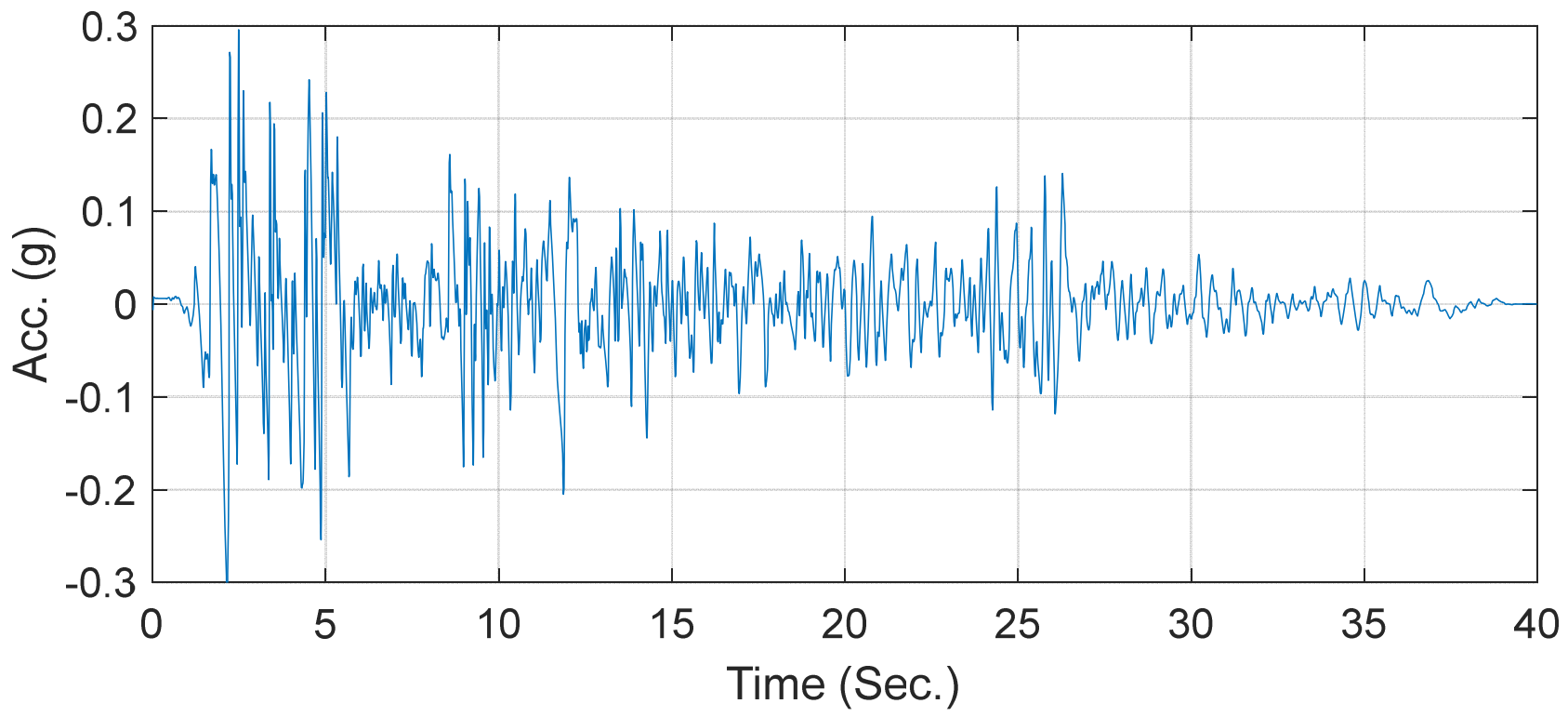

**Figure 4-5: North-South component of the ground motion recorded in the 1940 El Centro earthquake.**

Figure 4-6 shows the calculated displacement responses of the model at all four DOFs. As you can see, although this is a multi-degree-of-freedom model, the calculated responses are dominated by a single (first) mode because other modes have large damping values.

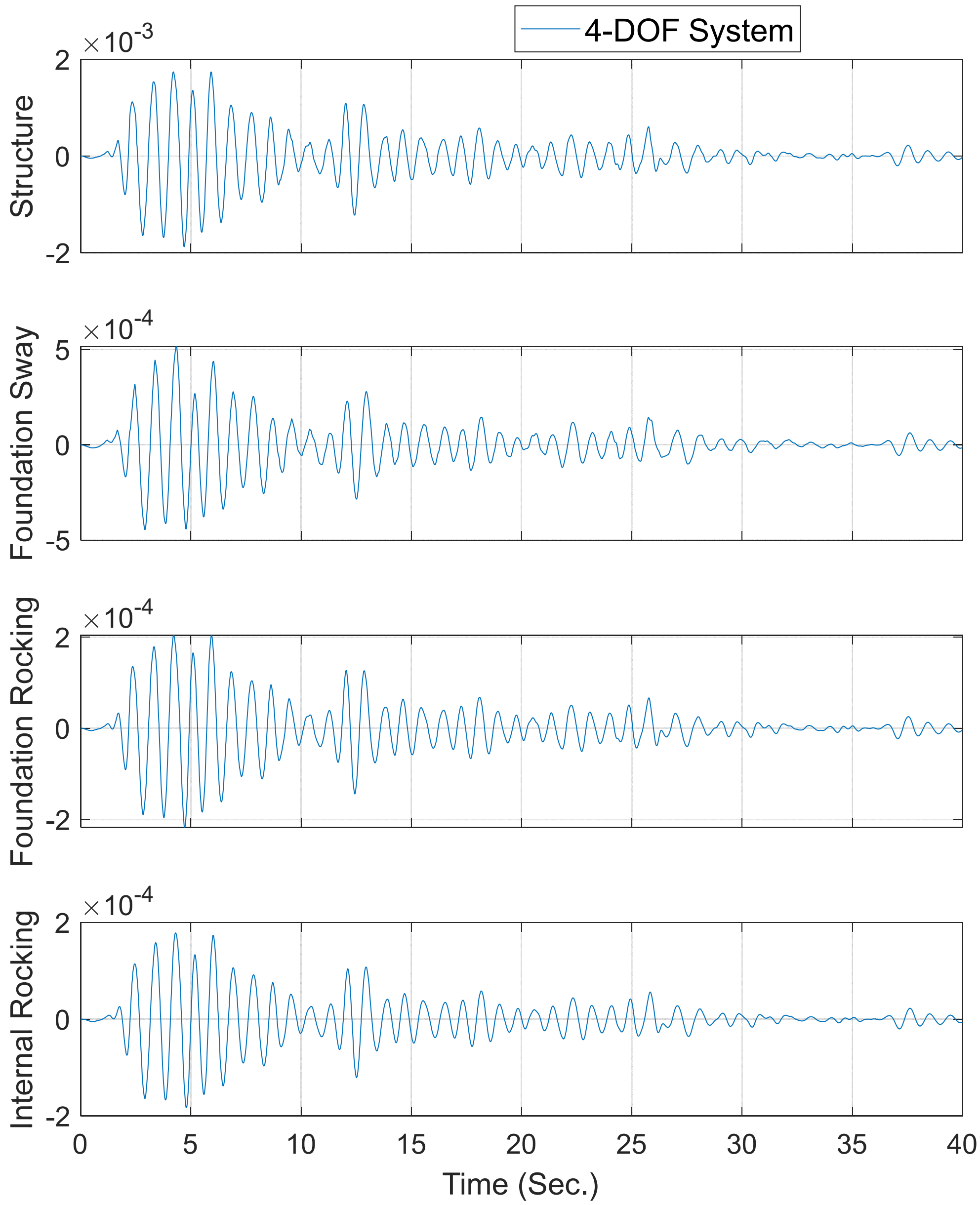

**Figure 4-6: Displacement responses (in meters) of the 4-DOF physical model under El Centro ground motion.**

## 4.5. LINEAR-ELASTIC RESPONSE OF THE PHYSICAL MODEL IN THE FREQUENCY DOMAIN

In addition to the time domain, the response of this model can also be calculated in the frequency domain, because its behavior is linear-elastic. Herein, we do such analysis and compare the response with the time domain to further verify both time and frequency domain solutions. It also helps to make sure the frequency domain implementation, which will be used later, is accurate enough. To solve the problem in the frequency domain, Eq. (4-18) is transformed to the frequency domain using Fourier Transform as

$$(-\mathbf{M}\omega^2 + \mathbf{C}i\omega + \mathbf{K})\hat{\mathbf{u}} = -\mathbf{L}\hat{\ddot{\mathbf{u}}}_g, \tag{4-23}$$

where the overbar denotes the Fourier Transform of the variable. Eq. (4-23) can be rewritten as

$$\hat{\mathbf{u}} = -(-\mathbf{M}\omega^2 + \mathbf{C}i\omega + \mathbf{K})^{-1}\mathbf{L}\hat{\ddot{\mathbf{u}}}_g. \tag{4-24}$$

The equation above is solved at every frequency in Matlab (the code is avilable upon request from the first author.) To carry out frequency domain calculation, two conditions must be satisfied [91]:

- The input excitation must be zero at both tails to prevent leakage;
- To be able to calculate linear convolution using circular convolution, the input excitation must be sufficiently zero-padded to prevent aliasing. In other words, because the linear convolution of two signals with length $m$ and $n$ in the time domain would be a signal with length $m + n - 1$, the input excitation must be zero-padded to have at least $m + n - 1$ data points in which $m$ and $n$ are, respectively, lengths of the original input excitation and impulse response function of the system.

To satisfy the first condition, a cubic spline is employed to smoothly taper the end of the input excitation. Subsequently, enough zeros are added to meet the second condition mentioned above. As mentioned above, the number of zeros needed to be used should be larger than the length of the impulse response function of the system which can be approximated as $\frac{-\ln 0.01}{dt\xi_1\omega_1}$ to make sure the first mode's impulse response function dies out below 1% of its initial value. Figure 4-7 repeats Figure 4-6 on which displacement responses calculated through the mentioned above frequency-domain solution are overlayed. As seen, the frequency-domain implementation is accurate and able to reproduce time-domain responses.

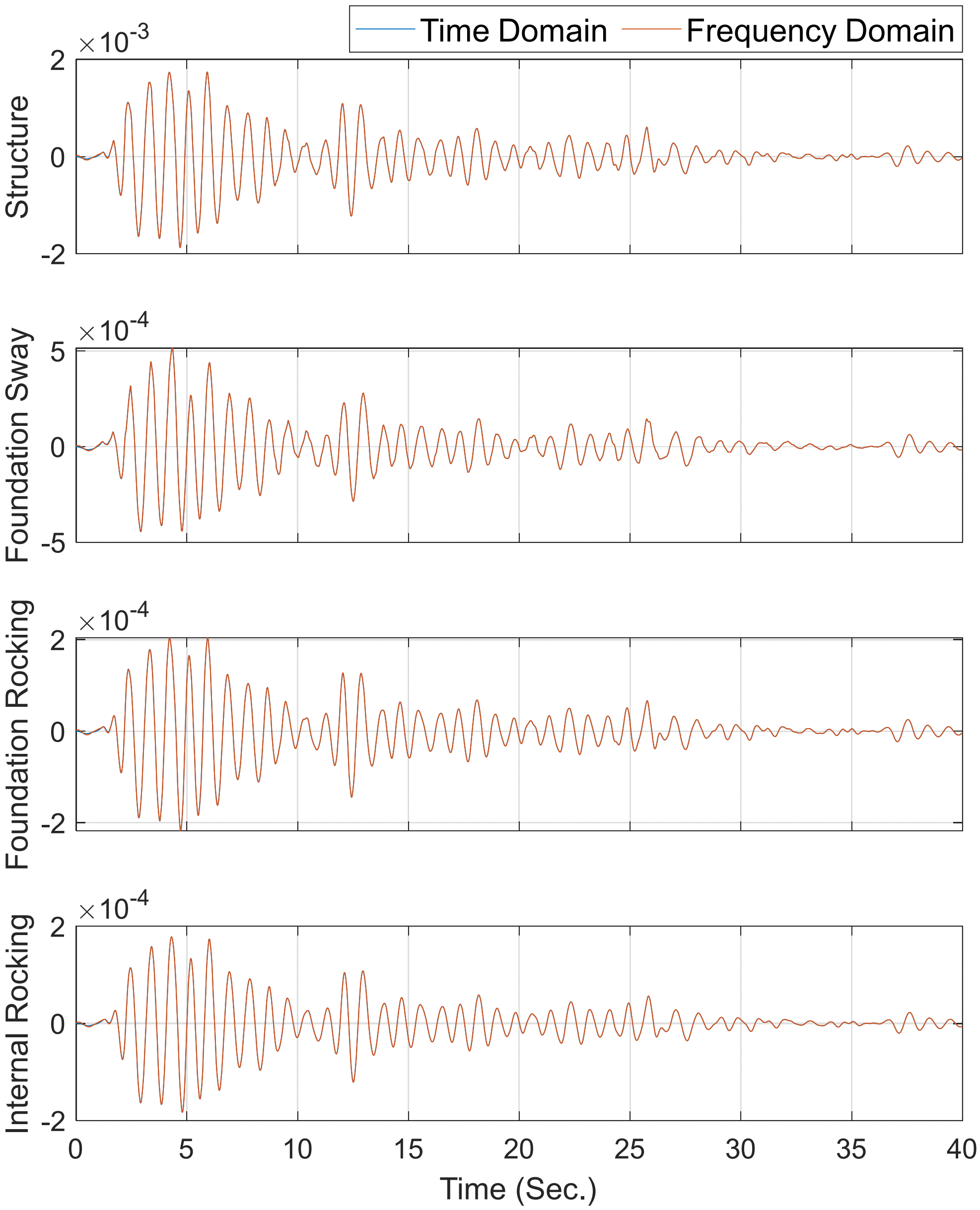

**Figure 4-7: Comparison between displacement responses (in meters) of the physical model under El Centro ground motion obtained through the time (blue) and frequency (red) domain analyses.**

### 4.6. LINEAR-ELASTIC RESPONSE OF THE ACTUAL MODEL IN THE TIME DOMAIN

In the linear-elastic regime the response of the actual system, i.e., linear-elastic SDOF on top of the frequency-dependent 2-DOF sway-rocking soil-foundation substructure, can be computed in the frequency domain without introducing the extra internal DOF. To do so, we need to generate a frequency-dependent impedance function of the rocking DOF. Neglecting the rocking component of the input excitation we repeat the equation of motion of the internal DOF (Eq. (4-17)) here:

$$I_{1r}\ddot{\varphi}_1 + c_{1r}\,(\dot{\varphi}_1 - \dot{\varphi}) = 0. \tag{4-25}$$

By using the Fourier Transform, the equation above can be transformed to the frequency domain as

$$-\omega^2 I_{1r}\hat{\varphi}_1 + i\omega c_{1r}\,(\hat{\varphi}_1 - \hat{\varphi}) = 0. \tag{4-26}$$

from which the internal DOF can be expressed using the foundation rocking DOF as

$$\hat{\varphi}_1 = \frac{c_{1r}^2 - I_{1r}c_{1r}i\omega}{c_{1r}^2 + I_{1r}^2\omega^2}\hat{\varphi}. \tag{4-27}$$

The frequency-dependent rocking impedance function can now be calculated as the ratio of an applied moment $M$ to the induced foundation rotation $\varphi$. Neglecting coupling effects ($f_k = f_c = 0$) and by looking at Figure 4-2, we can write the following equation of motion

$$M = I_f\ddot{\varphi} + c_{0r}\dot{\varphi} + k_{0r}\varphi + c_{1r}\,(\dot{\varphi} - \dot{\varphi}_1). \tag{4-28}$$

which can be expressed in the frequency-domain as

$$\hat{M} = \left(k_{0r} - \omega^2 I_f\right)\hat{\varphi} + i\omega c_{0r}\hat{\varphi} + i\omega c_{1r}\,(\hat{\varphi} - \hat{\varphi}_1). \tag{4-29}$$

In the equation above, the internal rocking DOF can be replaced by its equivalence from Eq. (4-27), so

$$\hat{M} = \left[\left(k_{0r} - \omega^2 I_f - \frac{I_{1r}c_{1r}^2\omega^2}{c_{1r}^2 + I_{1r}^2\omega^2}\right) + i\omega\left(\frac{I_{1r}^2 c_{1r}\omega^2}{c_{1r}^2 + I_{1r}^2\omega^2} + c_{0r}\right)\right]\hat{\varphi}, \tag{4-30}$$

from which the frequency-dependent rocking impedance function can be written as

$$\hat{S}(\omega) = \left[\left(k_{0r} - \omega^2 I_f - \frac{I_{1r}c_{1r}^2\omega^2}{c_{1r}^2 + I_{1r}^2\omega^2}\right) + i\omega\left(\frac{I_{1r}^2 c_{1r}\omega^2}{c_{1r}^2 + I_{1r}^2\omega^2} + c_{0r}\right)\right]. \tag{4-31}$$

It is a common practice to represent the frequency-dependent impedance function as $\hat{S}(a_0) = K_{st}[k(a_0) + ia_0c(a_0)]$ where $a_0 = \omega h/V_s$ and $K_{st} = \hat{S}(0)$ [53], [55]. $k(a_0)$ and $c(a_0)$ show the frequency-dependency of the impedance function and can be extracted from Eq. (4-31) as

$$k(a_0) = 1 - \frac{I_{1r}c_{1r}^2(a_0V_s/h)^2}{k_{0r}(c_{1r}^2 + I_{1r}^2(a_0V_s/h)^2)}, \tag{4-32}$$

$$c(a_0) = \frac{V_s}{hk_{0r}}\left[\frac{I_{1r}^2 c_{1r}(a_0 V_s/h)^2}{c_{1r}^2 + I_{1r}^2(a_0 V_s/h)^2} + c_{0r}\right]. \tag{4-33}$$

It is noteworthy that part of the frequency-dependency of the impedance function which comes from the foundation mass, $-\omega^2 I_f$, can be explicitly modeled, so it is removed from Eq. (4-32). Figure 4-8 shows the stiffness $k(a_0)$ and damping $c(a_0)$ parts of the soil-foundation rocking impedance function versus $a_0$.

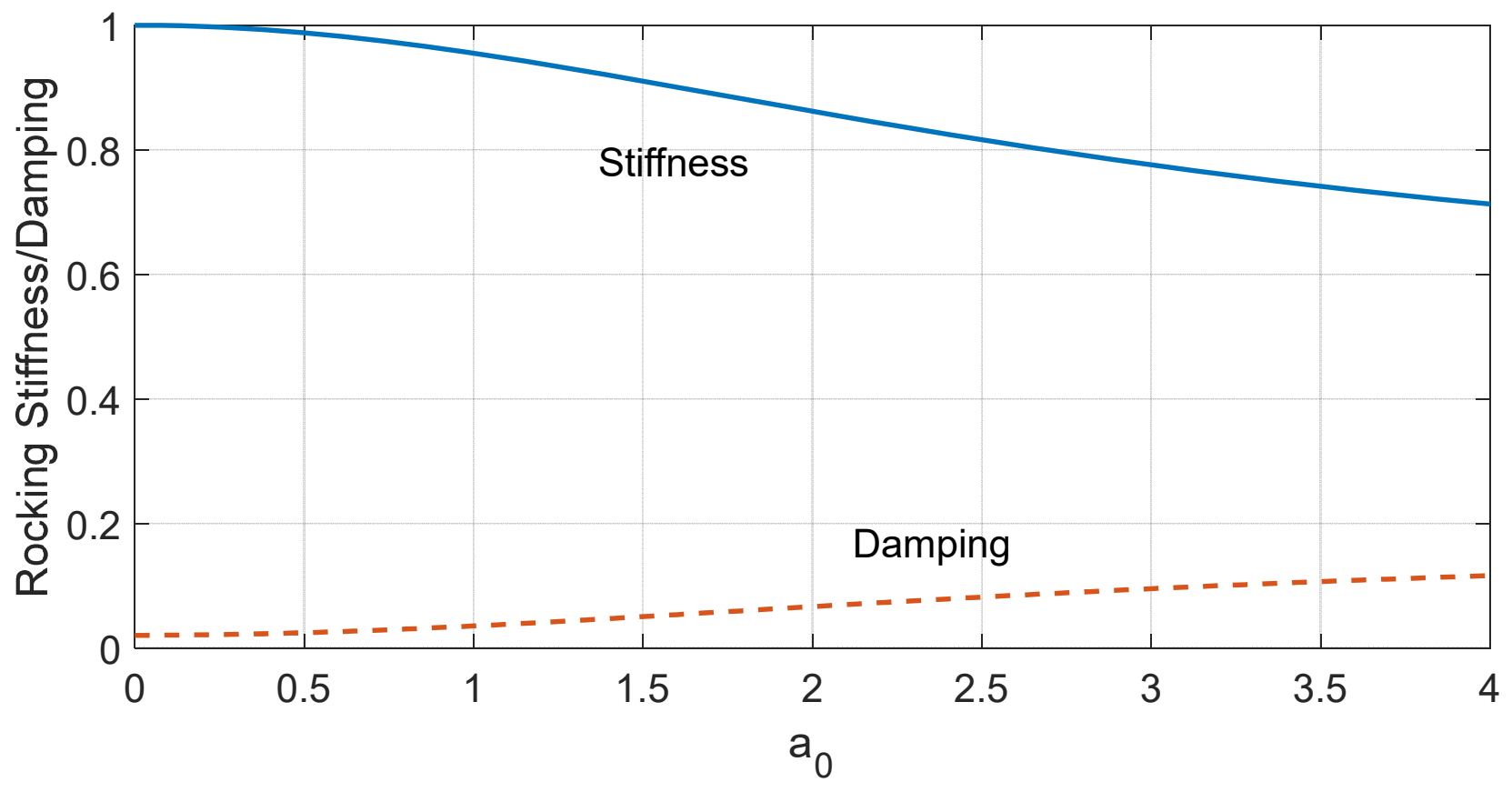

**Figure 4-8: Stiffness and damping parts of the rocking impedance function.**

Having a closed-form representation of the frequency-dependent rocking impedance function, we can easily write the equation of motion of the actual system as:

$$\mathbf{M}\ddot{\mathbf{u}} + \mathbf{C}(\omega)\dot{\mathbf{u}} + \mathbf{K}(\omega)\mathbf{u} = -\mathbf{L}\ddot{u}_g, \tag{4-34}$$

where system matrices are

$$\mathbf{M} = \begin{bmatrix} m & m & m(h+e) \\ m & m+m_f & m(h+e)+m_f\frac{e}{2} \\ m(h+e) & m(h+e)+m_f\frac{e}{2} & m(h+e)^2+m_f\left(\frac{e}{2}\right)^2+I+I_f \end{bmatrix}, \tag{4-35}$$

$$\mathbf{C}(\omega) = \begin{bmatrix} c & 0 & 0 \\ 0 & c_{0h} & c_{0h}f_c \\ 0 & c_{0h}f_c & c_{0h}{f_c}^2 + c_{0r}(\omega) \end{bmatrix}, \tag{4-36}$$

$$\mathbf{K}(\omega) = \begin{bmatrix} k & 0 & 0 \\ 0 & k_{0h} & k_{0h}f_k \\ 0 & k_{0h}f_k & k_{0h}{f_k}^2 + k_{0r}(\omega) \end{bmatrix}, \tag{4-37}$$

$$\mathbf{L} = \begin{bmatrix} m \\ m + m_f \\ m(h+e) + m_f \dfrac{e}{2} \end{bmatrix}, \tag{4-38}$$

where both stiffness and damping matrices are functions of the excitation frequency because of the frequency-dependent rocking impedance functions $k_{0r}(\omega)$ and $c_{0r}(\omega)$ which are

$$k_{0r}(\omega) = k_{0r} - \frac{I_{1r} c_{1r}^2 \omega^2}{c_{1r}^2 + I_{1r}^2 \omega^2}, \tag{4-39}$$

$$c_{0r}(\omega) = c_{0r} + \frac{I_{1r}^2 c_{1r} \omega^2}{c_{1r}^2 + I_{1r}^2 \omega^2}. \tag{4-40}$$

Note that like Figure 4-8, the term $-\omega^2 I_f$ is excluded from $k_{0r}(\omega)$ because it is already considered in the mass matrix. Now, the frequency-domain solution similar to Section 4.5 can be employed to obtain the response of the system.

Figure 4-9 shows a comparison between displacement responses (in meters) of the physical (blue) and actual (red) models under El Centro ground motion obtained through the time and frequency domain analyses, respectively. As it can be seen, these two models are equivalent which means the physical model is, indeed, able to provide frequency-dependency of the rocking impedance function through that additional internal DOF. Please note that the so-called actual model has only 3 DOFs, so no response is reported for the internal DOF of this model in Figure 4-9.

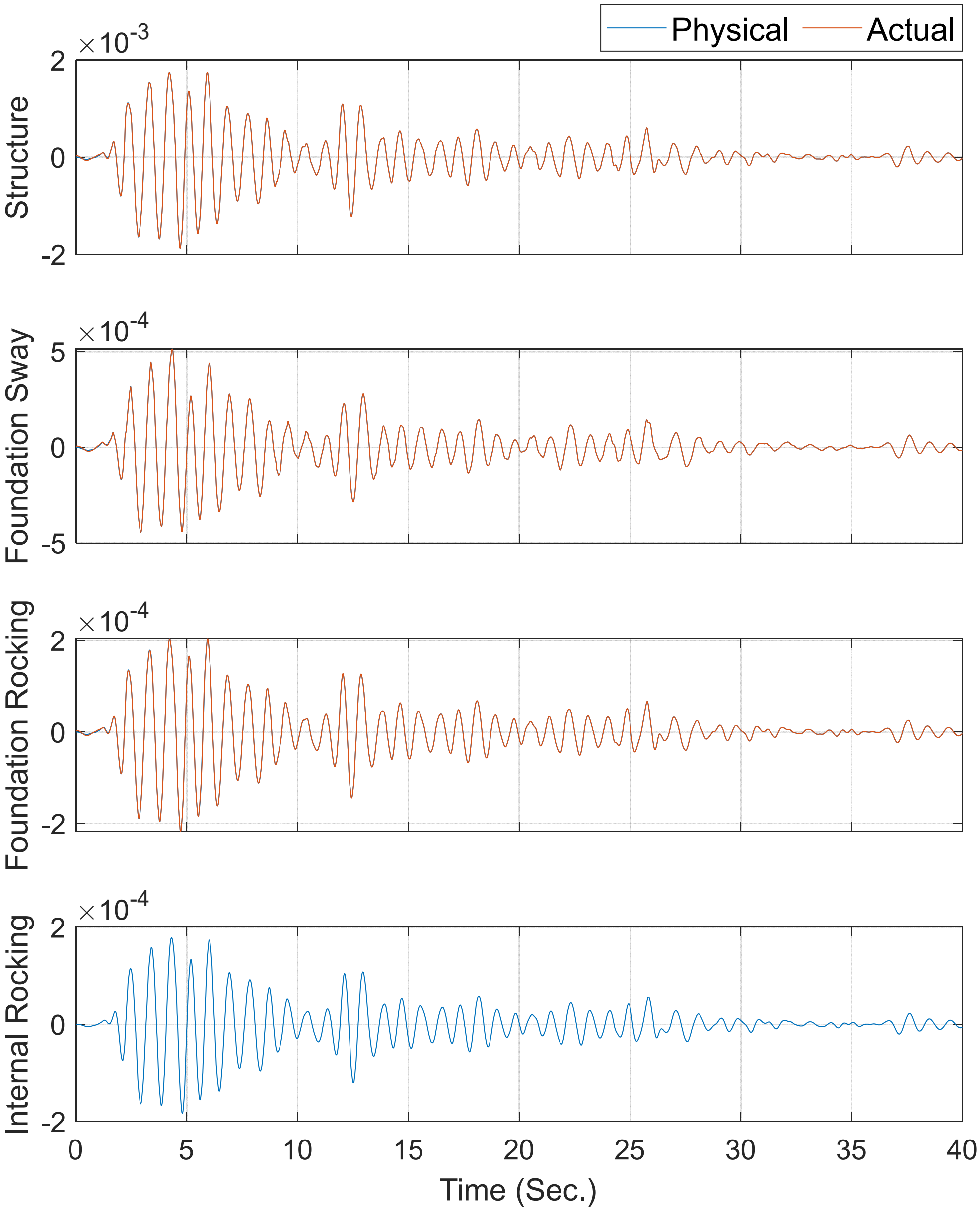

**Figure 4-9: Comparison between displacement responses (in meters) of the physical (blue) and actual (red) models under El Centro ground motion obtained through the time and frequency domain analyses, respectively.**

## 4.7. LINEAR-ELASTIC RESPONSE OF THE APPROXIMATE MODEL IN THE FREQUENCY DOMAIN

As discussed in Chapter 2, in design applications, the frequency-dependent impedance function is set at a single frequency, which should be the fundamental frequency of the soil-structure system, i.e., flexible-base natural frequency. Herein, we evaluate this approach by solving the 3-DOF system with the rocking stiffness set at the flexible-base natural frequency, that is,

$$k_{0r}(\omega) = k_{0r} - \frac{I_{1r} c_{1r}^2 \omega_n^2}{c_{1r}^2 + I_{1r}^2 \omega_n^2}, \tag{4-41}$$

$$c_{0r}(\omega) = c_{0r} + \frac{I_{1r}^2 c_{1r} \omega_n^2}{c_{1r}^2 + I_{1r}^2 \omega_n^2}. \tag{4-42}$$

Figure 4-10 shows a comparison between displacement responses (in meters) of the physical (blue) and approximate (red) models under El Centro ground motion obtained through the time and frequency domain analyses, respectively. Theoretically, we expect to see differences between these two systems, because the impedance function is set at a single frequency in the approximate model, while it is frequency-dependent in the physical model. However, as seen, the approximate model can very accurately represent the physical/actual system because 1- the impedance function is set at the flexible-base natural frequency, and 2- the system is a single-mode system. As will be shown later, even for this single-mode system, the frequency dependency could be very important if the superstructure goes beyond linear-elastic response.

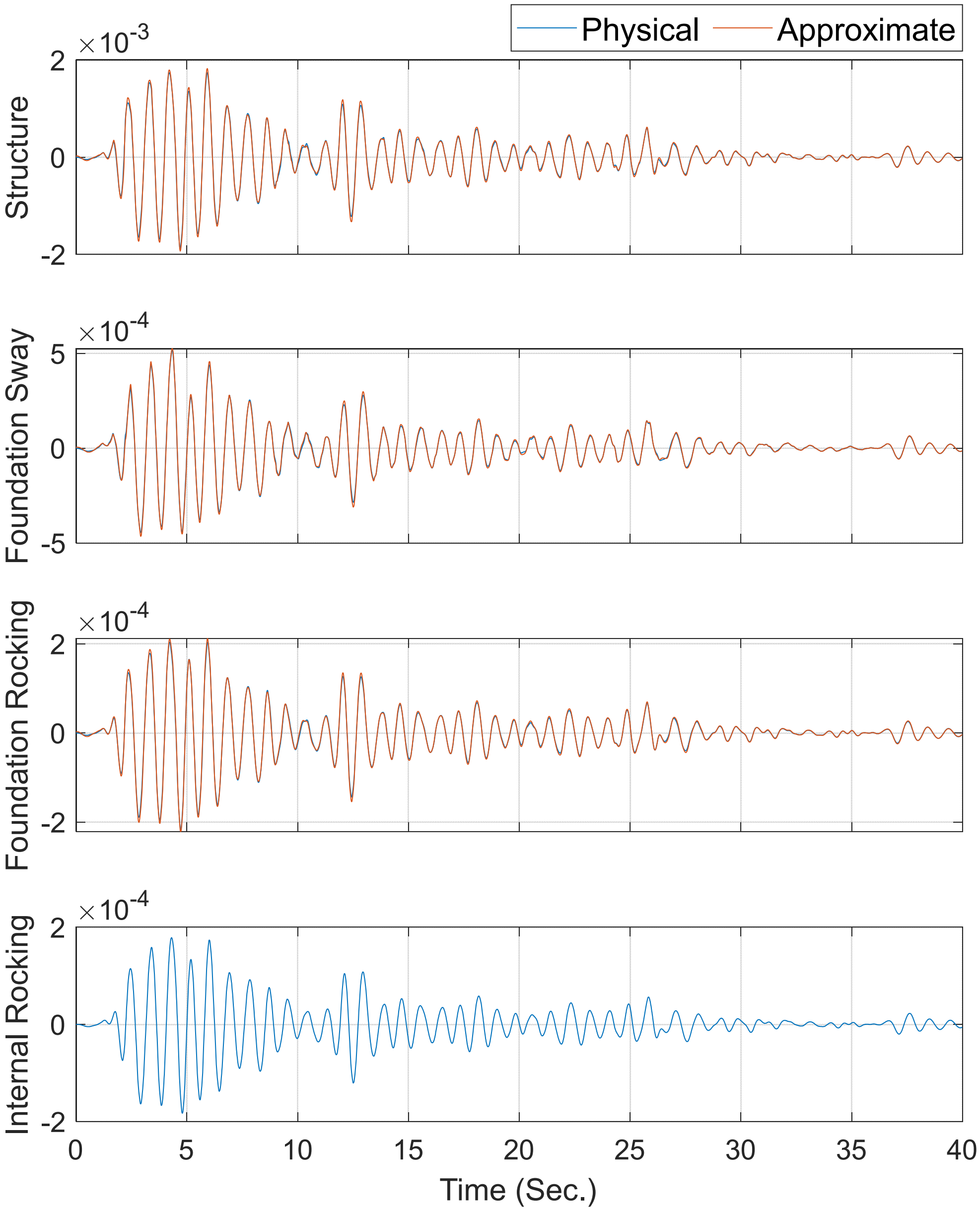

**Figure 4-10: Comparison between displacement responses (in meters) of the physical (blue) and approximate (red) models under El Centro ground motion obtained through the time and frequency domain analyses, respectively.**

## 4.8. NONLINEAR RESPONSE OF THE PHYSICAL MODEL IN THE TIME DOMAIN

The present project is all about the methods to consider soil-structure interaction frequency-dependency effects in the presence of superstructure's nonlinearity. So, herein, we extend the soil-structure system shown in Figure 4-3 to a system in which the superstructure can behave nonlinearly according to the kinematic hardening model presented in Figure 4-11. This nonlinear model is controlled by two additional parameters relative to the original linear-elastic model: 1- yield strength $F_y$, and 2- post-yield stiffness ratio $\alpha$.

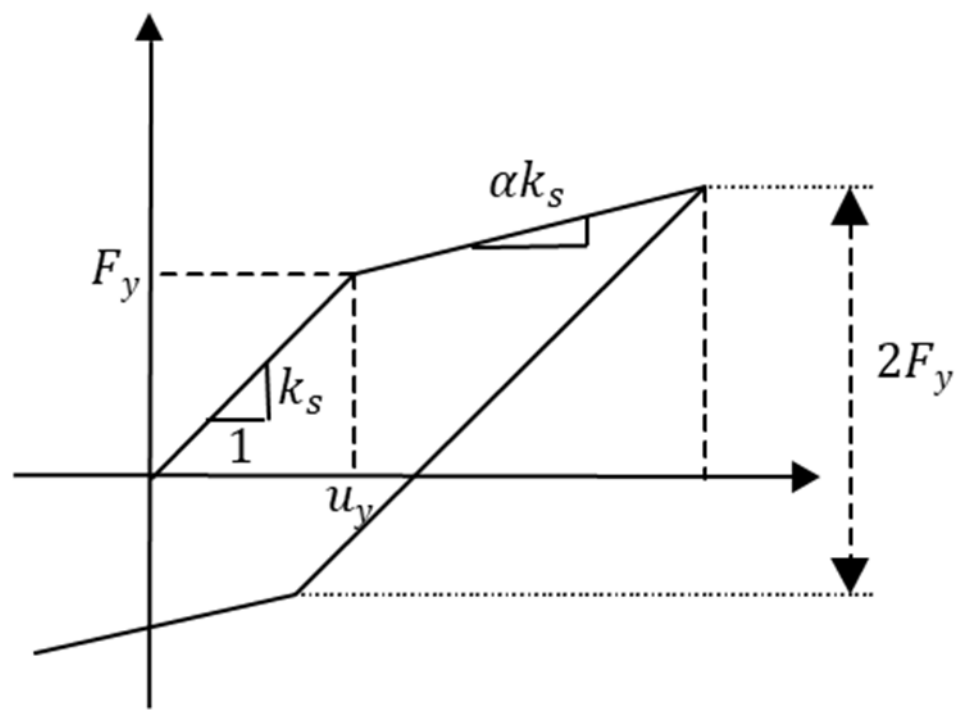

**Figure 4-11: Nonlinear model used for the superstructure.**

We first investigate how much differently the nonlinear system would behave relative to the original linear system. Figure 4-12 shows a comparison between the linear and nonlinear responses of the physical model obtained through time-domain analysis. For the nonlinear model, a post-yield stiffness ratio of zero ($\alpha = 0$) and yield strength equal to half of the maximum elastic strain force is considered, i.e., $F_y = \bar{f}_y F_e$ where $\bar{f}_y = 0.5$ and $F_e = k_s\, u_{max}$. As seen in this figure, the response of the nonlinear system at all DOFs is significantly different from the linear system. A permanent displacement is even observed in the structural DOF of the nonlinear system.

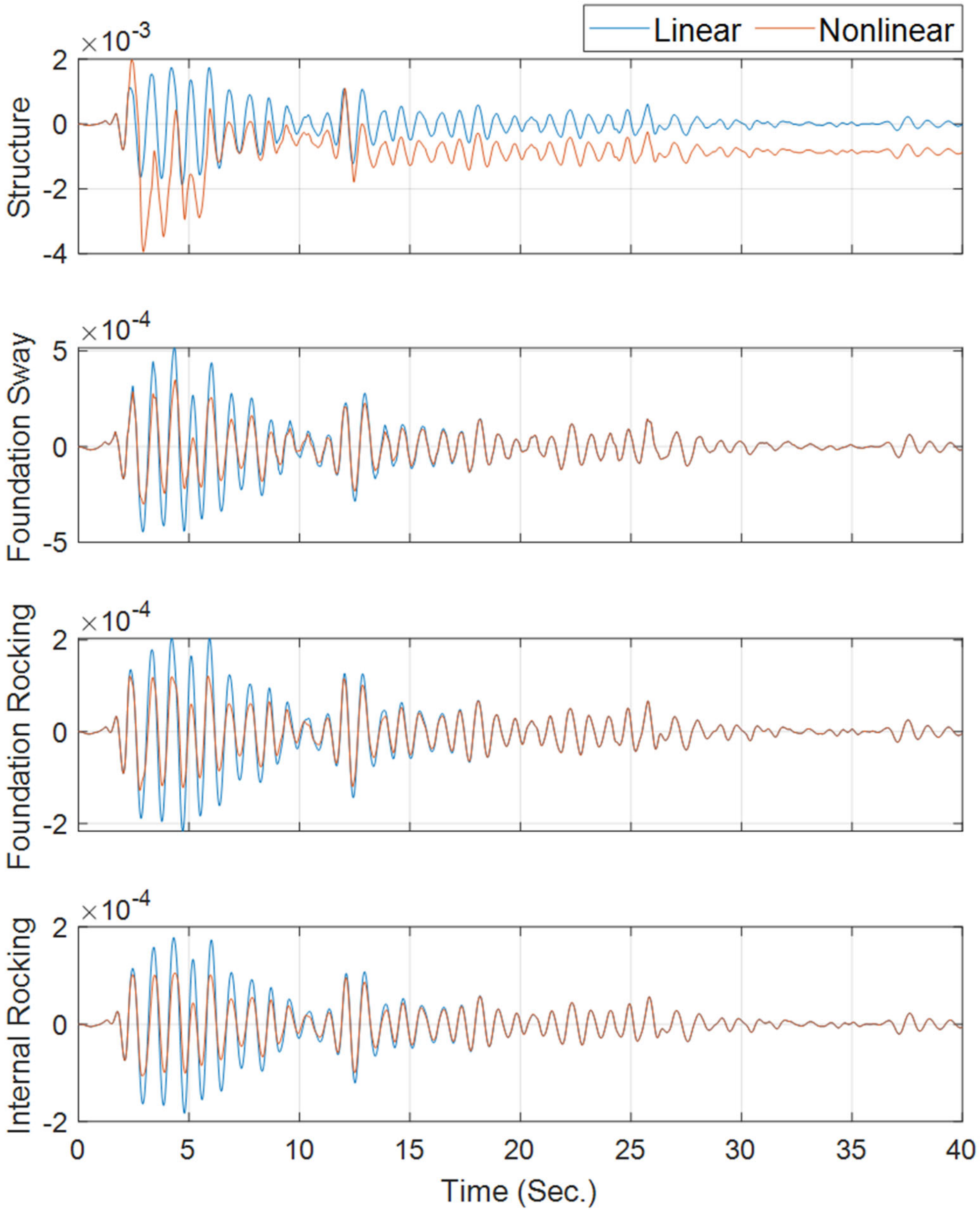

**Figure 4-12: Comparison between displacement responses (in meters) of the linear (blue) and nonlinear (red) physical models under El Centro ground motion obtained through the time domain analyses.**

## 4.9. NONLINEAR RESPONSE OF THE APPROXIMATE MODEL IN THE TIME DOMAIN

As shown before, the linear response of the approximate model in which the rocking impedance function is set at the flexible-base natural frequency is almost identical to the actual/physical models because the system is a single-mode system. However, if this single-mode system varies in time, then it is expected to see differences between the actual and approximate responses. Herein, we show this difference by comparing the nonlinear response of the physical model, shown in Figure 4-12, and the approximate model. As seen in Figure 4-13, the response of the approximate model is significantly different from the physical model in the superstructure DOF where nonlinearity is concentrated. In other words, the traditional substructure approach with a constant (frequency-independent) impedance function is not able to predict the response of the system if the superstructure is allowed to behave nonlinearly even if the impedance function is set at the initial flexible-base natural frequency.

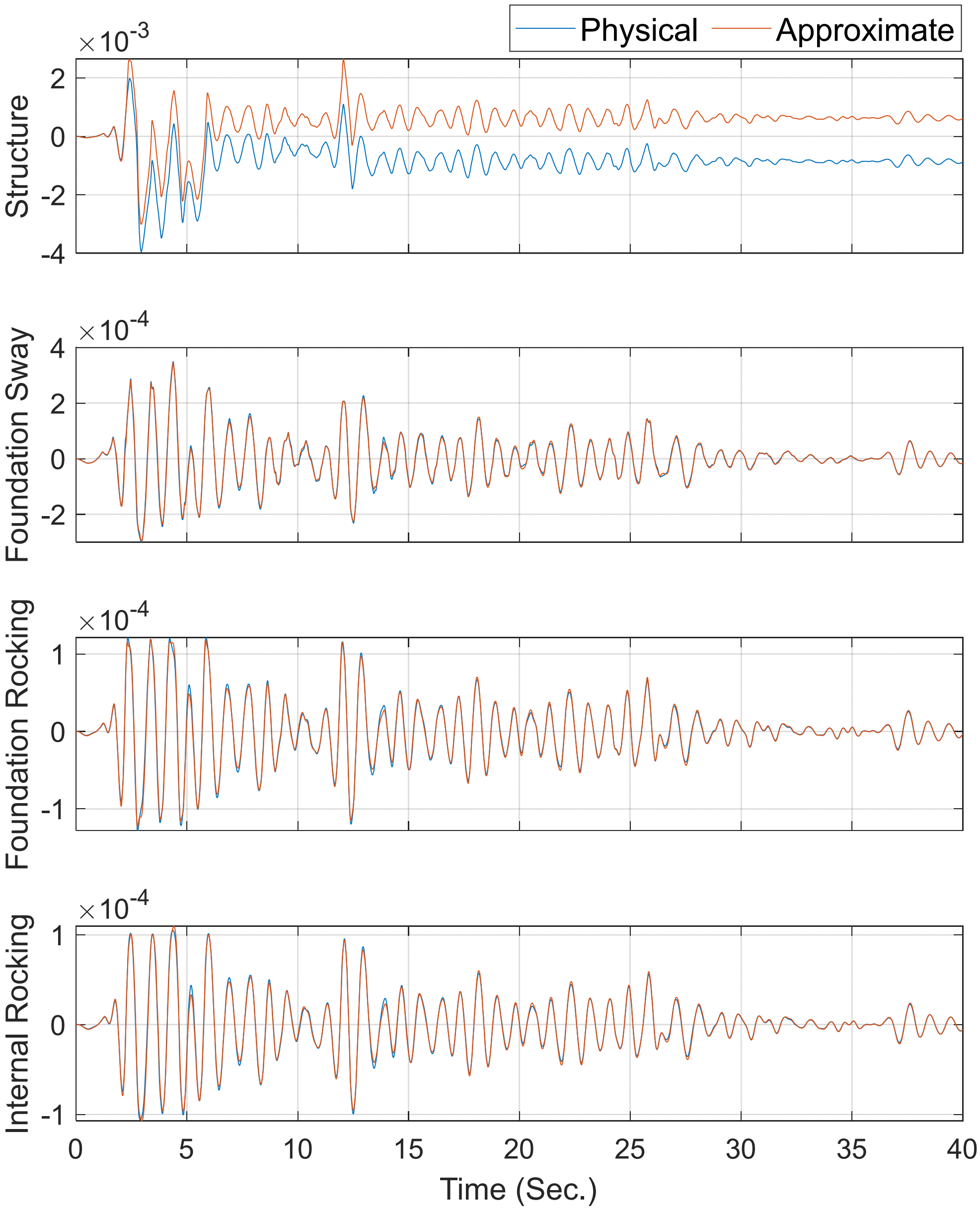

**Figure 4-13: Comparison between displacement responses (in meters) of the physical (blue) and approximate (red) models under El Centro ground motion obtained through the time domain analyses.**

## 4.10. NONLINEAR RESPONSE IN THE TIME DOMAIN USING THE HTFD METHOD

In this section, the HTFD method is employed to obtain the nonlinear time history response of the analytical model (3-DOF model with frequency-dependent soil-foundation impedance function) in the time domain.

As discussed in the HTFD section, We first need to set up a base model and then add a reference substructure at soil-foundation nodes, and finally analyze it through the iterative HTDF approach.

The system matrices of the base model are shown below in which elements related to the soil-foundation domain are removed, while the mass properties of the soil-foundation system are already included in the mass matrix.

$$\mathbf{M} = \begin{bmatrix} m & m & m(h+e) \\ m & m+m_f & m(h+e)+m_f\frac{e}{2} \\ m(h+e) & m(h+e)+m_f\frac{e}{2} & m(h+e)^2+m_f\left(\frac{e}{2}\right)^2+I_s+I_f \end{bmatrix}, \tag{4-43}$$

$$\mathbf{C} = \begin{bmatrix} c & 0 & 0 \\ 0 & 0 & 0 \\ 0 & 0 & 0 \end{bmatrix}, \tag{4-44}$$

$$\mathbf{K} = \begin{bmatrix} k & 0 & 0 \\ 0 & 0 & 0 \\ 0 & 0 & 0 \end{bmatrix}, \tag{4-45}$$

$$\mathbf{L} = \begin{bmatrix} m \\ m+m_f \\ m(h+e)+m_f\frac{e}{2} \end{bmatrix}. \tag{4-46}$$

The next step is to find the singular part of the impedance function. As shown earlier, the analytical impedance function is

$$\overline{\mathbf{S}}_{bb}^{g}(\omega) = \begin{bmatrix} k_{0h}+i\omega c_{0h} & k_{0h}f_k+i\omega c_{0h}f_c \\ k_{0h}f_k+i\omega c_{0h}f_c & k_{0h}{f_k}^2+k_{0r}(\omega)+i\omega\left[c_{0h}{f_c}^2+c_{0r}(\omega)\right] \end{bmatrix}, \tag{4-47}$$

where

$$k_{0r}(\omega) = k_{0r} - \frac{I_{1r}c_{1r}^2\omega^2}{c_{1r}^2+I_{1r}^2\omega^2}, \tag{4-48}$$

$$c_{0r}(\omega) = c_{0r} + \frac{I_{1r}^2c_{1r}\omega^2}{c_{1r}^2+I_{1r}^2\omega^2}. \tag{4-49}$$

So using Eqs. (3-98) to (3-100), we have

$$\mathbf{M}_{\infty} = \begin{bmatrix} 0 & 0 \\ 0 & 0 \end{bmatrix}, \tag{4-50}$$

$$\mathbf{C}_{\infty} = \begin{bmatrix} c_{0h} & c_{0h} f_c \\ c_{0h} f_c & c_{0h} {f_c}^2 + c_{0r} + c_{1r} \end{bmatrix}, \tag{4-51}$$

$$\mathbf{K}_{\infty} = \begin{bmatrix} k_{0h} & k_{0h} f_k \\ k_{0h} f_k & k_{0h} {f_k}^2 + k_{0r} + k_{0r} - \dfrac{c_{1r}^2}{I_{1r}} \end{bmatrix}, \tag{4-52}$$

In real-life applications, an analytical representation of the impedance function might not be available, and numerical differentiation should be used. Then, the evaluation can be carried out at the Nyquist frequency.

To be able to use Eq. (3-112) and (3-113) for stability criteria, the regular part of the impedance function at time zero, $\mathbf{S}_{bb}^{g,r}[0]$ is needed. Using inverse Fourier Transform, we can write

$$\mathbf{S}_{bb}^{g,r}[0] = \frac{1}{2\pi} \int_{-\infty}^{\infty} \bar{\mathbf{S}}_{bb}^{g,r}(\omega) d\omega. \tag{4-53}$$

As $\bar{\mathbf{S}}_{bb}^{g,r}(\omega)$ is an odd function and assuming that $\mathbf{S}_{bb}^{g,r}(t)$ is causal, we have

$$\mathbf{S}_{bb}^{g,r}[0] = \frac{2}{\pi} \int_{0}^{\infty} Real\{\bar{\mathbf{S}}_{bb}^{g,r}(\omega)\} d\omega. \tag{4-54}$$

For this example, it is easy to show that

$$\mathbf{S}_{bb}^{g,r}[0] = \begin{bmatrix} 0 & 0 \\ 0 & \lim_{\omega \to \infty} \dfrac{2}{\pi} \dfrac{c_{1r}^3}{I_{1r}^2} \tan^{-1}\left(\dfrac{I_{1r}\omega}{c_{1r}}\right) \end{bmatrix}. \tag{4-55}$$

**Remark:** The HTFD method converges in a time-progressive manner. In other words, convergence at a time $t$ is achieved provided that the solution has already converged at previous time steps. So, we carry out the analysis in successive short time windows. A Matlab code for performing the window-based HTFD in the example introduced in this chapter is available upon request from the first author.

#### 4.10.1. Case 1

In this case, we run the HTFD analysis using the parameters shown in Table 4-2. As seen, a suitable value is chosen for the rocking stiffness of the reference model. Also, the damping is calculated through Eq. (3-114), so the maximum eigenvalue is zero and we expect to see convergence. So, we use a large window size of 10 seconds. As shown in Figure 4-14, the response of the system obtained through the HTFD solution is identical to one obtained using the ground truth 4-DOF physical model. A video showing the progress of the analysis will be provided upon request from the first author.

**Table 4-2: Values of parameters set in Case 1.**

| Parameter | Value |
|---|---|
| $\mathbf{M}_{ref}$ | $\begin{bmatrix} 0 & 0 \\ 0 & 0 \end{bmatrix}$ |
| $\mathbf{K}_{ref}$ | $\begin{bmatrix} k_{0h} & 0 \\ 0 & k_{0r} \end{bmatrix}$ |
| $\mathbf{C}_{ref}$ | Eq. (3-114) |
| $N_{iter}$ | 1000 |
| $\epsilon$ | 0.001 |
| $L_{window}$ | 10 sec. |
| $\max\left\| Eig\left(\mathrm{A}_0{}^{-1}\Delta\mathrm{A}\right)\right\|$ | 0 |

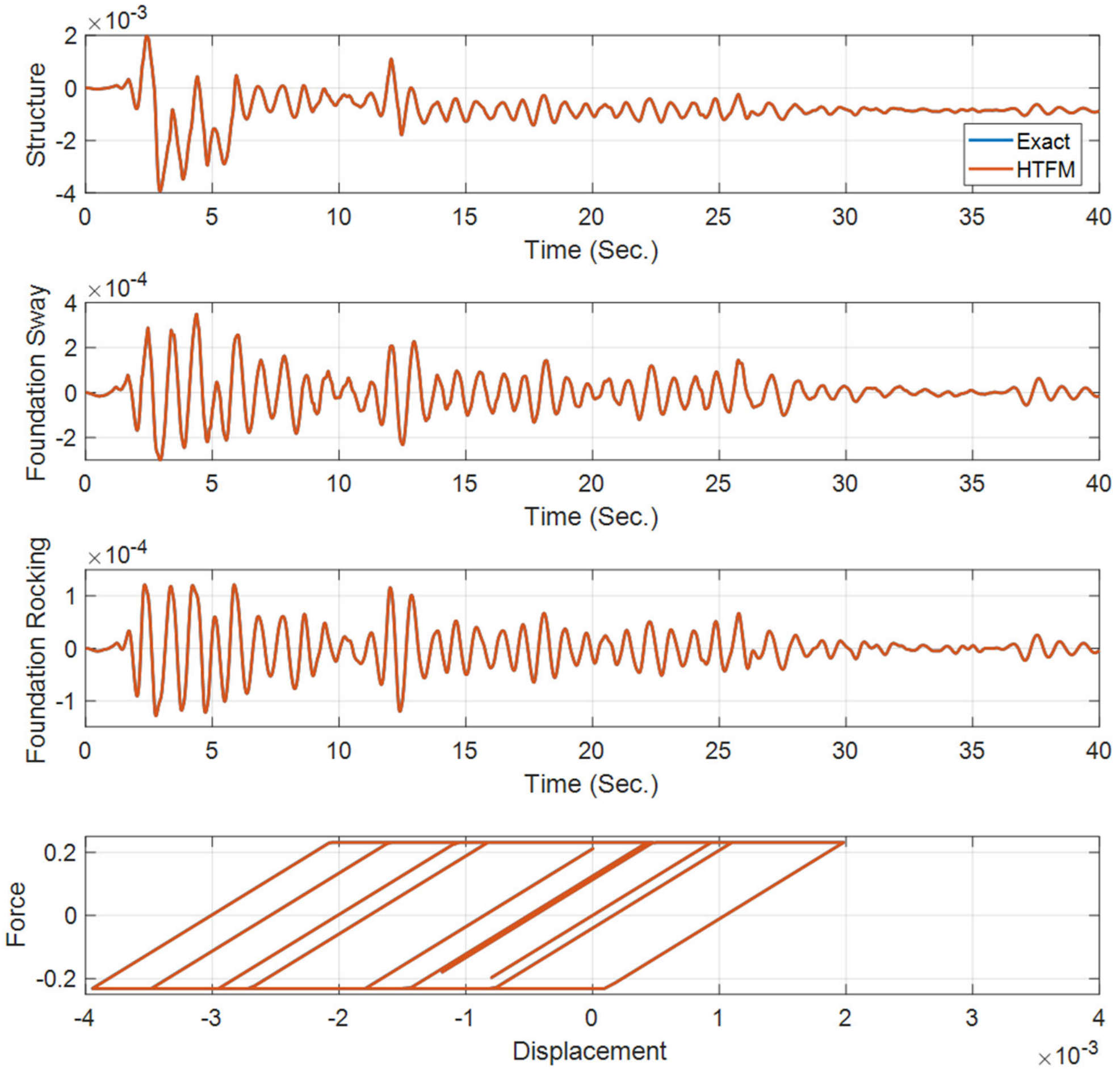

**Figure 4-14: Comparison between exact and HTFD solutions in Case 1.**

### 4.10.2. Case 2

In this case, we run the HTFD analysis using the parameters shown in Table 4-3. As seen, a zero value is chosen for the rocking stiffness of the reference model. However, the damping is calculated through Eq. (3-114), so the maximum eigenvalue is zero and we expect to see convergence. But, due to the large window size, convergence is not achieved for this case.

**Table 4-3: Values of parameters set in Case 2.**

| Parameter | Value |
|---|---|
| $\mathbf{M}_{ref}$ | $\begin{bmatrix} 0 & 0 \\ 0 & 0 \end{bmatrix}$ |
| $\mathbf{K}_{ref}$ | $\begin{bmatrix} k_{0h} & 0 \\ 0 & 0 \end{bmatrix}$ |
| $\mathbf{C}_{ref}$ | Eq. (3-114) |
| $\boldsymbol{N}_{iter}$ | 1000 |
| $\boldsymbol{\epsilon}$ | 0.001 |
| $\boldsymbol{L}_{window}$ | 10 sec. |
| $\mathbf{max}\left\|\boldsymbol{Eig}\left(\mathbf{A_0}^{-1}\Delta\mathbf{A}\right)\right\|$ | 0 |

### 4.10.3. Case 3

We now repeat Case 2 by using a smaller window size as reported in Table 4-4. By reducing the window size, the response of the system obtained through the HTFD solution converges to one obtained using the ground truth 4-DOF physical model. However, as the choice for the $\mathbf{K}_{ref}$ is not suitable, the convergence is slow. A video showing the progress of the analysis will be provided upon request from the first author.

### 4.10.4. Case 4

Finally, we test easy setting by setting $\mathbf{M}_{ref} = \mathbf{0}$, $\mathbf{K}_{ref} = \mathbf{K}_{static}$, and $\mathbf{C}_{ref} = \mathbf{C}_{\infty}$. As shown in Table 4-5, the convergence criteria, i.e., $\max\left|Eig\left(\mathbf{A_0}^{-1}\Delta\mathbf{A}\right)\right|$, is still very small supporting convergence observed in the results. A video showing the progress of the analysis will be provided upon request from the first author.

**Table 4-4: Values of parameters set in Case 3.**

| Parameter | Value |
|---|---|
| $\mathbf{M}_{ref}$ | $\begin{bmatrix} 0 & 0 \\ 0 & 0 \end{bmatrix}$ |
| $\mathbf{K}_{ref}$ | $\begin{bmatrix} k_{0h} & 0 \\ 0 & 0 \end{bmatrix}$ |
| $\mathbf{C}_{ref}$ | Eq. (3-114) |
| $\boldsymbol{N}_{iter}$ | 1000 |
| $\boldsymbol{\epsilon}$ | 0.001 |
| $\boldsymbol{L}_{window}$ | 1 sec. |
| $\mathbf{max}\left\|\boldsymbol{Eig}\left(\mathbf{A_0}^{-1}\Delta\mathbf{A}\right)\right\|$ | 0 |

**Table 4-5: Values of parameters set in Case 4.**

| Parameter | Value |
|---|---|
| $\mathbf{M}_{ref}$ | $\begin{bmatrix} 0 & 0 \\ 0 & 0 \end{bmatrix}$ |
| $\mathbf{K}_{ref}$ | $\begin{bmatrix} k_{0h} & 0 \\ 0 & k_{0r} \end{bmatrix}$ |
| $\mathbf{C}_{ref}$ | $\mathbf{C}_{\infty}$ |
| $\boldsymbol{N}_{iter}$ | 1000 |
| $\boldsymbol{\epsilon}$ | 0.001 |
| $\boldsymbol{L}_{window}$ | 1 sec. |
| $\mathbf{max}\left\|\boldsymbol{Eig}\left(\mathbf{A_0}^{-1}\Delta\mathbf{A}\right)\right\|$ | 0.0016 |

# CHAPTER 5: OPENSEES MODELING

## 5.1. INTRODUCTION

In this chapter, it is shown how the available tools in Opensees [49] can be used to model frequency-dependency through the HTFD method. We use OpenseesPy [92] for the examples in this chapter.

## 5.2. SINGLE-DEGREE-OF-FREEDOM SUPERSTRUCTURE

The first example studied in this chapter is the model which was extensively used in Chapter 4:Chapter 5 to verify the HTFD method in Matlab. The ground truth model, which we call the physical model, is shown in Figure 5-1(left) in which the frequency-dependent soil-foundation rocking stiffness is modeled through the additional internal rotational DOF. The equivalent model in which the frequency-dependency is being modeled through applying pseudo-force, $f_{ps}$, is shown in Figure 5-1(right). As shown in this figure, rocking dynamic stiffness (spring and dashpot) is set at some reference values. The values of the parameters used for the modeling are reported in Table 5-1 and are identical to those used in Chapter 5.

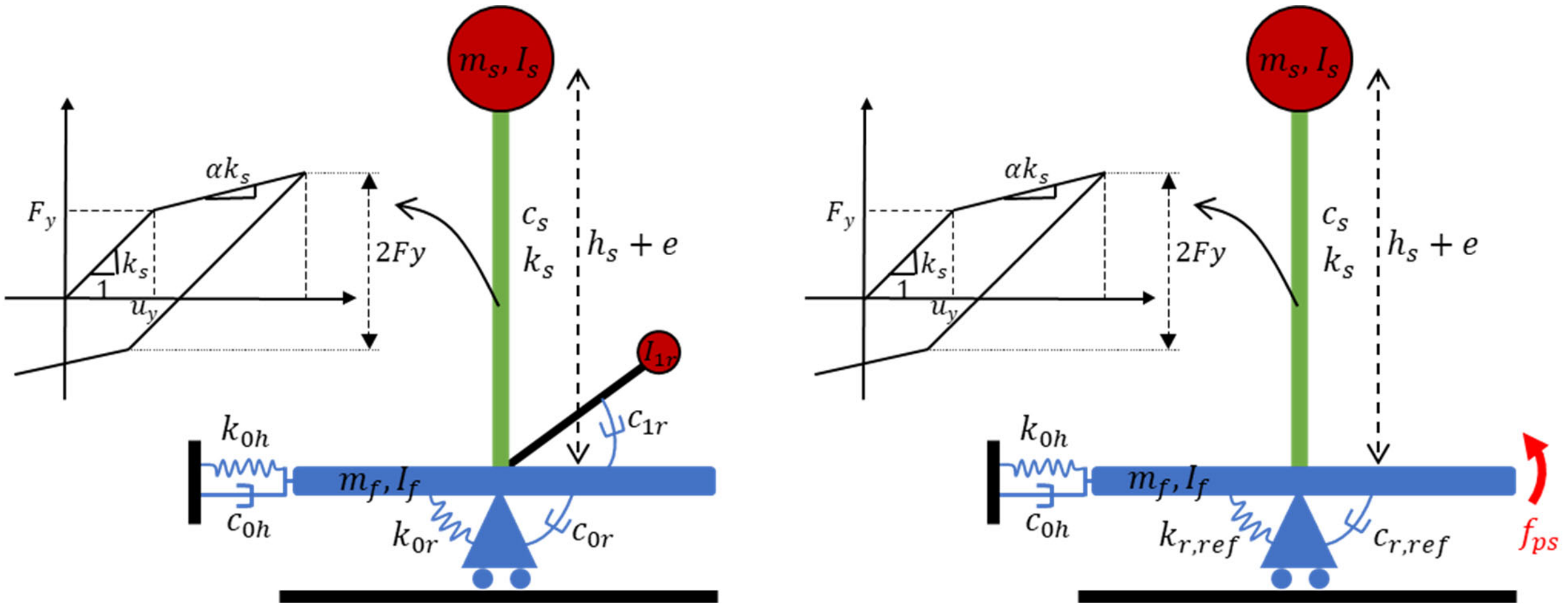

**Figure 5-1: The physical (left) and equivalent (right) models.**

**Table 5-1: Properties of the physical and equivalent models.**

| Parameter | Physical Model | Equivalent Model |
|---|---|---|
| $m_f$ | 0.5 | 0.5 |
| $I_f$ | 8.0 | 8.0 |
| $e$ | 8.0 | 8.0 |
| $m_s$ | 1.0 | 1.0 |
| $I_s$ | 16.0 | 16.0 |
| $k_s$ | 247.0 | 247.0 |
| $c_s$ | 0.63 | 0.63 |
| $h_s$ | 24.0 | 24.0 |
| $k_{0h}$ | 846.0 | 846.0 |
| $c_{0h}$ | 90.0 | 90.0 |
| $k_{0r}$ | 78310.0 | Not Applicable |
| $c_{0r}$ | 406.0 | Not Applicable |
| $I_{1r}$ | 253.0 | Not Applicable |
| $c_{1r}$ | 2982.0 | Not Applicable |
| $m_{r,ref}$ | Not Applicable | 0.0 |
| $k_{r,ref}$ | Not Applicable | 78310.0 |
| $c_{r,ref}$ | Not Applicable | 3227.0 |
| $u_y$ | 0.0009 | 0.0009 |

Figure 5-2 shows the Python code to run the physical model under El Centro ground motion. To understand the lines of this code and the way the physical model is modeled in Opensees, Figure 5-3 is presented. As seen in this figure, 5 nodes are defined for ground, the bottom of the foundation, the middle of the foundation, the superstructure, and internal rotational DOF. The ground node is connected to the bottom of the foundation with a sway-rocking zero-length element composed of springs and dashpots. Then, the bottom of the foundation is rigidly connected to the foundation node which is further connected to the superstructure through a "twoNodeLink" element with an elastic Perfectly-Plastic spring and a linear dashpot. The internal DOF is also connected to the bottom of the foundation using a zero-length linear dashpot. The ground node is fully fixed, while only the vertical DOF of all nodes as well as the translational DOF of the internal DOF are restrained. The system is supposed to have one rotational DOF (in addition to the internal rotation introduced for the frequency-dependency), so the bottom of the foundation and superstructure are connected by an "equalDOF" constraint to make sure their rotation is the same.

```
import openseespy.opensees as ops
import numpy as np
# wipe model
ops.wipe()
# create model
ops.model('basic', '-ndm', 2, '-ndf', 3)
# PARAMETERS
# Superstructures
ns = 1
ms = np.zeros((ns,1))
ks = np.zeros((ns,1))
cs = np.zeros((ns,1))
Is = np.zeros((ns,1))
h = np.zeros((ns,1))
epsy = np.zeros((ns,1))
ms[0,0] = 1.0;
cs[0,0] = 0.6283
ks[0,0] = 246.7401
Is[0,0] = 16
h[0,0] = 24
epsy[0,0] = 9.3722e-04
# Foundation
mf = 0.5
If = 8.0
e = 8.0
# Soil-foundation physical model
m0r = 0.0
c0h = 89.7589
k0h = 845.9661
c0r = 405.7319
k0r = 7.8310e+04
I1r = 253.1025
c1r = 2.9818e+3
# NODE
ops.node(1001, 0.0, 0.0)
ops.node(1002, 0.0, 0.0)
ops.node(1003, 0.0, e/2.0)
ops.node(1, 0.0, h[0,0]+e)
ops.node(100, 0.0, 0.0)
# MASS
ops.mass(1003, mf, 0.0, If)
ops.mass(1, ms[0,0], 0.0, Is[0,0])
ops.mass(100, 0.0, 0.0, I1r)
# FIXITY
ops.fix(1001, 1, 1, 1)
ops.fix(1002, 0, 1, 0)
ops.fix(1003, 0, 1, 0)
ops.fix(1, 0, 1, 0)
ops.fix(100, 1, 1, 0)
ops.rigidLink('beam', 1002, 1003)
ops.equalDOF(1002, 1, 3)
# MATERIAL
ops.uniaxialMaterial('ElasticPP', 1, ks[0,0], epsy[0,0])
ops.uniaxialMaterial('Viscous', 2, cs[0,0], 1)
ops.uniaxialMaterial('Parallel', 3, 1, 2)
ops.uniaxialMaterial('Elastic', 101, k0h)
ops.uniaxialMaterial('Viscous', 102, c0h, 1)
ops.uniaxialMaterial('Elastic', 201, k0r)
ops.uniaxialMaterial('Viscous', 202, c0r, 1)
ops.uniaxialMaterial('Parallel', 103, 101, 102)
ops.uniaxialMaterial('Parallel', 203, 201, 202)
ops.uniaxialMaterial('Viscous', 1002, c1r, 1)
# ELEMENTS
ops.element('twoNodeLink', 1, 1003, 1, '-mat', 3, '-dir', 2, '-orient', 1, 0, 0, '-shearDist', 0.0)
ops.element('zeroLength', 2, 1001, 1002, '-mat', 103, 203, '-dir', 2, 6, '-orient', 0, 1, 0, 1, 0, 0)
ops.element('zeroLength', 3, 1002, 100, '-mat', 1002, '-dir', 6, '-orient', 0, 1, 0, 1, 0, 0); t
# BASE EXCITATION
N = 4000
dt = 0.01
ops.timeSeries('Path', 1, '-dt', dt, '-filePath', 'Record.txt', '-factor', 1.0)
ops.pattern('UniformExcitation', 1, 1, '-accel', 1)
# RECORDER (ABSOLUTE DISPLACEMENT)
tagRc = ops.recorder('Node', '-file', 'Response_Ref.out', '-time', '-dT', dt,'-node', 1001, 1002, 1003, 1, 100, '-dof', 1, 3, 'disp')
# ANALYSIS
ops.constraints('Transformation')
ops.test('RelativeNormUnbalance', 1e-8, 50, 0)
ops.algorithm('Newton')
ops.numberer('Plain')
ops.system('FullGeneral')
ops.integrator('Newmark', 0.5, 0.25)
ops.analysis('Transient')
Run = ops.analyze(N, dt)
# REMOVE RECORDER
ops.remove('recorders')
ops.wipe()
```

**Figure 5-2: A Python code to run the physical model under El Centro ground motion.**

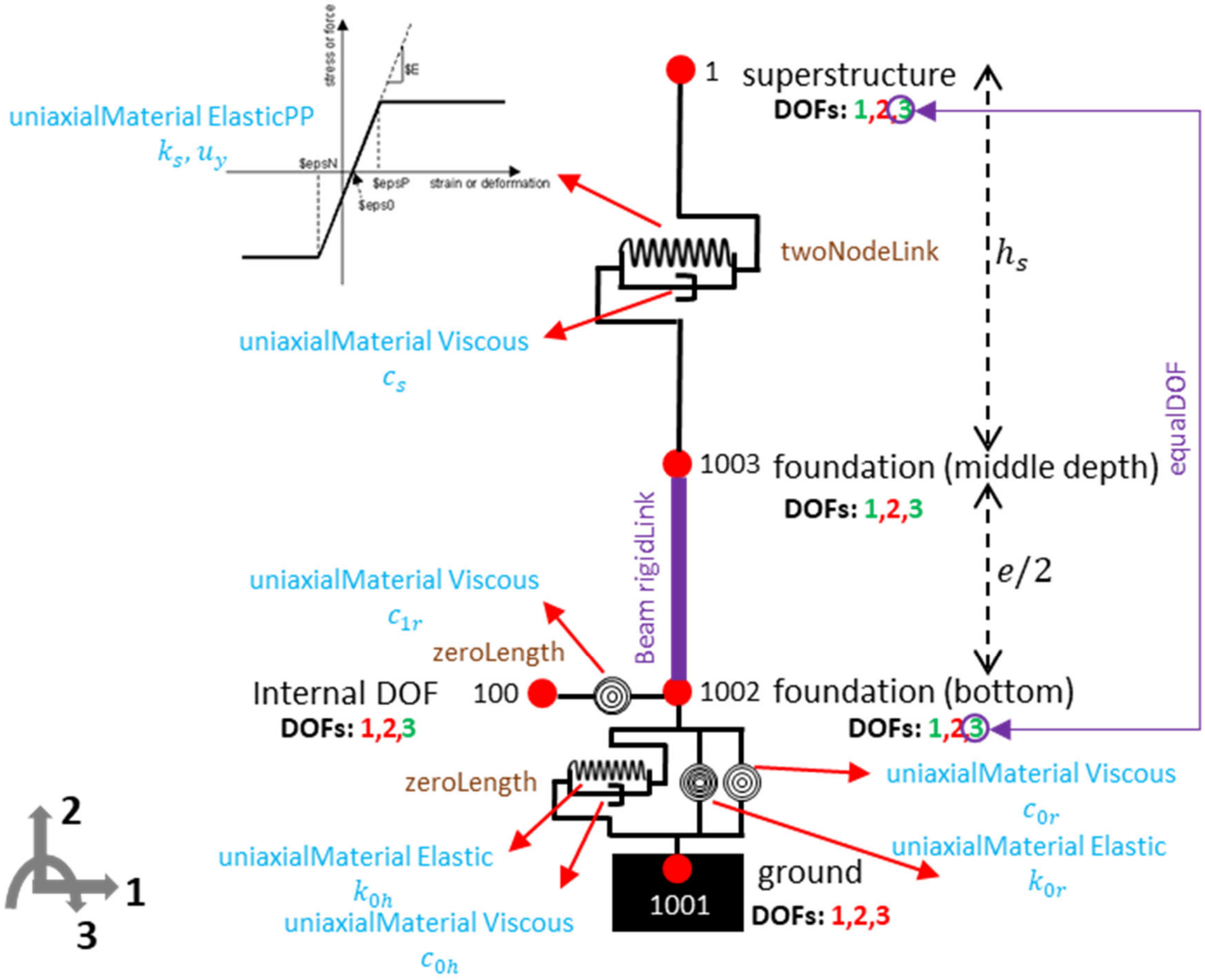

**Figure 5-3: Opensees model of the physical model.**

To verify the Opensees model of the physical system, Figure 5-4 is shown. As shown in this figure, the displacement responses obtained from the Opensees model are identical to the corresponding responses obtained from the Matlab model used in Chapter 4.

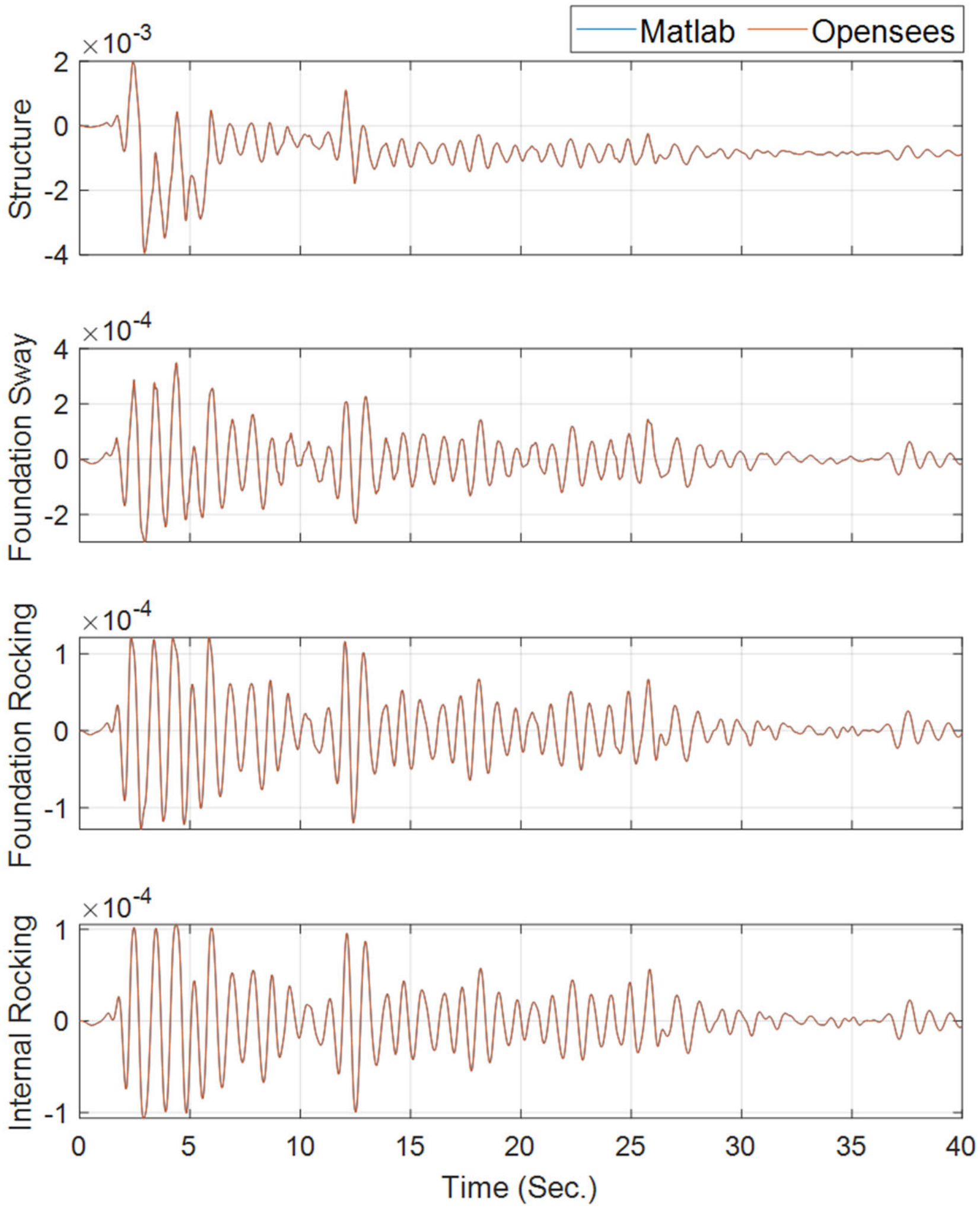

**Figure 5-4: Comparison between responses of the physical model in Opensees and Matlab.**

Figure 5-5 shows how the equivalent model is modeled in Opensees. As seen, it is almost identical to the physical model with three main differences:

- the internal rotational mass and its connected rotational dashpot are removed;
- the coefficients of rocking springs and dashpots are changed from $k_{0r}$ and $c_{0r}$ to $k_{r,ref}$ and $c_{r,ref}$, respectively;
- an unknown pseudo-force (moment in this example) is applied to the foundation node;

The Python code for this equivalent model is shown in Figure 5-6. The first half of this code is similar to Figure 5-2 except few differences mentioned above, but the second half is new and contains the HTFD implementation. Here are the steps carried out to run the model using the HTFD method:

1. A load time series, the pseudo-force, is defined with zero values and applied to node 1002;

2. The model is run in the first time window under the load above and ground motion;
3. rocking displacement (rotation at node 1002) is recorded;
4. The recorded displacement is transformed to the frequency domain after proper zero padding;
5. The true impedance function is multiped by the recorded response;
6. The reference impedance function (impedance function of the equivalent system) is multiplied by the recorded response;
7. The difference between the two moments above, which is the pseudo-force, is calculated;
8. The error between this new force and the initial guess, which is zero at the first iteration, is compared with the threshold;
9. If the error is less than the threshold, the convergence at this window is achieved and steps 1 to 8 are repeated for the next window. If the convergence is not achieved, the pseudo-force is replaced by the new force and steps 2 to 8 are repeated.

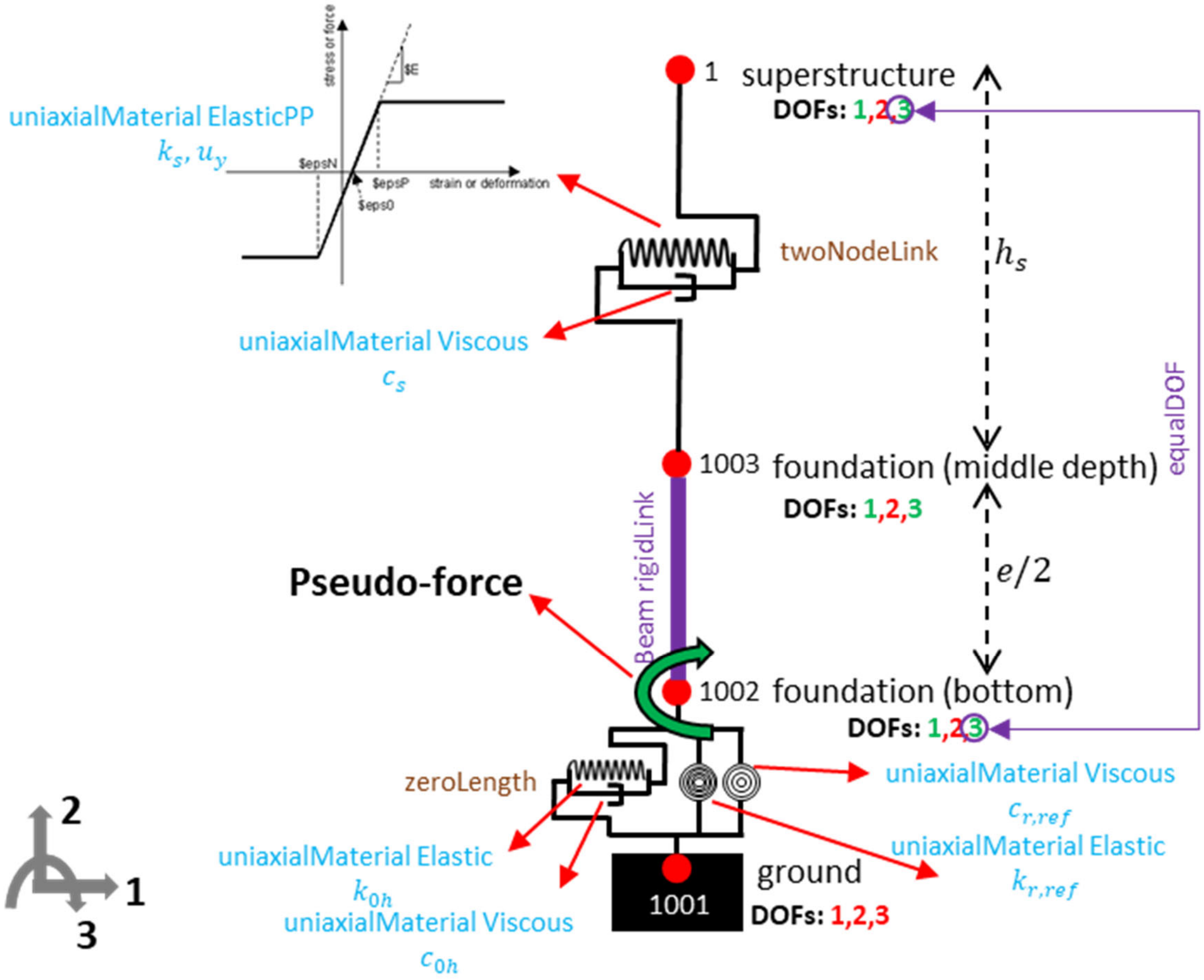

**Figure 5-5: Opensees model of the equivalent model.**

```
import openseespy.opensees as ops
import numpy as np
import matplotlib.pyplot as plt
import time
import os
plt.rcParams['figure.dpi'] = 300
plt.rcParams['savefig.dpi'] = 300
##################################  FUNCTIONS ######################################
def Forward_FFT(x,N1,N2):
   Temp = np.fft.fft(x,N1)/N1; X  = Temp[0:N2];
   return X
def Inverse_FFT(X,N1,N2):
   temp1 = N1*X;  temp2 = np.flip(np.conjugate(temp1))
   if N1%2==0:
      Temp = np.real(np.fft.ifft(np.append(temp1,temp2[1:]),N1))
   else:
      Temp = np.real(np.fft.ifft(np.append(temp1,temp2),N1))
   x = Temp[0:N2]
   return x
##################################################################################
# wipe model
ops.wipe()
# create model
ops.model('basic', '-ndm', 2, '-ndf', 3)
# plot model
# opsplt.plot_model()
# PARAMETERS
# Superstructures
ns = 1; ms = np.zeros((ns,1)); ks = np.zeros((ns,1)); cs = np.zeros((ns,1)); Is = np.zeros((ns,1)); h = np.zeros((ns,1)); epsy = np.zeros((ns,1))
ms[0,0] = 1.0; cs[0,0] = 0.6283; ks[0,0] = 246.7401; Is[0,0] =  16.0; h[0,0] = 24.0; epsy[0,0] = 9.3722e-04;
# Foundation
mf = 0.5; If = 8.0; e = 8.0; c0h = 89.759790102565490; k0h = 8.459660915219447e+02;
# Reference  impedance function for the frequency-dependent part
m_ref = 0.0; k_ref =  7.831014112986252e+04; c_ref = 3.227678266639782e+03;
# NODE
ops.node(1001, 0.0, 0.0); ops.node(1002, 0.0, 0.0); ops.node(1003, 0.0, e/2.0); ops.node(1, 0.0, h[0,0]+e);
# MASS
ops.mass(1003, mf, 0.0, If);ops.mass(1, ms[0,0], 0.0, Is[0,0]);
# FIXITY
ops.fix(1001, 1, 1, 1);ops.fix(1002, 0, 1, 0);ops.fix(1003, 0, 1, 0);ops.fix(1, 0, 1, 0);ops.rigidLink('beam', 1002, 1003);ops.equalDOF(1002, 1, 3);
# MATERIAL
ops.uniaxialMaterial('ElasticPP', 1, ks[0,0], epsy[0,0]); ops.uniaxialMaterial('Viscous', 2, cs[0,0], 1); ops.uniaxialMaterial('Parallel', 3, 1, 2); ops.uniaxialMaterial('Elastic', 101, k0h);
ops.uniaxialMaterial('Viscous', 102, c0h, 1)
ops.uniaxialMaterial('Elastic', 201, k_ref); ops.uniaxialMaterial('Viscous', 202, c_ref, 1); ops.uniaxialMaterial('Parallel', 103, 101, 102); ops.uniaxialMaterial('Parallel', 203, 201, 202);
# ELEMENTS
ops.element('twoNodeLink', 1, 1003, 1, '-mat', 3, '-dir', 2, '-orient', 1, 0, 0, '-shearDist', 0.0); ops.element('zeroLength',  2, 1001, 1002, '-mat', 103, 203, '-dir', 2, 6, '-orient', 0, 1, 0, 1, 0, 0)
# BASE EXCITATION
N = 4000; dt = 0.01; ops.timeSeries('Path', 1, '-dt', dt, '-filePath', 'Record.txt', '-factor', 1.0); ops.pattern('UniformExcitation', 1, 1, '-accel', 1)
# ANALYSIS
ops.constraints('Transformation'); ops.test('RelativeNormUnbalance', 1e-8, 50, 1); ops.algorithm('Newton'); ops.numberer('Plain'); ops.system('FullGeneral');
ops.integrator('Newmark', 0.5, 0.25); ops.analysis('Transient')
######################################  HTFD ########################################
# SETTING BY THE USER
L_W = 1000; tol = 0.001; MaxIter = 1000; NDecay = 100; NZero = 100
# FREQUENCY DOMAIN SETTING
N_E = int(N+NDecay+NZero)
if N_E%2==0:
   Nf = int(1+N_E/2)
else:
   Nf = int(1+(N_E-1)/2)
fs = 1/dt # Sampling frequency;df = fs/N_E; f = np.arange(0,Nf*df,df);
# FREQUENCY DEPENDENT IMPEDANCE FUNCTION (IT CAN BE IMPORTED FROM AN EXTERNAL FILE)
w = 2*np.pi*f
S = (7.8310e+04-2.981e+03**2*2.53e+02*w**2/(2.981e+03**2+2.53e+02**2*w**2))+1j*w*(4.05e+02+2.981e+03*2.53e+02**2*w**2/(2.981e+03**2+2.53e+02**2*w**2))
# FREQUENCY INDEPENDENT IMPEDANCE FUNCTION
S_Ref = k_ref-m_ref*w**2+1j*w*c_ref
# INITIALIZATION
NDOF = 3
d_E = np.zeros((NDOF,N_E));PsF = np.zeros((1,N));F_Int = np.zeros((1,N));PsF_old = np.zeros((1,N));N_W = round(N/L_W);
Windows = np.append(np.arange(0,N_W*L_W+1,L_W),N); Windows = np.unique(Windows);
np.savetxt('Load.txt', PsF.T); ops.timeSeries('Path', 10, '-dt', dt, '-filePath', 'Load.txt', '-factor', 1.0); ops.pattern('Plain', 101, 10); ops.load(1002, 0.0, 0.0, 1.0)
for k_window in range(1,np.size(Windows)):
   N_R = Windows[k_window]
   # Convergence loop
   iteration = 0;   convlock = 0;
   while (convlock == 0) and (iteration < MaxIter):
      iteration = iteration+1
      # RESPONSE CALCULATION OF THE REFERENCE SYSTEM IN TIME DOMAIN
      ops.reset();  ops.setTime(0.0); d = np.zeros((NDOF,N_R))
      for n in range(1,N_R): # Note: Opensees does not output response at t=0, so n starts at 1 and t=0 is implicitly considered by the definition of d!
         ops.analyze(1, dt); d[0,n] = ops.nodeResponse(1, 1, 1); d[1,n] = ops.nodeResponse(1002, 1, 1); d[2,n] = -ops.nodeResponse(1002, 3, 1
      # TRANSFERING RESPONSES TO THE FREQUENCY DOMAIN
      DISP = np.zeros((NDOF,Nf),dtype=complex
      for IDOF in range(0,NDOF):
         y0 = d[IDOF,N_R-1];  yd0 = d[IDOF,N_R-1]-d[IDOF,N_R-2]; y1 = 0; yd1 = 0;
         Coeff = np.matmul(np.linalg.inv(np.matrix([[0,0,0,1],[0,0,1,0],[NDecay**3,NDecay**2,NDecay,1],[3*NDecay**2,2*NDecay,1,0]])),(np.array([y0,yd0,y1,yd1])).T)
         E_Part = np.matmul(np.column_stack(((np.arange(1,NDecay+1,1)**3).T, (np.arange(1,NDecay+1,1)**2).T,(np.arange(1,NDecay+1,1)).T, np.ones((NDecay)))),Coeff.T)
         d_E[IDOF,0:N_R-1+NDecay+1] = np.append(d[IDOF,:],E_Part); DISP[IDOF,:] = Forward_FFT(d_E[IDOF,:],N_E,Nf)
      # UPDATE PSEUDO-FORCE
      F_Exp = Inverse_FFT(S*DISP[2,:],N_E,N)
      F_Ref = k_ref*Inverse_FFT(DISP[2,:],N_E,N)+c_ref*Inverse_FFT(w*1j*DISP[2,:],N_E,N)+m_ref*Inverse_FFT(-w**2*DISP[2,:],N_E,N)
      PsF[0,Windows[k_window-1]:N_R-1] = F_Exp[Windows[k_window-1]:N_R-1]-F_Ref[Windows[k_window-1]:N_R-1]
      # CONVERGENCE CHECK
      PsF_Er = np.linalg.norm(PsF[0,Windows[k_window-1]:N_R-1]-PsF_old[0,Windows[k_window-1]:N_R-1])/np.linalg.norm(PsF[0,Windows[k_window-1]:N_R-1])
      PsF_old = np.copy(PsF)
      if PsF_Er<=tol:
         convlock=1
# REPLACING OLD FORCE WITH NEW ONE
      ops.remove ('timeSeries',10); ops.remove ('loadPattern',101); np.savetxt('Load.txt', PsF.T); ops.timeSeries('Path', 10, '-dt', dt, '-filePath', 'Load.txt', '-factor', 1.0);
ops.pattern('Plain', 101, 10); ops.load(1002, 0.0, 0.0, 1.0);
```

**Figure 5-6: A Python code to run the equivalent model under El Centro ground motion through the HTFD method.**

Figure 5-7 shows the comparison between exact responses and the final responses obtained through the HTFD method. As seen in this figure, the Opensees implementation shows perfect performance similar to the Matlab implementation presented in Chapter 5. A video showing the progress of the analysis is available upon request from the first author.

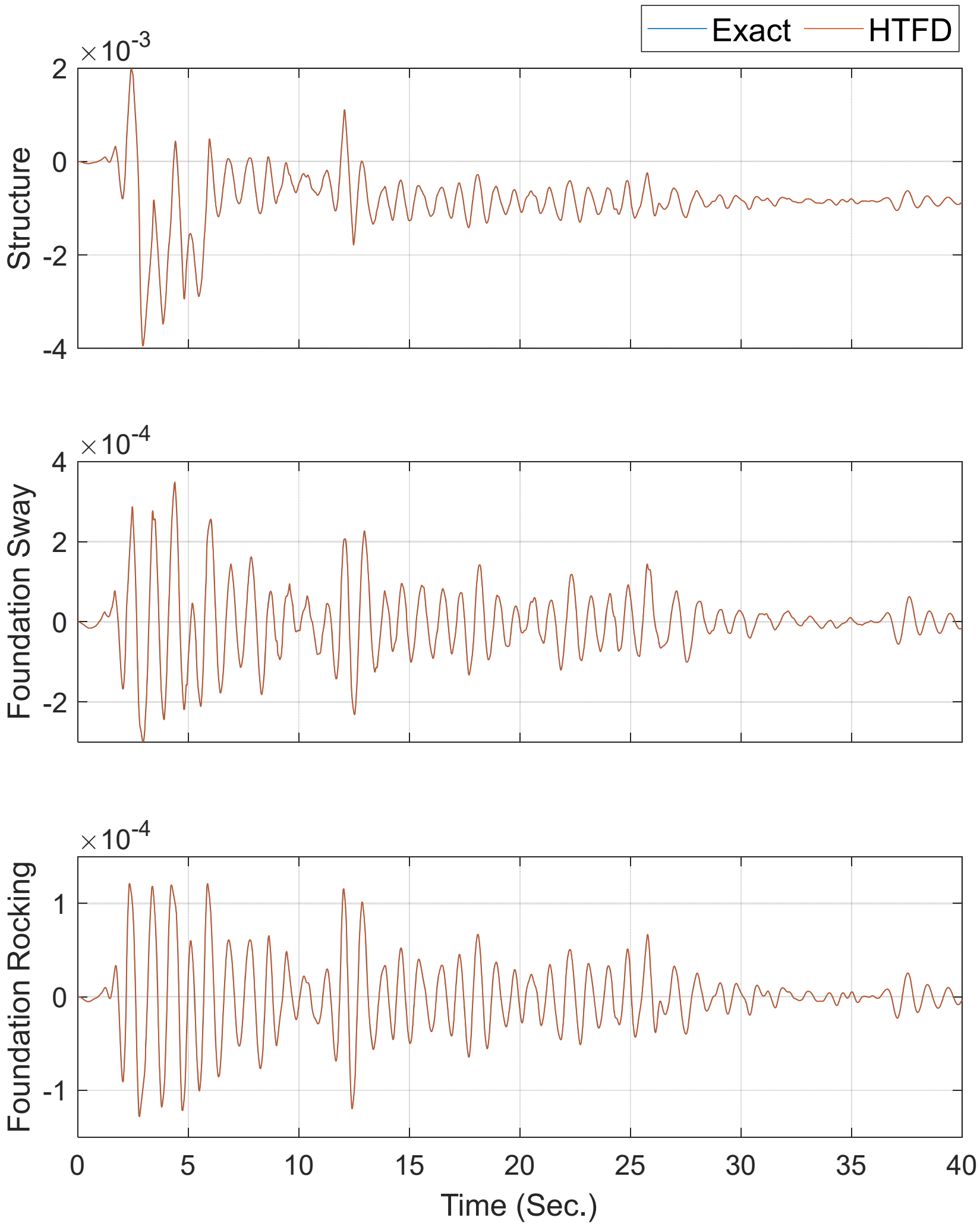

**Figure 5-7: Comparison between responses of the physical and equivalent (HTFD) models in Opensees.**

## 5.3. MULTI-DEGREE-OF-FREEDOM SUPERSTRUCTURE

The previous example does not show the importance of the HTFD implementation in Opensees because it was a simple example that can be easily solved in Matlab as shown in Chapter 5. Herein, we extend the example by replacing the SDOF superstructure with a Multi-Degree-Of-Freedom (MDOF) superstructure as displayed in Figure 5-8. Similar to the previous case, the superstructure is on top of a sway-rocking soil-foundation substructure in which the frequency-dependent rocking impedance function is modeled through the internal rotational DOF in the physical model Figure 5-8(left). The equivalent model in which frequency-dependency is replaced by an unknown pseudo-force (moment) at the foundation node is shown in Figure 5-8(right). The properties of both systems are reported in Table 5-2. Note that the mass, stiffness, and height of all stories are the same, while their strength, represented by yield interstory displacement capacity is different along the height of the building. To model the damping in the superstructure, a Rayleigh model with mass- and stiffness-proportional damping of 0.78 and 0.0024, respectively, is used in both physical and equivalent models. As shown in Table 5-2, we run the equivalent model using two different reference properties. In both models we assume $m_{r,ref} = 0$ and $c_{r,ref} = c_{r,inf} = 9.37 \times 10^8$; however, in Model 1 we set $k_{r,ref} = k_{0r}$ and in Model 2 we set $k_{r,ref} = 0$.

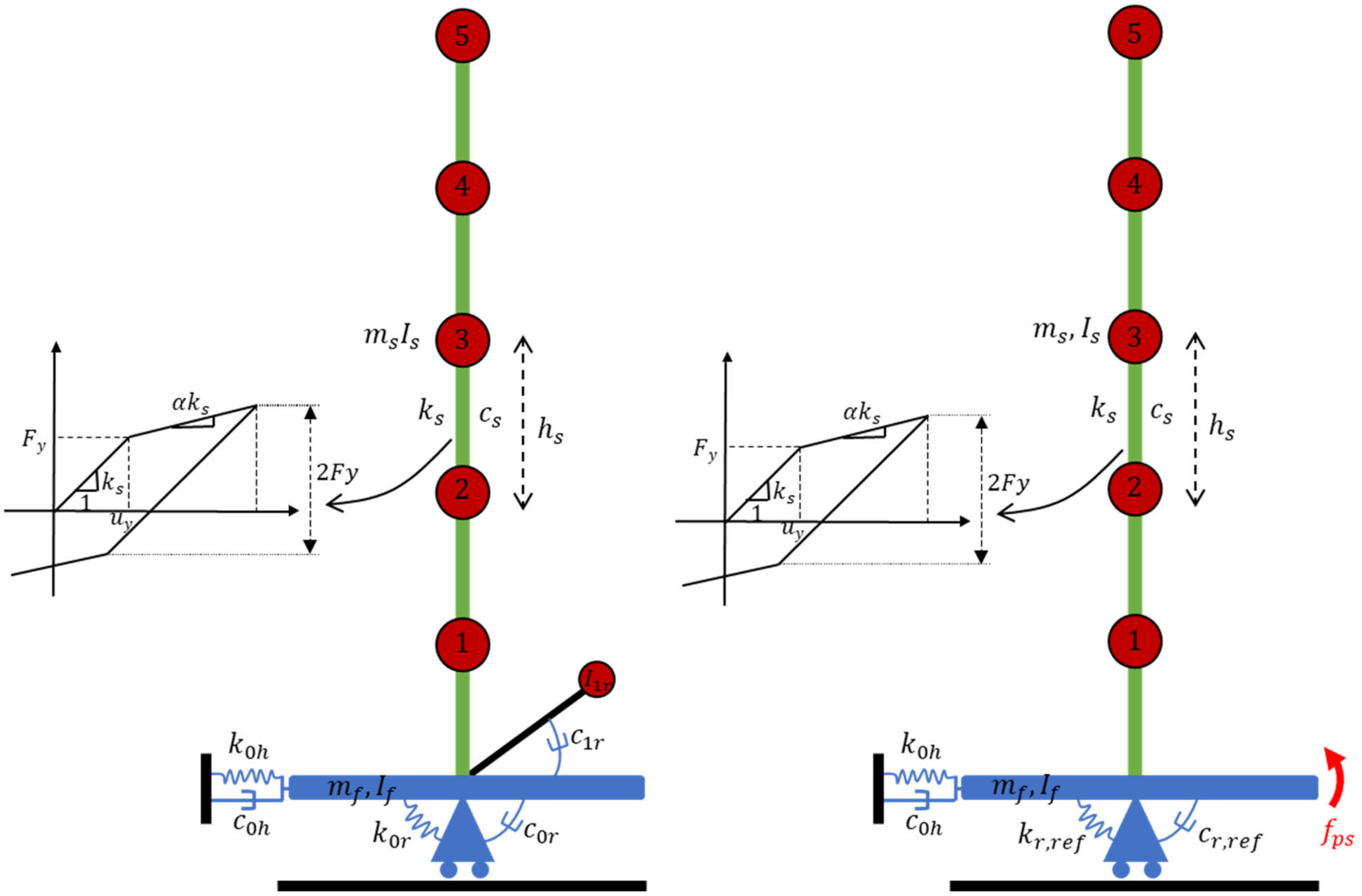

**Figure 5-8: The physical (left) and equivalent (right) models.**

Table 5-2: Properties of the physical and equivalent models.

| Parameter | Physical Model | Equivalent Model 1 | Equivalent Model 2 |
|---|---|---|---|
| $\boldsymbol{m_f}$ | $4.85 \times 10^3$ | $4.85 \times 10^3$ | $4.85 \times 10^3$ |
| $\boldsymbol{I_f}$ | $4.58 \times 10^4$ | $4.58 \times 10^4$ | $4.58 \times 10^4$ |
| $\boldsymbol{e}$ | 3.07 | 3.07 | 3.07 |
| $\boldsymbol{m_s}$ | $9.70 \times 10^3$ | $9.70 \times 10^3$ | $9.70 \times 10^3$ |
| $\boldsymbol{I_s}$ | $9.17 \times 10^4$ | $9.17 \times 10^4$ | $9.17 \times 10^4$ |
| $\boldsymbol{k_s}$ | $1.31 \times 10^7$ | $1.31 \times 10^7$ | $1.31 \times 10^7$ |
| $\boldsymbol{h_s}$ | 3.5 | 3.5 | 3.5 |
| $\boldsymbol{k_{0h}}$ | $5.53 \times 10^8$ | $5.53 \times 10^8$ | $5.53 \times 10^8$ |
| $\boldsymbol{c_{0h}}$ | $2.14 \times 10^7$ | $2.14 \times 10^7$ | $2.14 \times 10^7$ |
| $\boldsymbol{k_{0r}}$ | $3.26 \times 10^{10}$ | Not Applicable | Not Applicable |
| $\boldsymbol{c_{0r}}$ | $1.11 \times 10^8$ | Not Applicable | Not Applicable |
| $\boldsymbol{I_{1r}}$ | $4.44 \times 10^7$ | Not Applicable | Not Applicable |
| $\boldsymbol{c_{1r}}$ | $8.28 \times 10^8$ | Not Applicable | Not Applicable |
| $\boldsymbol{m_{r,ref}}$ | Not Applicable | 0.0 | 0.0 |
| $\boldsymbol{k_{r,ref}}$ | Not Applicable | $k_{0r}$ | 0.0 |
| $\boldsymbol{c_{r,ref}}$ | Not Applicable | $c_{r,inf} = 9.37 \times 10^8$ | $c_{r,inf} = 9.37 \times 10^8$ |
| $\boldsymbol{u_{y,1}}$ | 0.0099 | 0.0099 | 0.0099 |
| $\boldsymbol{u_{y,2}}$ | 0.0092 | 0.0092 | 0.0092 |
| $\boldsymbol{u_{y,3}}$ | 0.0079 | 0.0079 | 0.0079 |
| $\boldsymbol{u_{y,4}}$ | 0.0059 | 0.0059 | 0.0059 |
| $\boldsymbol{u_{y,5}}$ | 0.0033 | 0.0033 | 0.0033 |

The Python codes to run the physical and equivalent models are shown in Figure 5-9 and Figure 5-10, respectively. The details of the models are very similar to the SDOF case and are not repeated here.

```
import openseespy.opensees as ops
import numpy as np
# wipe model
ops.wipe()
# create model
ops.model('basic', '-ndm', 2, '-ndf', 3)
# PARAMETERS
# Superstructures
ns = 5
ms = 9.7014e3*np.ones((ns,1))
ks = 1.3132e7*np.ones((ns,1))
cs = np.zeros((ns,1))
Is = 4.5842e5/ns*np.ones((ns,1))
h = 3.5*np.ones((ns,1))
H = np.cumsum(h)
alphaM = 0.78
betaKinit = 0.0024
epsy = np.array(([0.0099,0.0092,0.0079,0.0059,0.0033]))
# Soil-foundation physical model
deltaM = 0.0
m0r = 0.0
c0h = 2.1377e7
k0h = 5.5335e8
c0r = 1.1138e+08
k0r = 3.2611e+10
I1r = 4.4365e+07
c1r = 8.2829e+08
mf = 4.8507e3
If = 4.5842e4+deltaM
e = 3.0742
# NODE
ops.node(1001, 0.0, 0.0)
ops.node(1002, 0.0, 0.0)
ops.node(1003, 0.0, e/2.0)
ops.node(100, 0.0, 0.0)
for i in range(ns):
    ops.node(i+1, 0.0, H[i]+e)
# MASS
ops.mass(1003, mf, 0.0, If)
ops.mass(100, 0.0, 0.0, I1r)
for i in range(ns):
    ops.mass(i+1, ms[i,0], 0.0, Is[i,0])
# FIXITY
ops.fix(1001, 1, 1, 1)
ops.fix(1002, 0, 1, 0)
ops.fix(1003, 0, 1, 0)
ops.fix(100, 1, 1, 0)
ops.rigidLink('beam', 1002, 1003)
for i in range(ns):
    ops.fix(i+1, 0, 1, 0)
    ops.equalDOF(1002, i+1, 3)
# MATERIAL
for i in range(ns):
    ops.uniaxialMaterial('ElasticPP', i+1, ks[i,0], epsy[i])
ops.uniaxialMaterial('Elastic', 101, k0h)
ops.uniaxialMaterial('Viscous', 102, c0h, 1)
ops.uniaxialMaterial('Elastic', 201, k0r)
ops.uniaxialMaterial('Viscous', 202, c0r, 1)
ops.uniaxialMaterial('Parallel', 103, 101, 102)
ops.uniaxialMaterial('Parallel', 203, 201, 202)
ops.uniaxialMaterial('Viscous', 1002, c1r, 1)
# ELEMENTS
ops.element('twoNodeLink', 1, 1003, 1, '-mat', 1, '-dir', 2, '-orient', 1, 0, 0, '-shearDist', 0.0)
for i in range(ns-1):
    ops.element('twoNodeLink', i+2, i+1, i+2, '-mat', i+2, '-dir', 2, '-orient', 1, 0, 0, '-shearDist', 0.0)
ops.element('zeroLength', 200, 1001, 1002, '-mat', 103, 203, '-dir', 2, 6, '-orient', 0, 1, 0, 1, 0, 0)
ops.element('zeroLength', 300, 1002, 100, '-mat', 1002, '-dir', 6, '-orient', 0, 1, 0, 1, 0, 0);
# BASE EXCITATION
N = 4000
dt = 0.01
ops.timeSeries('Path', 1, '-dt', dt, '-filePath', 'Record.txt', '-factor', 9.806)
ops.pattern('UniformExcitation', 1, 1, '-accel', 1)
# RECORDER (ABSOLUTE DISPLACEMENT)
tagRc = ops.recorder('Node', '-file', 'Response_Ref.out', '-time', '-dT', dt,'-node', 1001, 1002, 1003, 1, 2, 3, 4, 5, '-dof', 1, 3, 'disp')
# DAMPING
ops.region(1,'-node', 1, 2, 3, 4, 5, '-rayleigh', alphaM, 0.0, betaKinit, 0.0)
# EIGEN ANALYSIS
lambda2 = ops.eigen('-fullGenLapack', ns+3)
# ANALYSIS
ops.constraints('Transformation')
ops.test('RelativeNormUnbalance', 1e-12, 50, 0)
ops.algorithm('Newton')
ops.numberer('Plain')
ops.system('FullGeneral')
ops.integrator('Newmark', 0.5, 0.25)
ops.analysis('Transient')
Run = ops.analyze(N, dt)
# REMOVE RECORDER
ops.remove('recorders')
ops.wipe()
```

**Figure 5-9: A Python code to run the physical model under El Centro ground motion.**

```
import openseespy.opensees as ops
import numpy as np
import matplotlib.pyplot as plt
import time
import os
# wipe model
ops.wipe()
# create model
ops.model('basic', '-ndm', 2, '-ndf', 3)
# PARAMETERS
# Superstructures
ns = 5; ms = 9.7014e3*np.ones((ns,1)); ks = 1.3132e7*np.ones((ns,1)); cs = np.zeros((ns,1)); Is = 4.5842e5/ns*np.ones((ns,1)); h = 3.5*np.ones((ns,1)); H = np.cumsum(h)
alphaM = 0.78
betaKinit = 0.0024
epsy = np.array(([0.0099,0.0092,0.0079,0.0059,0.0033]))
# Soil-foundation physical model
deltaM = 0; c0h = 2.1377e7; k0h = 5.5335e8; mf = 4.8507e3; If = 4.5842e4+deltaM; e = 3.0742
# BASE EXCITATION
N = 4000; dt = 0.01; ops.timeSeries('Path', 1, '-dt', dt, '-filePath', 'Record.txt', '-factor', 9.806); ops.pattern('UniformExcitation', 1, 1, '-accel', 1)
# INITIAL SETTINGS OF HTFD METHOD
# Parameters must be adjusted by user
L_W = 1000; tol = 0.001; MaxIter = 100; NDecay = 100;NZero = 100
# FFT parameters
N_E = int(N+NDecay+NZero)
if N_E%2==0:
    Nf = int(1+N_E/2)
else:
    Nf = int(1+(N_E-1)/2)
fs = 1/dt; df = fs/N_E; f = np.arange(0,Nf*df,df); w = 2*np.pi*f; dw = 2*np.pi*df
# True frequency-dependent impedance function (it can be imported from a given data file)
c0r = 1.1138e+08; k0r = 3.2611e+10; I1r = 4.4365e+07; c1r = 8.2829e+08; S = (k0r-c1r**2*I1r*w**2/(c1r**2+I1r**2*w**2))+1j*w*(c0r+c1r*I1r**2*w**2/(c1r**2+I1r**2*w**2))
# Reference impedance function for the frequency-dependent DOF
gamma = 0.5; beta = 0.25; Temp = np.diff(np.diff(S))/(dw**2); m_inf = -0.5*np.real(Temp[len(Temp)-1]); Temp = S[len(w)-1]/w[len(w)-1]; c_inf = np.imag(Temp)
Temp = S+m_inf*w**2; k_inf = np.real(Temp[len(Temp)-1]); m_ref = 0.0; k_ref = k0r; Sr = S-(k_inf-m_inf*w**2+1j*w*c_inf); S0 = 2/np.pi*np.sum(np.real(Sr))*dw
c_ref = c_inf;#c_inf+beta*dt/gamma*(k_inf-k_ref+S0*dt)+1/gamma/dt*(m_inf-m_ref)
# NODE
ops.node(1001, 0.0, 0.0); ops.node(1002, 0.0, 0.0); ops.node(1003, 0.0, e/2.0)
for i in range(ns):
    ops.node(i+1, 0.0, H[i]+e)
# MASS
ops.mass(1003, mf, 0.0, If)
for i in range(ns):
    ops.mass(i+1, ms[i,0], 0.0, Is[i,0])
# FIXITY
ops.fix(1001, 1, 1, 1); ops.fix(1002, 0, 1, 0); ops.fix(1003, 0, 1, 0); ops.rigidLink('beam', 1002, 1003);
for i in range(ns):
    ops.fix(i+1, 0, 1, 0);     ops.equalDOF(1002, i+1, 3)
# MATERIAL
for i in range(ns):
    ops.uniaxialMaterial('ElasticPP', i+1, ks[i,0], epsy[i])  # Interstory behavior
ops.uniaxialMaterial('Elastic', 101, k0h); ops.uniaxialMaterial('Viscous', 102, c0h, 1); ops.uniaxialMaterial('Elastic', 201, k_ref); ops.uniaxialMaterial('Viscous', 202, c_ref, 1)
ops.uniaxialMaterial('Parallel', 103, 101, 102); ops.uniaxialMaterial('Parallel', 203, 201, 202)
# ELEMENTS
ops.element('twoNodeLink', 1, 1003, 1, '-mat', 1, '-dir', 2, '-orient', 1, 0, 0, '-shearDist', 0.0)
for i in range(ns-1):
    ops.element('twoNodeLink', i+2, i+1, i+2, '-mat', i+2, '-dir', 2, '-orient', 1, 0, 0, '-shearDist', 0.0)
ops.element('zeroLength', 200, 1001, 1002, '-mat', 103, 203, '-dir', 2, 6, '-orient', 0, 1, 0, 1, 0, 0)
# DAMPING
ops.region(1,'-node', 1, 2, 3, 4, 5, '-rayleigh', alphaM, 0.0, betaKinit, 0.0)
# ANALYSIS
ops.constraints('Transformation’); ops.test('RelativeNormUnbalance', 1e-12, 50, 1); ops.algorithm('Newton’); ops.numberer('Plain’); ops.system('FullGeneral')
ops.integrator('Newmark', 0.5, 0.25);ops.analysis('Transient')
####################################  HTFD ######################################
# FREQUENCY INDEPENDENT IMPEDANCE FUNCTION
S_Ref = k_ref-m_ref*w**2+1j*w*c_ref
# INITIALIZATION
NDOF = 3
d_E = np.zeros((NDOF,N_E)); PsF = np.zeros((1,N)); F_Int = np.zeros((1,N)); PsF_old = np.zeros((1,N)); N_W = round(N/L_W);
Windows = np.append(np.arange(0,N_W*L_W+1,L_W),N); Windows = np.unique(Windows); np.savetxt('Load.txt', PsF.T);
ops.timeSeries('Path', 10, '-dt', dt, '-filePath', 'Load.txt', '-factor', 1.0); ops.pattern('Plain', 101, 10); ops.load(1002, 0.0, 0.0, 1.0)
for k_window in range(1,np.size(Windows)):
    N_R = Windows[k_window]
    # Convergence loop
    iteration = 0; convlock = 0
    while (convlock == 0) and (iteration < MaxIter):
        iteration = iteration+1
        # RESPONSE CALCULATION OF THE REFERENCE SYSTEM IN TIME DOMAIN
        ops.reset(); ops.setTime(0.0); d = np.zeros((NDOF,N_R))
        for n in range(1,N_R):
            ops.analyze(1, dt); d[0,n] = ops.nodeResponse(5, 1, 1); d[1,n] = ops.nodeResponse(1002, 1, 1); d[2,n] = -ops.nodeResponse(1002, 3, 1);
        # TRANSFERING RESPONSES TO THE FREQUENCY DOMAIN
        DISP = np.zeros((NDOF,Nf),dtype=complex)
        for IDOF in range(0,NDOF):
            y0 = d[IDOF,N_R-1]; yd0 = d[IDOF,N_R-1]-d[IDOF,N_R-2]; y1 = 0; yd1 = 0
            Coeff = np.matmul(np.linalg.inv(np.matrix([[0,0,0,1],[0,0,1,0],[NDecay**3,NDecay**2,NDecay,1],[3*NDecay**2,2*NDecay,1,0]])),(np.array([y0,yd0,y1,yd1])).T)
            E_Part = np.matmul(np.column_stack(((np.arange(1,NDecay+1,1)**3).T, (np.arange(1,NDecay+1,1)**2).T,(np.arange(1,NDecay+1,1)).T, np.ones((NDecay)))),Coeff.T)
            d_E[IDOF,0:N_R-1+NDecay+1] = np.append(d[IDOF,:],E_Part); DISP[IDOF,:] = Forward_FFT(d_E[IDOF,:],N_E,Nf)
        # UPDATE PSEUDO-FORCE
        F_Exp = Inverse_FFT(S*DISP[2,:],N_E,N) ;        F_Ref = k_ref*Inverse_FFT(DISP[2,:],N_E,N)+c_ref*Inverse_FFT(w*1j*DISP[2,:],N_E,N)+m_ref*Inverse_FFT(-w**2*DISP[2,:],N_E,N)
        PsF[0,Windows[k_window-1]:N_R-1] = F_Exp[Windows[k_window-1]:N_R-1]-F_Ref[Windows[k_window-1]:N_R-1]
        # CONVERGENCE CHECK
        PsF_Er = np.linalg.norm(PsF[0,Windows[k_window-1]:N_R-1]-PsF_old[0,Windows[k_window-1]:N_R-1])/np.linalg.norm(PsF[0,Windows[k_window-1]:N_R-1])
        PsF_old = np.copy(PsF)
        if PsF_Er<=tol:
            convlock=1
 # REPLACING OLD FORCE WITH NEW ONE
        ops.remove ('timeSeries',10)
        ops.remove ('loadPattern',101)
        np.savetxt('Load.txt', PsF.T)
        ops.timeSeries('Path', 10, '-dt', dt, '-filePath', 'Load.txt', '-factor', 1.0)
        ops.pattern('Plain', 101, 10)
        ops.load(1002, 0.0, 0.0, 1.0)
```

**Figure 5-10: A Python code to run the equivalent model under El Centro ground motion.**

Figure 5-11 shows the comparison between exact responses and the final responses obtained through the HTFD method using Models 1 and 2. Note that a time window with a length of 10 seconds is used for Model 1, while a window with a length of 0.5 seconds is used for Model 2 which significantly increases computational time. As seen in this figure, the Opensees implementation of the HTFD method shows perfect performance in both cases, but considering the computational time, a reasonable setting for the reference model is recommended. A video showing the progress of the analysis for Model 1 is available upon request from the first author.

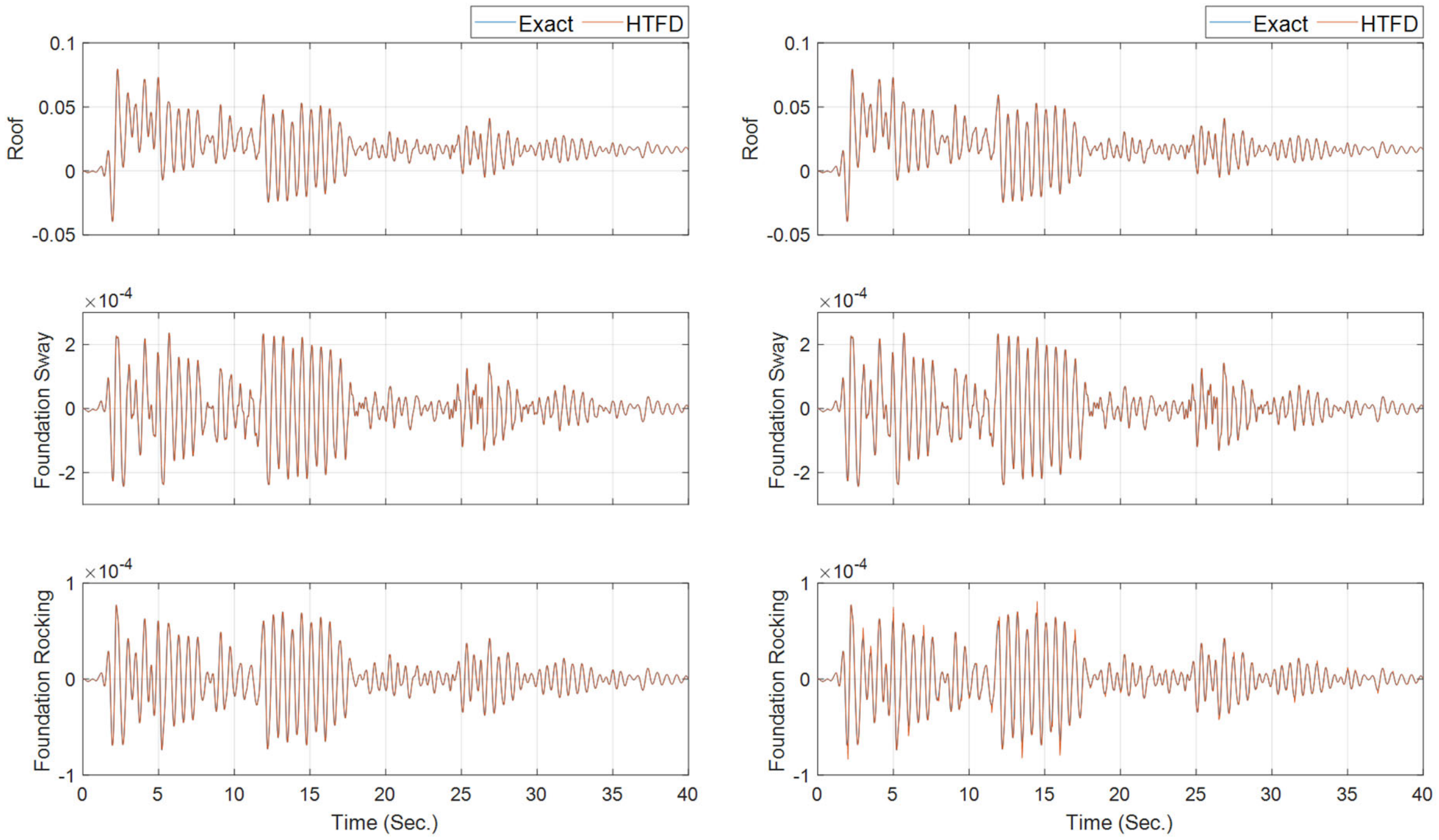

**Figure 5-11: Comparison between responses of the physical and equivalent (HTFD) models 1 (left) and 2 (right)in Opensees.**

# CHAPTER 6: BUILTIN OPENSEES IMPLEMENTATION

## 6.1. OPENSEES

### 6.1.1. Introduction

While commercial software packages offer various advantages, including computational stability and user-friendly graphic user interfaces, they also have notable drawbacks that require careful consideration. Firstly, the high cost of commercial software makes it less accessible to researchers, limiting its usage and hindering equal opportunities for exploration. Moreover, the unavailability of source codes in commercial software restricts researchers from comprehensively studying and discussing the internal mechanisms and functionalities of the software. This lack of transparency undermines the ability to gain a deeper understanding of the underlying algorithms. Additionally, the stringent control over secondary development functions in commercial software poses challenges in integrating the latest research outcomes and advancements into the existing source codes. This lack of flexibility hampers the ability to adapt and innovate based on cutting-edge research findings. Furthermore, computational models created with commercial software cannot be easily shared among the research community unless all participating institutes possess the same software, creating barriers to collaboration and impeding the free exchange of ideas and methodologies. In contrast, open-source software offers significant potential for greater benefits. Open-source software is freely available, eliminating financial barriers and ensuring broader accessibility. The availability of complete source codes empowers researchers to explore, modify, and enhance the software according to their specific needs. The flexibility and convenience of adding new modules to the codebase facilitate the incorporation of the latest research findings and the adaptation of the software to evolving requirements. Ultimately, the open-source software paradigm fosters a collaborative and inclusive research environment, enabling the research community to freely share computational models, methodologies, and insights, leading to mutual progress and advancement.

Opens System for Earthquake Engineering Simulation (OpenSEES) ) [49] is an open-source object-oriented Finite Element software framework for numerical simulation, which has increasingly become one of the most influential open platforms. OpenSees stands apart from conventional commercial software by offering versatility in terms of incorporating various materials, elements, and powerful algorithms. This versatility allows for the flexible definition of numerical models, accommodating the diverse requirements of different research projects. Additionally, OpenSees is designed with an advanced philosophy that promotes sustainability and seamless integration of the latest research outcomes. As a result, researchers are not only permitted but also encouraged to actively participate in the code development of OpenSees, contributing to its continuous improvement. Furthermore, the open exchange environment facilitated by OpenSees enables the reuse of previous achievements, including structural numerical models, which greatly benefits subsequent research activities. This facilitates the reproducibility of existing research outcomes and empowers other researchers to build upon them and make their valuable contributions. With these inherent advantages, OpenSees proves to be an ideal choice to be able to implement the HTFD methodology.

### 6.1.2. OpenSees Program Methodology

The OpenSees framework provides several modules to facilitate software development for finite element modeling and analysis tasks. The "Domain" module is the container that holds different components of the Finite Element model including the nodes, elements, materials, etc. The building blocks of the Finite Element model are added and accessed through the Domain module. The "ModelBuilder" module is responsible for building the model by adding the components of the Finite Element model to the "Domain" that it owns. This allows OpenSees user interface to be extended to a variety of types including reading from an input file or through a graphical user interface. "Node", "Element", "Material", "Constraint", and "Load" modules are the main building blocks of the Finite Element model in a "Domain". The "Recorder" module is responsible for monitoring, storing, and saving analysis results as the model is being analyzed, while the "Analysis" module handles forming and solving the system of equations that governs the behavior of the model. Finite Element analysis consists of several steps each encapsulated in separate modules in OpenSees. These steps include:

- defining the governing equations for analysis, e.g., static or dynamic (transient) equation of motion or a modal eigenvalue problem,
- defining the solution strategy for differential equations by specifying the numerical time integration scheme, e.g., Newmark method, Central Difference method, etc.,
- defining the solution strategy for a system of nonlinear equations including the nonlinear solver and the convergence criteria, e.g., Newton-Raphson method, BFGS, etc.,
- forming the linear system of equations for the model including efficient numbering of DOFs which leads to the rearrangement of the equations for more efficient solutions,
- handling the constraints defined by the model,
- and finally, solving the system of linear equations.

The "Analysis" module is also OpenSees's interface to different numerical computation frameworks which allows OpenSees to be used with a variety of different computing libraries.

### 6.1.3. Programming Strategy

OpenSees is developed in C++ which is an object-oriented programming language. Object-oriented programming (OOP) is a programming paradigm that uses objects and classes to organize code. Objects are data structures that contain both data and methods. Classes are blueprints for objects that define the data and methods that objects of that class will have. The OOP paradigm allows the design of OpenSees modules to follow well-known design patterns, making the framework very flexible for modification and extension.

Data abstraction plays a crucial role in the modeling process of OpenSees. It involves breaking down data, algorithms, and calculations into simplified procedures, showcasing the essential behavior of the related information. This approach effectively manages complexity and provides a user-friendly environment for modifying the software through data decomposition and algorithmic processes necessary for software analysis. The modular structure of OpenSees operates independently, thanks to the effective abstraction at the programming base levels. This abstraction facilitates the seamless addition or modification of the program framework. New codes can be implemented into the program with minimal changes required to other objects or modules, enabling easy integration and adaptability.

### 6.1.4. New SSI Element and HTFD Analysis Module

The implementation of the HTFD method in OpenSees, provided by the first author upon request, involves two separate tasks:

1. introducing a new element ("ZeroLengthSSI" in this case) which abstracts away the additional parameters for the element such as the definition of the dynamic stiffness for the element,

2. emulating the iterative process of HTFD analysis which involves solutions in both time and frequency domains.

The first task is done by extending the existing "ZeroLength" element to accept additional parameters defining the dynamic stiffness related to the frequency-dependent degree of freedom used for the HTFD analysis. The dynamic stiffness of the DOF is defined through a "Series" in OpenSees. A "Series" is a container for defining an x-y set of data, mainly used in OpenSees to define a time-series in which x is time and y is typically a load factor. In the context of "ZeroLengthSSI" element, two "Series" objects are used to define the real and imaginary parts of the dynamic stiffness in the frequency domain, i.e., x is frequency and y is the dynamic stiffness. The definition syntax for the new "ZeroLengthSSI" element is

```
element ZeroLengthSSI eleTag? iNode? jNode? -mat matID1? … -dir dirMat1? -freqDepDof dof1?
… -dynStiffReal series1? … -dynStiffImag series1? -refMass m_ref? -refDamping c_ref? -
refStiffness k_ref? <-orient x1? x2? x3? y1? y2? y3?>
```

Here the "-freqDepDof" specifies a list of the frequency-dependent DOFs of the element, "-dynStiffReal" specifies the series tag that defines the real part of the dynamic stiffness of the element, "-dynStiffImag" specifies the series tag that defines the imaginary part of the dynamic stiffness of the element, "-refMass", "-refDamping", and "-refStiffness" specify the stiffness characteristics of the reference physical model of the soil-foundation element. All other arguments are inherited from the "ZeroLength" element.

The second task abstracts away all the HTFD analysis subtasks into a command, which allows the user to perform the HTFD analysis very easily by calling the following "SSIAnalyze" command:

```
SSIAnalyze NumSteps? dt? -tol tol? -max_iter maxIter? -window_length Lw? -decay_length
dLen? -zero_pad_length zpLen? -log_level level?
```

The “NumSteps” specifies the total number of analysis steps in the time domain, “dt” specifies the time increment to be used for the analysis, “-tol” defines the tolerance value to be used for convergence checks between the time domain and the frequency domain, “-max_iter” specifies the maximum number of iterations allowed for performing the HTFD analysis, “-window_length” specifies the number of time steps used to define the window over which the HTFD analysis should be performed, “-decay_length” and “-zero_pad_length” are respectively the number of time steps that is used to decay the displacement to zero in time-domain, and the number of steps with zero displacement added to the end of time-domain displacements series, in order to smoothen the Gibbs effect during the Fourier transformation. Finally, the “-log_level” specifies the depth of details to be printed out while the HTFD analysis is being performed.

The “SSIAnalyze” command works as follows:

1. Gather all frequency-dependent degrees of freedom in the model
2. For each time interval window of length “windoew_length” perform the following:
    a. While the analysis has not converged and the number of iterations does not exceed “maxIter”, do:
        i. Reset the domain and set the time to 0
        ii. Analyze the model in the time domain to the end of the current window and store the displacement of all frequency-dependent DOFs, decay the displacements to zero and pad the end of the time series with zeros.
        iii. Transform the displacement to the frequency domain using Fast Fourier Transform. We added the third-party FFTW libraries (fftw3.dll) to OpenSees to perform the Fast Fourier Analysis.
        iv. For each of the DOFs calculate the dynamic force using the transformed displacement, and the element’s dynamic stiffness.
        v. For each of the DOFs calculate the reference linear force using the transformed displacement, “m_ref”, “c_ref”, and “k_ref”.
        vi. Transform both calculated forces back to the time domain using Inverse Fast Fourier Transform, calculate the error and check convergence.
        vii. Apply the residual load to the corresponding DOFs and go to the next step
3. Record the results of the converged analysis.

In our implementation of HTFD analysis in OpenSees, a model can have several “ZeroLengthSSI” elements each with one or more frequency-dependent degrees of freedom. A specific dynamic stiffness can be applied to each element. If separate dynamic stiffnesses are needed to be used for different degrees of freedom of one element, the user can define separate "ZeroLengthSSI" elements each associated with a specific dynamic stiffness for the desired degree of freedom. As such, a wide variety of models can be analyzed with the general interface of our HTFD implementation in OpenSees.

## 6.2. VERIFICATION EXAMPLES

### 6.2.1. SDOF Example

The first example studied in this chapter is the SDOF example used in the previous chapter to verify the HTFD implementation in Opensees. In this chapter, instead of explicitly implementing HTFD within Opensees script, we use the newly developed SSI elements and analysis module. For the benefit of the reader, the properties of both physical and equivalent models are repeated here in Table 6-1. The Python script to run the equivalent model under El Centro ground motion is shown in Figure 6-1.

**Table 6-1: Properties of the physical and equivalent models.**

| Parameter | Physical Model | Equivalent Model |
|---|---|---|
| $m_f$ | 0.5 | 0.5 |
| $I_f$ | 8.0 | 8.0 |
| $e$ | 8.0 | 8.0 |
| $m_s$ | 1.0 | 1.0 |
| $I_s$ | 16.0 | 16.0 |
| $k_s$ | 247.0 | 247.0 |
| $c_s$ | 0.63 | 0.63 |
| $h_s$ | 24.0 | 24.0 |
| $k_{0h}$ | 846.0 | 846.0 |
| $c_{0h}$ | 90.0 | 90.0 |
| $k_{0r}$ | 78310.0 | Not Applicable |
| $c_{0r}$ | 406.0 | Not Applicable |
| $I_{1r}$ | 253.0 | Not Applicable |
| $c_{1r}$ | 2982.0 | Not Applicable |
| $m_{r,ref}$ | Not Applicable | 0.0 |
| $k_{r,ref}$ | Not Applicable | 78310.0 |
| $c_{r,ref}$ | Not Applicable | 3227.0 |
| $u_y$ | 0.0009 | 0.0009 |

```
import os
import opensees as ops
import numpy as np

# wipe model
ops.wipe()

# create model
ops.model('basic', '-ndm', 2, '-ndf', 3)

# PARAMETERS
# Superstructures
ns = 1; ms = np.zeros((ns,1)); ks = np.zeros((ns,1)); cs = np.zeros((ns,1)); Is = np.zeros((ns,1)); h = np.zeros((ns,1)); epsy = np.zeros((ns,1))
ms[0,0] = 1.0; cs[0,0] = 0.6283; ks[0,0] = 246.7401; Is[0,0] =  16.0; h[0,0] = 24.0; epsy[0,0] = 9.3722e-04
# Foundation
mf = 0.5; If = 8.0; e = 8.0
# Frequency-independent impedance function
c0h = 89.759790102565490; k0h = 8.459660915219447e+02
# Reference  impedance function for the frequency-dependent part
m_ref = 0.0; k_ref =  7.831014112986252e+04; c_ref = 3.227678266639782e+03

# NODE
ops.node(1001, 0.0, 0.0)
ops.node(1002, 0.0, 0.0)
ops.node(1003, 0.0, e/2.0)
ops.node(1, 0.0, h[0,0]+e)

# MASS
ops.mass(1003, mf, 0.0, If)
ops.mass(1, ms[0,0], 0.0, Is[0,0])

# FIXITY
ops.fix(1001, 1, 1, 1)
ops.fix(1002, 0, 1, 0)
ops.fix(1003, 0, 1, 0)
ops.fix(1, 0, 1, 0)
ops.rigidLink('beam', 1002, 1003)
ops.equalDOF(1002, 1, 3)

# MATERIAL
ops.uniaxialMaterial('ElasticPP', 1, ks[0,0], epsy[0,0]); ops.uniaxialMaterial('Viscous', 2, cs[0,0], 1); ops.uniaxialMaterial('Parallel', 3, 1, 2) ;
ops.uniaxialMaterial('Elastic', 101, k0h); ops.uniaxialMaterial('Viscous', 102, c0h, 1); ops.uniaxialMaterial('Elastic', 201, k_ref); ops.uniaxialMaterial('Viscous', 202, c_ref, 1)
ops.uniaxialMaterial('Parallel', 103, 101, 102); ops.uniaxialMaterial('Parallel', 203, 201, 202);

# ELEMENTS
ops.element('twoNodeLink', 1, 1003, 1, '-mat', 3, '-dir', 2, '-orient', 1, 0, 0, '-shearDist', 0.0)

# BASE EXCITATION
N = 4000; dt = 0.01
ops.timeSeries('Path', 1, '-dt', dt, '-filePath', 'Record.txt', '-factor', 1.0); ops.pattern('UniformExcitation', 1, 1, '-accel', 1)

# ANALYSIS
ops.constraints('Transformation'); ops.test('RelativeNormUnbalance', 1e-8, 50, 0); ops.algorithm('Newton'); ops.numberer('Plain'); ops.system('FullGeneral');
ops.integrator('Newmark', 0.5, 0.25); ops.analysis('Transient')

####################################  HTFD ######################################
# SETTING ADJUSTED BY THE USER
L_W = 1000; tol = 0.001; MaxIter = 1000 ; NDecay = 100; NZero = 100

# FREQUENCY DOMAIN SETTING
N_E = int(N+NDecay+NZero)
if N_E%2==0:
   Nf = int(1+N_E/2)
else:
   Nf = int(1+(N_E-1)/2)
fs = 1/dt; df = fs/N_E; f = np.arange(0,Nf*df,df)

# FREQUENCY DEPENDENT IMPEDANCE FUNCTION
w = 2*np.pi*f
S_real = (7.831014112986252e+04-2.981793099803996e+03**2*2.531024592592593e+02*w**2/(2.981793099803996e+03**2+2.531024592592593e+02**2*w**2))
S_imag = w*(4.057318936500739e+02+2.981793099803996e+03*2.531024592592593e+02**2*w**2/(2.981793099803996e+03**2+2.531024592592593e+02**2*w**2))
S = S_real+1j*S_imag; S_real_list = list(S_real); S_imag_list = list(S_imag); w_list = list(w)
ops.timeSeries('Path', 20, '-values', *S_real_list, '-time', *w_list, '-factor', 1.0)
ops.timeSeries('Path', 30, '-values', *S_imag_list, '-time', *w_list, '-factor', 1.0)

# SSI ELEMENT
ops.element('zeroLengthSSI', 2, 1001, 1002, '-mat', 103, 203, '-dir', 2, 6, '-freqDepDof', 3,'-dynStiffReal',20,'-dynStiffImag',30, '-refMass', m_ref, '-refDamping', c_ref, '-
refStiffness', k_ref, '-orient', 0, 1, 0, 1, 0, 0)
ops.recorder('Node', '-file', 'Response_Builtin.out', '-time','-dT', dt, '-node', 1001, 1002, 1003, 1, '-dof', 1, 3, 'disp')

# ANALYSIS
ops.SSIAnalyze(N, dt, '-tol', tol, '-max_iter', MaxIter, '-window_length', L_W, '-decay_length', NDecay, '-zero_pad_length', NZero, '-log_level', 1)
```

**Figure 6-1: A Python code to run the equivalent model under El Centro ground motion.**

Figure 6-2 shows the comparison between exact responses (obtained using the physical model) and the final responses obtained through the HTFD method with the newly implemented SSI element and analysis module. As seen in this figure, the Opensees implementation shows perfect performance.

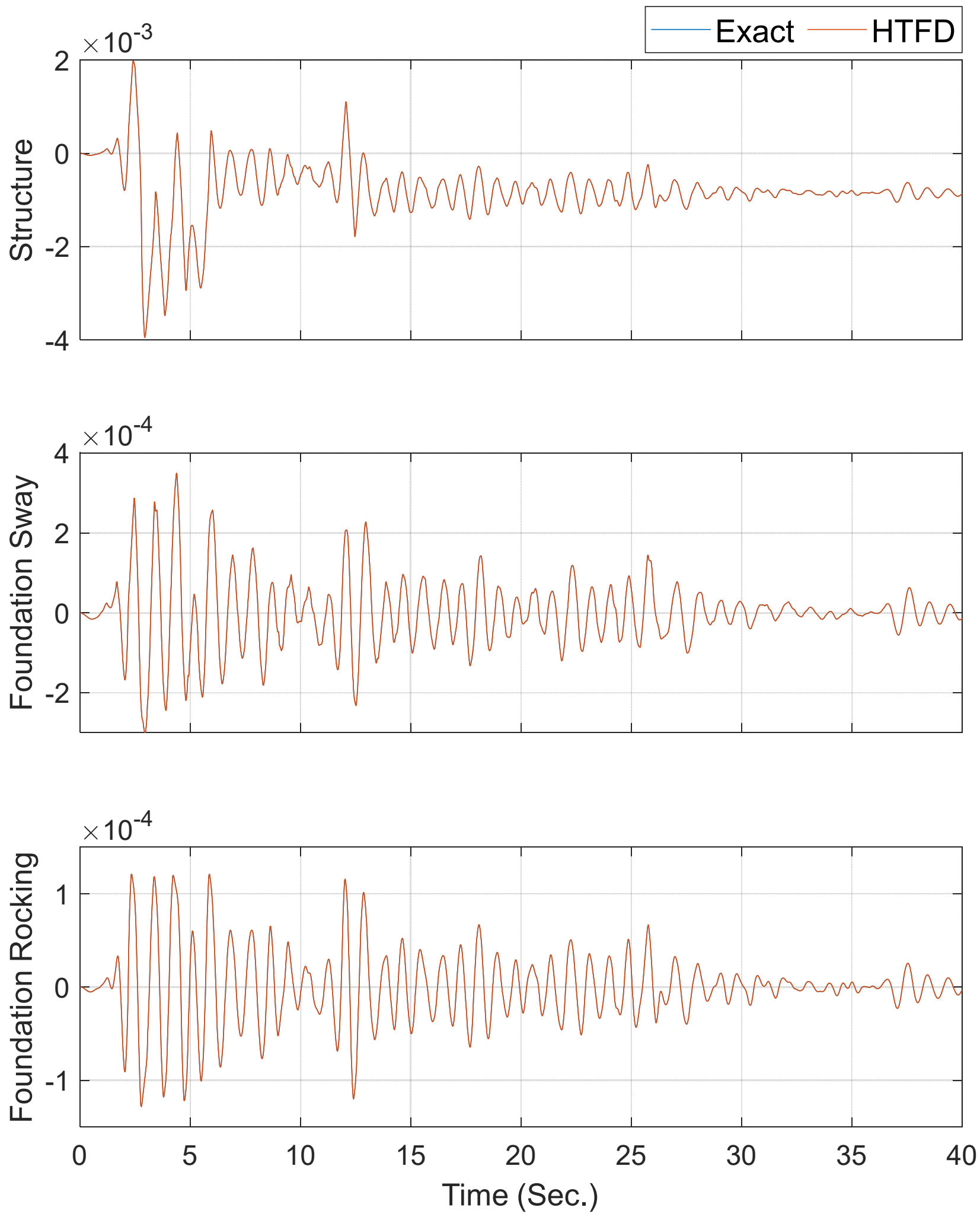

**Figure 6-2: Comparison between responses of the physical and equivalent (HTFD) models in Opensees. The equivalent model is constructed using a newly developed SSI element.**

### 6.2.2. MDOF Example

As the second example, the MDOF example presented in the previous chapter is used. Similar to the SDOF example, instead of explicitly implementing HTFD within the Opensees script, we use the newly developed SSI elements and analysis module. For the benefit of the reader, the properties of both physical and equivalent models are repeated here in Table 6-2. The Python script to run the equivalent model under El Centro ground motion is shown in Figure 6-3.

**Table 6-2: Properties of the physical and equivalent models.**

| Parameter | Physical Model | Equivalent Model 1 | Equivalent Model 2 |
|---|---|---|---|
| $m_f$ | $4.85 \times 10^3$ | $4.85 \times 10^3$ | $4.85 \times 10^3$ |
| $I_f$ | $4.58 \times 10^4$ | $4.58 \times 10^4$ | $4.58 \times 10^4$ |
| $e$ | 3.07 | 3.07 | 3.07 |
| $m_s$ | $9.70 \times 10^3$ | $9.70 \times 10^3$ | $9.70 \times 10^3$ |
| $I_s$ | $9.17 \times 10^4$ | $9.17 \times 10^4$ | $9.17 \times 10^4$ |
| $k_s$ | $1.31 \times 10^7$ | $1.31 \times 10^7$ | $1.31 \times 10^7$ |
| $h_s$ | 3.5 | 3.5 | 3.5 |
| $k_{0h}$ | $5.53 \times 10^8$ | $5.53 \times 10^8$ | $5.53 \times 10^8$ |
| $c_{0h}$ | $2.14 \times 10^7$ | $2.14 \times 10^7$ | $2.14 \times 10^7$ |
| $k_{0r}$ | $3.26 \times 10^{10}$ | Not Applicable | Not Applicable |
| $c_{0r}$ | $1.11 \times 10^8$ | Not Applicable | Not Applicable |
| $I_{1r}$ | $4.44 \times 10^7$ | Not Applicable | Not Applicable |
| $c_{1r}$ | $8.28 \times 10^8$ | Not Applicable | Not Applicable |
| $m_{r,ref}$ | Not Applicable | 0.0 | 0.0 |
| $k_{r,ref}$ | Not Applicable | $k_{0r}$ | 0.0 |
| $c_{r,ref}$ | Not Applicable | $c_{r,inf} = 9.37 \times 10^8$ | $c_{r,inf} = 9.37 \times 10^8$ |
| $u_{y,1}$ | 0.0099 | 0.0099 | 0.0099 |
| $u_{y,2}$ | 0.0092 | 0.0092 | 0.0092 |
| $u_{y,3}$ | 0.0079 | 0.0079 | 0.0079 |
| $u_{y,4}$ | 0.0059 | 0.0059 | 0.0059 |
| $u_{y,5}$ | 0.0033 | 0.0033 | 0.0033 |

```
import os
import opensees as ops
import numpy as np

# wipe model
ops.wipe()

# create model
ops.model('basic', '-ndm', 2, '-ndf', 3)

# PARAMETERS
# Superstructures
ns = 5; ms = 9.7014e3*np.ones((ns,1)); ks = 1.3132e7*np.ones((ns,1)); cs = np.zeros((ns,1)); Is = 4.5842e5/ns*np.ones((ns,1)); h = 3.5*np.ones((ns,1)); H = np.cumsum(h)
alphaM = 0.78; betaKinit = 0.0024; epsy = np.array(([0.0099,0.0092,0.0079,0.0059,0.0033]))
deltaM = 0; c0h = 2.1377e7; k0h = 5.5335e8; mf = 4.8507e3; If = 4.5842e4+deltaM; e = 3.0742; m_ref = 0.0; k_ref = 0.0;#3.2611e+10; c_ref = 936755020.9324069;

# NODE
ops.node(1001, 0.0, 0.0); ops.node(1002, 0.0, 0.0); ops.node(1003, 0.0, e/2.0)
for i in range(ns):
    ops.node(i+1, 0.0, H[i]+e)

# MASS
ops.mass(1003, mf, 0.0, If)
for i in range(ns):
    ops.mass(i+1, ms[i,0], 0.0, Is[i,0])

# FIXITY
ops.fix(1001, 1, 1, 1); ops.fix(1002, 0, 1, 0); ops.fix(1003, 0, 1, 0); ops.rigidLink('beam', 1002, 1003)
for i in range(ns):
    ops.fix(i+1, 0, 1, 0)
    ops.equalDOF(1002, i+1, 3)

# MATERIAL
for i in range(ns):
    ops.uniaxialMaterial('ElasticPP', i+1, ks[i,0], epsy[i])  # Interstory behavior

ops.uniaxialMaterial('Elastic', 101, k0h); ops.uniaxialMaterial('Viscous', 102, c0h, 1); ops.uniaxialMaterial('Elastic', 201, k_ref); ops.uniaxialMaterial('Viscous', 202, c_ref, 1)
ops.uniaxialMaterial('Parallel', 103, 101, 102); ops.uniaxialMaterial('Parallel', 203, 201, 202)

# ELEMENTS
ops.element('twoNodeLink', 1, 1003, 1, '-mat', 1, '-dir', 2, '-orient', 1, 0, 0, '-shearDist', 0.0)
for i in range(ns-1):
    ops.element('twoNodeLink', i+2, i+1, i+2, '-mat', i+2, '-dir', 2, '-orient', 1, 0, 0, '-shearDist', 0.0)

# DAMPING
ops.region(1,'-node', 1, 2, 3, 4, 5, '-rayleigh', alphaM, 0.0, betaKinit, 0.0)

# BASE EXCITATION
N = 4000; dt = 0.01
ops.timeSeries('Path', 1, '-dt', dt, '-filePath', 'Record.txt', '-factor', 9.806); ops.pattern('UniformExcitation', 1, 1, '-accel', 1)

# ANALYSIS
ops.constraints('Transformation’); ops.test('RelativeNormUnbalance', 1e-12, 50, 1); ops.algorithm('Newton’); ops.numberer('Plain’); ops.system('FullGeneral’);
ops.integrator('Newmark', 0.5, 0.25); ops.analysis('Transient')


################################### HTFD ######################################
# SETTING ADJUSTED BY THE USER
L_W = 50; tol = 0.001; MaxIter = 1000; NDecay = 100; NZero = 100
# FREQUENCY DOMAIN SETTING
N_E = int(N+NDecay+NZero)
if N_E%2==0:
    Nf = int(1+N_E/2)
else:
    Nf = int(1+(N_E-1)/2)
fs = 1/dt; df = fs/N_E; f = np.arange(0,Nf*df,df)

# FREQUENCY DEPENDENT IMPEDANCE FUNCTION
w = 2*np.pi*f; c0r = 1.1138e+08; k0r = 3.2611e+10; I1r = 4.4365e+07; c1r = 8.2829e+08
S = (k0r-c1r**2*I1r*w**2/(c1r**2+I1r**2*w**2))+1j*w*(c0r+c1r*I1r**2*w**2/(c1r**2+I1r**2*w**2))
S_real = np.real(S); S_imag = np.imag(S); S_real_list = list(S_real); S_imag_list = list(S_imag); w_list = list(w)
ops.timeSeries('Path', 20, '-values', *S_real_list, '-time', *w_list, '-factor', 1.0); ops.timeSeries('Path', 30, '-values', *S_imag_list, '-time', *w_list, '-factor', 1.0)

# SSI ELEMENT
ops.element('zeroLengthSSI', 200, 1001, 1002, '-mat', 103, 203, '-dir', 2, 6, '-freqDepDof', 3,'-dynStiffReal',20,'-dynStiffImag',30, '-refMass', m_ref, '-refDamping', c_ref, '-
refStiffness', k_ref, '-orient', 0, 1, 0, 1, 0, 0) ops.recorder('Node', '-file', 'Response_Builtin_MDOF.out', '-time','-dT', dt, '-node', 1001, 1002, 1003, 1, 2, 3, 4, 5, '-dof', 1, 3,
'disp')

# ANALYSIS
ops.SSIAnalyze(N, dt, '-tol', tol, '-max_iter', MaxIter, '-window_length', L_W, '-decay_length', NDecay, '-zero_pad_length', NZero, '-log_level', 1)
```

**Figure 6-3: Python code to run the equivalent model under El Centro ground motion.**

Figure 6-4 shows the comparison between exact responses (obtained using the physical model) and the final responses obtained through the HTFD method with the newly implemented SSI element and analysis module. Note that a time window with a length of 10 seconds is used for Model 1, while a window with a length of 0.5 seconds is used for Model 2 which significantly increases computational time. As seen in this figure, the Opensees implementation of the HTFD method shows perfect performance in both cases, but considering the computational time, a reasonable setting for the reference model is recommended.

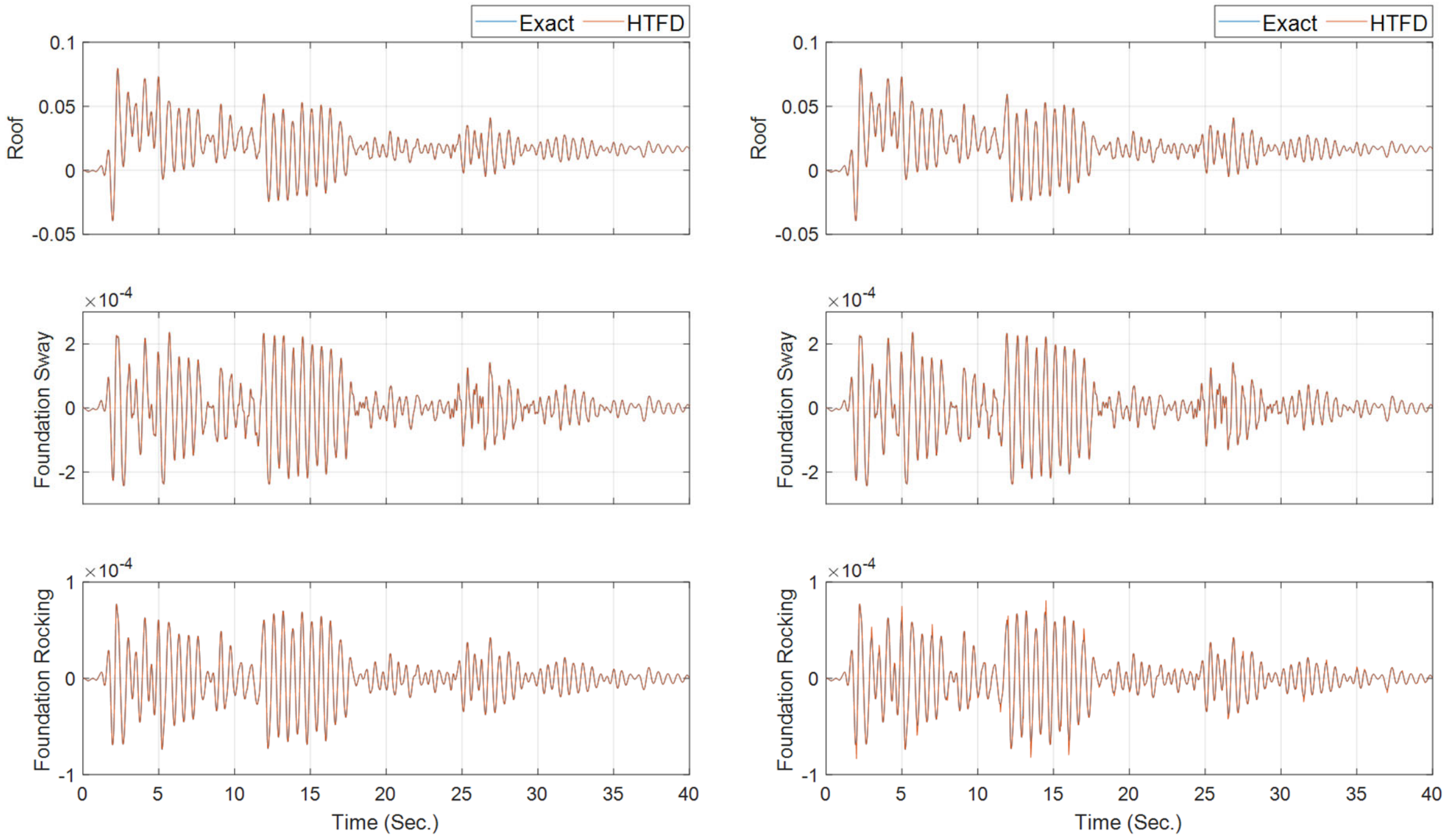

**Figure 6-4: Comparison between responses of the physical and equivalent (HTFD) models 1 (left) and 2 (right) in Opensees. The equivalent models are constructed and analyzed using a newly developed SSI element and analysis module.**

## 6.3. AN APPLICATION

In this section, the builtin Openssees implementation method/element is employed to analyze a nonlinear MDOF superstructure on a frequency-dependent sway-rocking soil-foundation substructure in which impedance functions are estimated through direct analysis.

Figure 6-5 shows the direct model which is representative of a strip foundation on a thin soil layer over a half-space. This example is chosen to have impedance functions with significant frequency dependency. Also, this example is similar to the one used in [93], [94] to make sure the impedance functions estimated through the direct method are reliable although the accuracy of the impedance functions is not the purpose of this example. The model is composed of a 16 m soil layer with a shear wave velocity of 75 $m/s$ over a half-space with a shear wave velocity of 150 $m/s$. Both soil layer and half-space have similar mass density and Poisson's ratio of 2200 $kg/m^3$ and 0.4, respectively. While Perfectly Matched Layer (PML) [95] has been recently implemented in Opensees [96] there is still no clear documentation to be able to use it. So, to prevent wave reflection at the boundaries of the domain, traditional parallel and normal viscous dashpots [17] are employed. Both TCL and Python versions of this model are developed and are available upon request from the first author. In these models, the soil is modeled by using 4-node "quad" elements with "PlaneStrain" behavior and "ElasticIsotropic" material. Element size is $\Delta L = 1\,m$ to have at least 5 elements with the smallest wavelength to be able to analyze the model up to 15 Hz frequency. A sampling frequency of 1000 Hz is used for the analysis to satisfy the rule of thumb of $\Delta t \leq \Delta L/3V_p$ [97] where $V_p$ is the largest soil's compressional wave velocity. To obtain the frequency-dependent impedance function, A harmonic force (moment) as $F(t) = A\sin(\omega t)$ with sufficient duration (5 seconds) is applied at the center of the foundation and its horizontal displacement (rotation) is recorded. Then, a sinusoidal function as $U(t) = D\sin(\omega t + \varphi)$ is fitted to the recorded response from which the complex-valued impedance function is calculated as

$$\hat{S}(\omega) = \frac{A}{D}[\cos(-\varphi) + i\sin(-\varphi)]. \tag{6-1}$$

This process is repeated for a range of frequencies to obtain the impedance function. Herein, we vary the dimensionless frequency $a_0 = \frac{\omega B}{V_{s2}}$ between 0.1 to 3 with steps of 0.1 to cover frequencies between 0.3 to 9 Hz. While the impedance function can be estimated more efficiently using a pulse excitation, we chose to use harmonic excitation to be able to consider frequency-independent material damping by setting the stiffness-proportional Rayleigh damping coefficient at $\xi/\omega$ where $\xi = 5\%$ for all frequencies.

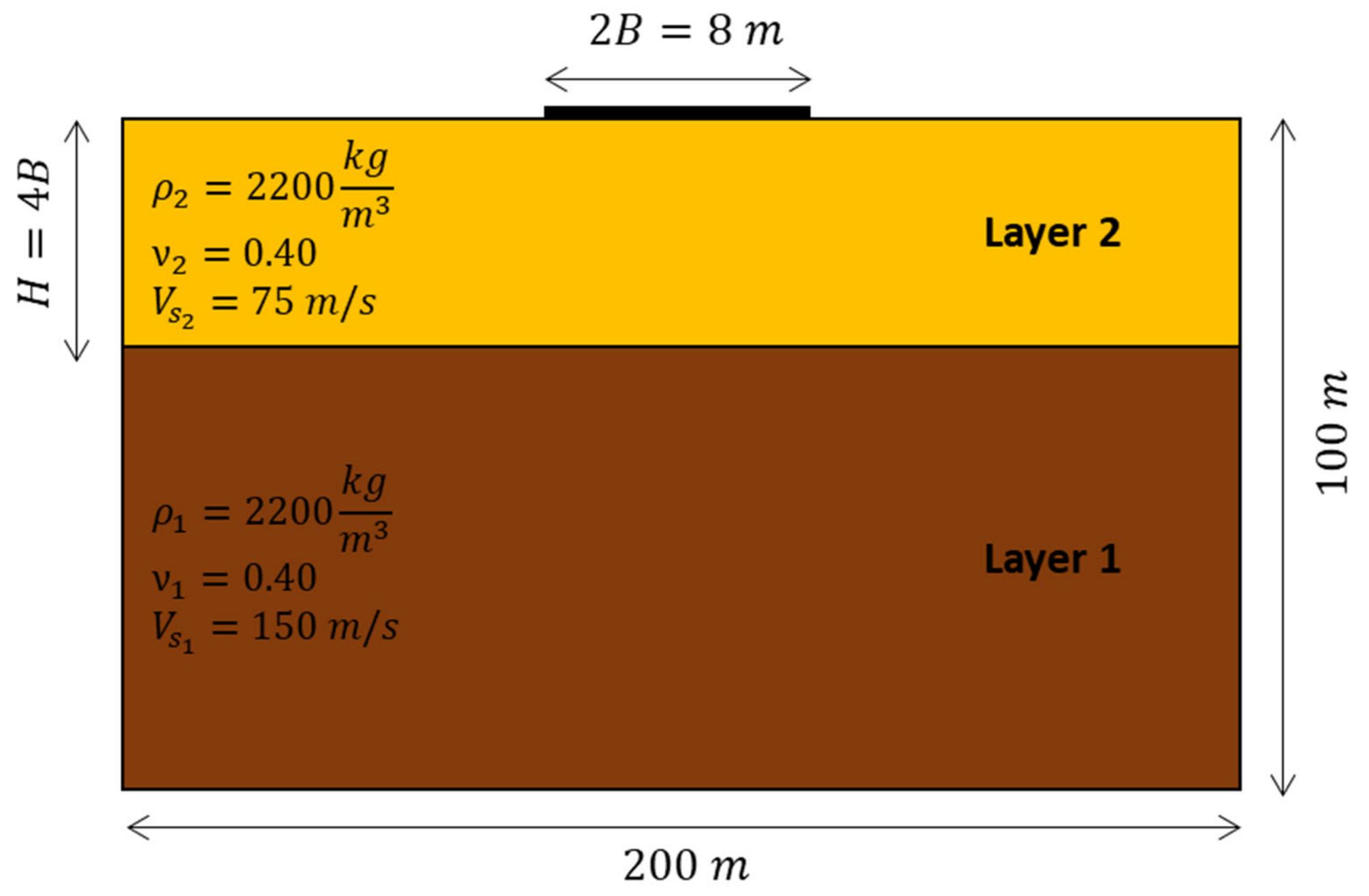

**Figure 6-5: Strip foundation on stratum over half-space.**

Figure 6-6 shows the estimated stiffness and damping as a function of dimensionless frequency for both horizontal and rocking degrees of freedom from which the complex-valued impedance function is constructed as

$$\hat{S}_j(a_0) = k_j(a_0) + i\omega c_j(a_0) \quad \text{for } j = H \text{ and } R. \tag{6-2}$$

To be able to use these impedance functions in the substructure analysis through the HTFD approach, they must be available in finer frequency resolution. So, cubic spline interpolation is employed as shown in Figure 6-6 by blue curves. To verify the accuracy of the estimated impedance functions, stiffness, and damping parts shown in (a) and (b) of Figure 6-6 can be directly compared against Figure 4 in [93] and stiffness and damping parts shown in parts (c) and (d) can be compared against Figure 7 in [94]. Two videos showing the response of the soil-foundation system under harmonic horizontal and rocking loading at $a_0 = 1$ are available upon request from the first author.

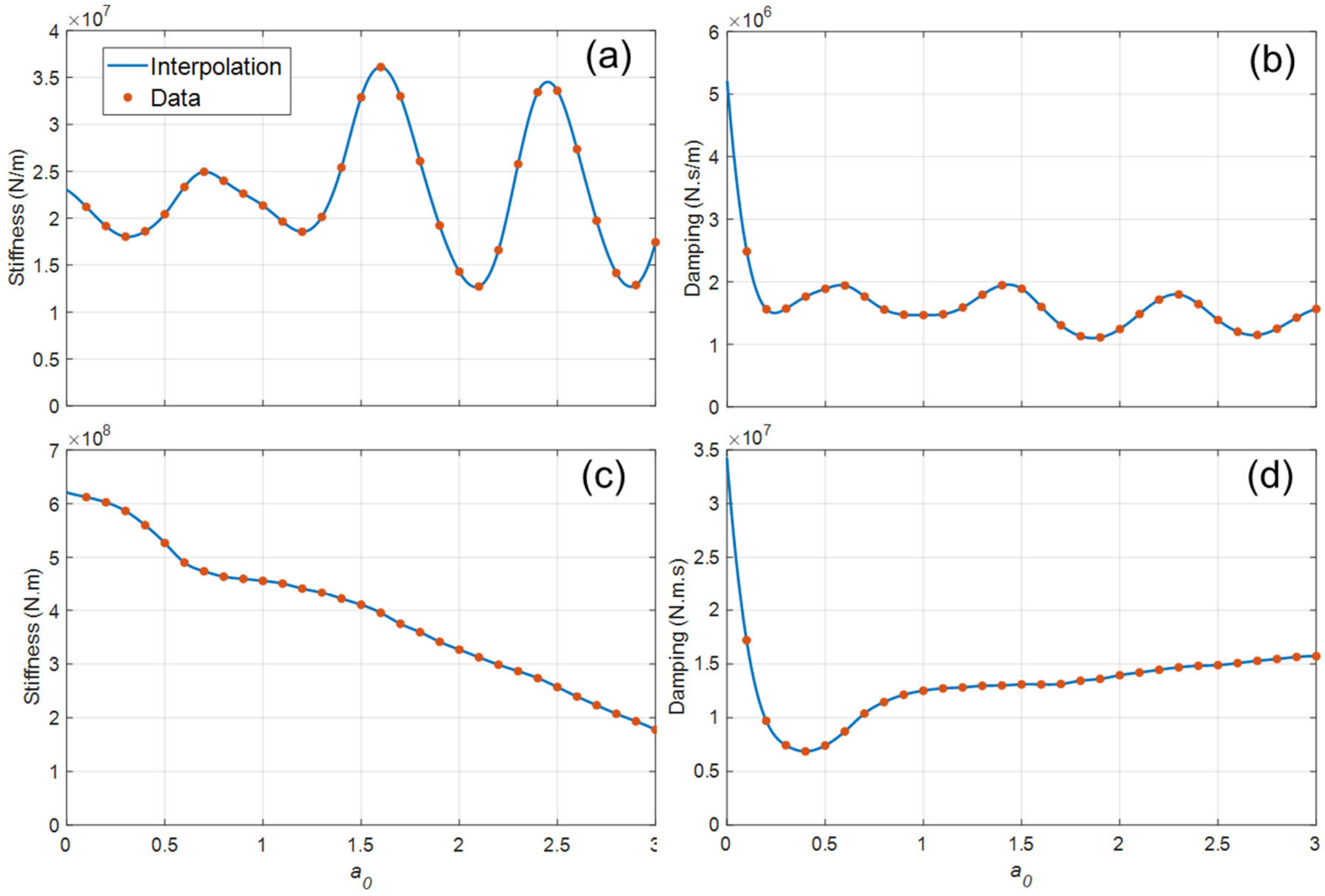

**Figure 6-6: Estimated impedance functions in the horizontal (a and b) and rocking (c and d) directions.**

Before using these estimated impedance functions with the built-in SSI element/analysis to analyze a structure under dynamic loading, it is worth verifying the new element/analysis implementation by repeating the impedance function estimation through the substructure approach. Openseespy codes for estimating horizontal and rocking impedance functions are shown in Figure 6-7 and Figure 6-8, respectively. The estimated impedance functions at discrete dimensionless frequencies as well as interpolated points are shown in Figure 6-9. As it can be seen, these plots are almost identical to the plots shown in Figure 6-6 confirming the equivalency between the direct and substructure models at the soil-foundation level.

```
import os
import opensees as ops
import numpy as np
A0 = np.arange(0.1, 3.1, 0.1);
################################ PARAMETERS ####################################
dz = 1.0;
B = 4.0;
Vs = 75;
dt = 0.001;
N = 5000;
L_W = 1000
tol = 1e-3
MaxIter = 100
NDecay = 500
NZero = 500
k_ref_H = 2.303815144145173e+07;
c_ref_H = 5.211476380043654e+06;
tsim = N*dt;
t = dt*np.arange(0.0, N, 1);
for k in range(len(A0)):
    ops.wipe()
    ops.model('basic', '-ndm', 2, '-ndf', 3)
    ############################ IMPEDANCE FUNCTION ################################
    with open('Impedance_H.txt', 'r') as KH:
        fH_list = []
        SH_real_list = []
        SH_imag_list = [];
        for line in KH:
            p = line.split()
            fH_list.append(float(p[0]));
            SH_real_list.append(float(p[1]));
            SH_imag_list.append(float(p[2]));
    KH.close()
    N_E = int(N+NDecay+NZero)
    if N_E%2==0:
        Nf = int(1+N_E/2)
    else:
        Nf = int(1+(N_E-1)/2)
    fs = 1/dt
    df = fs/N_E
    f = np.arange(0,Nf*df,df)
    indexesH = abs(f[:, None] - np.array(fH_list)).argmin(axis=1)
    outH = list(enumerate(indexesH))
    ############################## SUPERSTRUCTURE ##################################
    ops.node(197, 96.0, 100.0)
    ops.node(198, 97.0, 100.0)
    ops.node(199, 98.0, 100.0)
    ops.node(200, 99.0, 100.0)
    ops.node(1000, 100.0, 100.0)
    ops.node(202, 101.0, 100.0)
    ops.node(203, 102.0, 100.0)
    ops.node(204, 103.0, 100.0)
    ops.node(205, 104.0, 100.0)
    ops.fix(1000,0,1,1)
    ops.rigidLink('beam',1000,197)
    ops.rigidLink('beam',1000,198)
    ops.rigidLink('beam',1000,199)
    ops.rigidLink('beam',1000,200)
    ops.rigidLink('beam',1000,202)
    ops.rigidLink('beam',1000,203)
    ops.rigidLink('beam',1000,204)
    ops.rigidLink('beam',1000,205)
    ops.node(103, 100.0, 100.0)
    ops.fix(103,1,1,1)
    ############################ FOUNDATION ELEMENTS ###############################
    ops.uniaxialMaterial('Elastic', 2001, k_ref_H)
    ops.uniaxialMaterial('Viscous', 2002, c_ref_H, 1)
    ops.uniaxialMaterial('Parallel', 2003, 2001, 2002)
    ops.timeSeries('Path', 200, '-values', *[SH_real_list[x[1]]  for x in outH], '-time', *[2.0*np.pi*fH_list[x[1]]  for x in outH], '-factor', 1.0)
    ops.timeSeries('Path', 300, '-values', *[SH_imag_list[x[1]]  for x in outH], '-time', *[2.0*np.pi*fH_list[x[1]]  for x in outH], '-factor', 1.0)
    ops.element('zeroLengthSSI', 300, 103, 1000, '-mat', 2003, '-dir', 1, '-freqDepDof', 1,'-dynStiffReal',200,'-dynStiffImag',300, '-refMass', 0.0, '-
refDamping', c_ref_H,  '-refStiffness', k_ref_H)
    ################################### LOADING ####################################
    a0 = A0[k]
    f0 = a0*Vs/B/2.0/np.pi;
    t = np.arange(0.0, tsim+dt, dt);
    ForceX = 30000*np.sin(2.0*np.pi*f0*t);
    ops.timeSeries('Path', 100, '-values', *list(ForceX), '-time', *list(t))
    ops.pattern('Plain', 100, 100);
    ops.load(1000, 1.0, 0.0, 0.0);
    ################################### RECORDER ###################################
    np.savetxt('ForceX'+str(k)+'.txt',ForceX)
    ops.recorder('Node', '-file', 'dispX'+str(k)+'.out','-time', '-node' ,1000 , '-dof', 1, 'disp')
    ################################### ANALYSIS ###################################
    ops.constraints('Transformation')
    ops.algorithm('Linear',  'initial', 'factorOnce')
    ops.numberer('Plain')
    ops.system('BandSPD')
    ops.integrator('Newmark', 0.5, 0.25)
    ops.analysis('Transient')
    ops.SSIAnalyze(N, dt, '-tol', tol, '-max_iter', MaxIter, '-window_length', L_W, '-decay_length', NDecay, '-zero_pad_length', NZero, '-log_level', 1)
```

**Figure 6-7: A Python code to calculate frequency-dependent horizontal impedance function using the soil-foundation model with built-in SSI element/analysis.**

```
import os
import opensees as ops
import numpy as np
A0 = np.arange(0.1, 3.1, 0.1);
################################ PARAMETERS ####################################
dz = 1.0;
B = 4.0;
D = 0.1;
Vs = 75;
dt = 0.001;
N = 10000;
L_W = 10000 # Window length
tol = 1e-3 # Allowable pseudo-force error
MaxIter = 100 # Maximum number of iterations
NDecay = 500 # Decaying length
NZero = 500 # Zero padding length
k_ref_R = 6.2065315e+08;
c_ref_R = 3.3477e+07;
tsim = N*dt;
t = dt*np.arange(0.0, N, 1);
for k in range(len(A0)):
  ops.wipe()
  ops.model('basic', '-ndm', 2, '-ndf', 3)
  ############################ IMPEDANCE FUNCTION ################################
  with open('Impedance_R.txt', 'r') as KR:
    fR_list = []
    SR_real_list = []
    SR_imag_list = [];
    for line in KR:
      p = line.split()
      fR_list.append(float(p[0])); # f(Hz)
      SR_real_list.append(float(p[1])); # KR (N.m)
      SR_imag_list.append(float(p[2])); # w*CR (rad/sec*N.s/m)
  KR.close()
  N_E = int(N+NDecay+NZero)
  if N_E%2==0:
    Nf = int(1+N_E/2)
  else:
    Nf = int(1+(N_E-1)/2)
  fs = 1/dt
  df = fs/N_E
  f = np.arange(0,Nf*df,df)
  indexesR = abs(f[:, None] - np.array(fR_list)).argmin(axis=1)
  outR = list(enumerate(indexesR))
  ############################## SUPERSTRUCTURE ##################################
  ops.node(197, 96.0, 100.0)
  ops.node(198, 97.0, 100.0)
  ops.node(199, 98.0, 100.0)
  ops.node(200, 99.0, 100.0)
  ops.node(1000, 100.0, 100.0)
  ops.node(202, 101.0, 100.0)
  ops.node(203, 102.0, 100.0)
  ops.node(204, 103.0, 100.0)
  ops.node(205, 104.0, 100.0)
  ops.fix(1000,1,1,0)
  ops.rigidLink('beam',1000,197)
  ops.rigidLink('beam',1000,198)
  ops.rigidLink('beam',1000,199)
  ops.rigidLink('beam',1000,200)
  ops.rigidLink('beam',1000,202)
  ops.rigidLink('beam',1000,203)
  ops.rigidLink('beam',1000,204)
  ops.rigidLink('beam',1000,205)
  ops.node(103, 100.0, 100.0)
  ops.fix(103,1,1,1)
  ############################ FOUNDATION ELEMENTS ###############################
  ops.uniaxialMaterial('Elastic', 201, k_ref_R)
  ops.uniaxialMaterial('Viscous', 202, c_ref_R, 1)
  ops.uniaxialMaterial('Parallel', 203, 201, 202)
  ops.timeSeries('Path', 20, '-values', *[SR_real_list[x[1]]  for x in outR], '-time', *[2.0*np.pi*fR_list[x[1]]  for x in outR], '-factor', 1.0)
  ops.timeSeries('Path', 30, '-values', *[SR_imag_list[x[1]]  for x in outR], '-time', *[2.0*np.pi*fR_list[x[1]]  for x in outR], '-factor', 1.0)
  ops.element('zeroLengthSSI', 100, 103, 1000, '-mat', 203, '-dir', 6, '-freqDepDof', 3,'-dynStiffReal',20,'-dynStiffImag',30, '-refMass', 0.0, '-
refDamping', c_ref_R,  '-refStiffness', k_ref_R)
  #################################### LOADING #####################################
  a0 = A0[k]
  f0 = a0*Vs/B/2.0/np.pi;
  t = np.arange(0.0, tsim+dt, dt);
  ForceR = 30000*np.sin(2.0*np.pi*f0*t);
  ops.timeSeries('Path', 100, '-values', *list(ForceR), '-time', *list(t))
  ops.pattern('Plain', 100, 100);
  ops.load(197, 0.0, 1.0, 0.0);
  ops.load(205, 0.0, -1.0, 0.0);
  #################################### RECORDER ####################################
  np.savetxt('ForceR'+str(k)+'.txt',ForceR)
  ops.recorder('Node', '-file', 'dispYL'+str(k)+'.out','-time', '-node' ,197 , '-dof', 2, 'disp')
  ops.recorder('Node', '-file', 'dispYR'+str(k)+'.out','-time', '-node' ,205 , '-dof', 2, 'disp')
  #################################### ANALYSIS ####################################
  ops.constraints('Transformation')
  ops.algorithm('Linear',  'initial', 'factorOnce')
  ops.numberer('Plain')
  ops.system('BandSPD')
  ops.integrator('Newmark', 0.5, 0.25)
  ops.analysis('Transient')
  ops.SSIAnalyze(N, dt, '-tol', tol, '-max_iter', MaxIter, '-window_length', L_W, '-decay_length', NDecay, '-zero_pad_length', NZero, '-log_level', 1)
```

**Figure 6-8: A Python code to calculate frequency-dependent rocking impedance function using the soil-foundation model with built-in SSI element/analysis.**

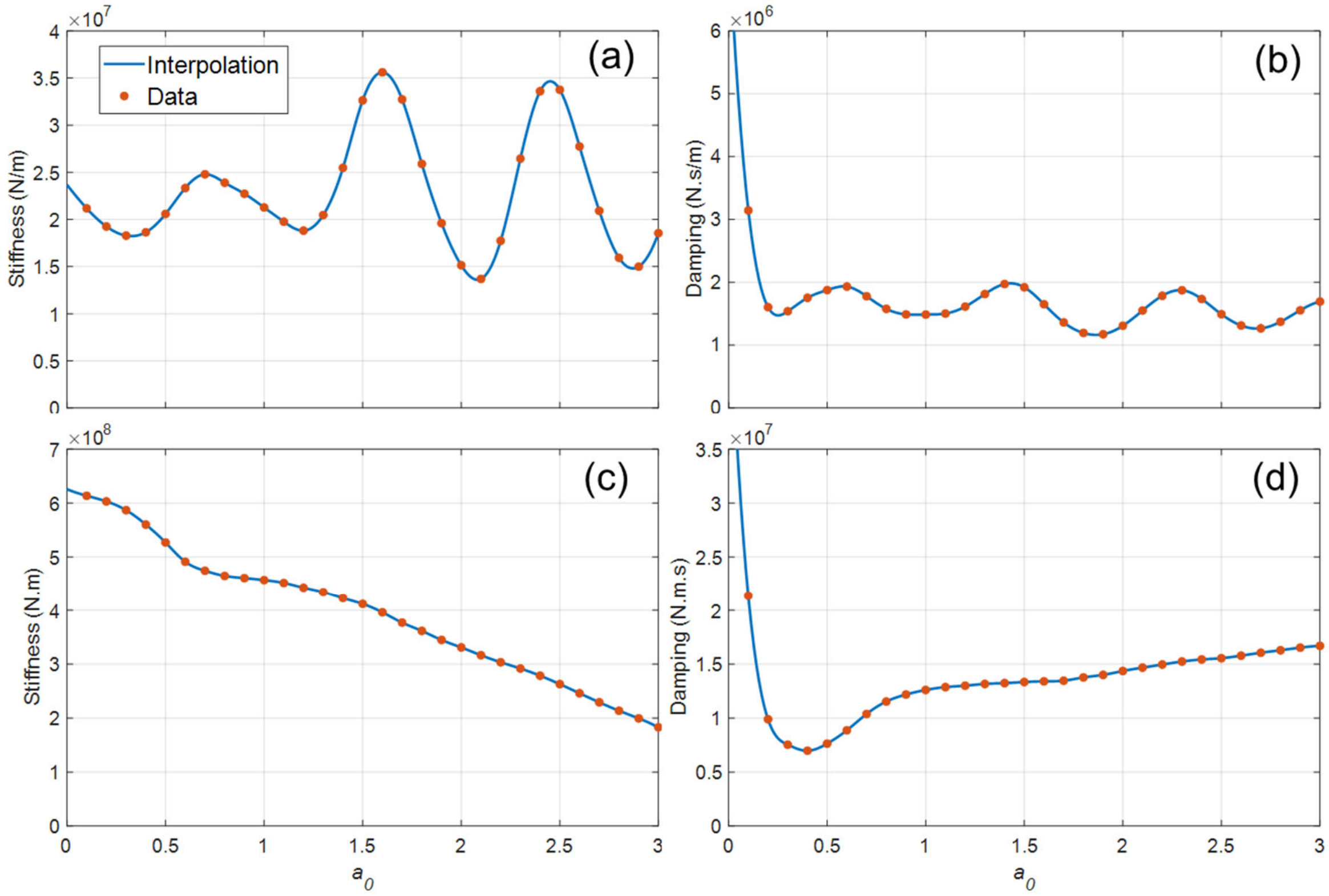

**Figure 6-9: Estimated impedance functions in the horizontal (a and b) and rocking (c and d) directions using built-in SSI element/analysis.**

Having frequency-dependent impedance functions, a superstructure is added to both direct and substructure models. The superstructure is a 5-DOF shear-building with properties presented in Table 6-3. To model inter-story behavior, a uniaxial material with kinematic hardening is used. The elastic stiffnesses and hardening ratios as well as yield displacement are shown in Table 6-3.

**Table 6-3: Properties of the superstructure.**

| Story | Height (m) | Mass (kg) | Stiffness (N/m) | Yield Disp. (mm) | Hardening |
|---|---|---|---|---|---|
| 1 | 3 | $1.0 \times 10^5$ | $1.0 \times 10^8$ | 1.0 | 0.1 |
| 2 | 3 | $1.0 \times 10^4$ | $1.0 \times 10^8$ | 0.1 | 0.1 |
| 3 | 3 | $1.0 \times 10^4$ | $1.0 \times 10^7$ | 1.0 | 0.1 |
| 4 | 3 | $1.0 \times 10^4$ | $1.0 \times 10^7$ | 1.0 | 0.1 |
| 5 | 3 | $1.0 \times 10^4$ | $1.0 \times 10^7$ | 1.0 | 0.1 |

The natural frequencies of the superstructure in its fixed-based condition are reported in Table 6-4. The example is designed to have multiple modes within the estimated impedance functions. Also, dynamic properties are designed to have significant soil-structure interaction effects. To see the level of period elongation caused by the soil-structure interaction, eigenanalysis is carried out for all points of the impedance functions and the fundamental frequency is plotted in Figure 6-10 (blue curve). The correct flexible-base natural frequency is the one that intersects with the 45-degree line (red line), which is 1.2 Hz. So, a period elongation of $\frac{2.1}{1.2} = 1.75$ is expected. A Matlab code to carry out such calculation is presented in Figure 6-11. Note that, the provided Matlab code carries out an approximate eigenanalysis because the effects of non-classical damping in the calculation are neglected [90].

**Table 6-4: Natural frequencies (Hz) of the superstructure.**

| Mode 1 | Mode 2 | Mode 3 | Mode 4 | Mode 5 |
|---|---|---|---|---|
| 2.11 | 4.82 | 6.35 | 9.04 | 17.50 |

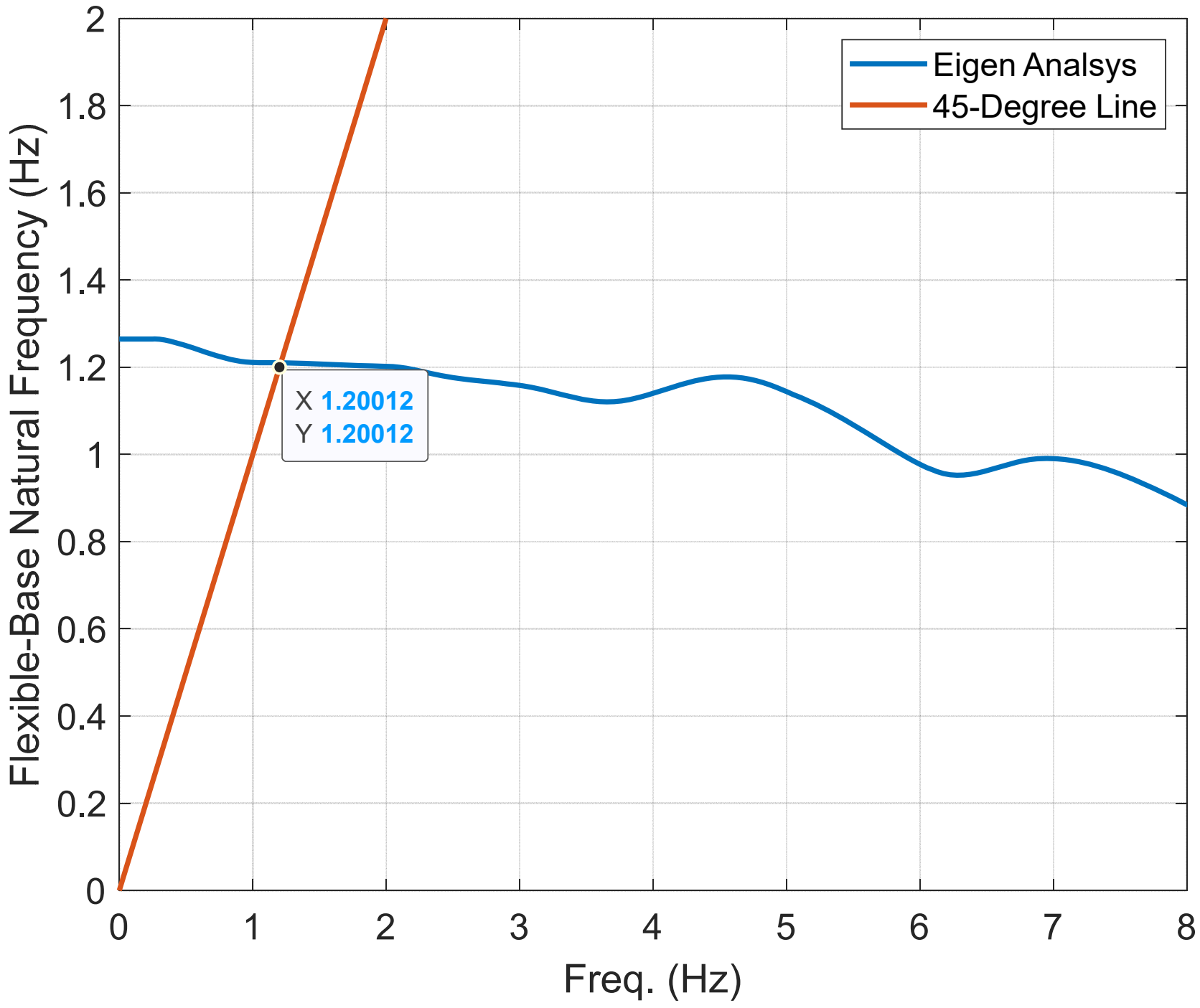

**Figure 6-10: Fundamental flexible-base natural frequency estimation.**

```
ns = 5;
ms = 1.0e2*[1000,100,100,100,100];
ks = 1.0e3*[100000,100000,10000,10000,10000];
hs = [3,3,3,3,3];
xi = 0.05;
mf = 0.1;
%% Fixed-base system matrices
Ms = zeros(ns,ns);
for p=1:ns
    for q=p:ns
        Ms(p,q) = sum(ms(q:ns));
    end
end
Ms = Ms+Ms'-diag(diag(Ms));
Ks = diag(ks(1:ns));
%% Fixed-base modal properties
[V,D] = eig(Ms\Ks);
ws = diag(sqrt(D));
[ws,id] = sort(ws);
fs = ws/2/pi;
Phis = V(:,id);
Phis = Phis./((diag(Phis'*Ms*Phis)).^(0.5))';
alpha = xi*2*fs(1)*fs(end)/(fs(1)+fs(end));
betta = xi*2/(fs(1)+fs(end));
%% Flexible-base mass matrix
H = cumsum(hs(1:ns));
m = zeros(ns+2,ns+2);
m(1,1) = mf+sum(ms(1:ns));
m(1,2) = 0;
for l=1:ns
     m(1,2) = m(1,2)+ms(1,l)*(sum(hs(1,1:l)));
end
m(2,1) = m(1,2);
m(2,2) = 0;
for l=1:ns
     m(2,2) = m(2,2)+(ms(1,l)*(sum(hs(1,1:l)))^2);
end
Mf = [m(1,1),m(1,2);m(1,2),m(2,2)];
Mfs = [];
for k=1:ns
     m(1,k+2) = sum(ms(1,k:ns));
     m(k+2,1) = m(1,k+2);
          for l=k:ns
               m(2,k+2) = m(2,k+2)+ms(1,l)*(sum(hs(1,1:l)));
               m(k+2,2) = m(2,k+2);
          end
          Mfs = [Mfs [m(1,k+2);m(2,k+2)]];
end
M = [Mf,Mfs;Mfs',Ms];
%% Frequency-dependent impedance functions
temp = dlmread('Impedance_H.txt');
f = temp(:,1);
Kh = temp(:,2);
temp = dlmread('Impedance_R.txt');
Kr = temp(:,2);
%% Iterative modal analysis
N = length(f);
for k=1:N
    K = diag([Kh(k) Kr(k) ks(1:ns)]);
    [V,D] = eig(-pinv(M)*K);
    lambda = diag(sqrt(D));
    temp = sort(abs(lambda)/2/pi);
    fn(k) = temp(2);
end
```

**Figure 6-11: A Matlab code to generate Figure 6-10.**

To obtain the ground truth responses, the superstructure is placed on top of the foundation in the Direct model. A Rayleigh damping is considered for the superstructure to have 5% modal damping ratios at the first and last fixed-base modes. To carry out a time-history analysis, a chirp signal as shown in Figure 6-12 is magnified by the weight of every floor and applied to every floor. Two cases are analyzed: one considering a linear superstructure (large yield displacement) and one with a nonlinear superstructure. Opensees models for these two cases along with videos showing the response the model are available upon request from the first author.

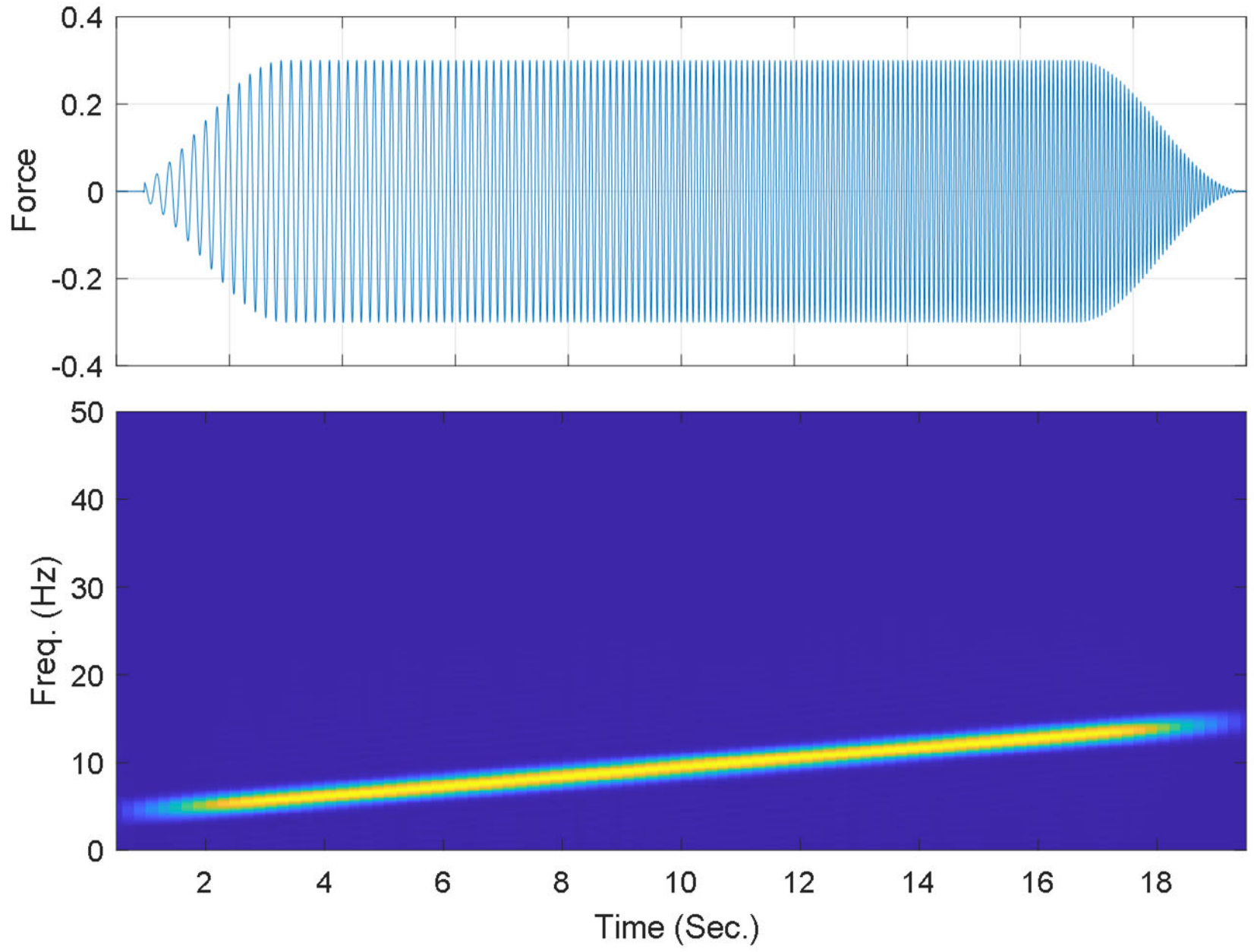

**Figure 6-12: Time (top) and time-frequency (bottom) representation of the input load.**

Finally, we run the substructure models with linear and nonlinear superstructure under the same lateral loads and compare the absolute displacements of the structural response with the ground truth response obtained from the corresponding direct models. To apply the HTFD method, parameters of the physical reference soil-foundation model are taken from impedance functions at zero frequency, as shown in Table 6-5. As there were no convergence issues, both models are analyzed using a single time window. Two OpenSeesPy models, incorporating the substructure method through the newly developed SSI element/analysis, are available upon request from the first author.

**Table 6-5: Properties of the physical reference soil-foundation model**

| $m_{h,ref}$ | $c_{h,ref}$ | $k_{h,ref}$ | $m_{r,ref}$ | $c_{r,ref}$ | $k_{r,ref}$ |
|---|---|---|---|---|---|
| **0** | $5.2 \times 10^6$ | $2.3 \times 10^7$ | 0 | $3.3 \times 10^7$ | $6.2 \times 10^8$ |

Figure 6-13 and Figure 6-14 show the comparison between the displacement response of all five stories obtained from the direct and substructure approach when the superstructure is linear and nonlinear, respectively. As these two figures show, the substructure approach in which the frequency-dependency is considered through newly implemented HTFD element/analysis is able to reproduce ground truth responses. It is important to mention that it takes 3 days to complete a nonlinear time-history analysis of this direct model on an ordinary computer, while the substructure model can be analyzed in under 5 minutes on the same computer.

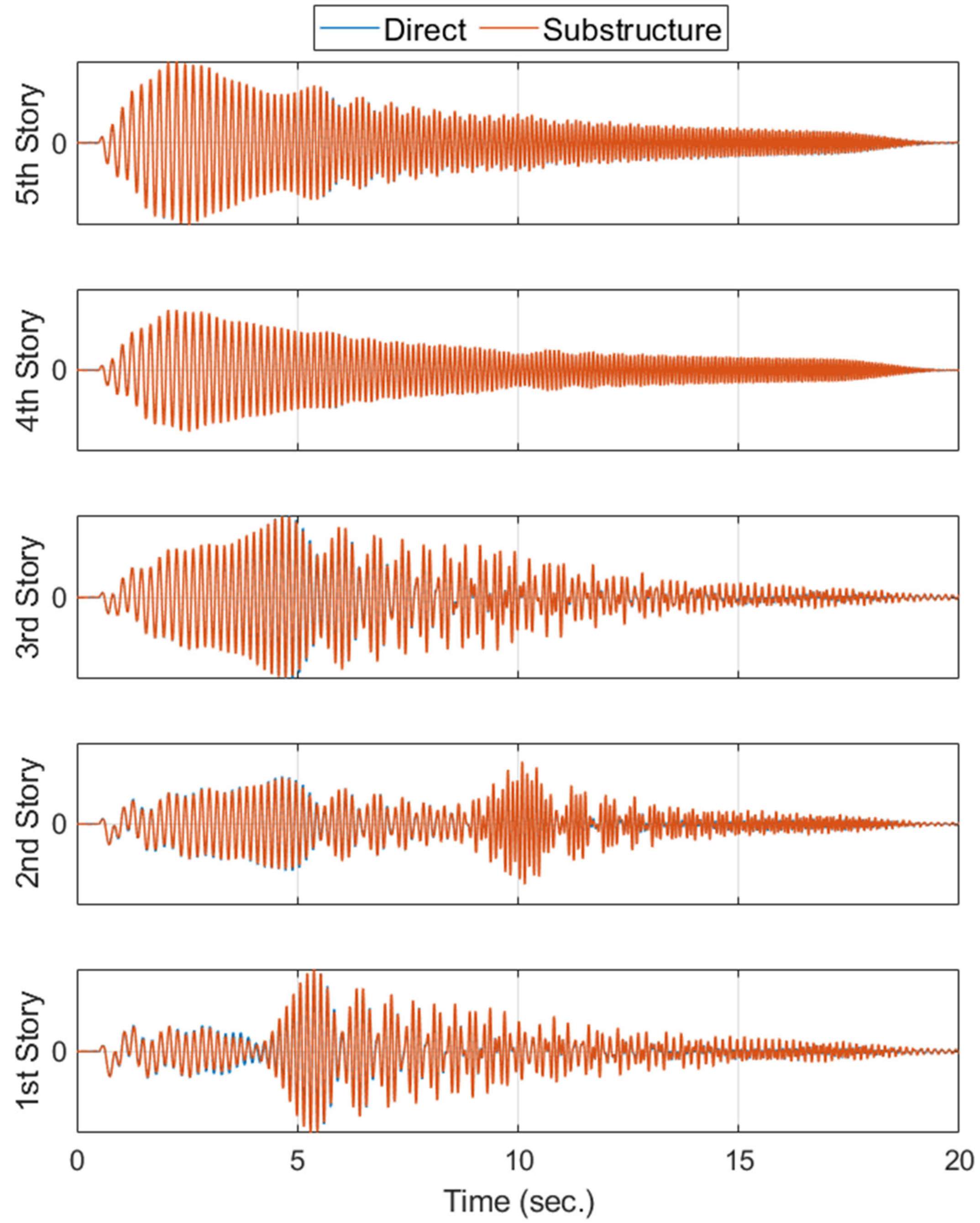

**Figure 6-13: Comparison between direct and substructure methods (linear superstructure).**

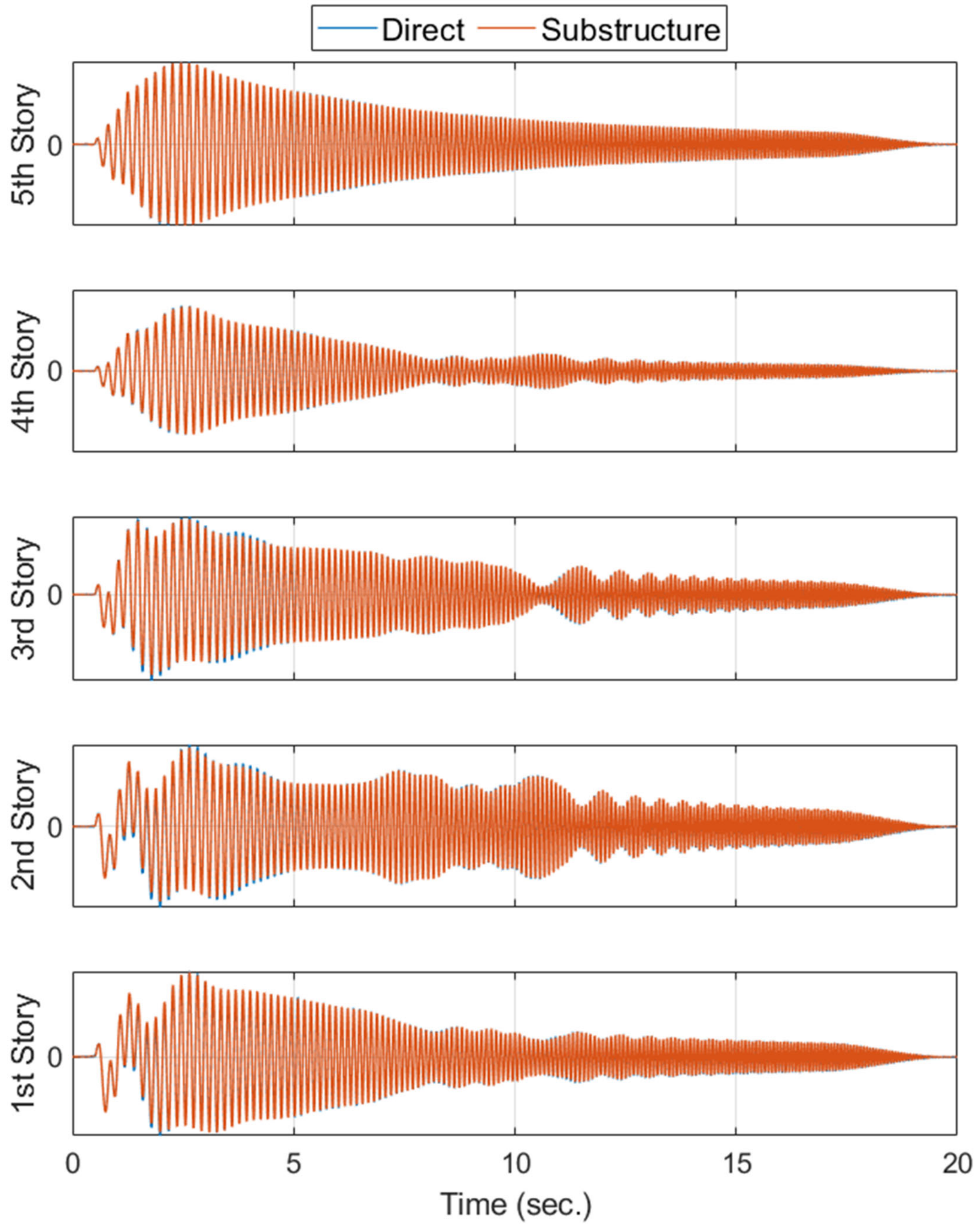

**Figure 6-14: Comparison between direct and substructure methods (nonlinear superstructure).**

# CHAPTER 7: CONCLUSIONS

## 7.1. CONCLUSIONS

To ensure accurate structural analysis, it is essential to consider Soil-Structure Interaction (SSI) effects. One approach to analyzing SSI effects involves creating and analyzing a comprehensive Finite Element Model (FEM) of the entire system, with the soil medium represented as a semi-infinite domain. This approach, known as the "direct" method, is commonly employed in research studies. However, in engineering practice, it is typically avoided due to the labor-intensive nature of developing finite element models and the associated high computational costs. Instead, SSI analysis is predominantly carried out using a substructure approach. In this approach, the superstructure is meticulously modeled using an FE model and placed on a soil-foundation substructure, represented by an Impedance Function (IF) system. The entire system is then analyzed considering Kinematic Interaction (KI) effects under Foundation Input Motions (FIMs) derived from Free-Field Motions (FFMs). While the method is theoretically designed for linear-elastic behavior based on the superposition assumption, the substructure method can be partially applied to nonlinear systems by condensing the viscous elastic soil-foundation only. Obtaining IFs for various soil and foundation configurations from analytical, numerical, or experimental analyses is possible, but implementing them in the time domain poses challenges as they exhibit frequency-dependent characteristics with unlimited bandwidth. One straightforward solution to this issue has been converting IFs into lumped-parameter physical models with frequency-independent components. However, connecting these components lacks a direct approach, and the coefficients of these components may entail non-physical parameters that cannot be modeled in FE software like OpenSEES, potentially resulting in an unstable lumped model. To address these issues associated with physical models, various alternative approaches have been proposed in the literature. In this project, we reviewed some of these existing solutions and evaluated their performance and stability through numerical examples. Following extensive review and evaluation, the Hybrid Time Frequency Domain (HTFD) method emerged as a more practical solution with fewer stability concerns. To facilitate its use by researchers and practitioners, this method was implemented in OpenSEES. This new implementation provides a new opportunity for researchers and engineers to consider both frequency-dependency from soil-foundation and nonlinearity from the superstructure in their analyses efficiently. Previously, a simple nonlinear time-history analysis using the direct method could take up to several days on an ordinary computer to run, while the new capability added to OpenSEES reduces this time to a few minutes. Verification examples presented in this report demonstrated the superb stability of the implementation, which is crucial when considering frequency-dependency in the time domain.

## 7.2. RECOMMENDATION FOR FUTURE STUDIES

Despite successfully passing verification using a diverse range of examples, the newly implemented approach necessitates additional validation and refinement to ensure its reliability across a wide spectrum of modeling conditions. For instance, although the new element/analysis demonstrates potential applicability in tackling 3D problems, rigorous testing specific to such scenarios is still pending. Moreover, the current implementation relies on a sequential non-overlapping windows approach for analysis, which, when compared to a methodology utilizing overlapping windows, may exhibit certain limitations, particularly in terms of achieving smoother behavior at the boundaries between windows. Consequently, exploring the integration of overlapping windows could potentially enhance the overall performance and continuity of the implemented approach. Additionally, while the current implementation assesses convergence solely based on the pseudo-force level, it is imperative to consider alternative options such as incorporating nodal displacements for convergence evaluation. By broadening the scope of convergence analysis techniques, a more comprehensive assessment of the implemented approach can be achieved. Consequently, further verification, particularly under diverse modeling conditions, coupled with the exploration of alternative convergence evaluation methods, will significantly contribute to the ongoing improvement and refinement of the implemented approach.

# APPENDIX A: THE IMPEDANCE FUNCTION OF A SEMI-INFINITE ROD WITH EXPONENTIALLY INCREASING AREA

Consider a foundation with an area $A_0$ resting on a semi-infinite rod with an exponentially increasing area as shown in Figure A-1(a). The area of the rod at any depth *x* can be expressed as [50]

$$A(x) = A_0 e^{\frac{x}{f}}, \tag{A-1}$$

where $f$ is a constant. The one-dimensional equation of motion is

$$\frac{\partial^2 u}{\partial x^2} + \frac{1}{A(x)}\frac{\partial A(x)}{\partial x}\frac{\partial u}{\partial x} = \frac{1}{c_1^2}\frac{\partial^2 u}{\partial t^2}, \tag{A-2}$$

where $c_1^2 = E/\rho$. By applying Fourier Transform, Eq. (A-2) can be written as

$$\frac{\partial^2 U}{\partial x^2} + \frac{1}{f}\frac{\partial U}{\partial x} + \frac{\omega^2}{c_1^2}U = 0, \tag{A-3}$$

To obtain the dynamic stiffness, a unit-impulse displacement is imposed in the frequency domain, i.e., $u(0,\omega) = 1$. The solution of the equation of motion under this boundary condition is

$$U(x, a_0) = e^{-\frac{x}{2f}\left(1+\sqrt{1-4a_0^2}\right)}, \tag{A-4}$$

with the dimensionless frequency $a_0 = \frac{\omega f}{c_1}$. Therefore, the dynamic stiffness is

$$S(a_0) = K\hat{S}(a_0), \tag{A-5}$$

where $K = \frac{EA_0}{f}$ is the static stiffness and $\hat{S}(a_0) = \frac{1}{2}\left(1 + \sqrt{1 - 4a_0^2}\right)$ is the dimensionless dynamic stiffness that can be split into real and imaginary parts as

$$\hat{S}(a_0) = k(a_0) + i a_0 c(a_0), \tag{A-6}$$

where $k(a_0)$ and $c(a_0)$ are frequency-dependent spring and dashpot coefficients. These coefficients are

$$a_0 < 0.5 \begin{cases} k(a_0) = \frac{1}{2}\left(1 + \sqrt{1 - 4a_0^2}\right) \\ c(a_0) = 0 \end{cases}$$
$$a_0 > 0.5 \begin{cases} k(a_0) = 0.5 \\ c(a_0) = \frac{i}{2}\sqrt{4a_0^2 - 1} \end{cases} \quad . \tag{A-7}$$

Introducing the dimensionless time parameter $\hat{t} = \frac{c_1 t}{f}$, time-domain representation of $\hat{S}(a_0)$ can be obtained through inverse Fourier transform as

$$\hat{s}(\hat{t}) = \frac{c_1}{2\pi f} \int_{-\infty}^{+\infty} \hat{S}(a_0)\, e^{ia_0\hat{t}} da_0. \tag{A-8}$$

However, $\hat{S}(a_0)$ becomes unbounded for $a_0 = \infty$ ($\hat{S}(a_0) \approx ia_0 + \frac{1}{2}$). It is thus necessary to decompose it into a regular and a singular part which results in the following time-domain representation

$$\hat{s}(\hat{t}) = \frac{c_1}{f}\left[\frac{1}{2}\delta(\hat{t}) + \frac{d\delta(\hat{t})}{d\hat{t}} + \frac{1}{2\hat{t}} J_1\left(\frac{\hat{t}}{2}\right)\right]. \tag{A-9}$$

where $J_1$ is the Bessel function of the first kind of the first order. The regular part in the time domain, $\frac{1}{2\hat{t}} J_1\left(\frac{\hat{t}}{2}\right)$, is shown in Figure A-1(b).

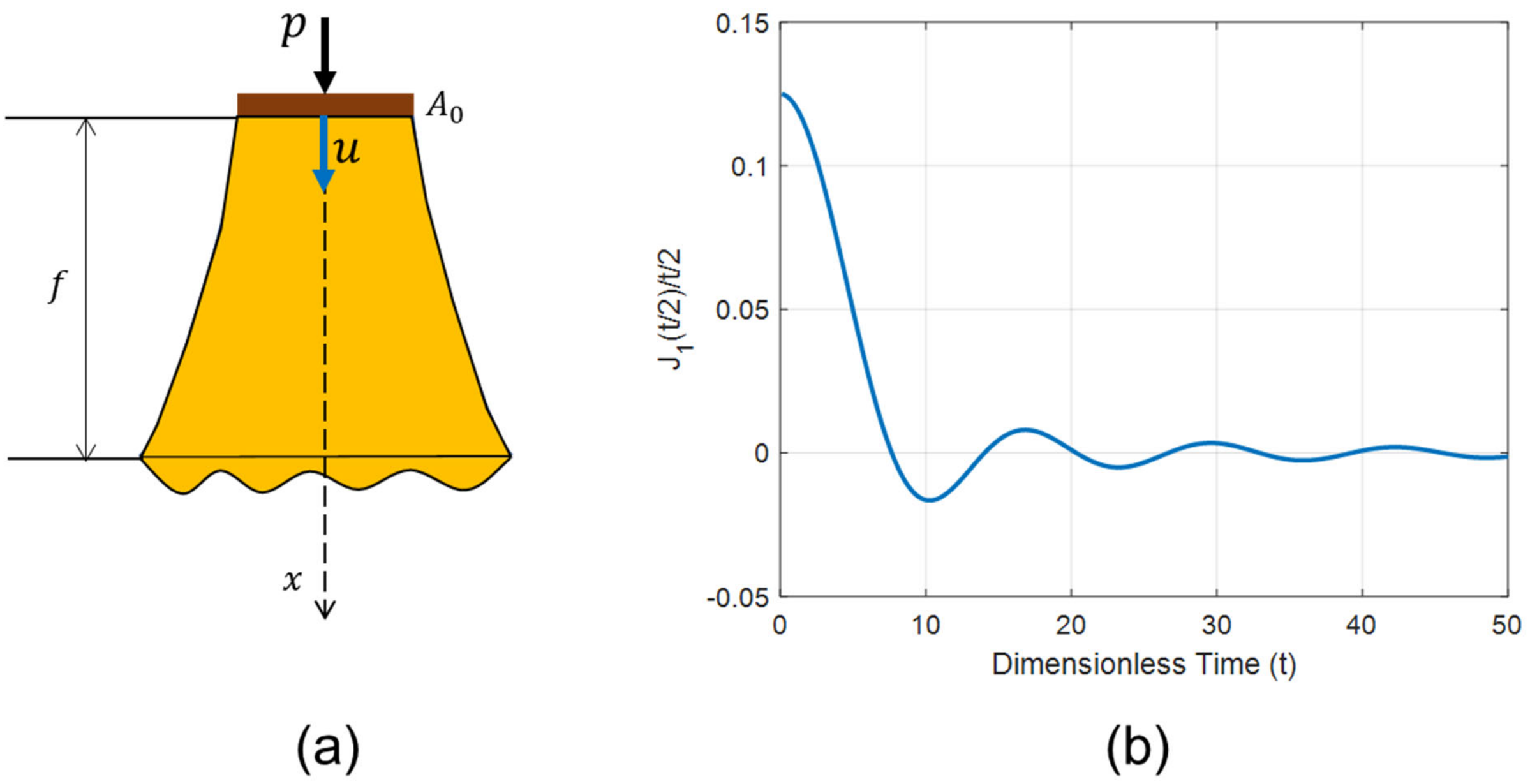

**Figure A-1: (a) A rigid foundation on a semi-infinite rod, (b) the time-domain representation of the regular part of its impedance function.**

In the unscaled time domain, the dynamic stiffness can be expressed as

$$\hat{s}(t) = \frac{1}{2}\delta(t) + \frac{f}{c_1}\frac{d\delta(t)}{d\hat{t}} + \frac{1}{2t}J_1\left(\frac{c_1}{2f}t\right), \tag{A-10}$$

by which the interaction force under any imposed displacement can be obtained through time convolution as

$$p(t) = K\left[\frac{1}{2}u(t) + \frac{f}{c_1}\dot{u}(t) + \frac{1}{2}\int_0^t \frac{1}{t-\tau}J_1\left(\frac{c_1}{2f}(t-\tau)\right)u(\tau)d\tau\right], \tag{A-11}$$

# APPENDIX B: THE IMPEDANCE FUNCTION OF A SEMI-INFINITE ROD ON AN ELASTIC FOUNDATION

## B.1. INTRODUCTION

Consider a semi-infinite rod resting on an elastic foundation with stiffness per length of $k_g$ as shown in Figure B-1(a) [50], [98]. The equation of motion of this system can be written as

$$\frac{\partial^2 u}{\partial x^2} - \alpha^2 u = \frac{1}{c_1^2}\frac{\partial^2 u}{\partial t^2}, \tag{B-1}$$

where

$$\alpha^2 = \frac{k_g}{EA}, \tag{B-2}$$

$$c_1^2 = \frac{E}{\rho}, \tag{B-3}$$

where $E$, $\rho$, and $A$ are elastic modulus, mass density, and section area, respectively. Introducing the dimensionless frequency $a_0 = \frac{\omega}{c_1 \alpha}$ and wave number $k = \frac{\omega}{c}$ where $c$ is the phase velocity, the solution in the frequency domain is given by

$$U(a_0) = ae^{-ikx} + be^{-ikx} \tag{B-4}$$

with $k = \alpha\sqrt{a_0^2 - 1}$ and $c = \frac{a_0}{\sqrt{a_0^2-1}} c_1$ the latter shows the system is dispersive. As seen, there is a cut-off frequency at $a_0 = 1$ below which the motion does not propagate, but decays exponentially. The group velocity $c_g = \frac{d\omega}{dk}$ is equal to $\frac{\sqrt{a_0^2-1}}{a_0} c_1$, so phase and group velocities converge to infinite and zero, respectively, at $a_0 = 1$, while both converge to $c_1$ for $a_0 = +\infty$.

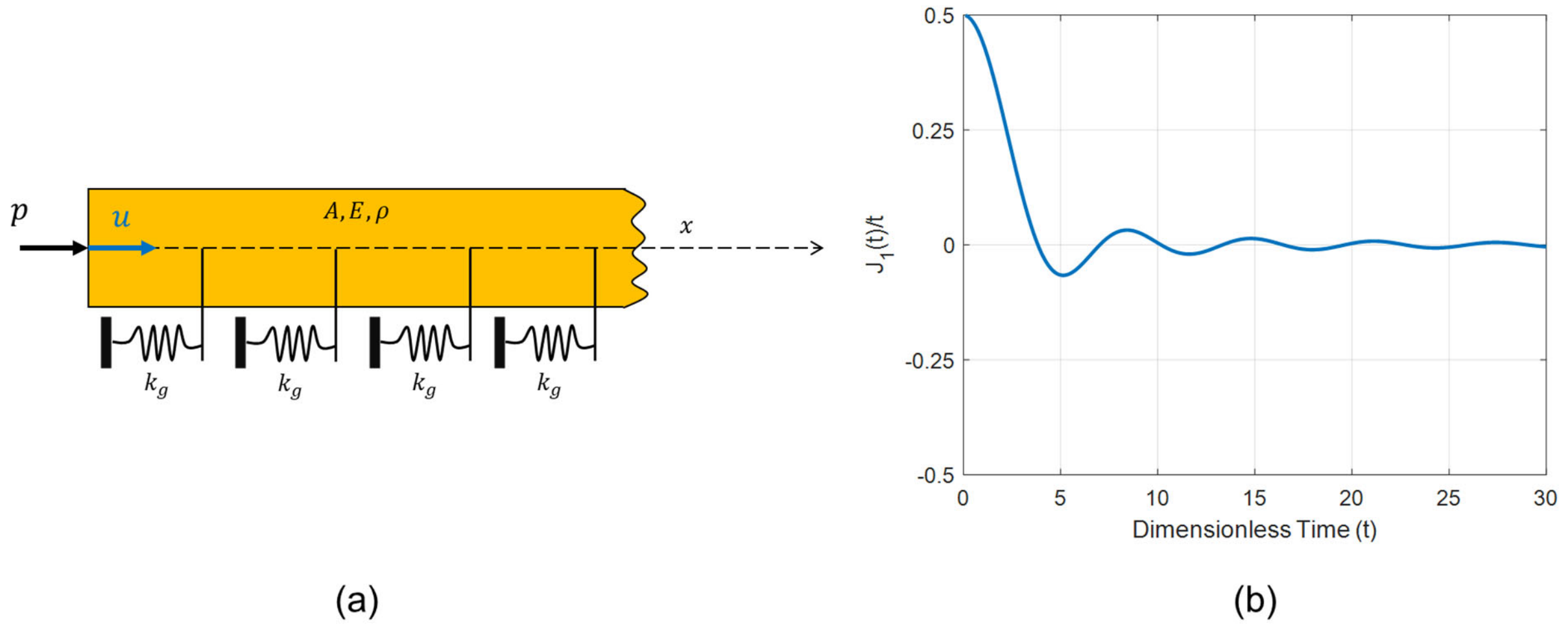

**Figure B-1: (a) A semi-infinite rod on an elastic half-space, and (b) the time-domain representation of the regular part of its impedance function.**

## B.2. THE IMPEDANCE FUNCTION IN THE FREQUENCY DOMAIN

With $p(a_0)$ denoting the amplitude of the force in the frequency domain, we can write

$$P(a_0) \; = S(a_0)U(a_0) \tag{B-5}$$

where $S(a_0) = K\hat{S}(a_0)$ is the impedance function in which $K = \sqrt{EAk_g}$ is the static stiffness and $\hat{S}(a_0) = i\sqrt{a_0^2 - 1}$ is the dimensionless dynamic stiffness that can be split into real and imaginary parts as

$$\hat{S}(a_0) \; = k_0(a_0) + ia_0c_0(a_0), \tag{B-6}$$

where $k_0(a_0)$ and $c_0(a_0)$ are frequency-dependent spring and dashpot coefficients which are defined as

$$\begin{aligned} a_0 < 1 & \begin{cases} k_0(a_0) = \sqrt{1 - a_0^2} \\ c_0(a_0) = 0 \end{cases} \\ a_0 > 1 & \begin{cases} k_0(a_0) = 0 \\ c_0(a_0) = \sqrt{1 - \dfrac{1}{a_0^2}} \end{cases} \end{aligned} \; . \tag{B-7}$$

## B.3. THE IMPEDANCE FUNCTION IN THE TIME DOMAIN

Introducing the dimensionless time parameter $\hat{t} = c_1\alpha t$, the time-domain representation of $\hat{S}(a_0)$ can be obtained through Inverse Fourier Transform as

$$\hat{s}(\hat{t}) = c_1\alpha \int_{-\infty}^{+\infty} \hat{S}(a_0)\, e^{ia_0\hat{t}} da_0. \tag{B-8}$$

However, $\hat{S}(a_0)$ becomes unbounded for $a_0 = \infty$ ( $\hat{S}(a_0) \approx ia_0$ ). So, we rewrite $\hat{S}(a_0) = i\sqrt{a_0^2 - 1} + ia_0 - ia_0$ which can be decomposed into regular $(i\sqrt{a_0^2 - 1} - ia_0)$ and singular $(ia_0)$ parts resulting in the following time-domain representation

$$\hat{s}(\hat{t}) = c_1 \left[ \frac{d\delta(\hat{t})}{d\hat{t}} + \frac{1}{\hat{t}} J_1(\hat{t}) \right], \tag{B-9}$$

where $J_1$ is the Bessel function of the first kind and of the first order. The regular part in the time domain is shown in Figure B-1(b). By convolving $\hat{s}$ with $u$, the following relationship can be written for the applied force

$$p(t) = K \left[ \sqrt{\frac{A\rho}{k_g}} \dot{u}(t) + \int_0^t \frac{1}{t-\tau} J_1 \left( \sqrt{\frac{k_g}{A\rho}} (t - \tau) \right) u(\tau) d\tau \right], \tag{B-10}$$

that shows the singular part can be interpreted as a viscous damper with a coefficient $c_1\alpha$.

# APPENDIX C: THE STABILITY OF THE NEWMARK METHOD

Let's assume that we have an SDOF system without any frequency-dependency. The equation of motion is shown in (C-1).

$$m\ddot{u}(t) + c\dot{u}(t) + ku(t) = p(t). \tag{C-1}$$

The discrete-time version of this equation at time instant $n + 1$ is shown below

$$m\ddot{u}_{n+1} + c\dot{u}_{n+1} + ku_{n+1} = p_{n+1}. \tag{C-2}$$

In the Newmark integration method, we assume a distribution for the acceleration during a time step and based on that distribution, we estimate velocity and displacement at the next time step as shown in the equation below.

$$\dot{u}_{n+1} = \dot{u}_n + (1-\gamma)\Delta t\ddot{u}_n + \gamma\Delta t\ddot{u}_{n+1}, \tag{C-3}$$

$$u_{n+1} = u_n + \dot{u}_n\Delta t + \left(\frac{1}{2} - \beta\right)\Delta t^2\ddot{u}_n + \beta\Delta t^2\ddot{u}_{n+1}, \tag{C-4}$$

where parameters $\gamma$ and $\beta$ and sampling time $\Delta t$ determine the accuracy and stability of the solution. We know that $\gamma$ equal to 0.5 is the common choice to have zero numerical dampings, and $\beta$ equal to $\frac{1}{6}$ and $\frac{1}{4}$ are two common choices [99]. These two equations combined with the equilibrium equation at time step $n + 1$ can be written in a matrix form as shown in Eq. (C-5) to compute displacement, velocity, and acceleration at time step $n + 1$.

$$\begin{bmatrix} k & c & m \\ 0 & 1 & -\gamma\Delta t \\ 1 & 0 & -\beta\Delta t^2 \end{bmatrix} \boldsymbol{x}_{n+1} = \begin{bmatrix} 0 & 0 & 0 \\ 0 & 1 & (1-\gamma)\Delta t \\ 1 & \Delta t & \left(\frac{1}{2} - \beta\right)\Delta t^2 \end{bmatrix} \boldsymbol{x}_n. \tag{C-5}$$

where $\boldsymbol{x}_{n+1} = [u_n, \dot{u}_{n+1}, \ddot{u}_{n+1}]^T$. Note that the external force is dropped as we are going to study the stability. Based on Eq. (C-5), the relationship between two consecutive time instants can be written as shown in (C-6).

$$\mathbf{H}_1\boldsymbol{x}_{n+1} = \mathbf{H}_0\boldsymbol{x}_n. \tag{C-6}$$

If matrix $\mathbf{H}_1$ can be written in a triangular form, this equation is explicit; otherwise, the formula is implicit. The Newmark method is implicit because, as seen here, $\mathbf{H}_1$ is not triangular. However, it is possible to convert this equation to an explicit form if the system is linear, but let's keep it in the current form. By defining the approximation operator matrix **A**, we can write Eq. (C-7)

$$\boldsymbol{x}_{n+1} = \mathbf{A}\boldsymbol{x}_n, \tag{C-7}$$

where $\mathbf{A} = \mathbf{H}_1{}^{-1}\mathbf{H}_0$ and can be expanded as [100]

$$\mathbf{A} = \frac{1}{1+\gamma b_2+\beta b_1}\begin{bmatrix} 1+\gamma b_2 & \Delta t[1+b_2(\gamma-\beta)] & \Delta t^2\left[\left(\frac{1}{2}-\beta\right)+b_2\left(\frac{\gamma^2}{2}-\beta\right)\right] \\ -\frac{\gamma b_1}{\Delta t} & 1+b_1(\beta-\gamma) & \Delta t\left[(1-\gamma)+b_1\left(\beta-\frac{\gamma}{2}\right)\right] \\ -\frac{b_1}{\Delta t^2} & -\frac{b_1+b_2}{\Delta t} & -\left[b_2(1-b_1)+b_1\left(\frac{1}{2}-\beta\right)\right] \end{bmatrix}, \tag{C-8}$$

where two dimensionless parameters $b_1$ and $b_1$ are

$$b_1 = \Delta t^2\omega_n^2, \tag{C-9}$$

$$b_2 = 2\xi\Delta t\omega_n, \tag{C-10}$$

in which $\omega_n$ and $\xi$ are natural frequency and damping ratio, respectively.

An integration method is unconditionally stable if the solution for any initial conditions does not grow without bound for any time step $\Delta t$, especially when $\Delta t\ /T$ is large where the $T$ is the period of interest (here $T = \frac{2\pi}{\omega_n}$). The method is only conditionally stable if the same holds provided $\frac{\Delta t}{T}$ is smaller than a certain threshold [101]. So, the stability would be guaranteed if the absolute value of the largest eigenvalue (spectral radius) of matrix **A** is less than one. For the common $\gamma = 0.5$, the non-trivial eigenvalues of matrix **A** are shown below.

$$\lambda_{1,2} = \frac{1+b_1\left(\beta-\frac{1}{2}\right) \pm i\sqrt{b_1^2\left(\beta-\frac{1}{4}\right)+b_1-\frac{b_2^2}{4}}}{1+\frac{b_2}{2}+\beta b_1}. \tag{C-11}$$

Figure C-1 shows the spectral radius of the matrix **A** obtained from equation above for a range of $\beta$ and $\Delta t\ /T$ assuming zero damping. As seen, when $\beta \geq \frac{1}{4}$ the integration method is unconditionally stable, otherwise, the stability is a function of $\Delta t\ /T$ and goes toward unstable solution when $\Delta t\ /T$ increases. The Matlab code to reproduce Figure C-1 is presented in Figure C-2.

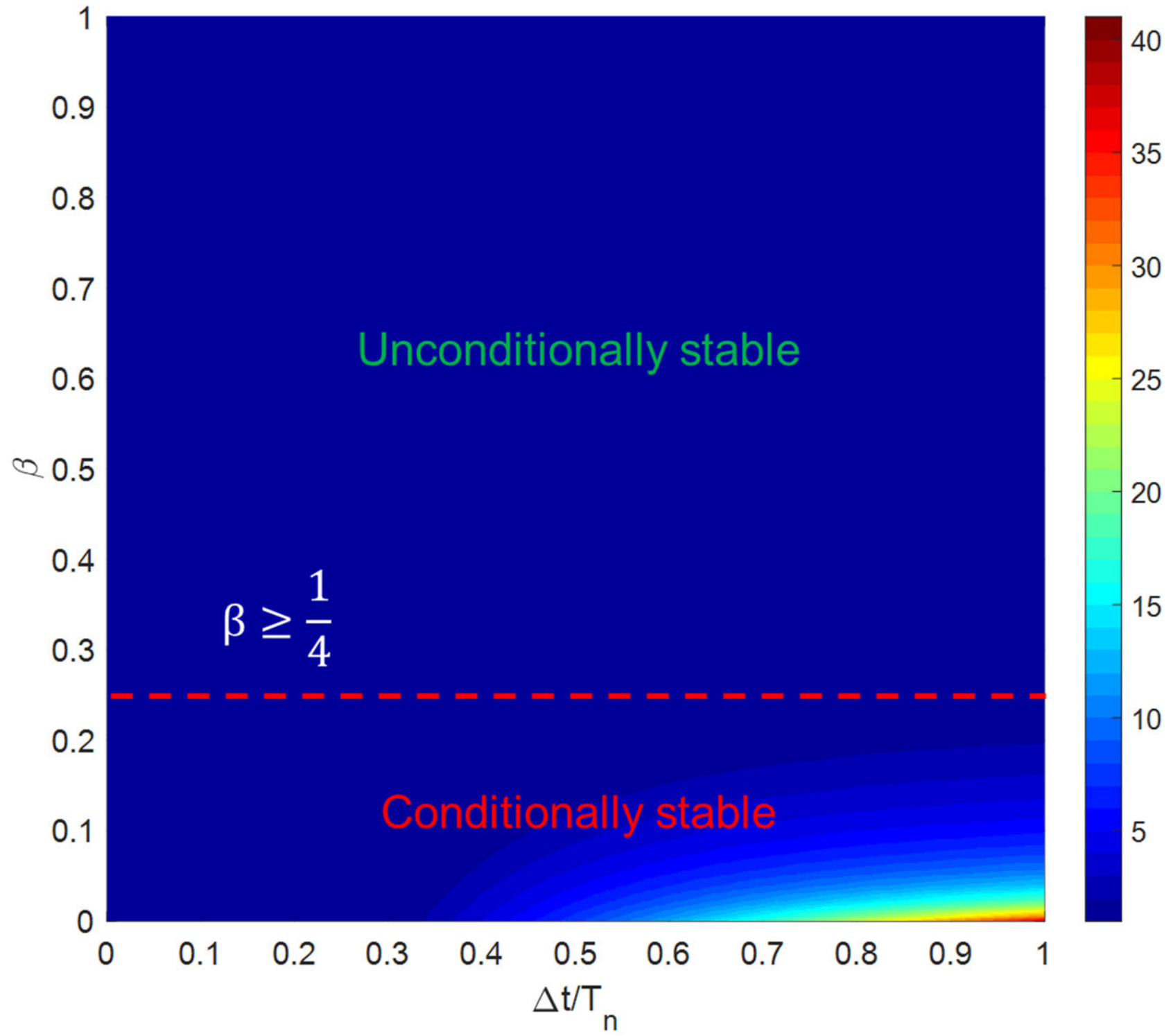

**Figure C-1: The spectral radius of matrix A**.

```matlab
g = 1/2;
syms dt b b1 b2

A = 1/(1+g*b2+b*b1)*[1+g*b2 dt*(1+b2*(g-b)) dt^2*((1/2-b)+b2*(g^2/2-b));...
-g*b1/dt 1+b1*(b-g) dt*((1-g)+b1*(b-g/2));...
-b1/dt^2 -(b1+b2)/dt -(b2*(1-b1)+b1*(1/2-b))];
A_Func = matlabFunction(A);
LAMBDA = matlabFunction(eig(A));
LAMBDA2 = @(b,b1,b2)[0;(1+b1*(b-1/2)+1i*sqrt(b1^2*(b-1/4)+b1-b2^2/4))/(1+b2/2+b*b1);(1+b1*(b-1/2)-
1i*sqrt(b1^2*(b-1/4)+b1-b2^2/4))/(1+b2/2+b*b1)];

%% Analytical formula
dt_T = 0.001:0.001:1;
betta = 0.001:0.001:1;
for ii = 1:length(dt_T)
        for jj = 1:length(betta)
                Temp = LAMBDA(betta(jj),4*pi^2*dt_T(ii)^2,0);
                lambda(ii,jj) = max(abs(Temp));
        end
end

figure1 = figure('Position',[971 407 779 650]);
axes1 = axes('Parent',figure1);
hold(axes1,'on');
image(dt_T,betta,lambda','CDataMapping','direct','Parent',axes1);
ylabel('\beta','FontSize',20);
xlabel('{\Delta}t/T_n','FontSize',20);
xlim(axes1,[0 1]);
ylim(axes1,[0 1]);
box(axes1,'on');
set(axes1,'CLim',[0 40],'FontSize',14,'XTick',[0 0.1 0.2 0.3 0.4 0.5 0.6 0.7 0.8 0.9 1]);
colorMap = jet(40);
colormap(colorMap);
colorbar('peer',axes1);

%% Analytical free vibration (under damped)
syms t
wn = 2*pi*2;
xi = 0.05;
wd = wn*sqrt(1-xi^2);
we = wn*sqrt(xi^2-1);
u0 = 1;
v0 = 0;
u = exp(-xi.*wn.*t).*(u0.*cos(wd.*t)+(v0+xi.*wn.*u0)./wd.*sin(wd.*t));
a = diff(diff((u)));
dt = 0.001;
T = 10;
N = fix(T/dt);
Time = 0:dt:(N-1)*dt;
x = zeros(3,N);
x(1) = u0;
for k=2:N
        x(:,k) = A_Func(0.25,(dt*wn)^2,2*xi*dt*wn,dt)*x(:,k-1);
end
figure; fplot(a,[0 Time(end)])
hold on;
plot(Time,x(3,:))
```

**Figure C-2: A Matlab code to produce Figure C-1.**

# APPENDIX D: THE STABILITY OF THE NAKAMURA'S METHOD

## D.1. STABILITY CONDITION

Herein the stability of Nakamura's method[15] [69] is investigated [102]. Let's assume the simple SDOF model on the frequency-dependent sway-rocking soil-foundation model shown in Figure D-1. The equation of motion governing this system can be written as

$$\mathbf{M}\ddot{\mathbf{u}}(t) + \mathbf{C}\dot{\mathbf{u}}(t) + \mathbf{K}\mathbf{u}(t) = \mathbf{p}(t) - \mathbf{r}(t), \tag{D-1}$$

where $\mathbf{u}(t) = [u_f(t) \quad \theta_f(t) \quad u_s(t)]^T$ and mass, damping, and stiffness matrices are defined as

$$\mathbf{M} = \begin{bmatrix} m_f + m_s & m_f h_f + m_s(2h_f + h_s) & m_s \\ m_f h_f + m_s(2h_f + h_s) & I_f + m_f h_f^2 + I_s + m_s(2h_f + h_s)^2 & m_s(2h_f + h_s) \\ m_s & m_s(2h_f + h_s) & m_s \end{bmatrix}, \tag{D-2}$$

$$\mathbf{C} = \begin{bmatrix} 0 & 0 & 0 \\ 0 & 0 & 0 \\ 0 & 0 & c_s \end{bmatrix}, \tag{D-3}$$

$$\mathbf{K} = \begin{bmatrix} 0 & 0 & 0 \\ 0 & 0 & 0 \\ 0 & 0 & k_s \end{bmatrix}, \tag{D-4}$$

and the external force vector caused by the horizontal ground acceleration $\ddot{\mathbf{u}}_g(t)$ is

$$\mathbf{p}(t) = \mathbf{M}\mathbf{s}\ddot{\mathbf{u}}_g(t). \tag{D-5}$$

in which the influence vector is $\mathbf{s} = [1 \quad 0 \quad 0]^T$. The soil-foundation reaction force vector is $\mathbf{r}(t) = [F(t) \quad M(t) \quad 0]^T$ in which $F(t)$ and $M(t)$ are calculated using the Nakamura's method as

$$F(t) = \sum_{j=0}^{n_j-1} k_{h,j}\, u_f(t - t_j) + \sum_{j=0}^{n_j-2} c_{h,j}\, \dot{u}_f(t - t_j) + m_{h,0}\ddot{u}_f(t), \tag{D-6}$$

$$M(t) = \sum_{j=0}^{n_j-1} k_{r,j}\, \theta_f(t - t_j) + \sum_{j=0}^{n_j-2} c_{r,j}\, \dot{\theta}_f(t - t_j) + m_{r,0}\ddot{\theta}_f(t). \tag{D-7}$$

[15] Herein we use Method B which includes stiffness, damping, and mass terms.

The instantaneous terms in Eqs. (D-6) and (D-7) can be moved to the left-hand-side of Eq. (D-1), so we can rewrite the equations as

$$\bar{\mathbf{M}}\ddot{\mathbf{u}}(t) + \bar{\mathbf{C}}\dot{\mathbf{u}}(t) + \bar{\mathbf{K}}\mathbf{u}(t) = \mathbf{p}(t) - \bar{\mathbf{r}}(t), \tag{D-8}$$

with

$$\bar{\mathbf{M}} = \begin{bmatrix} m_{h,0} + m_f + m_s & m_f h_f + m_s(2h_f + h_s) & m_s \\ m_f h_f + m_s(2h_f + h_s) & m_{r,0} + I_f + m_f h_f^2 + I_s + m_s(2h_f + h_s)^2 & m_s(2h_f + h_s) \\ m_s & m_s(2h_f + h_s) & m_s \end{bmatrix}, \tag{D-9}$$

$$\bar{\mathbf{C}} = \begin{bmatrix} c_{h,0} & 0 & 0 \\ 0 & c_{r,0} & 0 \\ 0 & 0 & c_s \end{bmatrix}, \tag{D-10}$$

$$\bar{\mathbf{K}} = \begin{bmatrix} k_{h,0} & 0 & 0 \\ 0 & k_{r,0} & 0 \\ 0 & 0 & k_s \end{bmatrix}, \tag{D-11}$$

$$\bar{\mathbf{r}}(t) = [\bar{F}(t) \quad \bar{M}(t) \quad 0]^T, \tag{D-12}$$

where

$$F(t) = \sum_{j=1}^{n_j-1} k_{h,j}\, u_f(t - t_j) + \sum_{j=1}^{n_j-2} c_{h,j}\, \dot{u}_f(t - t_j), \tag{D-13}$$

$$M(t) = \sum_{j=1}^{n_j-1} k_{r,j}\, \theta_f(t - t_j) + \sum_{j=1}^{n_j-2} c_{r,j}\, \dot{\theta}_f(t - t_j). \tag{D-14}$$

Now, we can use the Newmark integration method to solve the equation in the discrete-time domain. That is,

$$\bar{\mathbf{M}}\ddot{\mathbf{u}}_{n+1} + \bar{\mathbf{C}}\dot{\mathbf{u}}_{n+1} + \bar{\mathbf{K}}\mathbf{u}_{n+1} = \mathbf{p}_{n+1} - \bar{\mathbf{r}}_{n+1}, \tag{D-15}$$

$$\dot{\boldsymbol{u}}_{n+1} = \frac{\gamma}{\beta\Delta t}(\boldsymbol{u}_{n+1} - \boldsymbol{u}_n) + \left(1 - \frac{\gamma}{\beta}\right)\dot{\boldsymbol{u}}_n + \Delta t\left(1 - \frac{\gamma}{2\beta}\right)\ddot{\boldsymbol{u}}_n, \tag{D-16}$$

$$\ddot{\boldsymbol{u}}_{n+1} = \frac{1}{\beta\Delta t^2}(\boldsymbol{u}_{n+1} - \boldsymbol{u}_n) - \frac{1}{\beta\Delta t}\dot{\boldsymbol{u}}_n - \left(\frac{1}{2\beta} - 1\right)\ddot{\boldsymbol{u}}_n. \tag{D-17}$$

Because we are dealing with a linear system for the stability analysis, it is possible to modify the original equations (see Appendix B) to convert them to the explicit version. By replacing velocity and acceleration at time $n + 1$ into the equation of motion, we get the equation below with modified mass, damping, and stiffness matrices shown with over hat.

$$\hat{\mathbf{K}}\mathbf{u}_{\boldsymbol{n}+1} = -\hat{\mathbf{M}}\ddot{\mathbf{u}}_{\boldsymbol{n}} - \hat{\mathbf{C}}\dot{\boldsymbol{u}}_n + (\hat{\mathbf{K}} - \bar{\mathbf{K}})\mathbf{u}_n + \mathbf{p}_{n+1} - \bar{\mathbf{r}}_{n+1}, \tag{D-18}$$

$$\widehat{\mathbf{M}} = -\left(\frac{1}{2\beta} - 1\right)\overline{\mathbf{M}} + \Delta t\left(1 - \frac{\gamma}{2\beta}\right)\overline{\mathbf{C}}, \tag{D-19}$$

$$\widehat{\mathbf{C}} = -\frac{1}{\beta\Delta t}\overline{\mathbf{M}} + \left(1 - \frac{\gamma}{\beta}\right)\overline{\mathbf{C}}, \tag{D-20}$$

$$\widehat{\mathbf{K}} = \frac{1}{\beta\Delta t^2}\overline{\mathbf{M}} + \frac{\gamma}{\beta\Delta t}\overline{\mathbf{C}} + \overline{\mathbf{K}}. \tag{D-21}$$

Putting displacement, velocity, and acceleration into an extended vector $\mathbf{x}_{n+1} = [\mathbf{u}_{n+1} \quad \dot{\mathbf{u}}_{n+1} \quad \ddot{\mathbf{u}}_{n+1}]^T$, we can write the previous equation of motion along with the Newmark equations in a matrix form as shown below.

$$\begin{bmatrix} \widehat{\mathbf{K}} & \mathbf{0} & \mathbf{0} \\ -\frac{\gamma}{\beta\Delta t}\mathbf{I} & \mathbf{I} & \mathbf{0} \\ -\frac{1}{\beta\Delta t^2}\mathbf{I} & \mathbf{0} & \mathbf{I} \end{bmatrix}\mathbf{x}_{n+1} = \begin{bmatrix} \widehat{\mathbf{K}} - \overline{\mathbf{K}} & -\widehat{\mathbf{C}} & -\widehat{\mathbf{M}} \\ -\frac{\gamma}{\beta\Delta t}\mathbf{I} & \left(1 - \frac{\gamma}{\beta}\right)\mathbf{I} & \Delta t\left(1 - \frac{\gamma}{2\beta}\right)\mathbf{I} \\ -\frac{1}{\beta\Delta t^2}\mathbf{I} & -\frac{1}{\beta\Delta t}\mathbf{I} & -\left(\frac{1}{2\beta} - 1\right)\mathbf{I} \end{bmatrix}\mathbf{x}_n - \sum_{j=1}^{n_j}\begin{bmatrix} \mathbf{K}_j & \mathbf{C}_j & \mathbf{0} \\ \mathbf{0} & \mathbf{0} & \mathbf{0} \\ \mathbf{0} & \mathbf{0} & \mathbf{0} \end{bmatrix}\mathbf{x}_{n+1-j}, \tag{D-22}$$

where $\mathbf{I}$ and $\mathbf{0}$ are $\mathbf{3} \times \mathbf{3}$ identity and zero matrices and

$$\mathbf{K}_j = \begin{bmatrix} k_{h,j} & 0 & 0 \\ 0 & k_{r,j} & 0 \\ 0 & 0 & 0 \end{bmatrix}, \tag{D-23}$$

$$\mathbf{C}_j = \begin{bmatrix} c_{h,j} & 0 & 0 \\ 0 & c_{r,j} & 0 \\ 0 & 0 & 0 \end{bmatrix}. \tag{D-24}$$

Note that external force is dropped to study the stability. As seen, in comparison to the classic Newmark integration (Appendix C), a set of new diagonal matrices $\mathbf{K}_j$and $\mathbf{C}_j$ with filter coefficients on their diagonals appear here.

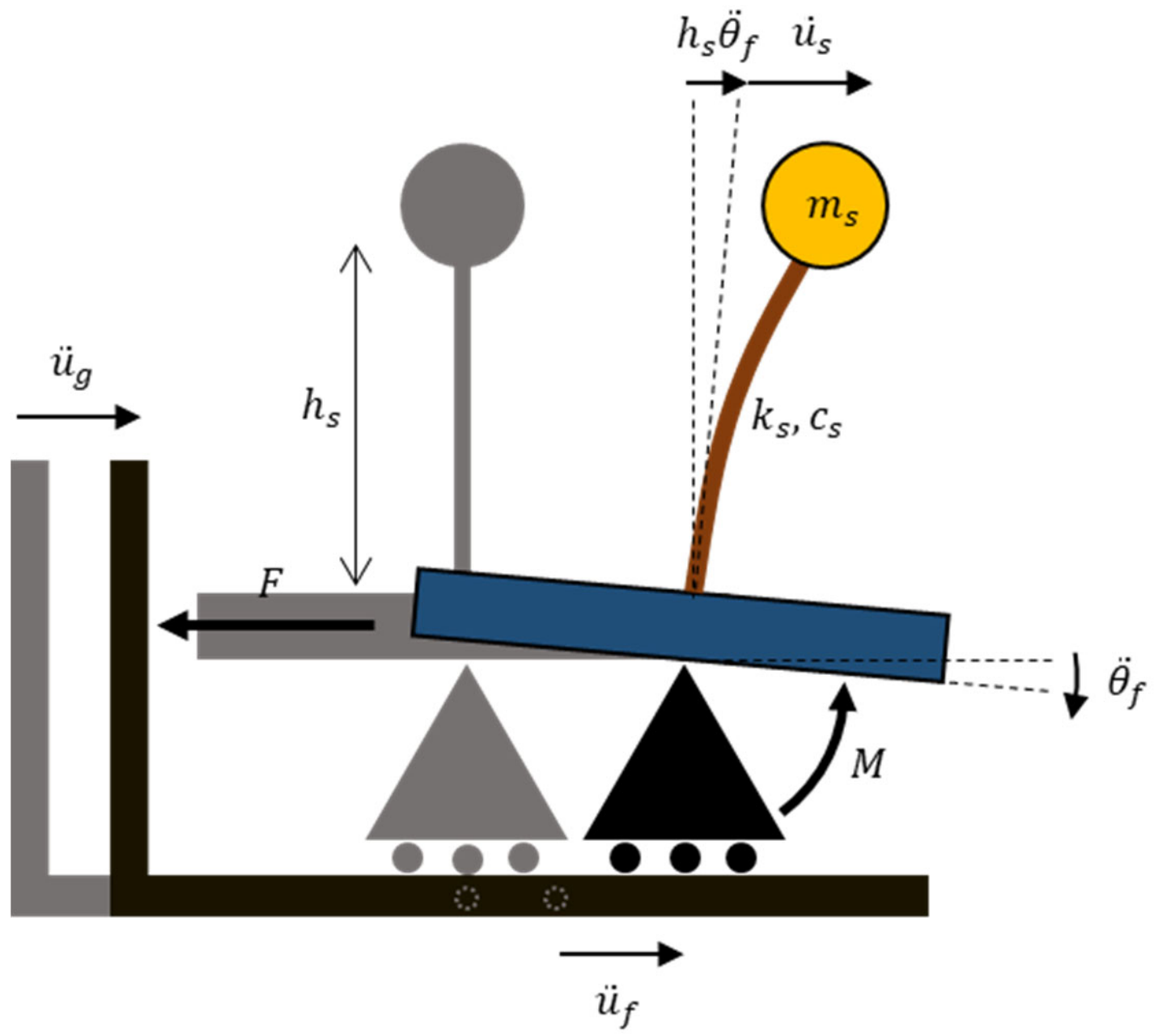

**Figure D-1: The soil-structure model.**

Similar to Appendix B where the stability of the classical Newmark integration is studied, we can rewrite Eq. (D-22) as

$$\mathbf{H}_1\mathbf{x}_{n+1} = \sum_{j=1}^{n_j} \mathbf{H}_{0,j}\mathbf{x}_{n+1-j}, \tag{D-25}$$

where

$$\mathbf{H}_1 = \begin{bmatrix} \widehat{\mathbf{K}} & \mathbf{0} & \mathbf{0} \\ -\frac{\gamma}{\beta\Delta t}\mathbf{I} & \mathbf{I} & \mathbf{0} \\ -\frac{1}{\beta\Delta t^2}\mathbf{I} & \mathbf{0} & \mathbf{I} \end{bmatrix}, \tag{D-26}$$

and

$$\mathbf{H}_{0,1} = \begin{bmatrix} \widehat{\mathbf{K}} - \overline{\mathbf{K}} - \mathbf{K}_1 & -\widehat{\mathbf{C}} - \mathbf{C}_1 & -\widehat{\mathbf{M}} \\ -\frac{\gamma}{\beta\Delta t}\mathbf{I} & \left(1 - \frac{\gamma}{\beta}\right)\mathbf{I} & \Delta t\left(1 - \frac{\gamma}{2\beta}\right)\mathbf{I} \\ -\frac{1}{\beta\Delta t^2}\mathbf{I} & -\frac{1}{\beta\Delta t}\mathbf{I} & -\left(\frac{1}{2\beta} - 1\right)\mathbf{I} \end{bmatrix}, \tag{D-27}$$

$$\boldsymbol{H}_{0,j>1} = \begin{bmatrix} -\mathbf{K}_j & -\mathbf{C}_j & \mathbf{0} \\ \mathbf{0} & \mathbf{0} & \mathbf{0} \\ \mathbf{0} & \mathbf{0} & \mathbf{0} \end{bmatrix}. \tag{D-28}$$

Eq. (D-25) can be solved as

$$\mathbf{x}_{n+1} = \mathbf{H}_1^{-1} \sum_{j=1}^{n_j} \mathbf{H}_{0,j} \mathbf{x}_{n+1-j}, \tag{D-29}$$

in which, contrary to the classical Newmark integration, the approximation operator matrix is not constant and evolves in time. By some mathematical manipulation, it is easy to show that the response of any time instant can be written versus the initial response $\mathbf{x}_{n+1} = \mathbf{A}_n \mathbf{x}_0$ where the approximation operator matrix $\mathbf{A}_n$ is shown in Table D-1 for some time instants. Now, it is possible to use spectral radius stability criteria as follows:

- $max(|eig(\mathbf{A}_k)|) < 1 \quad \forall\, k = 0, \dots, n_j + m$
- $max(|eig(\mathbf{A}_k)|) < max(|eig(\mathbf{A}_{k-1})|) \quad \forall\, k = 1, \dots, n_j + m$

In other words, to have a stable solution for time instants much larger than $n_j$, eigenvalues of matrices $\mathbf{A}_k$ must be decreasing and they must be less than 1.

**Table D-1: Values of parameters set in the example.**

| $n$ | $\mathbf{A}_n$ |
|---|---|
| 0 | $\mathbf{H}_1^{-1}\mathbf{H}_{0,1}$ |
| 1 | $\mathbf{H}_1^{-1}(\mathbf{H}_{0,1}\mathbf{A}_0 + \mathbf{H}_{0,2})$ |
| 2 | $\mathbf{H}_1^{-1}(\mathbf{H}_{0,1}\mathbf{A}_1 + \mathbf{H}_{0,2}\mathbf{A}_0 + \mathbf{H}_{0,3})$ |
| ⋮ | ⋮ |
| $n_j - 1$ | $\mathbf{H}_1^{-1}\left(\mathbf{H}_{0,1}\mathbf{A}_{n_j-2} + \mathbf{H}_{0,2}\mathbf{A}_{n_j-3} + \cdots + \mathbf{H}_{0,n_j}\right)$ |
| $n_j$ | $\mathbf{H}_1^{-1}\left(\mathbf{H}_{0,1}\mathbf{A}_{n_j-1} + \mathbf{H}_{0,2}\mathbf{A}_{n_j-2} + \cdots + \mathbf{H}_{0,n_j}\mathbf{A}_0\right)$ |
| $n_j + 1$ | $\mathbf{H}_1^{-1}\left(\mathbf{H}_{0,1}\mathbf{A}_{n_j} + \mathbf{H}_{0,2}\mathbf{A}_{n_j-1} + \cdots + \mathbf{H}_{0,n_j}\mathbf{A}_1\right)$ |
| ⋮ | ⋮ |
| $n_j + m$ | $\mathbf{H}_1^{-1}\left(\mathbf{H}_{0,1}\mathbf{A}_{n_j+m-1} + \mathbf{H}_{0,2}\mathbf{A}_{n_j+m-2} + \cdots + \mathbf{H}_{0,n_j}\mathbf{A}_m\right)$ |

## D.2. EXAMPLE

To verify the formulation presented in the previous section, a system similar to Figure D-1 is analyzed under a synthetic ground acceleration. A chirp signal with a frequency linearly varying from 1 to 15 Hz is used as a ground motion to emphasize the frequency-dependency of the soil-foundation impedance function. The impedance functions are those used in the MDOF example in Chapter 3 to verify Nakamura's method. Figure D-2 shows a comparison between responses obtained from the frequency-domain approach (exact approach) and the time-domain approach using the time-domain representation of the impedance functions. As seen, the responses obtained from the approximate solution are very close to the exact responses, and obviously, there is no stability issue. The largest eigenvalue of the matrix $\mathbf{A}_k$ is also shown in Figure D-3(a). As seen and expected, it is always less than one and it is decreasing over time.

The example is now repeated by reducing the imaginary (damping) part of the sway impedance function by a factor of 10. in this case, the time-history solution did not converge and the reason can be observed in Figure D-3(b). Unfortunately, while we can check the stability before analysis there is no easy way to make the system stable. Also, the computational costs associated with this stability check.

The Matlab code to reproduce the results of this example is available upon request from the first author.

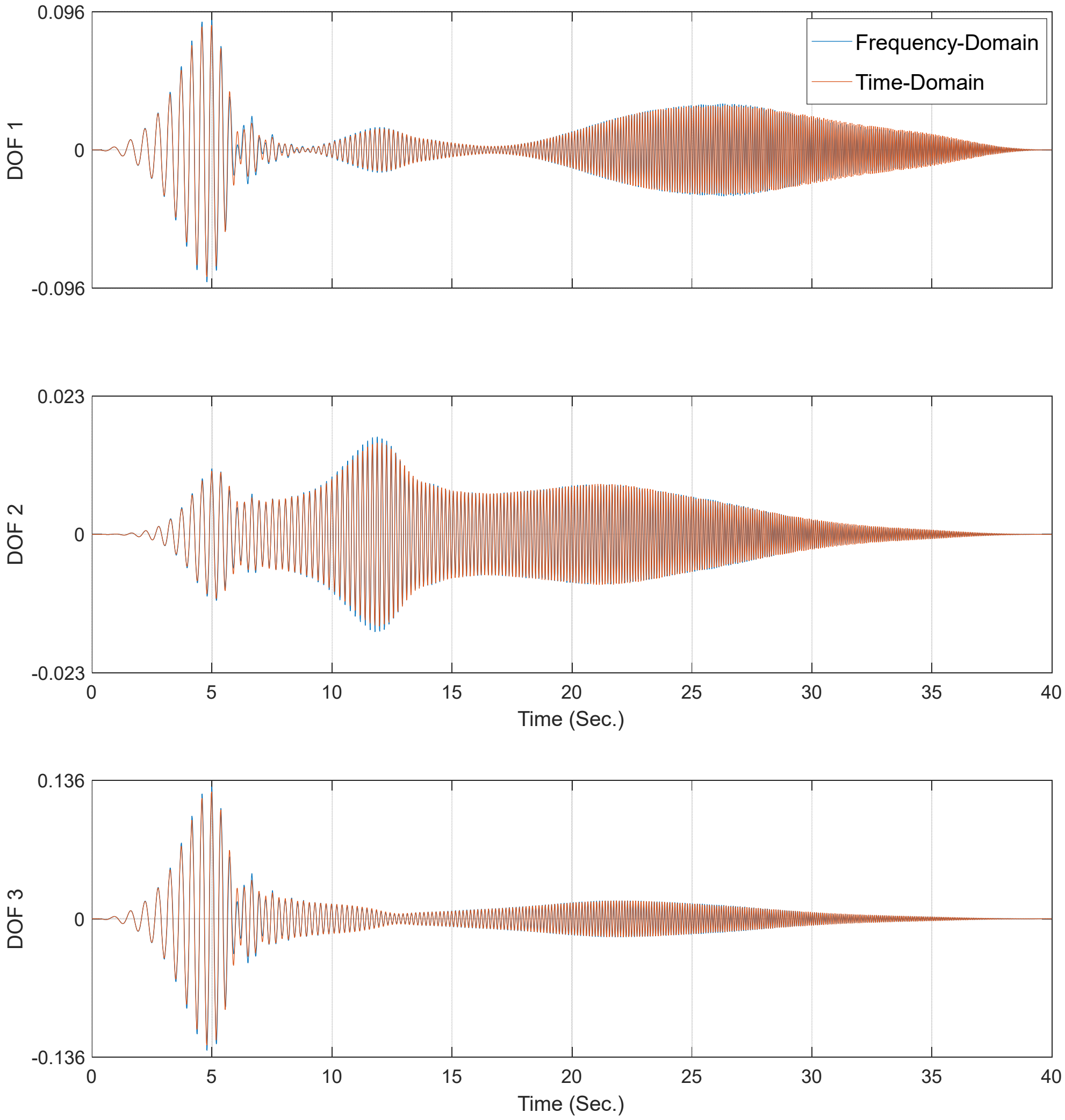

**Figure D-2: Responses obtained using the exact frequency- and approximate time-domain solutions.**

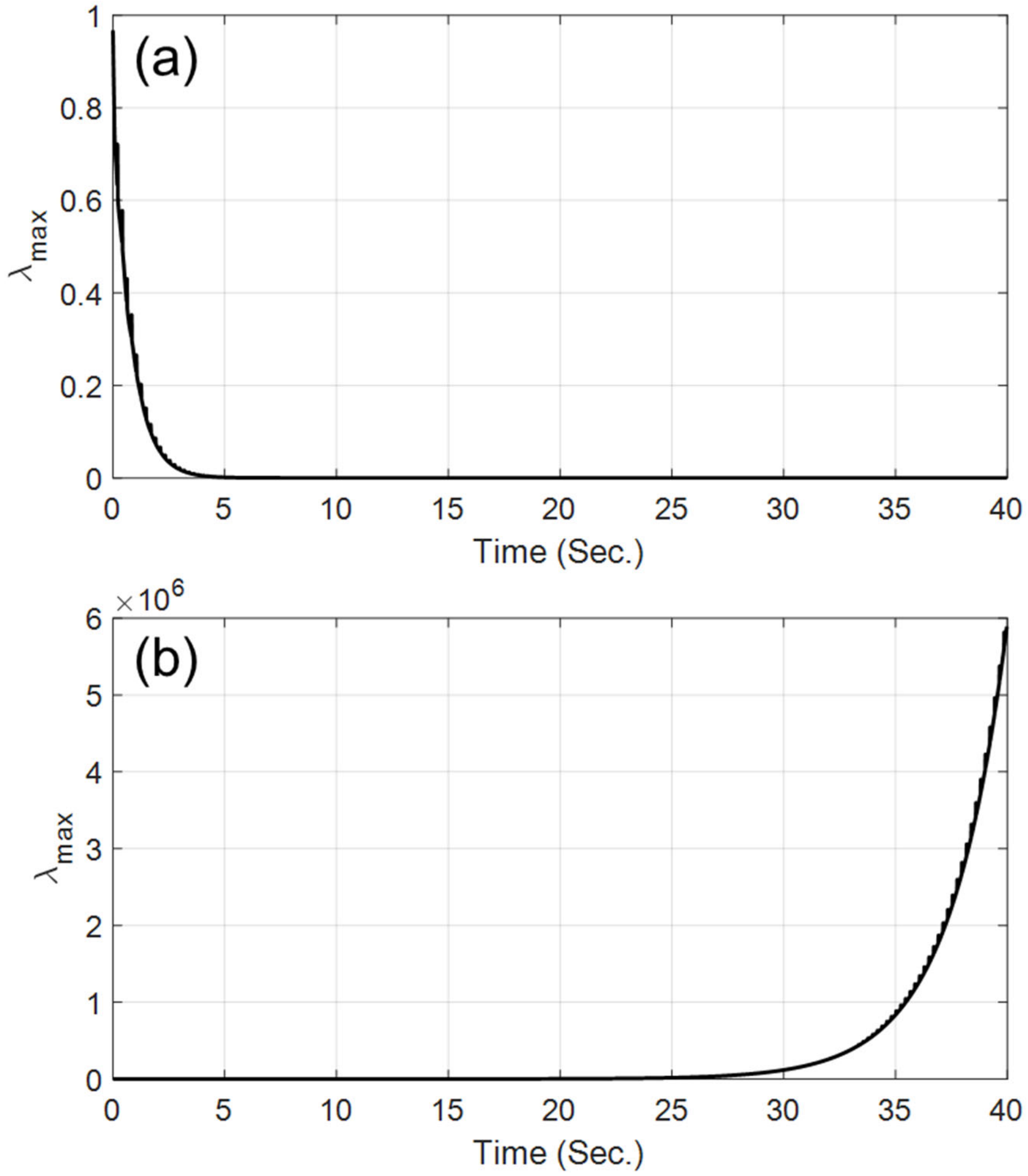

**Figure D-3: Largest eigenvalue of matrices $A_k$ in time for (a) the original example, and (b) the example with reduced sway impedance function.**